\newcommand{\sides}{oneside} % for single sided, change this to oneside
\newcommand{\userefs}{true} % for black and white only, change this to false

\documentclass[a4paper,11pt,\sides]{thesis_style}
\usepackage{fancyhdr,bm}
\usepackage{mathrsfs,amsmath,amssymb,amsthm,eucal,cancel,rotating}
\usepackage{ifthen,setspace,graphicx,verbatim,subfigure,booktabs}
\usepackage{multirow}
\usepackage{color}
\usepackage[usenames,dvipsnames]{xcolor}
\usepackage{captcont}
\usepackage[pagebackref=true]{hyperref}
\usepackage{wasysym}
\usepackage{marginnote}
\usepackage{ulem}
\usepackage{amsfonts,tensor}
\usepackage{float}

\DeclareMathOperator{\tr}{tr}
\newcommand{\qm}{\frac{q}{m}}
\newcommand{\xddd}{\dddot{x}}
\newcommand{\E}{{\cal E}}

\newcommand{\T}{{\cal T}}

\newcommand{\lpage}{\textbf{\textit{\thepage}}\ \ \textbar}
\newcommand{\rpage}{\textbar\ \ \textbf{\textit{\thepage}}}
\newcommand{\thesistitle}{Radiation reaction in strong fields from an alternative perspective}

\renewcommand{\geq}{\geqslant}
\renewcommand{\leq}{\leqslant}

\newcommand{\xd}{\dot{x}}
\newcommand{\xdd}{\ddot{x}}

\hypersetup{
    linktoc=page,%none,page,section,all
    pdftitle={\thesistitle},    % title
    pdfauthor={Yevgen Kravets},     % author
    pdfcreator={Yevgen Kravets}   % creator of the document
}
\ifthenelse{\equal{\userefs}{true}}{
\hypersetup{
    colorlinks=true,       % false: boxed links; true: colored links
    linkcolor=red,          % color of internal links
    citecolor=blue,        % color of links to bibliography
    urlcolor=OliveGreen           % color of external links
}
}{
\hypersetup{
    colorlinks=true,       % false: boxed links; true: colored links
    linkcolor=black,          % color of internal links
    citecolor=black,        % color of links to bibliography
    urlcolor=black           % color of external links
}
}

\renewcommand*{\backref}[1]{}
\renewcommand*{\backrefalt}[4]{%
  \ifcase #1 %
    \relax
  \or
$\langle\langle$~Cited~on~page~#2.~$\rangle\rangle$%
  \else
$\langle\langle$~Cited~on~pages~#2.~$\rangle\rangle$%
  \fi%
}

\begin{document}

\pagestyle{empty}

\begin{center}
\LARGE

\doublespacing 

% default size 0.3\linewidth
\includegraphics[width=0.5\linewidth]{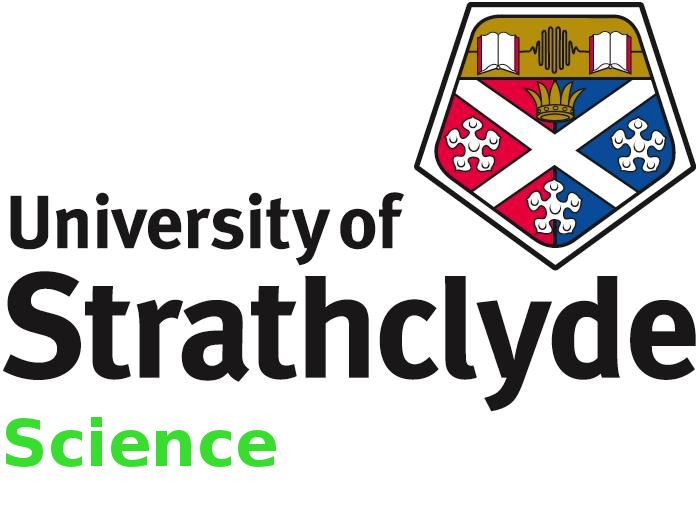}

\vspace{1cm}

\textbf{\thesistitle}

\vspace{1cm}

\doublespacing 

{\Large by Yevgen Kravets B. Sc., B. Eng., M. Sc.}\\
{\Large Supervisor: Prof. Dino Anthony Jaroszynski}\\

\vspace{1.25cm}

\normalsize
A thesis submitted in the partial fulfilment of the requirements for the degree of\\
\textit{Doctor of Philosophy in Physics}\\

\vspace{1cm}

\normalsize
Department of Physics\\
University \textit{of} Strathclyde\\
John Anderson Building\\
107 Rottenrow, G4 0NG, Glasgow\\
United Kingdom\\

\vspace{1cm}

2014

\end{center}
 % include the title page
\pagestyle{empty}

\frontmatter

\doublespacing

% header and footer
\renewcommand{\chaptermark}[1]{\markboth{#1}{}}
\renewcommand{\headheight}{28pt}

\renewcommand{\headrulewidth}{0pt}
\renewcommand{\footrulewidth}{0pt}

% Plain page style
\fancypagestyle{plain}{
 \fancyhf{} % clear all header and footer fields
 \renewcommand{\headrulewidth}{0pt}
 \ifthenelse{\equal{\sides}{oneside}}
 {
   \fancyfoot[R]{\rpage}
 }
 {
   \fancyfoot[RO]{\rpage}
   \fancyfoot[LE]{\lpage}
 }
}

% fancy header and footer for front matter
\pagestyle{fancy}
\fancyhf{}
\renewcommand{\chaptermark}[1]{\markboth{#1}{}}
\ifthenelse{\equal{\sides}{oneside}}
 {
   \fancyfoot[R]{\rpage}
 }
 {
   \fancyfoot[RO]{\rpage}
   \fancyfoot[LE]{\lpage}
 }

% include the front matter
\chapter{Declaration}

\doublespacing 

\vspace{-2em}

`I declare that this thesis is the result of author's original research. It has been composed by the author and has not been previously submitted for examination which has led to the award of a degree.'

\vspace{1em}
\noindent `The copyright of this thesis belongs to the author under the terms of the United Kingdom Copyright Acts as qualified by University of Strathclyde Regulation 3.50. Due acknowledgement must always be made of the use of any material contained in, or derived from, this thesis.'

\vspace{1in}
\begin{flushright}
\textit{Yevgen Kravets}\\
 April, 2014
\end{flushright}

\chapter*{}
\thispagestyle{empty}

\doublespacing 

 \vspace{5cm}
\begin{flushright}
 \Large\textit{I would like to dedicate this thesis to my gorgeous wife Renata, my beloved grandparents, all my friends who supported me in this endeavour and my cat Fluffy who drove me insane multiple times during my PhD.}
\end{flushright}

\chapter{Abstract}

\doublespacing 

\vspace{-2em}

Current classical theory of radiation reaction has several deficiencies such as ``runaway solutions'' and violation of causality. The Landau-Lifshitz approximation to the exact equation introduced by Lorentz, Abraham and Dirac is widely used, though questions remain regarding its domain of validity. 
This thesis explores an alternative treatment of the motion of a radiating electron, based on an equation first proposed by Ford and O'Connell. A general condition is found for solutions of this equation to deviate from those of Landau-Lifshitz. 

By exploring radiation reaction effects on a particle colliding with an ultra-intense laser pulse we show that the regime where there is a significant deviation of these two approaches can never be reached with existing or proposed laser facilities. 

The methods used to explore single particle interaction with an intense laser pulse are extended to describe the interaction of a particle bunch with various realistic laser pulses. We find that the interaction leads to a decrease in average momentum and relative momentum spread. However, the decrease appears to be independent of the length of the pulse and depends only on the energy in the pulse regardless of how it is distributed.

Radiation reaction effects occuring during the scattering of an electron by a heavy, highly-charged nucleus are studied. Radiation reaction is seen to affect the particle's motion. We find noticeable differences between the predictions of the Ford-O'Connell and Landau-Lifshitz equations, albeit in regimes where quantum effects would be important.
\chapter{Acknowledgements}

\doublespacing 

\vspace{-2em}

First and foremost, I wish to extend my gratitude to my supervisors, Prof. Dino Anthony Jaroszynski and Dr. Adam Noble, whom I also find to be a close friend. Without their continued support I would never have completed this thesis. I wish to thank Dr. Noble for his patient guidance and motivation towards research. I thank Prof. Jaroszynski for sharing his knowledge and enthusiasm with me, as well as for being approachable with any problems I had. I have learnt a lot from working with both of them.

Particular thanks are due to Dr. Samuel Yoffe, for his friendship and many stimulating discussions on the topic of radiation reaction and quantum corrections.

I cannot describe how indebted I am to my wonderful wife, Renata, whose love and encouragement will always motivate me to achieve all that I can. I could not have written this thesis without her support; in particular, my peculiar working hours and erratic behaviour towards the end could not have been easy to deal with!

Of course, I would never have made it this far without the love and support of my beloved grandparents. Their interest in my work and pride at my achievements has always been an inspiration.

I could also have not made it through without the many friends I have made along the way. I particularly wish to thank my colleagues at the University of Strathclyde and senior researchers at Lancaster University as working with them was a privilege. I also thank my cat Fluffy for keeping me smiling at the downfalls of my project. 

When I joined the SILIS group, I was instantly made to feel welcome and included, for which I owe additional thanks to Bernhard, Gaurav, Enrico, Silvia and Gregory. I extend my thanks and best wishes to all the students and post-docs I got to share lunch, coffee and/or (several) pints with.

I would like to thank Kirsten Munro, Catherine Cheshire and Lynn Gilmour for their kind approach in dealing with all the administrative matters and ensuring my PhD ran smoothly.

Additional thanks are due to Mr. Pawe\l\ Adamczyk for designing illustrations for the title pages of each Chapter of this thesis.

I gratefully acknowledge the generosity and support of the \textit{Scottish Universities Physics Alliance} (SUPA) and University \textit{of} Strathclyde, who provided me with a Prize Studentship enabling me to undertake this PhD. I am eternally grateful for this opportunity.

\chapter{The role of the Author}

\doublespacing 

\vspace{-2em}

Fundamental theoretical research is usually carried out by small teams of people. This section outlines the role of the author in the work presented in this thesis.

In Chapter 2, the development of the concept along with a complete derivation was done by the Author and Dr. Adam Noble. The initial $\verb!C++!$ code that produced results presented in Chapter 3 was designed and written by the Author, aided by a set of detailed discussions with Dr. Enrico Brunetti which led to further improvements and optimisation of the code.

The data presented in Chapters 4 and 5 are based on an extension of the computational routine developed by Dr. Samuel R. Yoffe. 

All the relevant simulations presented in this thesis were performed by the Author. Technical support and consultations were provided by Dr. Samuel R. Yoffe and Dr. Enrico Brunetti.
\chapter{List of publications}

\doublespacing 

\vspace{-2em}

Results presented in Chapter 2 and Chapter 3 have been published in 1 journal publication, 3 conference proceedings and 1 annual report. 

Results presented in Chapter 4 are currently being prepared for a journal publication. Detailed breakdown of these publications can be found below:

\begin{center}
{\bf Peer-reviewed journal articles:}
\end{center}
\vspace{-1.5em}
\begin{enumerate}
\item{{\bf Y. Kravets}, A. Noble, D. Jaroszynski; ``Radiation reaction effects on the interaction of an electron with an intense laser pulse''. \textit{Phys. Rev. E} {\bf 88}, 011201(R) (2013).}
%\href{http://link.aps.org/doi/10.1103/PhysRevE.88.011201}{$\triangleright$}
\end{enumerate}
\begin{center}
{\bf Conference proceedings:}
\end{center}
\vspace{-1.5em}
\begin{enumerate}
\item{{\bf Y. Kravets}, A. Noble and D. Jaroszynski; ``Validity of the Landau-Lifshitz approximation in an ultra-high intensity laser pulse''. \textit{Proc. EPS} {\bf P4.214} (2013).}
\item{{\bf Y. Kravets}, A. Noble and D. Jaroszynski; ``Energy losses due to radiation reaction in an intense laser pulse''. \textit{Proc. SPIE} {\bf 8779}, 87791X (2013).}
\item{A. Noble, {\bf Y. Kravets} et al.; ``Kinetic treatment of radiation reaction effects''. \textit{Proc. SPIE} {\bf 8079}, 80790L (2011).}
\end{enumerate}	
\begin{center}
{\bf Annual reports:}
\end{center}
\vspace{-1.5em}
\begin{enumerate}
\item{A. Noble, {\bf Y. Kravets}, S. Yoffe and D. Jaroszynski; ``Radiation damping of an electron in an intense laser pulse''. \textit{Central Laser Facility Annual Report} (2012--2013).}
\end{enumerate}

\doublespacing
\tableofcontents
\newpage

\listoffigures
%\listoftables

\mainmatter
\doublespacing % may need this to be double spacing

% Fancy header and footer for normal pages
\renewcommand{\chaptermark}[1]{\markboth{#1}{}}
\ifthenelse{\equal{\sides}{oneside}}
 {
   \renewcommand{\headrulewidth}{0.5pt}
   \fancyfoot[R]{\rpage}
   \fancyhead[R]{\textit{\nouppercase{\rightmark}}}
 }
 {
   \renewcommand{\headrulewidth}{0.5pt}
   \fancyhead[RO]{\textit{\nouppercase{\rightmark}}}
   \fancyhead[LE]{\textit{\chaptername\ \thechapter\ \ ---\ \ \leftmark}}
 }

%%% BEGIN INCLUDING CHAPTERS %%%

\chapter{Introduction}

\begin{figure}[H]
\vspace{-15em}
\centering
\includegraphics[width=\textwidth]{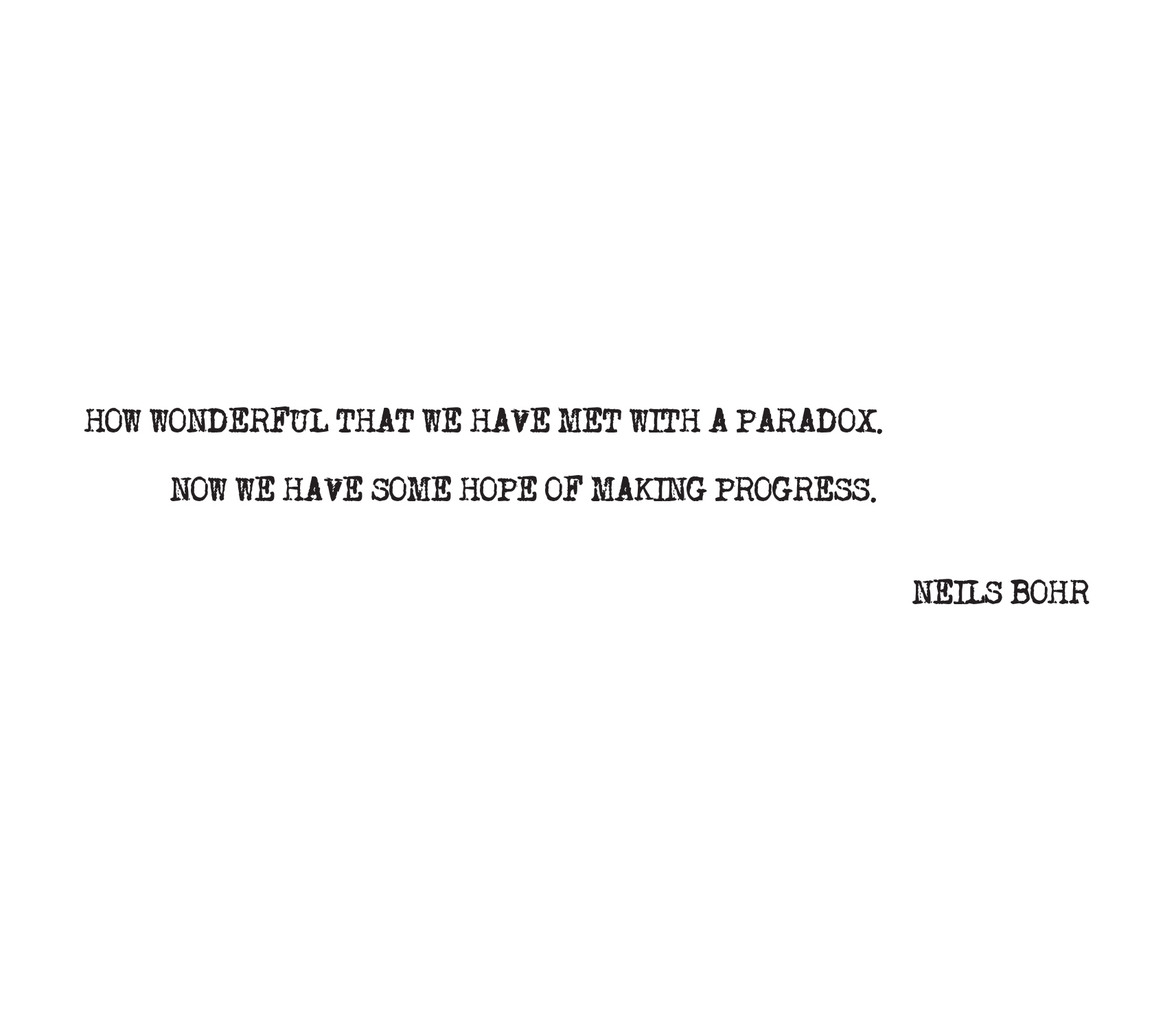}
\end{figure} 

\newpage 

\section{Thesis overview}
An accelerating charged particle emits radiation \cite{Adam9}. With this emission the particle loses energy and momentum, which influences its motion via the recoil force (so called ``radiation reaction'') in order to fulfill energy and momentum conservation. The question of how the particle interacts with the radiation it produces, however, remains unclear, despite investigations stretching back more than a century; see Refs. \cite{Burton2, Erber, Rohrlich2}. Until recently, interest has been motivated principally by theoretical curiosity, since the radiation reaction force is in general a negligible correction to the Lorentz force from the external fields. However, with the advent of the modern ultra-intense laser, this question is becoming relevant to experimental investigations.

This will be of particular relevance to research at the Extreme Light Infrastructure (ELI) \cite{ELI} facility that is currently under construction, which will be one of the leading high power laser facilities in the world. The work in this thesis will be most relevant to the fourth pillar of ELI (location to be decided), where the laser intensities are expected to reach $\sim 10^{25}$ $\text{W}\text{cm}^{-2}$ and the power to exceed that of the other three ELI pillars by at least one order of magnitude. 

The work presented in this thesis is focused on exploring the effects of classical radiation reaction in regimes where this effect is believed to be non-negligible. We present an alternative model for classical radiation reaction and compare its predictions with traditional approaches to help establish the limitations of commonly used methods. 

In the remaining part of this chapter we will discuss two common theoretical descriptions of radiation reaction: that of Lorentz \cite{Lorentz}, Abraham \cite{Abraham} and Dirac \cite{Dirac}, and that of Landau and Lifshitz \cite{Landau}, respectively. While the former leads to difficulties such as exponentially growing acceleration (``runaway solutions'') and violation of causality (``pre-acceleration''), the latter is perturbative, which raises questions on its validity in extremely high fields. 
 
In Chapter 2 we introduce an alternative approach based on the Ford-O'Connell equation \cite{OConnell, FO1, FO2}, which addresses these issues. The relation of this approach to other commonly used treatments is explored. We also introduce and discuss a condition allowing us to predict where the Landau-Lifshitz model is not valid. 
 
Chapter 3 is devoted to the study of the interaction of a high-energy electron with an intense laser pulse. By analysing this interaction we find that radiation reaction prevents the particle from accessing the regime where the Landau-Lifshitz approximation is not valid. The results presented in this chapter are summarised in Refs. \cite{ourPRE, ourEPS, ourSPIE, ourCLF}.

In Chapter 4 we present a study of how radiation reaction influences the momentum spread of a bunch of particles propagating through an intense laser pulse. 

In Chapter 5 we apply the model developed in Chapter 2 to the scattering of a particle off a heavy nucleus. We present a model that includes the classical radiation reaction corrections, and derive relevant equations of motion. Numerical simulations are presented and relevance of quantum effects during the interaction is discussed.

\section{Radiation reaction models}
\subsection{Lorentz-Abraham-Dirac equation}

An equation of motion for a \textit{non-relativistic} point particle of charge $q$ and mass $m$ in an external electromagnetic field is as follows:
\begin{equation}
\label{eq:1:LorentzforceonlyNonRel}
m\boldsymbol{a} = q\left(\boldsymbol{E} + \boldsymbol{v} \times \boldsymbol{B}\right)\ ,
\end{equation}
where $\boldsymbol{E}$ and $\boldsymbol{B}$ are the external electric and magnetic fields, respectively. 

Equation (\ref{eq:1:LorentzforceonlyNonRel}) can be written covariantly as:
\begin{equation}
\label{eq:1:LorentzforceonlyRel}
\ddot{x}^a = - \frac{q}{m} F^a{}_b \dot{x}^b\ ,
\end{equation}
where $F^a{}_b$ is the electromagnetic field tensor and the dot here denotes a derivative with respect to proper time\footnote{Here \textit{proper time} is the time elapsed between two events as seen by the particle in its rest frame. Throughout this thesis, we refer to proper time as $s$.}. Since $F^a{}_b$ is antisymmetric (\ref{eq:1:LorentzforceonlyRel}) preserves the 4-velocity normalisation condition:
\begin{equation}
\label{eq:1:norm}
\dot{x}^a\dot{x}_a = -1\ .
\end{equation}

We use the Einstein summation convention and indices are raised and lowered using the metric tensor $\eta_{ab} = diag\left(-1, 1, 1, 1\right)$. 

As mentioned earlier, a particle undergoing acceleration emits radiation, which produces a recoil force. For this to be taken into account, the above equation of motion must be modified to account for energy carried away by the radiation. This can be achieved by including a radiation reaction force in addition to the Lorentz force. Equation (\ref{eq:1:LorentzforceonlyRel}) then becomes:
\begin{equation}
\label{eq:1:withRR_1}
\ddot{x}^a = - \frac{q}{m} F^a{}_b \dot{x}^b + P^a + \dot{C}^a\ ,
\end{equation}
where $P^a$ is the recoil force or so called \textit{radiation reaction force}, which accounts for the emitted energy and momentum and $\dot{C}^a$ is the so-called \textit{Schott term} \cite{Schott, Ferris}, which ensures that the normalisation condition is preserved. 

The electromagnetic power emited by a relativistic particle can be obtained using the Larmor formula \cite{Adam9}:
\begin{equation}
\label{eq:1:Pa}
P^a = -\tau \ddot{x}^2 \dot{x}^a\ ,
\end{equation}
where $\tau = q^2/6\pi m = 2r_e/3c \sim 10^{-23}$s for an electron (with $r_e$ the classical electron radius). 

Differentiating (\ref{eq:1:norm}) leads to:
\begin{equation}
\label{eq:1:xddxd}
\dot{x}^a\ddot{x}_a=0\ .
\end{equation}

Contracting (\ref{eq:1:withRR_1}) with $\dot{x}_a$, (\ref{eq:1:Pa}) and (\ref{eq:1:xddxd}) yield:
\begin{equation}
%\label{eq:1:withRR_1}
\tau \ddot{x}^2 + \dot{C}^a\dot{x}_a = 0\ .
\end{equation}

Differentiating (\ref{eq:1:xddxd}) then leads to:
\begin{equation}
\label{eq:1:xdddxddxd}
\dot{x}^a\dddot{x}_a + \ddot{x}^2 = 0\ ,
\end{equation}
which suggests that
\begin{equation}
\label{eq:1:dotca}
\dot{C}^a = \tau\dddot{x}^a\ .
\end{equation}

Substituting (\ref{eq:1:dotca}) and (\ref{eq:1:Pa}) into (\ref{eq:1:withRR_1}) we obtain the relativistic Lorentz-Abraham-Dirac equation \cite{Lorentz, Abraham, Dirac}, which reads:
\begin{equation}
\label{eq:1:LADLAD}
\ddot{x}^a = - \frac{q}{m} F^a{}_b \dot{x}^b + \tau\left(\dddot{x}^a -  \ddot{x}^2 \dot{x}^a\right)\ .
\end{equation}
Using (\ref{eq:1:xdddxddxd}) the original Lorentz-Abraham-Dirac equation (\ref{eq:1:LADLAD}) can be written in the form:
\begin{equation}
\label{eq:1:LAD}
\ddot{x}^a = -\frac{q}{m} F^{ab} \dot{x}_b + \tau \Delta^a{}_b \dddot{x}^b\ .
\end{equation}
$\Delta^a{}_b$ here represents the projection operator and is given by:
\begin{equation}
\Delta^a{}_b = \delta^a{}_b + \xd^a\xd_b\ .
\end{equation}

A rigorous derivation of (\ref{eq:1:LADLAD}) has been obtained by Dirac \cite{Dirac} on the basis of energy and momentum conservation. It has subsequently been derived multiple times from different physical principles \cite{Schott, Ferris, Bhabha, Wheeler, Rohrlich1, Teitelboim, Barut, Gratus}.

It can be seen that the radiation reaction force involves the so-called ``\textit{jerk}'' term (time derivative of acceleration) making it a third order differential equation for the position of the particle. To solve this equation it is neccessary to specify initial conditions for position, velocity and acceleration. However the above equations (\ref{eq:1:LADLAD}, \ref{eq:1:LAD}) have a few important pathologies\footnote{The term \textit{pathologies} is widely used to describe the non-physical solutions of the Lorentz-Abraham-Dirac equation.}. 

Consider equation (\ref{eq:1:LAD}) in the absence of an external force:
\begin{equation}
\xdd^a=\tau\left(\delta^{a}{}_{b}+\xd^{a}\xd_{b}\right)\xddd^{b}\ .
\end{equation}
Assume motion only in one direction:
\begin{equation}
\xdd^1=\tau\xddd^{1}+\tau\xd^{1}\xd_{0}\xddd^{0}+\tau\xd^{1}\xd_{1}\xddd^{1}\ .
\end{equation}
Here $\xd^0=\cosh \alpha$ and $\xd^1=\sinh \alpha$ satisfy the normalisation condition (\ref{eq:1:norm}). An equation of motion for {\textit{proper acceleration}} can be produced. It has the form:
\begin{equation}
\tau\ddot{\alpha}=\dot{\alpha}\ ,
\end{equation}
which leads to the solution:
\begin{equation}
\alpha\left(s\right) \sim e^{s/\tau} \alpha\left(0\right)\ .
\end{equation}
This type of solution is known as a \textit{runaway solution} as the particle's acceleration increases exponentially with time, unless $\alpha(0)=0$. This solution is clearly unphysical and is one of the main pathologies of the Lorentz-Abraham-Dirac equation.

Another pathology of this equation \cite{Baylis2, Hammond1, Hammond2} can be most readily seen by examining the non-relativistic limit of (\ref{eq:1:LAD}):
\begin{equation}
\label{eq:1:LADnonrel}
m \boldsymbol{a} = \boldsymbol{F}_{ext} + m \tau \frac{\mathrm{d}\boldsymbol{a}}{\mathrm{d}t}\ ,
\end{equation}
which can be integrated to give:
\begin{equation}
\label{eq:1:rewritten}
m \boldsymbol{a}\left(t\right) = m \boldsymbol{a}\left(t_0\right) e^{\left(t - t_0\right)/\tau} - \frac{e^{t/\tau}}{\tau} \int_{t_0}^{t} \boldsymbol{F}_{ext} \left(t^\prime\right) e^{-t^\prime/\tau} \mathrm{d}t^\prime\ ,
\end{equation}
where $t_0$ is a constant and $\boldsymbol{F}_{ext}$ is given by the right hand side of (\ref{eq:1:LorentzforceonlyNonRel}). The previously discussed \textit{runaway solutions} can be eliminated by introducing a mix of initial and final conditions rather than just the initial ones. Demanding that the final acceleration is 0 once all the forces have finished acting we eliminate the previous pathology. 

This corresponds to $t_0 \to \infty$ and $a\left(t_0\right) \to 0$ in (\ref{eq:1:rewritten}). Applying the change of variable $z = \left(t^\prime - t\right)/\tau$, the original equation becomes:
\begin{equation}
\label{eq:1:preacceleration}
m \boldsymbol{a}\left(t\right) = \int_{0}^{\infty} \boldsymbol{F}_{ext} \left(t + \tau z\right) e^{-z} \mathrm{d}z\ .
\end{equation}
This solution uncovers yet another pathology of the Lorentz-Abraham-Dirac equation. From (\ref{eq:1:preacceleration}) it can be seen that the acceleration at time $t$ depends on the applied force still to come. Therefore removal of the \textit{runaway} pathology leads to the \textit{pre-acceleration} pathology, which is also unphysical. However, in this case, unlike the \textit{runaway} pathology, the influence of the unphysical part of the solution is very small as it is supressed by the $e^{-z}$ term. The same pathology remains in the relativistic equation of motion \cite{Dirac}.

\subsection{The Landau-Lifshitz equation}
In 1962 Landau and Lifshitz  \cite{Landau} proposed a perturbative method for removing runaway and pre-acceleration pathologies by reducing the order of the Lorentz-Abraham-Dirac equation (\ref{eq:1:LADLAD}). Assuming $\tau\simeq 6\cdot 10^{-24}$\ s is small compared to the timescale over which the Lorentz force varies (as measured in the rest frame of the particle) they considered the radiation reaction force (proportional to $\tau$) to be a small  perturbation about the Lorentz force:

%Because $\tau$ is small ($\tau\simeq 6\cdot 10^{-24}$s) they proposed to treat radiation reaction as a small perturbation about the Lorentz force:
\begin{equation}
\ddot{x}^a = -\frac{q}{m} F\indices{^a_b}\dot{x}^b + \mathcal{O}\left(\tau\right)\ ,
\end{equation}
allowing us to approximate the \textit{jerk} term:
\begin{align}
\label{eq:1:LLjerkapprox}
\nonumber \dddot{x}^a & = -\frac{q}{m} \dot{F}\indices{^a_b} \dot{x}^b -\frac{q}{m} F\indices{^a_b} \ddot{x}^b + \mathcal{O}\left(\tau\right) \\
& = - \frac{q}{m}\dot{x}^c\partial_{c} F^a{}_b \dot{x}^b + \frac{q^2}{m^2} F^a{}_b F^b{}_c \dot{x}^c + \mathcal{O}\left(\tau\right)\ .
\end{align}

If we now substitute (\ref{eq:1:LLjerkapprox}) into the original Lorentz-Abraham-Dirac equation (\ref{eq:1:LAD}) and drop terms of $\mathcal{O}\left(\tau^2\right)$ we obtain the \textit{Landau-Lifshitz equation}:
\begin{equation}
\label{eq:1:LLrel}
\ddot{x}^a = -\frac{q}{m}F^{ab}\dot{x}_b - \tau \frac{q}{m} \partial_c F^{ab}\dot{x}_b\dot{x}^c + \tau\frac{q^2}{m^2}\Delta^a{}_bF^{bc}F_{cd}\dot{x}^d\ .
\end{equation}
It has been shown \cite{Spohn} that (\ref{eq:1:LLrel}) approximates the non-runaway solutions of (\ref{eq:1:LADLAD}).

As this equation no longer depends on the derivative of the acceleration it is free from the pathologies of the original Lorentz-Abraham-Dirac equation (\ref{eq:1:LAD}) and is usually accepted to be the correct classical equation of motion for the relativistic charged point particle.

As noted earlier, the fields in forthcoming ultra-high intensity laser facilities will be sufficiently strong that the forces due to an electron's emission can exceed the Lorentz force of the electron due to the laser pulse, which raises questions regarding the domain of validity of the Landau-Lifshitz approach \cite{Griffiths, Bulanov, Pandit, Galley, Gron}.

\section{Alternative radiation reaction models}
Given the fundamental flaws of the Lorentz-Abraham-Dirac equation, a number of alternative approaches have been proposed throughout the last century. This section is devoted to a brief introduction of the most common of these  to compare attempts to solve this problem. Advantages and limitations of these approaches will be discussed.

\subsection{Mo-Papas equation}
In 1971 an alternative approach was presented by Mo and Papas \cite{MoPapas}. Instead of a rigorous derivation from first principles, they heurestically claimed that an equation of motion should balance inertia and radiation forces with the Lorentz force and an additional acceleration-dependent generalization of the Lorentz force. This led to an equation of the following form:
\begin{equation}
\ddot{x}^a - \tau \frac{q}{m} F^b{}_c \ddot{x}_b \dot{x}^c \dot{x}^a = - \frac{q}{m}F^a{}_b \dot{x}^b + g F^a{}_b \ddot{x}^b\ .
\end{equation}
It can be seen that the second term on the left-hand side is responsible for the compensation of the losses due to radiation, while the second term on the right-hand side is the new force. To fulfill the normalisation condition $g = -\tau q/m$ was defined.

The Mo-Papas equation has been criticised on a number of occasions. One of the most important criticisms \cite{Baylis3} is that for the case of linear motion this equation coincides with the Lorentz force describing the motion with no radiation reaction taken into account. The radiation losses in this case are small, but non-zero.

\subsection{Sokolov equation}
One of the most popular recent endeavours that have gained significant attention is the equation presented by Sokolov \cite{Sokolov}. The fundamental assumption of the derivation of this equation is that the 4-momentum does not have to be collinear with the 4-velocity. To justify this claim Sokolov uses the fact that part of the momentum of a charged particle may be regarded as distributed throughout space in its Coulomb field.

If we take the momentum and velocity to be parallel, $p^a = m \dot{x}^a$, the Einstein relation for energy and momentum corresponds to the normalisation condition with respect to proper time:
\begin{equation}
\label{eq:1:Einsteinrelation}
E^2 - \boldsymbol{p}^2 = m^2 \quad \Longleftrightarrow \quad \dot{x}^a\dot{x}_a = -1\ .
\end{equation}
If we now accept Sokolov's assumption and assume that due to acceleration momentum $p^a$ and velocity $\dot{x}^a$ are no longer parallel only one of (\ref{eq:1:Einsteinrelation}) holds. From the first equation in (\ref{eq:1:Einsteinrelation}) via some additional assumptions we acquire:
\begin{equation}
\label{eq:1:Sokolov_1}
\dot{x}^a = \left(\delta^a_b - \tau \frac{q}{m} F^a{}_b \right) \frac{p^b}{m}\ ,
\end{equation}
\begin{equation}
\label{eq:1:Sokolov_2}
\frac{\dot{p}^a}{m} = \frac{q}{m} F^a{}_b \frac{p^b}{m} + \tau \frac{q^2}{m^2} \left(F^a{}_b F^b{}_c \frac{p^c}{m} + F^b{}_d F^d{}_c \frac{p_b}{m}\frac{p^c}{m}\frac{p^a}{m} \right)\ .
\end{equation}

If we now substitute $p^a/m \longrightarrow \dot{x}^a$ into (\ref{eq:1:Sokolov_2}), apart from the field derivative terms it agrees with the Landau-Lifshitz equation. The novel feature of the Sokolov theory is given by (\ref{eq:1:Sokolov_1}), which describes the non-collinearity of $\dot{x}^a$ and $p^a$.

Using (\ref{eq:1:Sokolov_1}), the normalisation of velocities can be written as:
\begin{equation}
\label{eq:1:sokolovnorm}
- \dot{x}^a\dot{x}_a = \left( 1 - \tau^2 \frac{q^2}{m^2} F^a{}_b F_{ac} \frac{p^b}{m} \frac{p^c}{m} \right) \leq 1\ .
\end{equation}

It can be seen that for sufficiently large fields and/or high energies, it is possible that $\dot{x}^a\dot{x}_a \geq 0$. This would lead to the failure of the notion of proper time which would allow a massive particle to move with the speed of light or faster. This is one fundamental problem with the Sokolov theory indicating that it violates causality in extreme circumstances. 

\subsection{The Ford-O'Connell equation}
One of the first attempts to address the original issues related to the Lorentz-Abraham-Dirac equation was undertaken by Dirac's student Eliezer, in 1948 \cite{Eliezer}. He noticed that the equation of motion of a nonrelativistic \textit{non-pointlike} electron of radius $R$ can be written as:
\begin{equation}
m\boldsymbol{a} - m \tau \frac{\mathrm{d}\boldsymbol{a}}{\mathrm{d}t} + \sum\limits_{n=0}^\infty c_n R^n \frac{\mathrm{d}^n \boldsymbol{a}}{\mathrm{d}t^n} = \boldsymbol{F}_{ext}\ ,
\end{equation}
where the $c_n$ denote coefficients dependent upon the particle structure. If we now adjust the radius and charge density such that the $c_n$ give:
\begin{equation}
m \sum\limits_{n=0}^\infty \left(-\tau \frac{\mathrm{d}}{\mathrm{d}t}\right)^n \boldsymbol{a} = m \left(1 + \tau \frac{\mathrm{d}}{\mathrm{d}t}\right)^{-1} \boldsymbol{a} = \boldsymbol{F}_{ext}\ ,
\end{equation}
the equation of motion becomes:
\begin{equation}
\label{eq:1:FO}
m\boldsymbol{a} = \boldsymbol{F}_{ext} + \tau \frac{\mathrm{d}\boldsymbol{F}_{ext}}{\mathrm{d}t}\ .
\end{equation}
The relativistic form of (\ref{eq:1:FO}) can be written as follows:
\begin{equation}
\label{eq:1:EFO}
m\ddot{x} = f^a_{ext} + \tau \Delta^a{}_b \dot{f}^b_{ext}\ ,
\end{equation}
where the \textit{dot} denotes the derivative with respect to the proper time $s$. 

Equation (\ref{eq:1:EFO}) bears a certain similarity to the Landau-Lifshitz equation. However, it is worth emphasising that while the Landau-Lifshitz and Ford-O'Connell equations agree to order $\tau$, that of Ford-O'Connell includes corrections to all orders in $\tau$ and hence goes beyond it.

The equation (\ref{eq:1:FO}) was rediscovered some 40 years later by Ford and O'Connell \cite{FO1,FO2} while looking at the classical limit of a quantum theory for \textit{non-pointlike} electrons. In their work, they chose a form factor for the electron to be the following:
\begin{equation}
\rho\left(\omega\right) = \frac{\Omega^2}{\omega^2 + \Omega^2}
\end{equation}
with $\Omega$ a cut-off frequency. The $pointlike$ electron arises from $\Omega \rightarrow \infty$ giving rise to the Abraham-Lorentz equation \cite{FO1}. However, for $\Omega = \tau^{-1}$ we recover (\ref{eq:1:FO}), a pathology-free equation of motion for a \textit{non-pointlike} particle. 

Since the Ford-O'Connell equation also arises as an intermediate step in the derivation of the Landau-Lifshitz equation from the Lorentz-Abraham-Dirac equation, it can be used to benchmark the former: where the Landau-Lifshitz and Ford-O'Connell equations disagree, Ford-O'Connell might or might not be correct, but Landau-Lifshitz must be wrong. For this reason, in the remainder of this thesis, we focus on exploring the Ford-O'Connell treatment for various setups where effects of classical radiation reaction become significant.

\chapter{Ford-O'Connell equation as an alternative treatment of radiation reaction}

\begin{figure}[H]
\vspace{-21.3em}
\centering
\includegraphics[width=\textwidth]{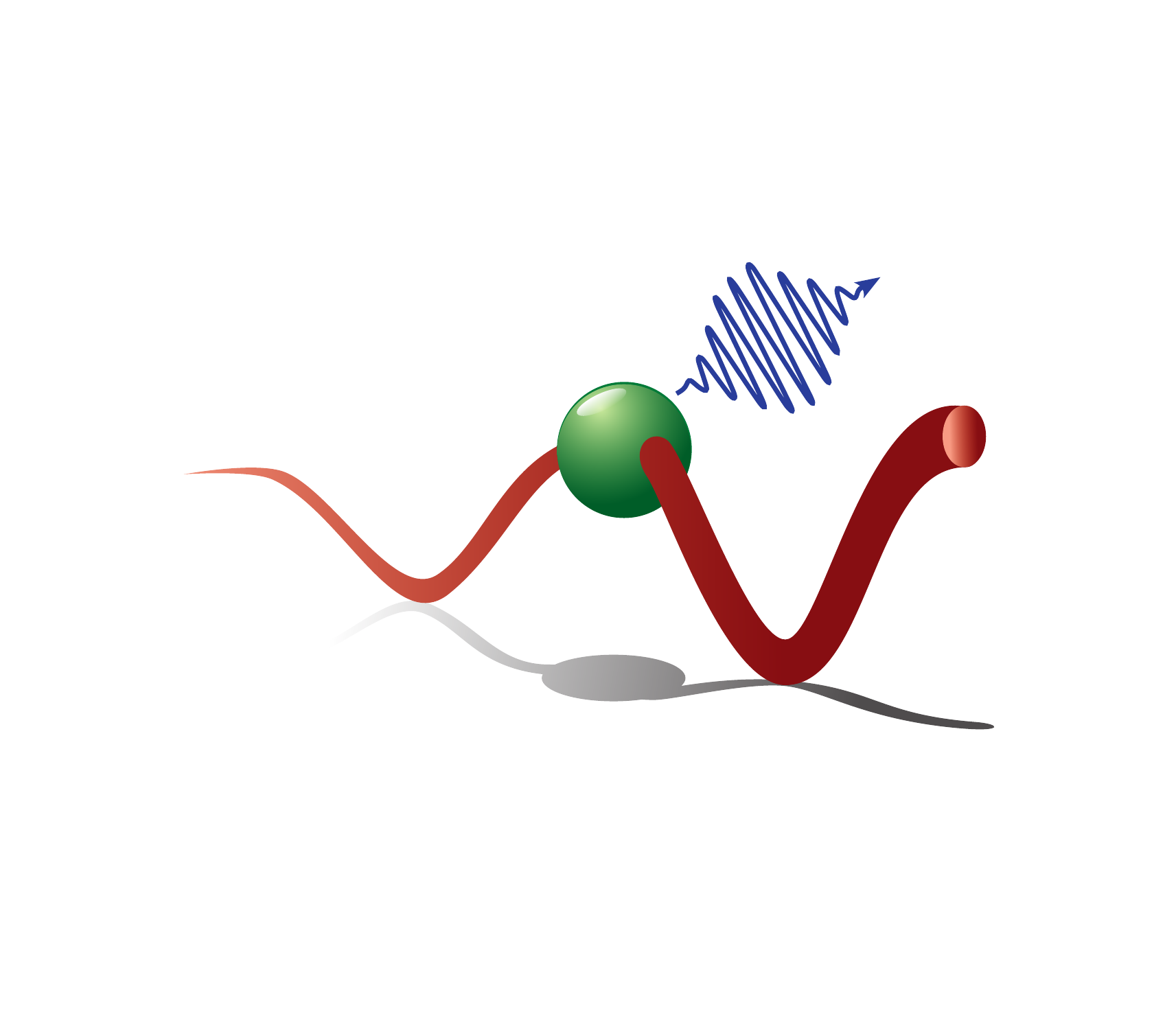}
\end{figure} 

\newpage
\section{Ford-O'Connell equation in an electromagnetic field}

In the presence of an arbitrary external force $f^a$, the Ford-O'Connell equation \cite{FO1,FO2} reads
\begin{equation}
\xdd^a= f^a + \tau \Delta^a{}_b \dot{f}^b\ , \label{eq:2:FOini}
\end{equation}
where $\Delta^a{}_b=\delta^a_b+\xd^a\xd_b$ is the projection operator. 

Consider the normalisation condition of the 4-velocity:
\begin{equation}
\label{eq:2:norm}
\xd\cdot \xd=-1\ . 
\end{equation}
Differentiating (\ref{eq:2:norm}):
\begin{equation}
\label{eq:2:FOdifferentiated}
\frac{d}{ds}\left(\xd\cdot \xd\right) = 2\xd \cdot \xdd = 0\ , 
\end{equation}
shows that $\dot{x}$ must be $\perp$ to $\ddot{x}$.

Combining (\ref{eq:2:FOdifferentiated}) with (\ref{eq:2:FOini}) it can be seen that:
\begin{equation}
2\xd \cdot \xdd=\underbrace{2f \cdot \xd}_{=0} + \underbrace{2\tau\Delta\dot{f} \cdot \xd}_{=0}=0\ .
\end{equation}

The first component $2f \cdot \xd=0$ is due to the property of an external force being orthogonal to the 4-velocity and the second component reflects the property of the projection operator $\Delta$ to annihilate the $\xd$ component of a vector, shown below:
\begin{equation}
\label{eq:2:Deltaannigh}
\Delta^f{}_b\xd_f=\left(\delta^f_b+\xd^f\xd_b\right)\xd_f=\delta^f_b\xd_f+\xd^f\xd_f\xd_b=\xd_b-\xd_b=0\ .
\end{equation}
These coupled together ensure that the normalisation condition (\ref{eq:2:norm}) is preserved.
 
The dominant forces on a classical charged particle are electromagnetic, so we use the Lorentz force: 
\begin{equation}
f^a=-\qm F^a{}_b \xd^b
\end{equation}
as the applied 
force in (\ref{eq:2:FOini}), leading to:
\begin{equation}
\xdd^a = -\qm F^a{}_b \xd^b - \tau \qm F^a{}_c \xdd^c - \tau \qm 
\frac{\partial F^a{}_c}{\partial x^d} \xd^d\xd^c - \tau \qm F^b{}_c 
\xd^a\xd_b\xdd^c - \tau \qm \frac{\partial F^b{}_c}{\partial x^d} 
\xd^a\xd_b\xd^d\xd^c\ . \label{eq:2:FO} 
\end{equation}
Because $F$ is antisymmetric $\frac{\partial F^b{}_c}{\partial x^d}\xd_b\xd^c = 0$, so (\ref{eq:2:FO}) can be rearranged as follows:
\begin{equation}
N^a{}_b\xdd^b = -\qm \left(F^a{}_b + \tau \xd^c \frac{\partial F^a{}_b}{\partial 
x^c}\right)\xd^b\ , \label{eq:2:FONmatrix}
\end{equation}
where $N^a{}_b = \delta^a_b + \tau \qm \Delta^a{}_c F^c{}_b$. Apart from the term 
involving derivatives of the fields, this coincides with the Mo-Papas equation \cite{MoPapas}, which reads:
\begin{equation}
\xdd^a = -\qm F^a{}_b \xd^b - \tau \qm F^a{}_c \xdd^c - \tau \qm F^b{}_c 
\xd^a\xd_b\xdd^c\ .
\end{equation}
The latter was derived heuristically, rather than either from first principles or as an approximation to the Lorentz-Abraham-Dirac equation.

To acquire a proper equation of motion we need to invert $N^a{}_b$. However, (\ref{eq:2:FONmatrix}) does not uniquely define $N$:
it is clear that the contraction of $\xdd^b$ with $N^a{}_b$ 
produces an identical result to the contraction with 
$M^a{}_b = N^a{}_b + V^a\xd_b$. The $M^a{}_b$ matrix is required to act only on vectors that are orthogonal to $\dot{x}$ and will produce vectors orthogonal to $\dot{x}$. Choosing $V$ to be:
\begin{equation}
\label{eq:2:Vcomp}
V^a=\xd^a + \tau \qm \Delta^a{}_c F^c{}_d\xd^d\ ,
\end{equation}
ensures that $M$ annihilates $\dot{x}$. This allows us to uniquely invert $M$ \textit{on the space of vectors orthogonal to $\dot{x}$}.
 
Incorporating (\ref{eq:2:Vcomp}) in equation (\ref{eq:2:FONmatrix}) the Ford-O'Connell equation takes the form
\begin{equation}
\left(\Delta^a{}_b+ \tau G^a{}_b\right)\xdd^b=-\qm\left(F^a{}_b+\tau \xd^c\partial_c 
F^a{}_b\right)\xd^b\ , \label{eq:2:FOMmatrix}
\end{equation}
where $G^a{}_b= \qm \Delta^a{}_c F^c{}_d \Delta^d{}_b$ is the ``sandwiched'' tensor, representing the magnetic field \textit{as seen by the particle}. 

\section{Matrix form of the Ford-O'Connell equation}

For (\ref{eq:2:FOMmatrix}) to be a valid equation of motion, it is necessary 
that it can be solved algebraically for the acceleration $\xdd$. By writing it as 
\begin{equation}
M^a{}_b \xdd^b= -\qm\left(F^a{}_b+\tau\xd^c\partial_c F^a{}_b\right)\xd^b\ ,
\end{equation}
where 
\begin{equation}
M^a{}_b=\Delta^a{}_b + \tau G^a{}_b\ , \label{eq:2:Mdef}
\end{equation}
we need to show that $M$ can be inverted. However, care must be taken in defining the inverse: taken as a matrix acting on \textit {all} 4-vectors, 
$M$ annihilates $\xd$, and therefore cannot be inverted. However, from 
(\ref{eq:2:FOMmatrix}), we only require $M$ to act on (and produce) vectors orthogonal 
to $\xd$. 

In the particle rest frame, denoted by $\star$, the 4-velocity is given by $\dot{x}^a \stackrel{\star}{=} \delta^a_0$ and the projection operator is given by:
\begin{equation}
\label{eq:2:Delta_time}
\Delta^0{}_a \stackrel{\star}{=} 0 \stackrel{\star}{=} \Delta^a{}_0
\end{equation}
and its spatial components
\begin{equation}
\label{eq:2:Delta_space}
\Delta^\mu{}_\nu \stackrel{\star}{=} \delta^\mu_\nu\ .
\end{equation}
In this section Greek indices run from 1 to 3.

Based on (\ref{eq:2:Mdef}) and (\ref{eq:2:Delta_time}) the time components of the matrix $M$ in the particle's rest frame are:
\begin{equation}
M^0{}_a \stackrel{\star}{=} 0 \stackrel{\star}{=} M^a{}_0
\end{equation}
and, according to (\ref{eq:2:Delta_space}) the spatial components are:
\begin{equation}
\label{eq:2:M_spatial_rest}
M^\mu{}_\nu \stackrel{\star}{=} \delta^\mu_\nu + \tau G^\mu{}_\nu\ .
\end{equation}
The definition of the inverse of $M$ in this frame is given by:
\begin{equation}
\left(M^{-1}\right)^\mu{}_\nu M^\nu{}_\lambda \stackrel{\star}{=} \delta^\mu_\lambda \stackrel{\star}{=} \Delta^\mu{}_\lambda\ .
\end{equation}
It follows that:
\begin{equation}
\left(M^{-1}\right)^\mu{}_a M^a{}_\lambda \stackrel{\star}{=} \left(M^{-1}\right)^\mu{}_0 M^0{}_\lambda + \left(M^{-1}\right)^\mu{}_\nu M^\nu{}_\lambda \stackrel{\star}{=} \Delta^\mu{}_\lambda\ ,
\end{equation}
which is true since in the rest frame of the particle, $M^0{}_\lambda = 0$. Also, because both $\left(M^{-1}\right)^0{}_a$ and $M^a{}_0$ are 0, the following holds:
\begin{equation}
\left(M^{-1}\right)^0{}_a M^a{}_b \stackrel{\star}{=} 0 \stackrel{\star}{=} \Delta^0{}_b\ ,
\end{equation}
\begin{equation}
\left(M^{-1}\right)^b{}_a M^a{}_0 \stackrel{\star}{=} 0 \stackrel{\star}{=} \Delta^b{}_0\ .
\end{equation}
Therefore, the inverse can be defined as
\begin{equation}
\label{eq:2:invcondition}
\left(M^{-1}\right)^a{}_b M^b{}_c \stackrel{\star}{=} M^a{}_b \left(M^{-1}\right)^b{}_c \stackrel{\star}{=} \Delta^a{}_c\ .
\end{equation}
Since (\ref{eq:2:invcondition}) is covariant it is true in all frames.

To specify its determinant, consider the 3 $\times$ 3 matrix of spatial components $M^\mu{}_\nu$ in the rest frame of the particle (\ref{eq:2:M_spatial_rest}). Its determinant is given by:
\begin{equation}
\det M \stackrel{\star}{=} \frac{1}{3!} \varepsilon^{\mu\nu\lambda} \varepsilon_{\alpha\beta\gamma} M^\alpha{}_\mu M^\beta{}_\nu M^\gamma{}_\lambda\ ,
\end{equation}
where $\varepsilon_{\alpha\beta\gamma}$ is the completely antisymmetric tensor of rank 3 with $\varepsilon_{123} = 1$. 

This can also be written as
\begin{align}
\label{eq:2:det_rest_step3}
\nonumber \det M & \stackrel{\star}{=} \frac{1}{3!} \varepsilon^{0\mu\nu\lambda} \varepsilon_{0\alpha\beta\gamma} M^\alpha{}_\mu M^\beta{}_\nu M^\gamma{}_\lambda \dot{x}^0 \dot{x}_0 \\
& \stackrel{\star}{=} \frac{1}{3!} \varepsilon^{a\mu\nu\lambda} \varepsilon_{b\alpha\beta\gamma} M^\alpha{}_\mu M^\beta{}_\nu M^\gamma{}_\lambda \dot{x}^b \dot{x}_a\ ,
\end{align}
with $\varepsilon_{abcd}$ the completely antisymmetric tensor of rank 4 with $\varepsilon_{0123} = 1$. 

Due to the antisymmetric property of $\varepsilon$, (\ref{eq:2:det_rest_step3}) can be written covariantly as
\begin{equation}
\label{eq:2:determinant}
\det M = \frac{1}{3!} \varepsilon^{abcd} \varepsilon_{efgh} M^f{}_b M^g{}_c M^h{}_d \dot{x}^e \dot{x}_a
\end{equation}

The combination $\varepsilon_{abcd}\varepsilon^{efgh}$ can be expanded as follows:
\begin{align}
\nonumber \varepsilon_{abcd}\varepsilon^{efgh} 
=&\delta^e_a\delta^f_b\delta^g_d\delta^h_c-\delta^e_a\delta^f_b\delta^g_c\delta^h_d+\delta^e_a\delta^f_c\delta^g_b\delta^h_d-\delta^e_a\delta^f_c\delta^g_d\delta^h_b-\delta^e_a\delta^f_d\delta^g_b\delta^h_c+\\
\nonumber 
&\delta^e_a\delta^f_d\delta^g_c\delta^h_b+\delta^e_b\delta^f_a\delta^g_c\delta^h_d-\delta^e_b\delta^f_a\delta^g_d\delta^h_c-\delta^e_b\delta^f_c\delta^g_a\delta^h_d+\delta^e_b\delta^f_c\delta^g_d\delta^h_a-\\
\nonumber 
&\delta^e_b\delta^f_d\delta^g_c\delta^h_a+\delta^e_b\delta^f_d\delta^g_a\delta^h_c+\delta^e_c\delta^f_b\delta^g_a\delta^h_d-\delta^e_c\delta^f_b\delta^g_d\delta^h_a+\delta^e_c\delta^f_d\delta^g_b\delta^h_a-\\
\nonumber 
&\delta^e_c\delta^f_d\delta^g_a\delta^h_b+\delta^e_c\delta^f_a\delta^g_d\delta^h_b-\delta^e_c\delta^f_a\delta^g_b\delta^h_d+\delta^e_d\delta^f_b\delta^g_c\delta^h_a-\delta^e_d\delta^f_b\delta^g_a\delta^h_c+\\
&\delta^e_d\delta^f_a\delta^g_b\delta^h_c-\delta^e_d\delta^f_a\delta^g_c\delta^h_b+\delta^e_d\delta^f_c\delta^g_a\delta^h_b-\delta^e_d\delta^f_c\delta^g_b\delta^h_a\ .
\end{align}
To produce general expression for the determinant the 
respective contractions of $\varepsilon_{abcd}\varepsilon^{efgh}$ with $M$'s 
must be carried out. To simplify the calculation we start with 
the contraction with $\xd^d\xd_h$, which yields:
\begin{align}
\label{eq:2:eexx}
\nonumber \varepsilon_{abcd}\varepsilon^{efgh}\xd^d\xd_h 
=&\delta^e_a\delta^f_b\delta^g_c+\delta^e_a\delta^f_b\xd^g\xd_c-\delta^e_a\delta^f_c\delta^g_b-\delta^e_a\delta^f_c\xd^g\xd_b-\delta^e_a\delta^g_b\xd^f\xd_c+\\
%%%%%%%%%%%
\nonumber 
&\delta^e_a\delta^g_c\xd^f\xd_b-\delta^e_b\delta^f_a\delta^g_c-\delta^e_b\delta^f_a\xd^g\xd_c+\delta^e_b\delta^f_c\delta^g_a+\delta^e_b\delta^f_c\xd^g\xd_a-\\
%%%%%%%%%%%
\nonumber 
&\delta^e_b\delta^g_c\xd^f\xd_a+\delta^e_b\delta^g_a\xd^f\xd_c-\delta^e_c\delta^f_b\delta^g_a-\delta^e_c\delta^f_b\xd^g\xd_a+\delta^e_c\delta^g_b\xd^f\xd_a-\\
%%%%%%%%%%%%
\nonumber 
&\delta^e_c\delta^g_a\xd^f\xd_b+\delta^e_c\delta^f_a\xd^g\xd_b+\delta^e_c\delta^f_a\delta^g_b+\delta^f_b\delta^g_c\xd^e\xd_a-\delta^f_b\delta^g_a\xd^e\xd_c+\\
&\delta^f_a\delta^g_b\xd^e\xd_c-\delta^f_a\delta^g_c\xd^e\xd_b+\delta^f_c\delta^g_a\xd^e\xd_b-\delta^f_c\delta^g_b\xd^e\xd_a\ .
\end{align}
Considering that $M$ is orthogonal to $\xd$ it can be seen that all the terms of $\varepsilon_{abcd}\varepsilon^{efgh}\xd^d\xd_h$ involving $\xd$ will vanish after all the contractions have taken place leaving only terms that contribute to the final result:
\begin{equation}
\label{eq:2:eexxsimplified}
\varepsilon_{abcd}\varepsilon^{efgh}\xd^d\xd_h\approx\delta^e_a\delta^f_b\delta^g_c-\delta^e_a\delta^f_c\delta^g_b-\delta^e_b\delta^f_a\delta^g_c+\delta^e_b\delta^f_c\delta^g_a-\delta^e_c\delta^f_b\delta^g_a+\delta^e_c\delta^f_a\delta^g_b\ .
\end{equation}

Based on the result in (\ref{eq:2:eexxsimplified}) and proceeding with the remaining 
contractions we obtain:
\begin{equation}
\label{eq:2:eexxmmmm}
\varepsilon_{abcd}\varepsilon^{efgh} M^a{}_e M^b{}_f M^c{}_g \xd^d\xd_h =  
M^3 - 3M M^a{}_b M^b{}_a + 2M^a{}_b M^b{}_c M^c{}_a\ ,
\end{equation}
where $M=M^a{}_a=\tr\left(M\right)$.

Considering the form and properties of $M$ (\ref{eq:2:Mdef}) all the required terms involving $M$ can be written out in terms of ``sandwiched'' electromagnetic field tensors $G$ as follows:
\begin{equation}
M^a{}_a=3\ , \label{eq:2:Maa}
\end{equation}
\begin{align}
\nonumber M^a{}_b M^b{}_a & = \left(\Delta^a{}_b+\tau G^a{}_b\right)\left(\Delta^b{}_a+\tau G^b{}_a\right) = 3 + \tau \Delta^a{}_b G^b{}_a + \tau \Delta^b{}_a G^a{}_b + \tau^2 G^a{}_b G^b{}_a\\
& = 3 + \tau^2 G^a{}_b G^b{}_a\ , \label{eq:2:MabMba}
\end{align}
and
\begin{align}
\nonumber M^a{}_b M^b{}_c M^c{}_a & = \left(\Delta^a{}_b+\tau G^a{}_b\right)\left(\Delta^b{}_c+\tau G^b{}_c\right)\left(\Delta^c{}_a+\tau G^c{}_a\right)\\
& = 3 + 3 \tau^2 G^a{}_b G^b{}_a\ . \label{eq:2:MabMbcMca}
\end{align}
Substituting (\ref{eq:2:Maa}, \ref{eq:2:MabMba}, \ref{eq:2:MabMbcMca}) into (\ref{eq:2:eexxmmmm}) leads to:
\begin{align}
\nonumber &\varepsilon_{abcd}\varepsilon^{efgh} M^a{}_e M^b{}_f M^c{}_g 
\xd^d\xd_h = 6 - 3\tau^2 G^a{}_b G^b{}_a\ ,
\end{align}
allowing the determinant from (\ref{eq:2:determinant}) to be specified as:
\begin{equation}
\label{eq:2:detfinal}
\det M = 1 + \frac{\tau^2}{2} G^{ab} G_{ab}\ .
\end{equation}
Expressing the electromagnetic field tensor in terms of the electric and magnetic fields \textit{as seen by the particle}:
\begin{equation}
\label{eq:2:fabdefinition}
F^a{}_b=E^a\xd_b-E_b\xd^a-\varepsilon^a{}_{bcd}B^c\xd^d
\end{equation}
the property of the projection operator $\Delta$ will lead to the annihilation of the $\xd$ components, such that $\Delta^a{}_b F^b{}_c \Delta^c{}_d$ does not involve $E$ in (\ref{eq:2:fabdefinition}).

Combined with the Lorentz invariant $F_{ab}F^{ab}=2\left(B^2-E^2\right)$, the form of the determinant (\ref{eq:2:detfinal}) in terms of the fields measured by an observer comoving with the particle can then be found:
\begin{equation}
\det M= 1+ \tau^2 \frac{q^2}{m^2}B^2>0\ .
\end{equation} 
It follows that $M$ is invertible, and the Ford-O'Connell equation is a viable description for the motion of a 
charged particle.

\section{Inversion of the $M$ matrix}

To consider the Ford-O'Connell equation as an equation of motion written in 
the following form:
\begin{equation}
\label{eq:2:FOwithinverse}
\xdd^a=-\qm\left(M^{-1}\right)^a{}_b \left(F^b{}_d+\tau \xd^c\partial_c 
F^b{}_d\right)\xd^d\ ,
\end{equation}
we require an analytical form for the inverse $\left(M^{-1}\right)^a{}_b$.

To solve (\ref{eq:2:invcondition}) for the inverse it is convenient to express $\left(M^{-1}\right)$ as a linear combination of a set of known matrices. 

A convenient choice of basis matrices can be constructed using the fields seen by the particle:
\begin{equation}
\label{eq:2:basis}
\left(\Gamma_\lambda\right)^c{}_a \in \left[ \begin{array}{c} S^c \\ E^c \\ B^c \end{array} \right]
\otimes \left[\begin{array}{ccc} S_a & E_a & B_a \end{array} \right]\ ,
\end{equation}
where $\lambda$ runs from 1 to 9. In terms of the electromagnetic field tensor, the electric field $E$, magnetic field $B$ and Poynting vector $S$ are as follows:
\begin{equation}
E^a = - F^a{}_b \dot{x}^b\ ,
\end{equation}
\begin{equation}
B^a = - \frac{1}{2} \varepsilon^{abcd}F_{cd} \dot{x}_b\ ,
\end{equation}
and
\begin{equation}
S^a = F^a{}_b F^b{}_c \dot{x}^c - F_{cb} F^c{}_d \dot{x}^b \dot{x}^d \dot{x}^a\ .
\end{equation}

The tensor product in (\ref{eq:2:basis}) leads to a combination of 9 linearly independent\footnote{For special cases $E$ may be parallel to $B$ and the frame collapses, however generically $E$, $B$ and $S$ are independent.} components: 
\begin{equation}
\label{eq:2:basislist}
\left(\Gamma_\lambda\right)^c{}_a\in\left\{{\Delta^c{}_a, S^c E_a, S^c B_a, E^c S_a, E^c E_a, E^c B_a, 
B^c S_a, B^c E_a, B^c B_a}\right\}\ .
\end{equation}
The $S^cS_a$ matrix in (\ref{eq:2:basis}) can be expanded as:
\begin{equation}
\label{eq:2:SScontr}
S^c S_a = -B^2 E^c E_a + \left(E\cdot B\right)E^c B_a + \left(E\cdot B\right)B^c E_a - E^2 B^c 
B_a+S^2\Delta^c{}_a\ ,
\end{equation}
which explains the presence of $\Delta^c{}_a$ in (\ref{eq:2:basislist}) instead of $S^cS_a$.

The matrix inverse can then be described as a sum:
\begin{equation}
\label{eq:2:invsumm}
\left(M^{-1}\right)^c{}_a = \displaystyle\sum\limits_\lambda \alpha_\lambda 
\left(\Gamma_\lambda\right)^c{}_a\ ,
\end{equation}
where $\alpha_\lambda$ is a set of coefficients we need to determine in order to obtain an expression for 
the inverse in terms of the electric and magnetic fields.

Taking into account the properties of the inverse (\ref{eq:2:invcondition}) and 
the above relation (\ref{eq:2:invsumm}) it can be straightforwardly seen that:
\begin{equation}
\label{eq:2:invconditionsumm}
\left(M^{-1}\right)^c{}_a M^a{}_b=\displaystyle\sum\limits_\lambda \alpha_\lambda 
\left(\Gamma_\lambda\right)^c{}_a M^a{}_b=\Delta^c{}_b\ .
\end{equation}

The matrix to be inverted can also be 
rewritten in terms of electric and magnetic fields:
\begin{equation}
M^a{}_b = \Delta^a{}_b+\tau\frac{q}{m}\varepsilon^a{}_b{}^{mn}B_m\xd_n\ . \label{eq:2:mabEB}
\end{equation}
Equation (\ref{eq:2:mabEB}) can be written explicitly as a linear combination of $\Gamma_\lambda$ components.
Because $\varepsilon^{abcd}$ is antisymmetric the $\varepsilon^{abcd}B_c\xd_d$ term can be expanded as:
\begin{equation}
\label{eq:2:bits}
\varepsilon^{abcd}B_c\xd_d=\zeta_1\left(E^aB^b-E^bB^a\right)+\zeta_2\left(E^aS^b-E^bS^a\right)+\zeta_3\left(B^aS^b-B^bS^a\right)\ .
\end{equation}
Contracting equation (\ref{eq:2:bits}) with $B_b$ leads to the left hand side 
vanishing due to properties of the completely antisymmetric tensor 
$\varepsilon^{abcd}$, therefore:
\begin{equation}
0=\zeta_1 E^aB^bB_b-\zeta_1 E^bB^aB_b+\zeta_2 E^aS^bB_b-\zeta_2 
E^bS^aB_b+\zeta_3 B^aS^bB_b-\zeta_3 B^bS^aB_b\ ,
\end{equation}
which is equivalent to:
\begin{equation}
\label{eq:2:bbcomponent}
0=\zeta_1 B^2E^a-\zeta_1 \left(E\cdot B\right)B^a-\zeta_2 \left(E\cdot B\right)S^a-\zeta_3 
B^2S^a\ .
\end{equation}
Because of the linear independence of $E$, $B$ and $S$ it can be clearly seen that:
\begin{equation}
\label{eq:2:zeta23relation1}
\zeta_1 = 0\ , \qquad \zeta_3=-\frac{\left(E\cdot B\right)}{B^2}\zeta_2\ .
\end{equation}

The analogous contraction of equation (\ref{eq:2:bits}) with $E_b$,  taking into account (\ref{eq:2:zeta23relation1}), turns the left 
hand side into the Poynting vector $S^a$:
\begin{equation}
S^a= \zeta_2 E^aS^bE_b-\zeta_2 E^bE_bS^a+\zeta_3 B^aS^bE_b-\zeta_3 B^bE_bS^a\ ,
\end{equation}
which is equivalent to:
\begin{equation}
\label{eq:2:sa}
S^a= - \zeta_2 E^2S^a - \zeta_3 \left(E\cdot 
B\right)S^a\ .
\end{equation}
From (\ref{eq:2:sa}) it can be seen that:
\begin{equation}
\label{eq:2:zeta23relation2}
\zeta_3=-\frac{1+\zeta_2 
E^2}{\left(E\cdot B\right)}\ .
\end{equation}
Equating (\ref{eq:2:zeta23relation1}) with (\ref{eq:2:zeta23relation2}) leads to:
\begin{equation}
\label{eq:2:zeta2}
\zeta_2=\frac{B^2}{\left(E\cdot B\right)^2-E^2B^2}=-\frac{B^2}{S^2}\ ,
\end{equation}
where $S^2 = E^2 B^2 - \left(E \cdot B\right)^2$ has been used.

After substitution of this result into (\ref{eq:2:zeta23relation1}), we obtain:
\begin{equation}
\label{eq:2:zeta3}
\zeta_3=-\frac{\left(E\cdot B\right)}{\left(E\cdot B\right)^2-E^2B^2}=\frac{\left(E\cdot 
B\right)}{S^2}\ .
\end{equation}
Recombining the $\zeta$ coefficients with (\ref{eq:2:bits}) we obtain an expression for $\varepsilon^a{}_b{}^{mn}B_m\xd_n$ in terms of the electric and magnetic fields, and the Poynting vector:
\begin{equation}
\label{eq:2:sandwichedtensor}
\varepsilon^a{}_b{}^{mn}B_m\xd_n = \frac{B^2}{S^2}\left(S^a 
E_b - E^a S_b\right)+\frac{E\cdot B}{S^2}\left(B^a S_b - S^a B_b\right)\ .
\end{equation}
This allows us to rewrite (\ref{eq:2:mabEB}) in the form of:
\begin{equation}
M^a{}_b = \Delta^a{}_b+\tau\frac{q}{m}\frac{B^2}{S^2}\left(S^a 
E_b - E^a S_b\right)+\tau\frac{q}{m}\frac{E\cdot B}{S^2}\left(B^a S_b - S^a B_b\right)\ . \label{eq:2:mabEBS}
\end{equation}

To obtain the inverse we can write out each $\lambda$-component of (\ref{eq:2:invconditionsumm}) separately, giving us the form of the inverse in terms of $E$, $B$ and $S$:
%% Sam: maybe organise as a list?
%% First element
%%\begin{center}
%%For $\lambda=1$ corresponding to $\Delta^c_a$ in the $B^c{}_a$ basis
%%\end{center}
%\begin{align}
%B^c{}_a M^a{}_b\nonumber &=\Delta^c_b+\tau\frac{q}{m}\frac{B^2}{E^2 
%B^2-(E\cdot B)^2}\Delta^c_a S^a E_b-\tau\frac{q}{m}\frac{B^2}{E^2 
%B^2-(E\cdot B)^2}\Delta^c_a E^a S_b\\ &+\tau\frac{q}{m}\frac{E\cdot B}{E^2 
%B^2-(E\cdot B)^2}\Delta^c_a B^a S_b-\tau\frac{q}{m}\frac{E\cdot B}{E^2 
%B^2-(E\cdot B)^2}\Delta^c_a S^a B_b.
%\end{align}
%simplified
\begin{enumerate}
\item $\lambda=1$ with $\left(\Gamma_1\right)^c{}_a = \Delta^c{}_a$:
\begin{align}
\left(\Gamma_1\right)^c{}_a M^a{}_b\nonumber =&\Delta^c{}_b+\tau\frac{q}{m}\frac{B^2}{E^2 
B^2-\left(E\cdot B\right)^2}S^c E_b-\tau\frac{q}{m}\frac{B^2}{E^2 B^2-\left(E\cdot B\right)^2}E^c 
S_b+\\ &\tau\frac{q}{m}\frac{E\cdot B}{E^2 B^2-\left(E\cdot B\right)^2}B^c 
S_b-\tau\frac{q}{m}\frac{E\cdot B}{E^2 B^2-\left(E\cdot B\right)^2}S^c B_b\ .
\end{align}
%% Second element
%\begin{center}
%For $\lambda=2$ corresponding to $S^c E_a$ in the $B^c{}_a$ basis
%\end{center}
%\begin{align}
%B^c{}_a M^a{}_b\nonumber &=\Delta^a_b S^c E_a+\tau\frac{q}{m}\frac{B^2}{E^2 
%B^2-(E\cdot B)^2}S^c E_a S^a E_b-\tau\frac{q}{m}\frac{B^2}{E^2 B^2-(E\cdot 
%B)^2}S^c E_a E^a S_b\\ &+\tau\frac{q}{m}\frac{E\cdot B}{E^2 B^2-(E\cdot 
%B)^2}S^c E_a B^a S_b-\tau\frac{q}{m}\frac{E\cdot B}{E^2 B^2-(E\cdot 
%B)^2}S^c E_a S^a B_b.
%\end{align}
%simplified
\item $\lambda=2$ with $\left(\Gamma_2\right)^c{}_a = S^c E_a$:
\begin{align}
\left(\Gamma_2\right)^c{}_a M^a{}_b=\nonumber &S^c E_b+\tau\frac{q}{m}\frac{B^2}{E^2 B^2-\left(E\cdot 
B\right)^2}E^2 S^c S_b-\\
&\tau\frac{q}{m}\frac{E\cdot B}{E^2 B^2-\left(E\cdot B\right)^2}\left(E\cdot 
B\right)S^c S_b\ .
\end{align}
%% Third element
%\begin{center}
%For $\lambda=3$ corresponding to $S^c B_a$ in the $B^c{}_a$ basis
%\end{center}
%\begin{align}
%B^c{}_a N^a{}_b\nonumber &=\Delta^a_b S^c B_a-\tau\frac{q}{m}\frac{B^2}{E^2 
%B^2-(E\cdot B)^2}S^c B_a S^a E_b+\tau\frac{q}{m}\frac{B^2}{E^2 B^2-(E\cdot 
%B)^2}S^c B_a E^a S_b\\ &-\tau\frac{q}{m}\frac{E\cdot B}{E^2 B^2-(E\cdot 
%B)^2}S^c B_a B^a S_b+\tau\frac{q}{m}\frac{E\cdot B}{E^2 B^2-(E\cdot 
%B)^2}S^c B_a S^a B_b.
%\end{align}
%simplified
\item $\lambda=3$ with $\left(\Gamma_3\right)^c{}_a = S^c B_a$:
\begin{align}
\left(\Gamma_3\right)^c{}_a M^a{}_b=\nonumber &S^c B_b+\tau\frac{q}{m}\frac{B^2}{E^2 B^2-\left(E\cdot 
B\right)^2}\left(E\cdot B\right) S^c S_b-\\
&\tau\frac{q}{m}\frac{E\cdot B}{E^2 B^2-\left(E\cdot 
B\right)^2}B^2 S^c S_b\ .
\end{align}
%% Fourth element
%\begin{center}
%For $\lambda=4$ corresponding to $E^c S_a$ in the $B^c{}_a$ basis
%\end{center}
%\begin{align}
%B^c{}_a N^a{}_b\nonumber &=\Delta^a_b E^c S_a-\tau\frac{q}{m}\frac{B^2}{E^2 
%B^2-(E\cdot B)^2}E^c S_a S^a E_b+\tau\frac{q}{m}\frac{B^2}{E^2 B^2-(E\cdot 
%B)^2}E^c S_a E^a S_b\\ &-\tau\frac{q}{m}\frac{E\cdot B}{E^2 B^2-(E\cdot 
%B)^2}E^c S_a B^a S_b+\tau\frac{q}{m}\frac{E\cdot B}{E^2 B^2-(E\cdot 
%B)^2}E^c S_a S^a B_b.
%\end{align}
%simplified
\item $\lambda=4$ with $\left(\Gamma_4\right)^c{}_a = E^c S_a$:
\begin{align}
\left(\Gamma_4\right)^c{}_a M^a{}_b=\nonumber &E^c S_b-\tau\frac{q}{m}\frac{B^2}{E^2 B^2-\left(E\cdot 
B\right)^2}S^2 E^c E_b+\\
&\tau\frac{q}{m}\frac{E\cdot B}{E^2 B^2-\left(E\cdot B\right)^2}S^2 E^c 
B_b\ .
\end{align}
%% Fifth element
%\begin{center}
%For $\lambda=5$ corresponding to $E^c E_a$ in the $B^c{}_a$ basis
%\end{center}
%\begin{align}
%B^c{}_a N^a{}_b\nonumber &=\Delta^a_b E^c E_a-\tau\frac{q}{m}\frac{B^2}{E^2 
%B^2-(E\cdot B)^2}E^c E_a S^a E_b+\tau\frac{q}{m}\frac{B^2}{E^2 B^2-(E\cdot 
%B)^2}E^c E_a E^a S_b\\ &-\tau\frac{q}{m}\frac{E\cdot B}{E^2 B^2-(E\cdot 
%B)^2}E^c E_a B^a S_b+\tau\frac{q}{m}\frac{E\cdot B}{E^2 B^2-(E\cdot 
%B)^2}E^c E_a S^a B_b.
%\end{align}
%simplified
\item $\lambda=5$ with $\left(\Gamma_5\right)^c{}_a = E^c E_a$:
\begin{align}
\left(\Gamma_5\right)^c{}_a M^a{}_b=\nonumber &E^c E_b+\tau\frac{q}{m}\frac{B^2}{E^2 B^2-\left(E\cdot 
B\right)^2}E^2 E^c S_b-\\
&\tau\frac{q}{m}\frac{E\cdot B}{E^2 B^2-\left(E\cdot B\right)^2}\left(E\cdot 
B\right) E^c S_b\ .
\end{align}
%% Sixth element
%\begin{center}
%For $\lambda=6$ corresponding to $E^c B_a$ in the $B^c{}_a$ basis
%\end{center}
%\begin{align}
%B^c{}_a N^a{}_b\nonumber &=\Delta^a_b E^c B_a-\tau\frac{q}{m}\frac{B^2}{E^2 
%B^2-(E\cdot B)^2}E^c B_a S^a E_b+\tau\frac{q}{m}\frac{B^2}{E^2 B^2-(E\cdot 
%B)^2}E^c B_a E^a S_b\\ &-\tau\frac{q}{m}\frac{E\cdot B}{E^2 B^2-(E\cdot 
%B)^2}E^c B_a B^a S_b+\tau\frac{q}{m}\frac{E\cdot B}{E^2 B^2-(E\cdot 
%B)^2}E^c B_a S^a B_b.
%\end{align}
%simplified
\item $\lambda=6$ with $\left(\Gamma_6\right)^c{}_a = E^c B_a$:
\begin{align}
\left(\Gamma_6\right)^c{}_a M^a{}_b=\nonumber &E^c B_b+\tau\frac{q}{m}\frac{B^2}{E^2 B^2-\left(E\cdot 
B\right)^2}\left(E\cdot B\right) E^c S_b-\\
&\tau\frac{q}{m}\frac{E\cdot B}{E^2 B^2-\left(E\cdot 
B\right)^2}B^2 E^c S_b\ .
\end{align}
%% Seventh element
%\begin{center}
%For $\lambda=7$ corresponding to $B^c S_a$ in the $B^c{}_a$ basis
%\end{center}
%\begin{align}
%B^c{}_a N^a{}_b\nonumber &=\Delta^a_b B^c S_a-\tau\frac{q}{m}\frac{B^2}{E^2 
%B^2-(E\cdot B)^2}B^c S_a S^a E_b+\tau\frac{q}{m}\frac{B^2}{E^2 B^2-(E\cdot 
%B)^2}B^c S_a E^a S_b\\ &-\tau\frac{q}{m}\frac{E\cdot B}{E^2 B^2-(E\cdot 
%B)^2}B^c S_a B^a S_b+\tau\frac{q}{m}\frac{E\cdot B}{E^2 B^2-(E\cdot 
%B)^2}B^c S_a S^a B_b.
%\end{align}
%simplified
\item $\lambda=7$ with $\left(\Gamma_7\right)^c{}_a = B^c S_a$:
\begin{align}
\left(\Gamma_7\right)^c{}_a M^a{}_b=\nonumber &B^c S_b-\tau\frac{q}{m}\frac{B^2}{E^2 B^2-\left(E\cdot 
B\right)^2}S^2 B^c E_b+\\
&\tau\frac{q}{m}\frac{E\cdot B}{E^2 B^2-\left(E\cdot B\right)^2}S^2 B^c 
B_b\ .
\end{align}
%% Eights element
%\begin{center}
%For $\lambda=8$ corresponding to $B^c E_a$ in the $B^c{}_a$ basis
%\end{center}
%\begin{align}
%B^c{}_a N^a{}_b\nonumber &=\Delta^a_b B^c E_a-\tau\frac{q}{m}\frac{B^2}{E^2 
%B^2-(E\cdot B)^2}B^c E_a S^a E_b+\tau\frac{q}{m}\frac{B^2}{E^2 B^2-(E\cdot 
%B)^2}B^c E_a E^a S_b\\ &-\tau\frac{q}{m}\frac{E\cdot B}{E^2 B^2-(E\cdot 
%B)^2}B^c E_a B^a S_b+\tau\frac{q}{m}\frac{E\cdot B}{E^2 B^2-(E\cdot 
%B)^2}B^c E_a S^a B_b.
%\end{align}
%simplified
\item $\lambda=8$ with $\left(\Gamma_8\right)^c{}_a = B^c E_a$:
\begin{align}
\left(\Gamma_8\right)^c{}_a M^a{}_b=\nonumber &B^c E_b+\tau\frac{q}{m}\frac{B^2}{E^2 B^2-\left(E\cdot 
B\right)^2}E^2 B^c S_b-\\
&\tau\frac{q}{m}\frac{E\cdot B}{E^2 B^2-\left(E\cdot B\right)^2}\left(E\cdot 
B\right) B^c S_b\ .
\end{align}
%% Ninth element
\item $\lambda=9$ with $\left(\Gamma_9\right)^c{}_a = B^c B_a$:
\begin{align}
\left(\Gamma_9\right)^c{}_a M^a{}_b=\nonumber &B^c B_b+\tau\frac{q}{m}\frac{B^2}{E^2 B^2-\left(E\cdot 
B\right)^2}\left(E\cdot B\right) B^c S_b-\\
&\tau\frac{q}{m}\frac{E\cdot B}{E^2 B^2-\left(E\cdot 
B\right)^2}B^2 B^c S_b\ .
\end{align}
\end{enumerate}

The above combined with (\ref{eq:2:SScontr}) allows us to rewrite the sum in (\ref{eq:2:invconditionsumm}) in terms of the corresponding $\alpha_{\lambda}$ coefficients and basis elements, $\left(\Gamma_\lambda\right)^c{}_a$:
%\begin{align}
%\label{eq:2:summEMfields}
%\nonumber \displaystyle\sum\limits_\lambda \alpha_\lambda \left(B_\lambda\right)^c{}_a 
%M^a{}_b=&\alpha_1\Delta^c_b+\Big[\alpha_2-\alpha_1\tau\frac{q}{m}\frac{B^2}{S^2}\Big]S^c 
%E_b+\Big[\alpha_1\tau\frac{q}{m}\frac{B^2}{S^2}+\alpha_4+\alpha_5\tau\frac{q}{m}\Big]E^c 
%S_b+\\ \nonumber &\Big[\alpha_7-\alpha_1\tau\frac{q}{m}\frac{E\cdot 
%B}{S^2}+\alpha_8\tau\frac{q}{m}\Big]B^c 
%S_b+\Big[\alpha_1\tau\frac{q}{m}\frac{E\cdot B}{S^2}+\alpha_3\Big]S^c 
%B_b+\\ %% alpha_3 end
%\nonumber &\Big[\alpha_2\tau\frac{q}{m}\Big]S^c S_b+\Big[\alpha_5-\alpha_4\tau\frac{q}{m}B^2\Big]E^c 
%E_b+\Big[\alpha_4\tau\frac{q}{m}E\cdot B+\alpha_6\Big]E^c 
%B_b+\\ 
%&\Big[\alpha_8-\alpha_7\tau\frac{q}{m}B^2\Big]B^c E_b+\Big[\alpha_7\tau\frac{q}{m}E\cdot B+\alpha_9\Big]B^c B_b.
%\end{align}
\begin{align}
\label{eq:2:summEMfieldssimplified}
\nonumber \displaystyle\sum\limits_\lambda \alpha_\lambda \left(\Gamma_\lambda\right)^c{}_a 
M^a{}_b=&\left[\alpha_1+\alpha_2\tau\frac{q}{m}S^2\right]\Delta^c{}_b+\left[\alpha_1\tau\frac{q}{m}\frac{B^2}{S^2}+\alpha_4+\alpha_5\tau\frac{q}{m}\right]E^c S_b+\\
\nonumber 
&\left[\alpha_2-\alpha_1\tau\frac{q}{m}\frac{B^2}{S^2}\right]S^c 
E_b+\left[\alpha_7-\alpha_1\tau\frac{q}{m}\frac{E\cdot 
B}{S^2}+\alpha_8\tau\frac{q}{m}\right]B^c S_b+\\
\nonumber &\left[\alpha_1\tau\frac{q}{m}\frac{E\cdot B}{S^2}+\alpha_3\right]S^c 
B_b+\left[\alpha_5-\alpha_4\tau\frac{q}{m}B^2-\alpha_2\tau\frac{q}{m}B^2\right]E^c 
E_b+\\
\nonumber &\left[\alpha_4\tau\frac{q}{m}E\cdot 
B+\alpha_6+\alpha_2\tau\frac{q}{m}E\cdot B\right]E^c 
B_b+\Big[\alpha_8-\alpha_7\tau\frac{q}{m}B^2+\\
&\alpha_2\tau\frac{q}{m}E\cdot B\Big]B^c E_b+\left[\alpha_7\tau\frac{q}{m}E\cdot 
B+\alpha_9-\alpha_2\tau\frac{q}{m}E^2\right]B^c B_b\ .
\end{align}

Coupling (\ref{eq:2:summEMfieldssimplified}) with the main condition of inversion 
of the matrix (\ref{eq:2:invcondition}) we find the coefficient in front of 
$\Delta^c{}_b$ in (\ref{eq:2:summEMfieldssimplified}) to be equal to 1 and the 
remaining coefficients in front of every other element of the 
basis to be 0.  This leads to a set of 9 algebraic equations 
with 9 unknowns, that are the $\alpha_{\lambda}$ coefficients we require. These 
are:

% Delta^c_b
\begin{align}
\label{eq:2:eq1}
\alpha_1+\alpha_2\tau\frac{q}{m}S^2&=1\\
% S^c E_b
\label{eq:2:eq2}
\alpha_2-\alpha_1\tau\frac{q}{m}\frac{B^2}{S^2}&=0\\
% E^c S_b
\label{eq:2:eq3}
\alpha_1\tau\frac{q}{m}\frac{B^2}{S^2}+\alpha_4+\alpha_5\tau\frac{q}{m}&=0\\
% B^c S_b
\label{eq:2:eq4}
\alpha_7-\alpha_1\tau\frac{q}{m}\frac{\left(E\cdot 
B\right)}{S^2}+\alpha_8\tau\frac{q}{m}&=0\\
% S^c B_b
\label{eq:2:eq5}
\alpha_1\tau\frac{q}{m}\frac{\left(E\cdot B\right)}{S^2}+\alpha_3 &=0\\
% E^c E_b
\label{eq:2:eq6}
\alpha_5-\alpha_4\tau\frac{q}{m}B^2-\alpha_2\tau\frac{q}{m}B^2&=0\\
% E^c B_b
\label{eq:2:eq7}
\alpha_4\tau\frac{q}{m}\left(E\cdot B\right)+\alpha_6+\alpha_2\tau\frac{q}{m}\left(E\cdot B\right)&=0\\
% B^c E_b
\label{eq:2:eq8}
\alpha_8-\alpha_7\tau\frac{q}{m}B^2+\alpha_2\tau\frac{q}{m}\left(E\cdot B\right)&=0\\
% B^c B_b
\label{eq:2:eq9}
\alpha_7\tau\frac{q}{m}\left(E\cdot B\right)+\alpha_9-\alpha_2\tau\frac{q}{m}E^2&=0
\end{align}

To obtain these coefficients we need to solve the above set of equations. From equation (\ref{eq:2:eq2}) it can be seen that:
\begin{equation}
\label{eq:2:eq1proc}
\alpha_2=\alpha_1\tau\frac{q}{m}\frac{B^2}{S^2}\ .
\end{equation}
Substituting this result into equation (\ref{eq:2:eq1}) we obtain:
\begin{equation}
\label{eq2proc}
\alpha_1+\alpha_1\tau^2\frac{q^2}{m^2}B^2=\alpha_1\left(1+\tau^2\frac{q^2}{m^2}B^2\right)=1\ ,
\end{equation}
leading to:
\begin{equation}
\label{eq:2:eq3proc}
\alpha_1=\frac{1}{1+\tau^2\frac{q^2}{m^2}B^2}\ .
\end{equation}
Substituting this result into equation (\ref{eq:2:eq1proc}) we obtain:
\begin{equation}
\label{eq:2:eq4proc}
\alpha_2=\frac{\tau q B^2}{m S^2\left(1+\tau^2\frac{q^2}{m^2}B^2\right)}\ .
\end{equation}
Considering equation (\ref{eq:2:eq5}) we can write the following 
condition:
\begin{equation}
\label{eq:2:eq5proc}
\alpha_3=-\alpha_1\tau\frac{q}{mS^2}\left(E\cdot B\right)\ .
\end{equation}
Substituting the result from equation (\ref{eq:2:eq3proc}) into (\ref{eq:2:eq5proc}) 
we obtain:
\begin{equation}
\alpha_3=-\frac{\tau q}{mS^2\left(1+\tau^2\frac{q^2}{m^2}B^2\right)}\left(E\cdot B\right)\ .
\end{equation}
Considering equations (\ref{eq:2:eq3}, \ref{eq:2:eq6}) we can write out the equations 
combining $\alpha_4$ and $\alpha_5$ coefficients:
\begin{equation}
\label{eq:2:subset1}
\alpha_4+\alpha_5\tau\frac{q}{m}=-\alpha_1\tau\frac{q}{m}\frac{B^2}{S^2}\ ,
\end{equation}
\begin{equation}
\label{eq:2:subset2}
\alpha_5-\alpha_4\tau\frac{q}{m}B^2=\alpha_2\tau\frac{q}{m}B^2\ .
\end{equation}
Substituting equation (\ref{eq:2:eq4proc}) into (\ref{eq:2:subset2}) we obtain:
\begin{equation}
\alpha_5-\alpha_4\tau\frac{q}{m}B^2=\frac{\tau^2 q^2 B^4}{m^2 
S^2\left(1+\tau^2\frac{q^2}{m^2}B^2\right)}\ ,  
\end{equation}
leading to:
\begin{equation}
\label{eq:2:alpha5pre}
\alpha_5=\frac{\tau^2 q^2 B^4}{m^2 S^2\left(1+\tau^2\frac{q^2}{m^2}B^2\right)}+\alpha_4\tau\frac{q}{m}B^2\ ,
\end{equation}
which can then be substituted into (\ref{eq:2:subset1}), allowing us to obtain an equation for the $\alpha_4$ component:
\begin{equation}
\alpha_4\left(1+\tau^2\frac{q^2}{m^2}B^2\right)=-\frac{\tau^3 q^3 B^4}{m^3 S^2\left(1+\tau^2\frac{q^2}{m^2}B^2\right)}-\frac{\tau q B^2}{m 
S^2\left(1+\tau^2\frac{q^2}{m^2}B^2\right)}
\end{equation}
therefore
\begin{equation}
\label{eq:2:alpha4}
\alpha_4=-\frac{\tau^3 q^3 B^4}{m^3 S^2\left(1+\tau^2\frac{q^2}{m^2}B^2\right)^2}-\frac{\tau q B^2}{m S^2\left(1+\tau^2\frac{q^2}{m^2}B^2\right)^2}\ .
\end{equation}
Substituting (\ref{eq:2:alpha4}) into (\ref{eq:2:alpha5pre}) we obtain:
\begin{equation}
\alpha_5=\frac{\tau^2 q^2 B^4}{m^2 S^2\left(1+\tau^2\frac{q^2}{m^2}B^2\right)}-\frac{\tau^4 q^4 B^6}{m^4 S^2\left(1+\tau^2\frac{q^2}{m^2}B^2\right)^2}-\frac{\tau^2 q^2 B^4}{m^2 S^2\left(1+\tau^2\frac{q^2}{m^2}B^2\right)^2}\ .  
\end{equation}
Considering equation (\ref{eq:2:eq7}) and the results we already have, we can find $\alpha_6$:
\begin{equation}
\alpha_6=-\frac{\tau^2 q^2 B^2}{m^2 S^2\left(1+\tau^2\frac{q^2}{m^2}B^2\right)}E\cdot B+\frac{\tau^4 q^4 B^4}{m^4 S^2\left(1+\tau^2\frac{q^2}{m^2}B^2\right)^2}E\cdot B+\frac{\tau^2 q^2 B^2}{m^2 S^2\left(1+\tau^2\frac{q^2}{m^2}B^2\right)^2}E\cdot B
\end{equation}

From equations (\ref{eq:2:eq4}) and (\ref{eq:2:eq8}) an analogous relation between $\alpha_7$ and $\alpha_8$ can be produced:\\
\begin{equation}
\label{eq:2:subset2_1}
\alpha_7+\alpha_8\tau\frac{q}{m}=\alpha_1\tau\frac{q}{m}\frac{\left(E\cdot 
B\right)}{S^2}\ ,
\end{equation}
\begin{equation}
\label{eq:2:subset2_2}
\alpha_8-\alpha_7\tau\frac{q}{m}B^2=-\alpha_2\tau\frac{q}{m}\left(E\cdot B\right)\ .
\end{equation}
From (\ref{eq:2:subset2_2}) it can be seen:
\begin{equation}
\alpha_8=\alpha_7\tau\frac{q}{m}B^2-\alpha_2\tau\frac{q}{m}\left(E\cdot B\right)\ ,
\end{equation}
substituting this back into (\ref{eq:2:subset2_1}) we obtain:
\begin{equation}
\alpha_7\left(1+\tau^2\frac{q^2}{m^2}B^2\right)=\alpha_1\tau\frac{q}{m}\frac{\left(E\cdot 
B\right)}{S^2}+\alpha_2\tau^2\frac{q^2}{m^2}\left(E\cdot B\right)\ .
\end{equation}
As we already know $\alpha_1$ and $\alpha_2$ from equations (\ref{eq:2:eq3proc}) and (\ref{eq:2:eq4proc}) respectively we obtain:
\begin{equation}
\alpha_7\left(1+\tau^2\frac{q^2}{m^2}B^2\right)=\frac{\tau q}{m S^2\left(1+\tau^2\frac{q^2}{m^2}B^2\right)}E\cdot B+\frac{\tau^3 q^3 B^2}{m^3 
S^2\left(1+\tau^2\frac{q^2}{m^2}B^2\right)}E\cdot B
\end{equation}
and
\begin{equation}
\alpha_7=\frac{\tau q}{m S^2\left(1+\tau^2\frac{q^2}{m^2}B^2\right)^2}E\cdot B+\frac{\tau^3 q^3 B^2}{m^3 S^2\left(1+\tau^2\frac{q^2}{m^2}B^2\right)^2}E\cdot B\ .
\end{equation}
Considering (\ref{eq:2:subset2_2}) the $\alpha_8$ coefficient can be written as:
\begin{equation}
\alpha_8=\alpha_7\tau\frac{q}{m}B^2-\alpha_2\tau\frac{q}{m}\left(E\cdot B\right)\ ,  
\end{equation}
\begin{align}
\alpha_8=\nonumber &\frac{\tau^2 q^2 B^2}{m^2 S^2\left(1+\tau^2\frac{q^2}{m^2}B^2\right)^2}E\cdot B+\frac{\tau^4 q^4 B^4}{m^4 S^2\left(1+\tau^2\frac{q^2}{m^2}B^2\right)^2}E\cdot B-\\
&\frac{\tau^2 q^2 B^2}{m^2 S^2\left(1+\tau^2\frac{q^2}{m^2}B^2\right)}E\cdot B\ .  
\end{align}
Rearranging equation (\ref{eq:2:eq9}) we obtain $\alpha_9$:
\begin{equation}
\alpha_9=\alpha_2\tau\frac{q}{m}E^2-\alpha_7\tau\frac{q}{m}\left(E\cdot B\right)\ .
\end{equation}
Substituting the appropriate solutions for $\alpha_2$ and $\alpha_7$:
\begin{align}
\alpha_9=\nonumber &\frac{\tau^2 q^2 E^2 B^2}{m^2 
S^2\left(1+\tau^2\frac{q^2}{m^2}B^2\right)}-\frac{\tau^2 q^2}{m^2 
S^2\left(1+\tau^2\frac{q^2}{m^2}B^2\right)^2}\left(E\cdot B\right)^2-\\
&\frac{\tau^4 q^4 B^2}{m^4 
S^2\left(1+\tau^2\frac{q^2}{m^2}B^2\right)^2}\left(E\cdot B\right)^2
\end{align}
Following the above calculation, all of the $\alpha_{\lambda}$ coefficients have been obtained and after small simplifications can be written out as follows:
\begin{align}
\alpha_1&=\frac{1}{1+\tau^2\frac{q^2}{m^2}B^2}\ , 
\end{align}
\begin{align}
\alpha_2& = -\alpha_4 =\tau\frac{q}{m S^2\left(1+\tau^2\frac{q^2}{m^2}B^2\right)} B^2\ , 
\end{align}
\begin{align}
\alpha_3& = -\alpha_7 =-\tau\frac{q}{mS^2\left(1+\tau^2\frac{q^2}{m^2}B^2\right)}E\cdot B\ , 
\end{align}
%\alpha_4&=-\tau\frac{q}{m S^2 (1+\tau^2\frac{q^2}{m^2}B^2)}B^2 \sam[]{ = 
%-\alpha_2}\\
\begin{align}
\alpha_5& = \alpha_6 = \alpha_8 = 0\ ,
\end{align}
%\alpha_6&=0\\
%\alpha_7&=\tau\frac{q}{m S^2(1+\tau^2\frac{q^2}{m^2}B^2)}E\cdot B \sam[]{ = 
%-\alpha_3}\\
%\alpha_8&=0\\
\begin{align}
\alpha_9&=\tau^2\frac{q^2}{m^2\left(1+\tau^2\frac{q^2}{m^2}B^2\right)}\ .
\end{align}
Recombining these coefficients with the basis elements and defining $D:=1+\tau^2\frac{q^2}{m^2}B^2$, the inverted tensor can be explicitly written as: 
%%%THE RIGHT ONE%%%
\begin{align}
\label{eq:2:MinvEBS}
\nonumber 
\left(M^{-1}\right)^c{}_a=&\frac{1}{D}\Delta^c{}_a + \tau^2\frac{q^2}{m^2 D }B^c B_a - \tau\frac{q}{m S^2 D} B^2 \left[S^c E_a - E^c S_a\right] + \\
 & \tau\frac{q}{mS^2 D }\left(E\cdot B\right) \left[S^c 
B_a - B^c S_a\right]\ .
\end{align}

To be able to use this inverse to assemble an equation of motion in terms of the electromagnetic field tensor we require the inverse to be written in the same terms. 

From (\ref{eq:2:sandwichedtensor}):
\begin{equation}
\label{eq:2:GAB}
G^a{}_b = - \frac{q}{m}\frac{B^2}{S^2} \left[S^a E_b - E^a S_b\right] + \frac{q}{m}\frac{\left(E\cdot B\right)}{S^2} \left[S^a B_b - B^a S_b\right]\ ,
\end{equation}
and the only term still required is $\tau^2\frac{q^2}{m^2 D }B^c B_a$. 

If we now consider the square of (\ref{eq:2:GAB}):
\begin{align}
\label{eq:2:GG}
\nonumber G^c{}_e G^e{}_a & = \left( - \frac{q}{m}\frac{B^2}{S^2} S^c E_e + \frac{q}{m}\frac{B^2}{S^2} E^c S_e + \frac{q}{m}\frac{\left(E\cdot B\right)}{S^2} S^c B_e - \frac{q}{m}\frac{\left(E\cdot B\right)}{S^2} B^c S_e\right)\\
\nonumber &\quad\enskip\left(- \frac{q}{m}\frac{B^2}{S^2} S^e E_a + \frac{q}{m}\frac{B^2}{S^2} E^e S_a + \frac{q}{m}\frac{\left(E\cdot B\right)}{S^2} S^e B_a - \frac{q}{m}\frac{\left(E\cdot B\right)}{S^2} B^e S_a\right) \\
\nonumber & = \frac{q^2}{m^2}\frac{B^2}{S^4}\left[\left(E\cdot B\right)^2 - B^2 E^2\right] S^c S_a - \frac{q^2}{m^2}\frac{B^4}{S^2} E^c E_a + \frac{q^2}{m^2}\frac{B^2\left(E\cdot B\right)}{S^2}  E^c B_a + \\
& \quad \enskip \frac{q^2}{m^2}\frac{B^2 \left(E\cdot B\right)}{S^2} E_a B^c
- \frac{q^2}{m^2}\frac{\left(E\cdot B\right)^2}{S^2} B_a B^c\ .
\end{align}
If we now expand the $S^c S_a$ term in (\ref{eq:2:GG}) using (\ref{eq:2:SScontr}) we acquire:
\begin{align}
\label{eq:2:GGstep2}
\nonumber G^c{}_e G^e{}_a & = \frac{q^2}{m^2}\frac{B^4}{S^2} E^c E_a - \frac{q^2}{m^2}\frac{B^2\left(E\cdot B\right)}{S^2} E^c B_a - \frac{q^2}{m^2}\frac{B^2\left(E\cdot B\right)}{S^2} B^c E_a + \frac{q^2}{m^2}\frac{E^2 B^2}{S^2} B^c B_a - \\
\nonumber & \quad \enskip \frac{q^2}{m^2}\frac{B^4}{S^2} E^c E_a + \frac{q^2}{m^2}\frac{B^2\left(E\cdot B\right)}{S^2}  E^c B_a + \frac{q^2}{m^2}\frac{B^2 \left(E\cdot B\right)}{S^2} B^c E_a - \frac{q^2}{m^2}\frac{\left(E\cdot B\right)^2}{S^2} B^c B_a - \\ 
\nonumber & \quad \enskip \frac{q^2}{m^2} B^2 \Delta^c{}_a \\
&= \frac{q^2}{m^2} B^c B_a - \frac{q^2}{m^2} B^2 \Delta^c{}_a\ .
\end{align}

Combining results yields the inverse of $M$ (\ref{eq:2:MinvEBS}) in the form:
\begin{equation}
\label{eq:2:inversef}
\left(M^{-1}\right)^c{}_a=\Delta^c{}_a-\frac{\tau}{D}G^c{}_a+\frac{\tau^2}{D}G^c{}_e G^e{}_a\ ,
\end{equation}
where $D=1+\frac{\tau^2}{2}G^{ab}G_{ab}$.

Equation (\ref{eq:2:inversef}) provides us with the form of the inverse we require to 
write the Ford-O'Connell equation of motion in terms of electromagnetic 
field tensor.

Combining the inverse (\ref{eq:2:inversef}) with the form of the Ford-O'Connell equation used in (\ref{eq:2:FOwithinverse}) we acquire an equation of motion as follows:
\begin{equation}
\label{eq:2:FOwithinverse1}
\xdd^a=-\qm\left(\Delta^a{}_b-\frac{\tau}{D}G^a{}_b+\frac{\tau^2}{D}G^a{}_n 
G^n{}_b\right) \left(F^b{}_d\xd^d+\tau \xd^c\partial_c 
F^b{}_d\xd^d\right)\ .
\end{equation}

\section{Condition of divergence between Ford-O'Connell and Landau-Lifshitz approaches}

It can be seen that, ignoring terms of $\mathcal{O}\left(\tau^2\right)$, the Ford-O'Connell and Landau-Lifshitz equations agree.

Considering that $D=1+\frac{\tau^2}{2}G^{ab}G_{ab}$ if we ignore terms of $\mathcal{O}\left(\tau^2\right)$ the inverse (\ref{eq:2:inversef}) becomes:
\begin{equation}
\label{eq:2:inverseftau}
\left(M^{-1}\right)^c{}_a = \Delta^c{}_a-\tau G^c{}_a\ ,
\end{equation}
which, after substitution in the equation of motion (\ref{eq:2:FOwithinverse1}) becomes:
\begin{equation}
\label{eq:2:FOwithinversenewer}
\xdd^a=-\qm\left(\Delta^a{}_b-\tau G^a{}_b\right) \left(F^b{}_d\xd^d+\tau \xd^c\partial_c 
F^b{}_d\xd^d\right)\ ,
\end{equation}
where $G^a{}_b = \qm\Delta^a{}_e F^e{}_f \Delta^f{}_b$.
After elimination of the remaining terms of $\mathcal{O}\left(\tau^2\right)$ we obtain the Landau-Lifshitz equation:
\begin{equation}
\xdd^a = -\qm F^a{}_d\xd^d - \tau\qm\xd^c\partial_c F^a{}_d\xd^d + \tau\frac{q^2}{m^2} \Delta^a{}_b F^b{}_f F^f{}_d\xd^d\ .
\end{equation}
 
For the Landau-Lifshitz equation to be a good approximation to the Ford-O'Connell equation it is necessary 
that
\begin{equation}
\mathcal{T} := \tau \sqrt{G^{ab}G_{ab}/2}\ll 1\ , \label{eq:2:approxcondition}
\end{equation}
though this involves only the magnetic field \textit{seen by the particle}, which 
does not contribute to the applied force. The condition 
(\ref{eq:2:approxcondition}) is necessary, but not sufficient. However, we focus on this 
scalar condition, as it is more readily applicable than the somewhat vague 
requirement that $M^a{}_b$  is ``close'' to the unit matrix.

\section{Ford-O'Connell vs. Landau-Lifshitz approaches for the case of linear motion}

We now consider linear motion in a field:
\begin{equation}
\label{eq:2:linearmotion}
F_{ab}=E\varepsilon_{ab}\ ,
\end{equation}
where $\varepsilon_{ab}$ is the antisymmetric tensor:
\begin{equation}
\label{eq:2:varepsilonlinear}
\varepsilon_{01}=-\varepsilon_{10}=1\ .
\end{equation}
Consider the ``sandwiched'' electromagnetic tensor $G^{ab}$ constructed under the above conditions: 
\begin{align}
\label{eq:2:gablinear}
G^{ab}=\nonumber \Delta^{ac}F_{cd}\Delta^{db} & = E\varepsilon_{cd}\delta^{ac}\delta^{db}+E\varepsilon_{cd}\delta^{ac}\xd^d\xd^b+E\varepsilon_{cd}\delta^{db}\xd^a\xd^c+\underbrace{E\varepsilon_{cd}\xd^a\xd^c\xd^d\xd^b}_{=0}\\
\nonumber & = E\varepsilon^{ab}+E\varepsilon^a{}_d\xd^d\xd^b+E\varepsilon_c{}^b\xd^a\xd^c\\
& = E\varepsilon^{ab}\left(1+\xd_1\xd^1+\xd_0\xd^0\right)\ .
\end{align}

During the linear motion in the $x^1$ direction we can assume $x^2=x^3=0$, therefore the normalisation condition becomes:
\begin{equation}
\label{eq:2:normalisationlinearmotion}
\xd_1\xd^1+\xd_0\xd^0=-1\ .
\end{equation}
If we now combine (\ref{eq:2:normalisationlinearmotion}) with (\ref{eq:2:gablinear}) it can be seen that for the case of linear motion:
\begin{equation}
G^{ab}=0\ .
\end{equation}
Therefore for this case Ford-O'Connell and Landau-Lifshitz equations are identical and it is not possible to enter the ``Ford-O'Connell regime''.

\chapter{Particle motion in a plane wave}\label{sec:benchmark}

\begin{figure}[H]
\vspace{-17em}
\centering
\includegraphics[width=\textwidth]{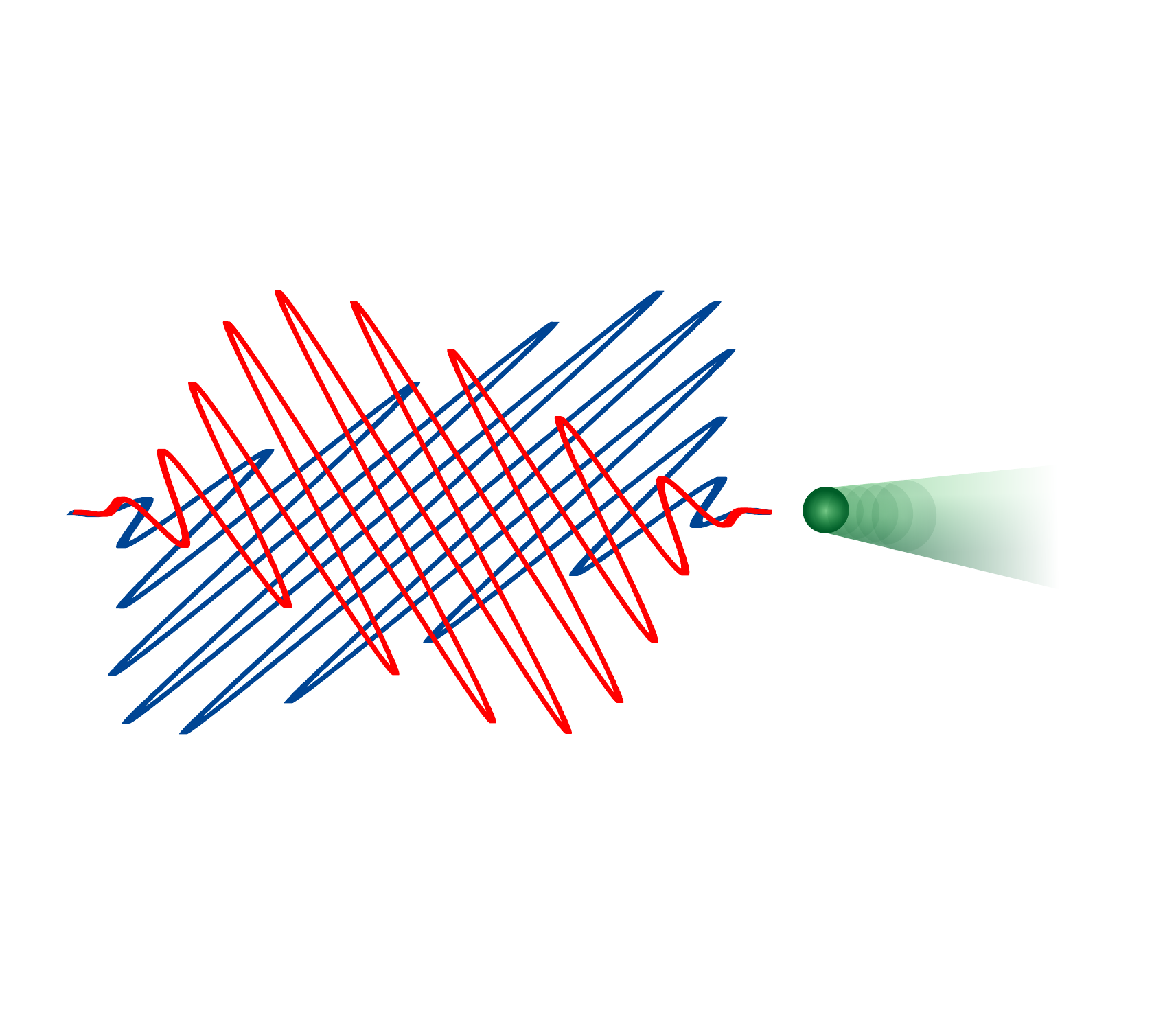}
\end{figure} 

\newpage 

The invention of the first laser in 1960 led to the possibility to concentrate the enormous power of intense light sources into a relatively small volume. This gave rise to the development of numerous research areas, one of which is laser-particle interactions.

The interaction between relativistic particles with intense counterpropagating laser pulses is relevant to high power laser facilities currently under construction, such as the Extreme Light Infrustructure (ELI) \cite{ELI} and Exawatt Center for Extreme Light Studies (XCELS) \cite{XCELS} facilities. The intensities to be achieved at these facilities require theoretical understanding of the influence of radiation reaction on the behaviour of particles to be considered. Many experimental research programmes will benefit from the theoretical description of this interaction by allowing us to better understand the limitations of theories describing the high intensity regime.

%A particle in a laser field emits electromagnetic radiation and the emission modifies both the particle trajectory and the emission spectrum. In the Landau-Lifshitz equation, which is commonly used, radiation reaction is included as an additional force acting on the electron (self-force). 

In this Chapter we investigate the effect of radiation reaction on particle motion in the radiation dominated regime using the Ford-O'Connell description, which has been described in the previous Chapter.

We explore the conditions under which the perturbative Landau-Lifshitz approximation breaks down, and whether this can be realised in the foreseeable future. To do that we compare the predictions of the Landau-Lifshitz and Ford-O'Connell methods for extreme cases.
 
\section{Introduction to the theoretical model}

In the absence of radiation reaction, the solution to the Lorentz force equation $\xdd^a=-\qm F^a{}_b \xd^b$ in a plane electromagnetic wave is well known: particularly lucid accounts may be found in \cite{Meyer, Heinzl}. Analytical solutions also exist for the Landau-Lifshitz radiation reaction correction \cite{Hadad, Harvey, DiPiazza}.

Radiation reaction will be most prominent for high energy electrons colliding with ultra-intense laser pulses which experience extremely large acceleration, and hence radiate most strongly according to (\ref{eq:1:Pa}). We therefore consider the Ford-O'Connell equations of motion in a laser pulse approximated by a plane wave with arbitrary shape and polarisation, moving in a given direction.

To keep the model as general as possible we do not specify any particular pulse parameters until we solve the equations of motion numerically. To derive these we introduce $n_a$ and $m_a$, which are null basis vectors, and the polarisation vectors $\epsilon$, $\lambda$ such that they satisfy the following conditions:
\begin{equation}
\label{eq:3:coordprop1}
 n\cdot n = m\cdot m = 0\ ,
\end{equation}
\begin{equation}
\label{eq:3:coordprop2}
n\cdot m = -2\ ,
\end{equation}
\begin{equation}
\label{eq:3:coordprop3}
\epsilon\cdot\epsilon = \lambda\cdot\lambda = 1\ ,
\end{equation}
\begin{equation}
\label{eq:3:coordprop4}
\epsilon\cdot\lambda = \epsilon\cdot n = \lambda\cdot n = \epsilon\cdot m = \lambda\cdot m = 0\ .
\end{equation}

%Assuming the electron's motion is in the spatial plane spanned by $\epsilon$ and $n$, 
We can define the coordinates as follows:
\begin{equation}
\label{eq:3:coordinates}
\phi = n\cdot x\ , \qquad \xi = \epsilon\cdot x\ , \qquad \sigma = \lambda\cdot x\ , \qquad \psi = m\cdot x\ .
\end{equation}
Note that $\xi$ and $\sigma$ are the transverse space-like coordinates whereas $\psi$ and $\phi$ (so called \textit{lightfront time}) are light-like coordinates.
 
Using the above specified set of coordinates, $x^a$ can be expressed as:
\begin{equation}
\label{eq:3:xacoord}
x^a = - \frac{1}{2}\phi m^a - \frac{1}{2}\psi n^a + \xi\epsilon^a + \sigma\lambda^a\ .
\end{equation}

The normalisation condition $\dot x^a \dot x_a = -1$ can be written in terms of these coordinates as follows:
\begin{equation}
\label{eq:3:normcond}
 \xd^a\xd_a = \left(- \frac{1}{2}\dot{\phi} m^a - \frac{1}{2}\dot{\psi} n^a + \dot{\xi}\epsilon^a + \dot{\sigma}\lambda^a\right)\left(- \frac{1}{2}\dot{\phi} m_a - \frac{1}{2}\dot{\psi} n_a + \dot{\xi}\epsilon_a + \dot{\sigma}\lambda_a\right) = -1\ .
\end{equation}

Combining (\ref{eq:3:normcond}) with the properties of the basis (\ref{eq:3:coordprop1}, \ref{eq:3:coordprop2}, \ref{eq:3:coordprop3}, \ref{eq:3:coordprop4}), the normalisation condition simplifies to:
\begin{equation}
\label{eq:3:norm}
 \dot\phi \dot\psi - \dot\xi^2 - \dot\sigma^2 = 1\ .
\end{equation}

The electron's energy, normalised to $mc^2$, is then given by
\begin{equation}
\label{eq:3:energy}
\gamma= \frac{1}{2}\left(\dot{\phi}+\dot{\psi}\right) = \frac{1+\dot{\phi}^2+\dot{\xi}^2+\dot{\sigma}^2}{2\dot{\phi}}\ ,
\end{equation}
with (\ref{eq:3:norm}) allowing us to eliminate $\dot{\psi}$.

Consider the Ford-O'Connell equation (\ref{eq:2:FOwithinverse}) with the Lorentz force $f^a = -\frac{q}{m} F\indices{^a_b} \dot x^b$ as an external force:
\begin{equation}
\label{eq:3:FOinv}
 \ddot x^a = -\frac{q}{m} \left(M^{-1}\right)\indices{^a_b} \left( F\indices{^b_c} + 
 \tau \dot x^d \partial_d F\indices{^b_c} \right) \dot x^c \ ,
\end{equation}
where the inverse $\left(M^{-1}\right)\indices{^a_b}$ is defined as:
\begin{equation}
 \left(M^{-1}\right)\indices{^a_b} = \frac{1}{\det M} \left(\left(\det M\right) 
 \Delta\indices{^a_b} - \tau G\indices{^a_b} + \tau^2 G\indices{^a_c} 
 G\indices{^c_b} \right) \ ,
\end{equation}
with $G^{ab}$ the ``sandwiched'' electromagnetic tensor $G^{ab} = \frac{q}{m}\Delta^{ac} F_{cd} \Delta^{db}$ and the determinant $\det M = 1 + (\tau^2/2) G\indices{^{ab}} 
G\indices{_{ab}}$.

We focus on deriving a general form for the equations of motion using the Ford-O'Connell radiation reaction force for the case of a particle interacting with an arbitrarily polarised laser pulse. Although realistic laser pulses have important transverse structure, for electrons co- or counter-propagating approximately through the centre of the pulse these are largely unimportant, unless the particle is deflected out of the pulse in the transverse direction. For simplicity then, we will consider a plane wave of the form:

\begin{equation}
\label{eq:3:pulseshape1pulse}
 \frac{q}{m} F\indices{_{ab}} = \mathcal{E}_1\left(\phi\right) \left( \epsilon_a n_b - 
 \epsilon_b n_a \right) + \mathcal{E}_2\left(\phi\right) \left( \lambda_a n_b - \lambda_b 
 n_a \right)\ ,
\end{equation}
where $\mathcal{E}_1$ and $\mathcal{E}_2$ correspond to electric fields of an arbitrary form in the $\epsilon$ and $\lambda$ directions respectively. Note the dependence on $\phi$ only to satisfy Maxwell's equations (see Appendix A).

To obtain the relevant equation of motion we need to substitute (\ref{eq:3:pulseshape1pulse}) into (\ref{eq:3:FOinv}).

Considering the form of the electromagnetic field tensor in equation (\ref{eq:3:pulseshape1pulse}), the sandwiched tensor $G\indices{_{ab}}$ can be written out:
\begin{align}
 G\indices{_{ab}} = &\mathcal{E}_1 \left[ \left(\epsilon_a + \dot\xi \dot x_a\right)\left(n_b 
 + \dot\phi \dot x_b\right) - \left(n_a + \dot\phi \dot x_a\right)\left(\epsilon_b + \dot\xi \dot 
 x_b\right) \right]+ \nonumber\\
 &\mathcal{E}_2 \left[ \left(\lambda_a + \dot\sigma \dot x_a\right)\left(n_b + 
 \dot\phi \dot x_b\right) - \left(n_a + \dot\phi \dot x_a\right)\left(\lambda_b + \dot\sigma \dot 
 x_b\right) \right]\ .
\end{align}

The $G\indices{^a_c} G\indices{^c_b}$ and $\dot{x}^d \partial_d F\indices{^b_c} \dot x^c$ are as follows:
\begin{align}
\label{eq:3:GGcontraction}
 \nonumber G\indices{^{ac}} G\indices{_{cb}} = &\dot\phi \left[ \mathcal{E}_1 \left( 
 \epsilon^a + \dot\xi\dot x^a \right) + \mathcal{E}_2 \left( \lambda^a + 
 \dot\sigma\dot x^a \right) \right] \Big[ \left(\mathcal{E}_1\dot\xi + 
 \mathcal{E}_2\dot\sigma\right) \left(n_b + \dot\phi\dot x_b\right) - \mathcal{E}_1 \dot\phi 
 \left(\epsilon_b + \dot\xi\dot x_b\right) - \\
 \nonumber &\mathcal{E}_2 \dot\phi \left(\lambda_b + \dot\sigma\dot x_b\right) \Big] + \left(n^a + \dot\phi\dot x^a\right) \Big[  \dot\phi \left(\mathcal{E}_1\dot\xi + 
 \mathcal{E}_2\dot\sigma\right) \left[ \mathcal{E}_1 \left(\epsilon_b + \dot\xi\dot x_b\right) 
 + \mathcal{E}_2 \left(\lambda_b + \dot\sigma\dot x_b\right) \right] - \\
 &\left[ \mathcal{E}_1^2\left(1+\dot\xi^2\right) + 
 \mathcal{E}_2^2\left(1+\dot\sigma^2\right) + 2\mathcal{E}_1 \mathcal{E}_2 
 \dot\xi\dot\sigma \right] \left(n_b + \dot\phi \dot x_b\right) \Big]\ ,
\end{align}
and
\begin{equation}
 \dot x^d \partial_d\mathcal{E}_i\left(\phi\right) = \dot x^d \mathcal{E}_i'\left(\phi\right) 
 \partial_d \phi = \dot x^d \mathcal{E}_i'\left(\phi\right) \partial_d \left(n_a x^a\right) = \dot 
 x^d n_d \mathcal{E}_i'\left(\phi\right) = \mathcal{E}_i' \dot\phi\ ,
\end{equation}
where prime represents differentiation with respect to $\phi$ and subscript $i \in {1,2}$.

Taking $a=b$ and summing in (\ref{eq:3:GGcontraction}) leads to the determinant of the form:
\begin{equation}
 \det M = 1 + \frac{\tau^2}{2} G\indices{^{ab}} 
G\indices{_{ab}} = 1 + \tau^2\left(\mathcal{E}_1^2 + \mathcal{E}_2^2\right) \dot\phi^2\ .
\end{equation}

%\begin{equation}
% \ddot x^a = -(M^{-1})\indices{^a_b} \Big[ (\mathcal{E}_1 + \tau \dot x^c 
% \partial_c\mathcal{E}_1) (\dot\phi \epsilon^b - \dot\xi n^b) + 
% (\mathcal{E}_2 + \tau \dot x^c \partial_c\mathcal{E}_2) (\dot\phi \lambda^b 
% - \dot\sigma n^b) \Big] \ .
%\end{equation}

Having all the necessary components of the Ford-O'Connell equation we can assemble the general equation of motion for a particle interacting with an arbitrarily polarised plane wave laser pulse, which reads:

\begin{align}
\label{eq:3:FO1pulse}
 \ddot x^a = & \left(\frac{\mathcal{E}_1 + \tau \mathcal{E}_1' \dot\phi}{1 + 
 \tau^2(\mathcal{E}_1^2 + \mathcal{E}_2^2) \dot\phi^2} \right) \Bigg[ \Big\{ 
   \dot\xi - \tau \mathcal{E}_1 \dot\phi - \tau^2 \mathcal{E}_2 
   \dot\phi^2(\mathcal{E}_1\dot\sigma - \mathcal{E}_2\dot\xi) \Big\} (n^a + 
   \dot\phi \dot x^a) \nonumber \\
   &\hspace{1.5in}- \dot\phi(1 + \tau^2 \mathcal{E}_2^2\dot\phi^2) (\epsilon^a 
   + \dot\xi \dot x^a) + \tau^2 \mathcal{E}_1 \mathcal{E}_2 \dot\phi^3 
   (\lambda^a + \dot\sigma\dot x^a) \Bigg] + \nonumber \\
   &\left(\frac{\mathcal{E}_2 + \tau \mathcal{E}_2' \dot\phi}{1 + 
   \tau^2(\mathcal{E}_1^2 + \mathcal{E}_2^2) \dot\phi^2} \right) \Bigg[ 
   \Big\{ \dot\sigma - \tau \mathcal{E}_2 \dot\phi - \tau^2 \mathcal{E}_1 
   \dot\phi^2(\mathcal{E}_2\dot\xi - \mathcal{E}_1\dot\sigma) \Big\} (n^a + 
   \dot\phi \dot x^a) \nonumber \\
   &\hspace{1.5in} - \dot\phi(1 + \tau^2 \mathcal{E}_1^2\dot\phi^2) 
   (\lambda^a + \dot\sigma \dot x^a) + \tau^2 \mathcal{E}_1 \mathcal{E}_2 
   \dot\phi^3 (\epsilon^a + \dot\xi\dot x^a) \Bigg] \ .
\end{align}

%Note that $\Delta\indices{^a_b} n^b = n^a + \dot\phi \dot x^a$, \\
%$\Delta\indices{^a_b} \epsilon^b = \epsilon^a + \dot\xi \dot x^a$, \\
%$\Delta\indices{^a_b} \lambda^b = \lambda^a + \dot\sigma \dot x^a$ and \\ 
%$\Delta\indices{^a_b} m^b = m^a + \dot\psi \dot x^a$.

Contracting equation (\ref{eq:3:FO1pulse}) with the basis vectors and using (\ref{eq:3:coordinates}) we obtain 4 separate equations of motion, one for each component, which are as follows:
\begin{equation}
 \ddot\phi = -\frac{\tau\dot\phi^3}{1 + \tau^2(\mathcal{E}_1^2 + 
 \mathcal{E}_2^2) \dot\phi^2} \Bigg[ \mathcal{E}_1 \Big( \mathcal{E}_1 + 
 \tau \mathcal{E}_1' \dot\phi \Big) + \mathcal{E}_2 \Big( \mathcal{E}_2 + 
 \tau \mathcal{E}_2' \dot\phi \Big) \Bigg]\ ,
\end{equation} 
\begin{align}
 \nonumber \ddot\xi = &-\frac{\dot\phi}{1 + \tau^2(\mathcal{E}_1^2 + \mathcal{E}_2^2) 
 \dot\phi^2} \Bigg[ \Big( \mathcal{E}_1 + \tau \mathcal{E}_1' \dot\phi \Big) 
 \Big( 1 + \tau \mathcal{E}_1 \dot\phi \dot\xi + \tau^2 \mathcal{E}_2^2 
 \dot\phi^2 \Big) \\
 &+\tau \mathcal{E}_2 \dot\phi \Big(\mathcal{E}_2 + \tau 
 \mathcal{E}_2' \dot\phi\Big) \Big( \dot\xi - \tau \mathcal{E}_1\dot\phi 
 \Big) \Bigg]\ ,
\end{align} 
\begin{align}
 \nonumber \ddot\sigma = &-\frac{\dot\phi}{1 + \tau^2(\mathcal{E}_1^2 + 
 \mathcal{E}_2^2) \dot\phi^2} \Bigg[ \Big( \mathcal{E}_2 + \tau 
 \mathcal{E}_2' \dot\phi \Big) \Big( 1 + \tau \mathcal{E}_2 \dot\phi 
 \dot\sigma + \tau^2 \mathcal{E}_1^2 \dot\phi^2 \Big) \\
 &+\tau \mathcal{E}_1 
 \dot\phi \Big(\mathcal{E}_1 + \tau \mathcal{E}_1' \dot\phi\Big) \Big( 
 \dot\sigma - \tau \mathcal{E}_2\dot\phi \Big) \Bigg]\ ,
\end{align} 
\begin{align}
 \ddot\psi & = -\frac{1}{1 + \tau^2(\mathcal{E}_1^2 + \mathcal{E}_2^2) 
 \dot\phi^2} \Bigg[ \Big( \mathcal{E}_1 + \tau \mathcal{E}_1' \dot\phi \Big) 
 \Big( 2\dot\xi + \tau \mathcal{E}_1\dot\phi (\dot\psi \dot\phi - 2) + 
 2\tau^2 \mathcal{E}_2 \dot\phi^2 (\mathcal{E}_2 \dot\xi - \mathcal{E}_1 
 \dot\sigma) \Big) \nonumber \\
 &+ \Big( \mathcal{E}_2 + \tau \mathcal{E}_2' \dot\phi \Big) 
 \Big( 2\dot\sigma + \tau \mathcal{E}_2\dot\phi (\dot\psi \dot\phi - 2) + 
 2\tau^2 \mathcal{E}_1 \dot\phi^2 (\mathcal{E}_1 \dot\sigma - \mathcal{E}_2 
 \dot\xi) \Big) \Bigg]\ .
\end{align} 

By dropping all the terms of $\mathcal{O}(\tau^2)$ we reduce to the Landau-Lifshitz set of equations, which reads:
\begin{equation}
 \ddot\phi = - \tau \dot\phi^3 \left( \mathcal{E}^2_1 + \mathcal{E}^2_2 \right)\ ,
\end{equation} 
\begin{align}
 \nonumber \ddot\xi = & - \dot\phi \left( \mathcal{E}_1 + \tau \mathcal{E}_1' \dot\phi \right) - \tau \dot\phi^2 \dot\xi \left( \mathcal{E}^2_1 + \mathcal{E}^2_2 \right)\ ,
\end{align} 
\begin{align}
 \nonumber \ddot\sigma = & - \dot\phi \left(\mathcal{E}_2 + \tau \mathcal{E}_2' \dot\phi \right) - \tau \dot\phi^2 \dot\sigma \left( \mathcal{E}^2_1 + \mathcal{E}^2_2 \right)\ ,
\end{align} 
\begin{equation}
 \ddot\psi = - 2 \left( \mathcal{E}_1 \dot\xi + \mathcal{E}_2 \dot\sigma \right) - 2 \tau \dot\phi \left( \mathcal{E}_1' \dot\xi + \mathcal{E}_2' \dot\sigma \right) - \tau \dot\phi \left(\dot \psi \dot\phi - 2\right) \left( \mathcal{E}^2_1 + \mathcal{E}^2_2 \right)\ .
\end{equation} 

If we now drop all the terms of $\mathcal{O}(\tau)$ from the Landau-Lifshitz equations of motion we end up with equations corresponding to the Lorentz force only with no radiation reaction taken into account:

\begin{equation}
 \ddot\phi = 0\ ,
\end{equation} 
\begin{equation}
 \ddot\xi = - \mathcal{E}_1 \dot\phi\ ,
\end{equation} 
\begin{equation}
\ddot\sigma = - \mathcal{E}_2 \dot\phi\ ,
\end{equation} 
\begin{equation}
 \ddot\psi = - 2 \left( \mathcal{E}_1 \dot\xi + \mathcal{E}_2 \dot\sigma \right)\ .
\end{equation} 

\section{Particle equations of motion in a linearly polarised plane wave}

We consider the simple case where we model the laser pulse by a linearly polarised (in $\epsilon$ direction) plane wave given by
\begin{equation}
\label{eq:3:pulseshape1pulseharmonic}
 \frac{q}{m} F\indices{_{ab}} = \mathcal{E}_1\left(\phi\right) \left( \epsilon_a n_b - 
 \epsilon_b n_a \right)\ .
\end{equation}
Applying the linearly polarised plane wave (\ref{eq:3:pulseshape1pulseharmonic}) to the set of Ford-O'Connell equations of motion, they simplify to:
\begin{equation}
\label{eq:3:FOhpw1}
 \ddot\phi = -\frac{\tau\dot\phi^3\mathcal{E}_1}{1 + \tau^2\mathcal{E}_1^2\dot\phi^2} \Bigg[ \mathcal{E}_1 + 
 \tau \mathcal{E}_1' \dot\phi \Bigg]\ ,
\end{equation} 
\begin{equation}
\label{eq:3:FOhpw2}
 \ddot\xi = -\frac{\dot\phi}{1 + \tau^2\mathcal{E}_1^2 
 \dot\phi^2} \Bigg[ \Big( \mathcal{E}_1 + \tau \mathcal{E}_1' \dot\phi \Big) 
 \Big( 1 + \tau \mathcal{E}_1 \dot\phi \dot\xi \Big) \Bigg]\ ,
\end{equation} 
\begin{equation}
\label{eq:3:FOhpw3}
 \ddot\sigma = -\frac{\dot\phi}{1 + \tau^2\mathcal{E}_1^2\dot\phi^2} \Bigg[ \tau \mathcal{E}_1 
 \dot\phi \dot\sigma \Big(\mathcal{E}_1 + \tau \mathcal{E}_1' \dot\phi\Big) \Bigg]\ ,
\end{equation} 
\begin{equation}
\label{eq:3:FOhpw4}
 \ddot\psi = -\frac{1}{1 + \tau^2\mathcal{E}_1^2 \dot\phi^2} \Bigg[ \Big( \mathcal{E}_1 + \tau \mathcal{E}_1' \dot\phi \Big) 
 \Big( 2\dot\xi + \tau \mathcal{E}_1\dot\phi (\dot\psi \dot\phi - 2) \Big) \Bigg]\ .
\end{equation} 

Neglecting terms of $\mathcal{O}(\tau^2)$, (\ref{eq:3:FOhpw1}, \ref{eq:3:FOhpw2}, \ref{eq:3:FOhpw3}, \ref{eq:3:FOhpw4}) reduce to their counterparts in the Landau-Lifshitz equation, as expected. The set of equations of motion then becomes:
\begin{equation}
\label{eq:3:LLhpw1}
 \ddot\phi = -\tau\mathcal{E}^2_1\dot\phi^3\ ,
\end{equation} 
\begin{equation}
\label{eq:3:LLhpw2}
 \ddot\xi = -\dot\phi \left(\mathcal{E}_1 + \tau \mathcal{E}^2_1 \dot\phi \dot\xi + \tau \mathcal{E}_1' \dot\phi \right)\ ,
\end{equation} 
\begin{equation}
\label{eq:3:LLhpw3}
 \ddot\sigma = - \tau \mathcal{E}^2_1 \dot\phi^2 \dot\sigma\ ,
\end{equation} 
\begin{equation}
\label{eq:3:LLhpw4}
 \ddot\psi = - 2\mathcal{E}_1\dot\xi - \tau \mathcal{E}^2_1\dot\phi^2 \dot\psi + 2 \tau \mathcal{E}^2_1\dot\phi - 2\tau \mathcal{E}_1' \dot\phi\dot\xi\ .
\end{equation} 

From (\ref{eq:2:approxcondition}), it follows that the Landau-Lifshitz equation should be reliable only when:
\begin{equation}
 \tau \sqrt{\frac{1}{2}G^{ab}G_{ab}}\ll 1\ , 
\end{equation} 
or when
\begin{equation}
\label{eq:3:Tcond}
\mathcal{T}:= \tau\dot{\phi}|\mathcal{E}_1|\ll 1\ ,
\end{equation}
as is clearly borne out by Eq. (\ref{eq:3:FOhpw1}--\ref{eq:3:FOhpw4}).

To compare the predictions of the influence of radiation reaction on the particle motion with both Landau-Lifshitz and Ford-O'Connell corrections, we solve the respective equations of motion numerically for the given configuration. A harmonic plane wave of the form:
\begin{equation}
\mathcal{E}_1 = \omega a_0 \sin \left( \omega \phi \right)
\end{equation}
is considered.

A particle with initial energy $\gamma_\text{in}$ is placed within the harmonic plane wave at the peak field. Numerical evaluation of the respective equations of motion allow us to visualise the evolution of the particle's energy $\gamma$ (normalised to $mc^2$) \textit{vs.} lightfront time $\phi$, and the evolution of lightfront time $\phi$ as a function of proper time, $s$. The transverse components $\dot{\sigma}$ and $\dot{\xi}$ are initially set to be 0, and $\dot{\phi}(0)$ corresponds to an initial energy $\gamma_\text{in}$ via (\ref{eq:3:energy}). We use units such that $\omega=1$, which for 790 nm wavelength gives $\tau = 1.5 \cdot 10^{-8}$.

The function $\phi(s)$, where $s$ is the proper time, represents the longitudinal position of the particle within the pulse and is therefore a useful measure of the rate at which the electron passes through the laser field, and thus, together with $\gamma$, is a good indication of the significance of radiation reaction. Note that $\dot{\phi}=const$ without radiation reaction. 
 
\begin{figure}
\centering
\includegraphics[width=0.85\textwidth]{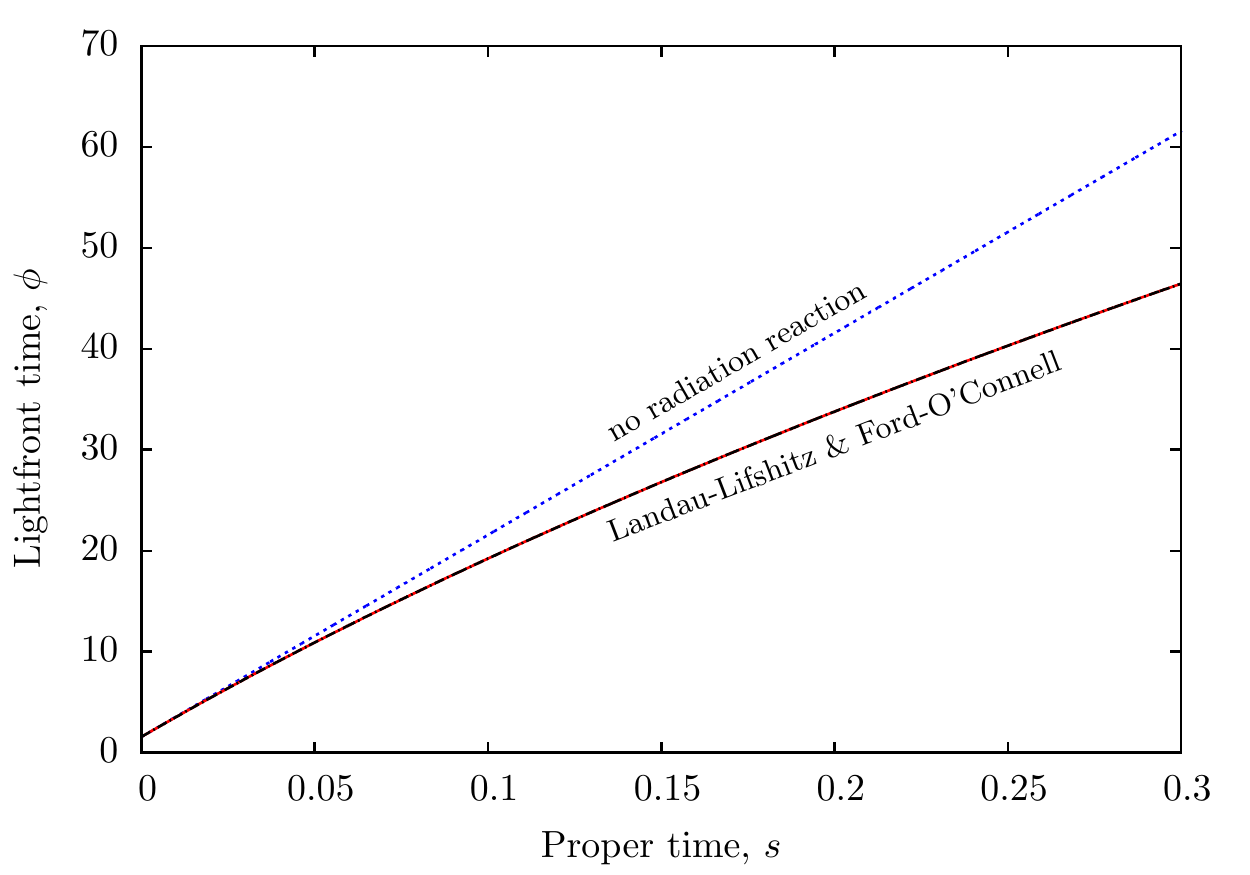}
\caption{\label{fig:3:phislowinfinite} 
Radiation reaction effects of a pulse with $a_0=100$ on an electron of initial energy $\gamma_\text{in}=100$: $\phi$ as a function of $s$.}
\end{figure}

As Fig.~\ref{fig:3:phislowinfinite}--\ref{fig:3:gammaphilowinfinite} show, for the highest currently attainable laser intensities ($a_0=100$, see Appendix B) and moderately high initial electron energies ($\gamma_\text{in}=100$), a particle starting at the peak of the laser field with initial $\phi=\pi/2\omega$ experiences significant radiation reaction, but Landau-Lifshitz and Ford-O'Connell are in good agreement, as expected.
\begin{figure}
\centering
\includegraphics[width=0.85\textwidth]{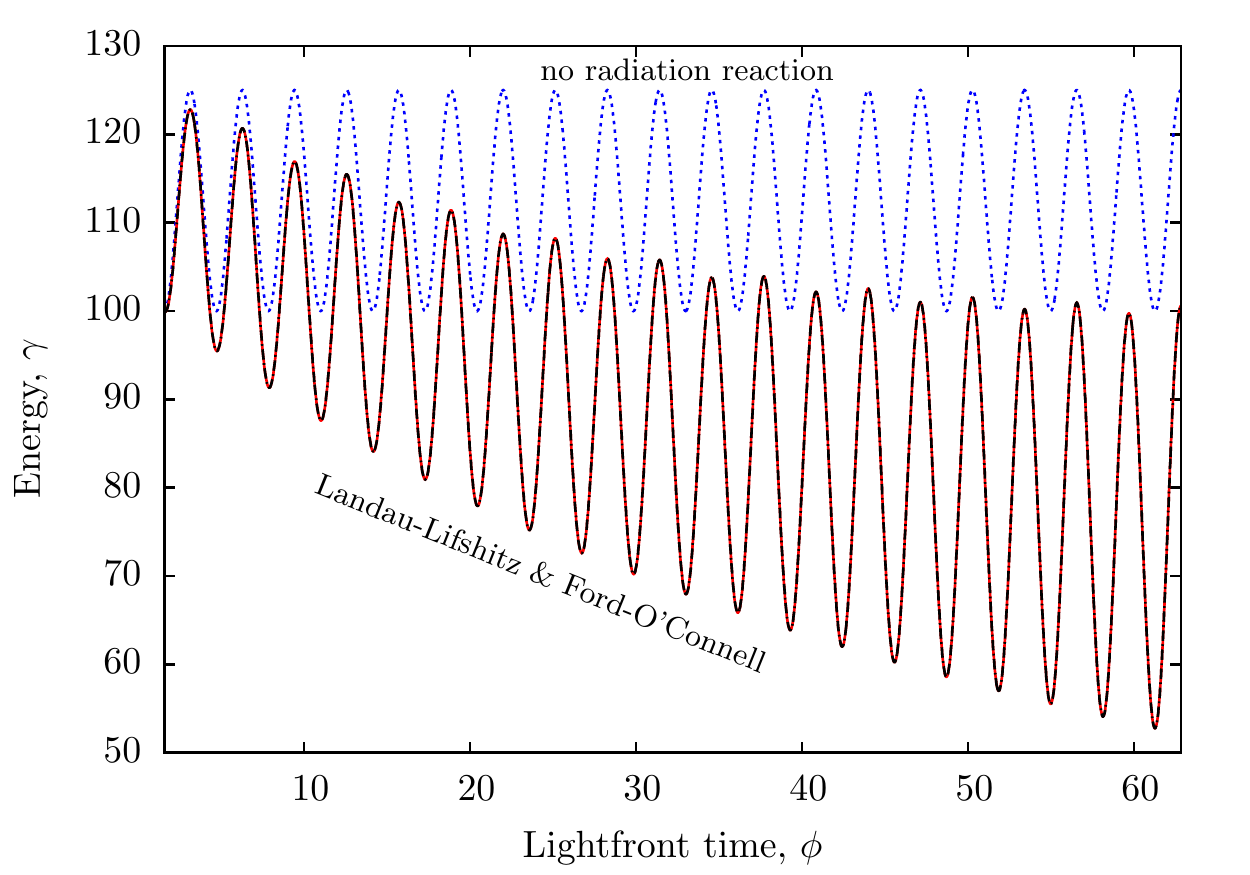}
\caption{\label{fig:3:gammaphilowinfinite} 
Radiation reaction effects of a pulse with $a_0=100$ on an electron of initial energy $\gamma_\text{in}=100$: $\gamma$ as a function of $\phi$. Dotted blue curves without radiation reaction; solid red curves with Landau-Lifshitz radiation reaction; double-dotted black curves with Ford-O'Connell radiation reaction.}
\end{figure} 

\begin{figure}
\centering
\includegraphics[width=0.85\textwidth]{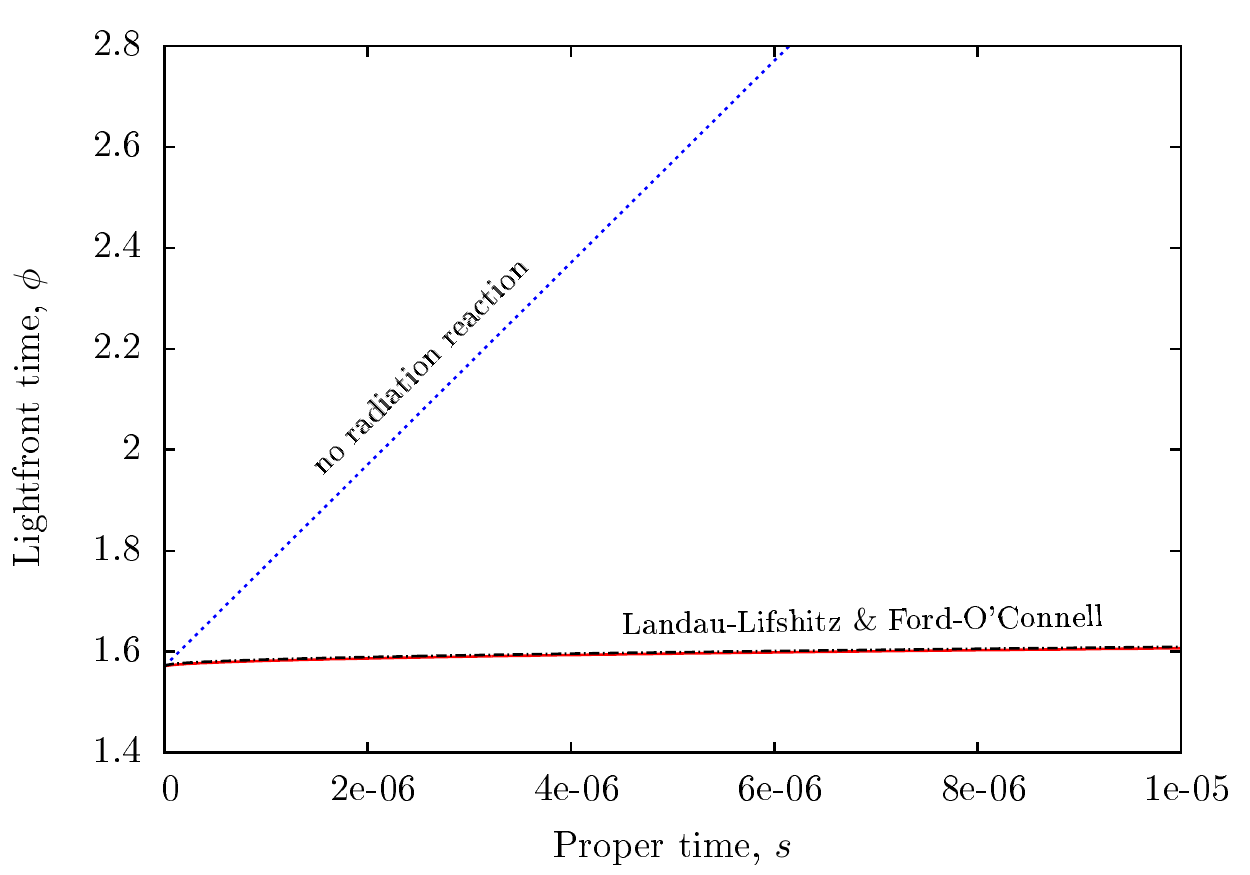}
\caption{\label{fig:3:phishighinfinite}
Radiation reaction effects of a pulse with $a_0=1000$ on an electron of initial energy $\gamma_\text{in}=10^5$: $\phi$ as a function of $s$.}
\end{figure}

\begin{figure}
\centering
\includegraphics[width=0.85\textwidth]{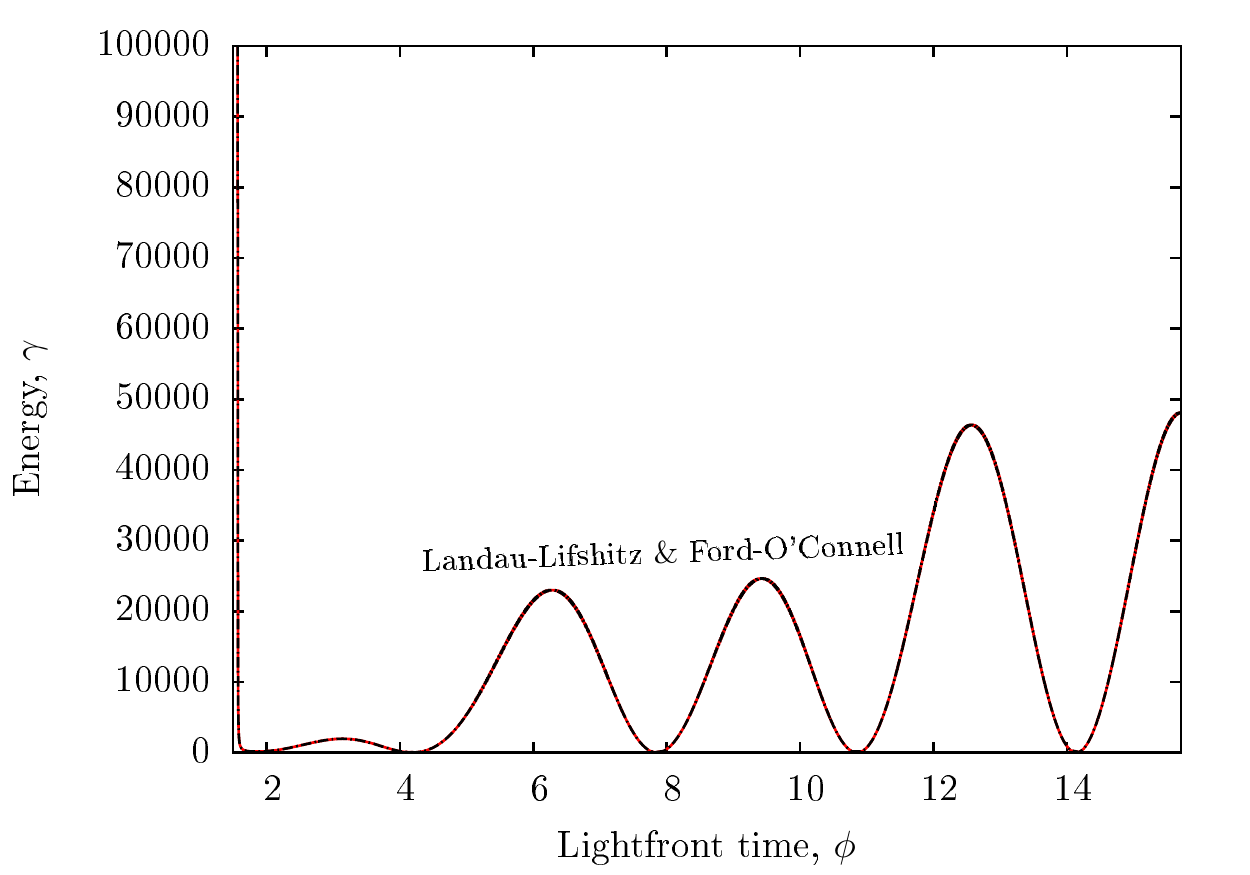}
\caption{\label{fig:3:gammaphihighinfinite}
Radiation reaction effects of a pulse with $a_0=1000$ on an electron of initial energy $\gamma_\text{in}=10^5$: $\gamma$ as a function of $\phi$. Solid red curves with Landau-Lifshitz radiation reaction; double-dotted black curves with Ford-O'Connell radiation reaction.}
\end{figure}

If we consider the most intense lasers under development, such as the ones to be used at the Extreme Light Infrustructure (ELI) facility, taking $a_0 = 1000$ (see Appendix B), and the highest energy electrons available, $\gamma_\text{in} = 10^5$ (electrons with this $\gamma$ were produced at the Large Electron-Positron Collider (LEP) at CERN), we appear to be in a regime where the condition (\ref{eq:3:Tcond}) is violated. We would therefore expect strong differences between Landau-Lifshitz and Ford-O'Connell. However, as shown in Fig.~\ref{fig:3:phishighinfinite}--\ref{fig:3:gammaphihighinfinite}, although the dynamics is dominated by radiation reaction, agreement between the two theories remains strong. How are we to explain this?

The condition (\ref{eq:3:Tcond}) refers to the instantaneous energy and field strength, whereas the previously quoted values of $a_0$ and $\gamma$ refer to the \textit{peak field} and the \textit{initial energy}. From Fig.~\ref{fig:3:gammaphihighinfinite}, it is clear that the electron almost instantaneously loses most of its energy to radiation. After this, it hardly radiates at all, and its evolution is well described by the Lorentz force alone, at a greatly reduced initial energy. 

\begin{figure}
\centering
\includegraphics[width=0.85\textwidth]{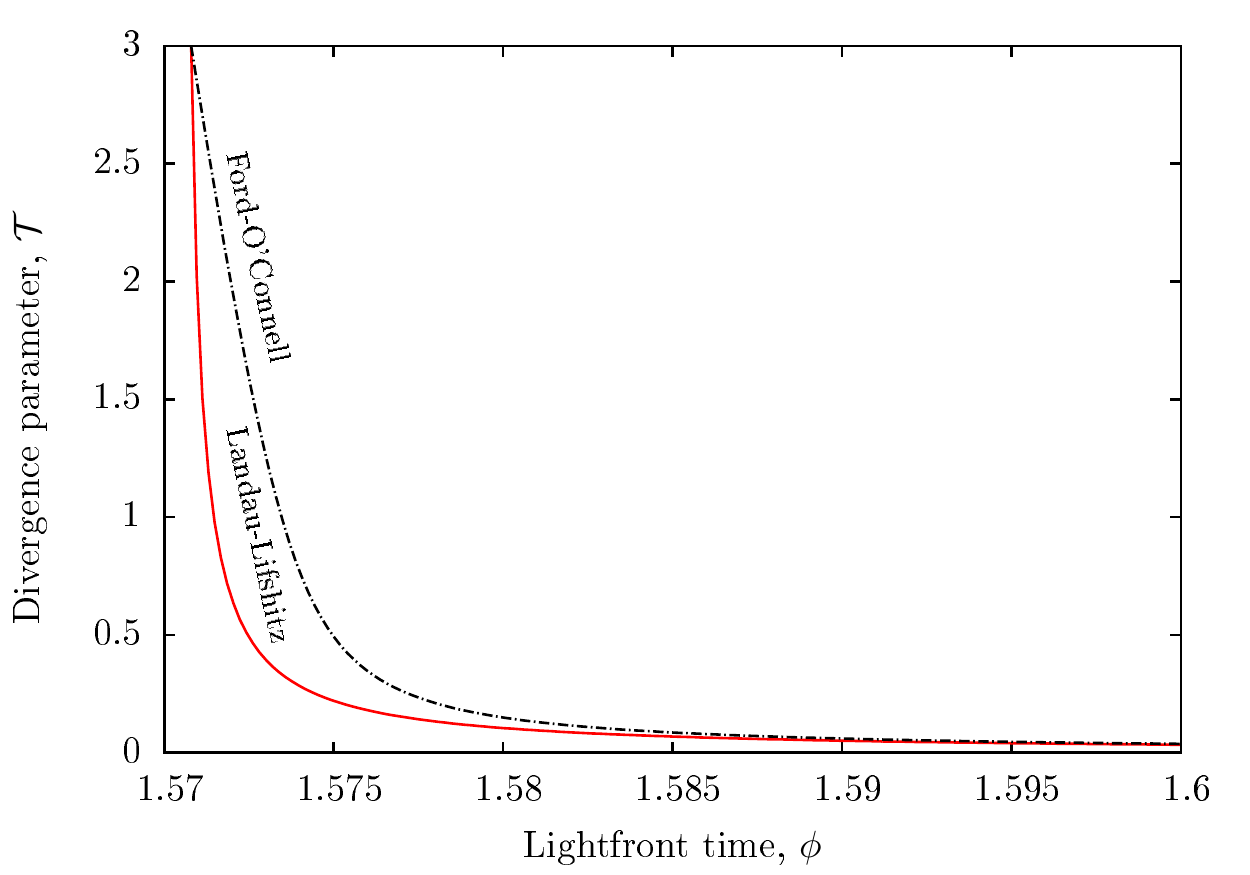}
\caption{\label{fig:3:Tpeakinfinite} 
$\T$ as a function of $\phi$, for $a_0=1000$ and $\gamma_\text{in}=10^5$. Solid red curves with Landau-Lifshitz radiation reaction; double-dotted black curves with Ford-O'Connell radiation reaction: particle beginning at field peak.}
\end{figure}

\begin{figure}
\centering
\includegraphics[width=0.85\textwidth]{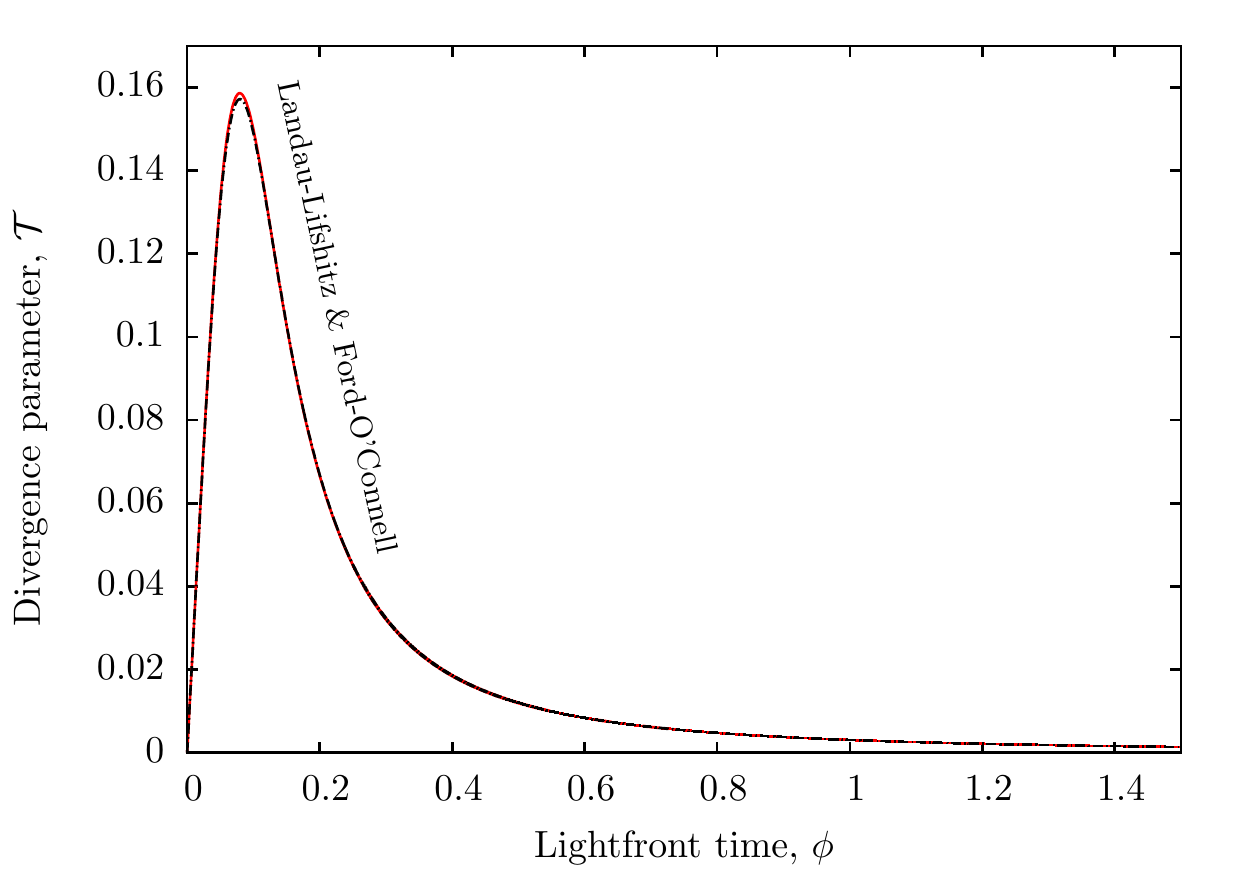}
\caption{\label{fig:3:Tnodeinfinite} 
$\T$ as a function of $\phi$, for $a_0=1000$ and $\gamma_\text{in}=10^5$. Solid red curves with Landau-Lifshitz radiation reaction; double-dotted black curves with Ford-O'Connell radiation reaction: particle beginning at field node.}
\end{figure}

As shown in Fig.~\ref{fig:3:Tpeakinfinite}, as the electron propagates through the laser field, a particle initially at the peak of the field starts with $\T>1$, and in the very early stages of its motion Ford-O'Connell predictions deviate from those of Landau-Lifshitz, indicating that breakdown of the Landau-Lifshitz approximation is possible if the peak field and peak energy coincide. However, after approximately $1/300$ of a cycle, $\T$ becomes sufficiently small that the two descriptions are indistinguishable. For a particle initially at a node of the wave, $\T$ begins at zero, and never approaches unity, as can be seen in Fig.~\ref{fig:3:Tnodeinfinite}.

The analysis presented above assumes that the laser pulse can be described by a harmonic plane wave. It has been shown that the radiation reaction effects within this approximation ensure that it cannot enter a regime where the Landau-Lifshitz description breaks down, even when {\it a priori} estimates would suggest otherwise, provided the particle does not start at the peak of the field. It is of interest to explore whether the results remain valid for a pulse with a more realistic structure, which we consider next.

\section{Particle motion in a finite laser pulse}

\begin{figure}
\centering
\includegraphics[width=0.85\textwidth]{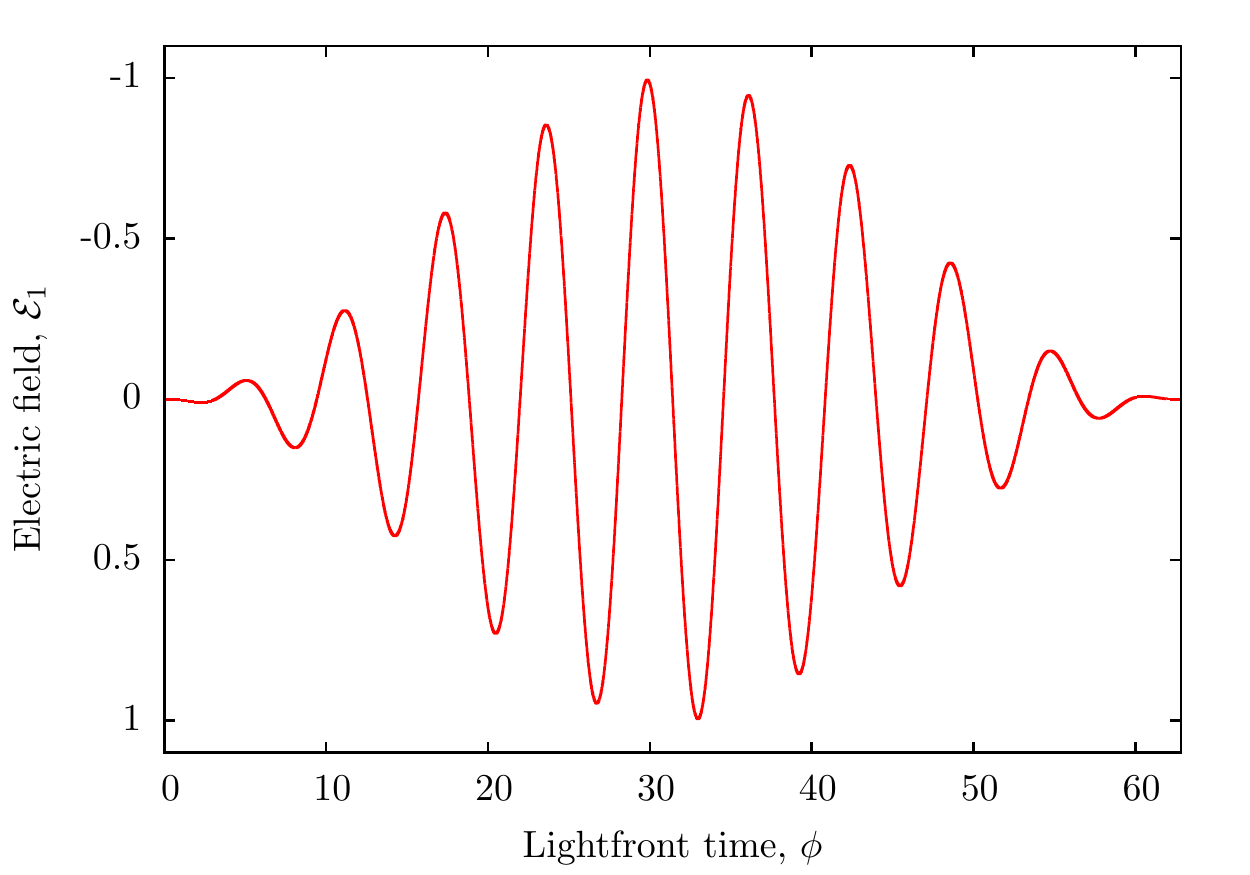}
\caption{\label{fig:pulse} 
Electric field $\E_1$ (Eq. \ref{eq:3:pulse_shape}) as a function of $\phi$, for $N=10$, $a_0=1$, in units such that $\omega=1$.}
\end{figure}

To compare the predictions of Ford-O'Connell and Landau-Lifshitz in a more realistic scenario, we need to specify a finite pulse shape for the profile of the electric field $\E_1$, though the specific choice does not significantly affect the results. It will be convenient to choose $\E_1$ to have compact support, so the electron can begin and end in vacuum. Furthermore, both $\E_1$ and its derivative should be continuous. We adopt the simple choice (related to profiles used in, for example, \cite{Harvey, Heinzl2, Mackenroth}),
\begin{equation}
\label{eq:3:pulse_shape}
\E_1= 
\begin{cases}
\omega a_0 \sin (\omega \phi) \sin^2(\omega \phi/ 2N) & \text{for } 0 < \phi < 2\pi N/\omega\ ,\\
0 & \text{otherwise}\ .
\end{cases}
\end{equation}
This represents an $N$-cycle pulse of central frequency $\omega$, modulated by a $\sin^2$-envelope. $a_0$ is the usual intensity parameter (sometimes called ``normalised vector potential''). Fig.~\ref{fig:pulse} shows~(\ref{eq:3:pulse_shape}) for $N=10$, in units such that $\omega=1$.

To describe the impact of radiation reaction we will again look at the evolution of $\phi(s)$, indicating how long the particle experiences the pulse with radiation reaction taken into account, along with the evolution of $\gamma$ in order to have a quantitive estimate of the energy lost to radiation. 

To do this we consider a laser pulse approximated by a plane wave with $N=10$ oscillations stretching from $\phi_\text{i} = 0$ to $\phi_\text{f} = 2\pi N$ in the lightfront time $\phi$. A particle with initial energy $\gamma_\text{in}$ is placed in front of the pulse at $\phi(0) = -\pi/2$. The transverse components $\dot{\sigma}$ and $\dot{\xi}$ are initially set to be 0, and $\dot{\phi}(0)$ corresponds to the initial energy $\gamma_\text{in}$ via (\ref{eq:3:energy}). We trace the particle motion while it collides with the laser pulse up until the point where it exits the pulse.

\begin{figure}
\centering
\includegraphics[width=0.85\textwidth]{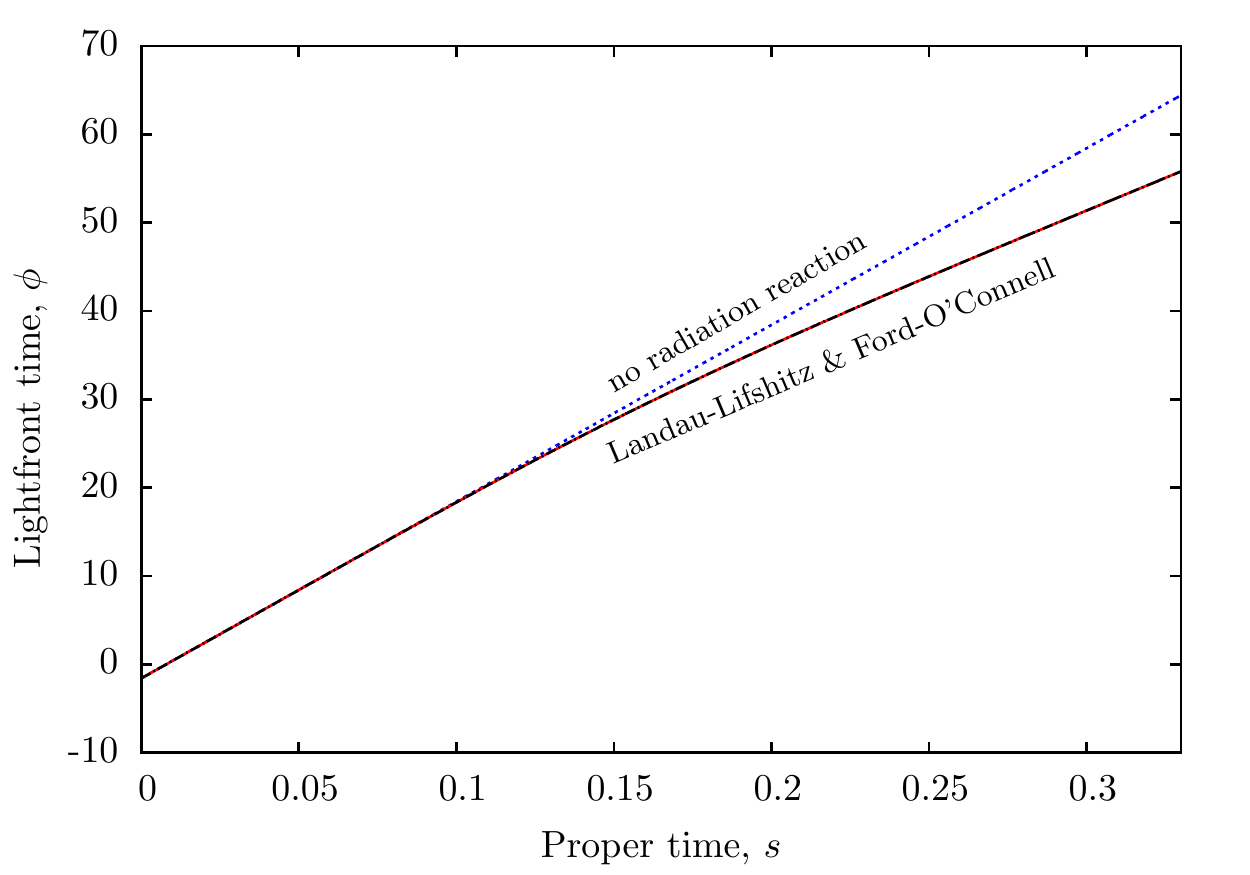}
\caption{\label{fig:3:phislowfinite} 
Radiation reaction effects of a pulse with $a_0=100$ on an electron of initial energy $\gamma_\text{in}=100$: $\phi$ as a function of $s$. Dotted blue curves without radiation reaction; solid red curves with Landau-Lifshitz radiation reaction; double-dotted black curves with Ford-O'Connell radiation reaction.}
\end{figure}

\begin{figure}
\centering
\includegraphics[width=0.85\textwidth]{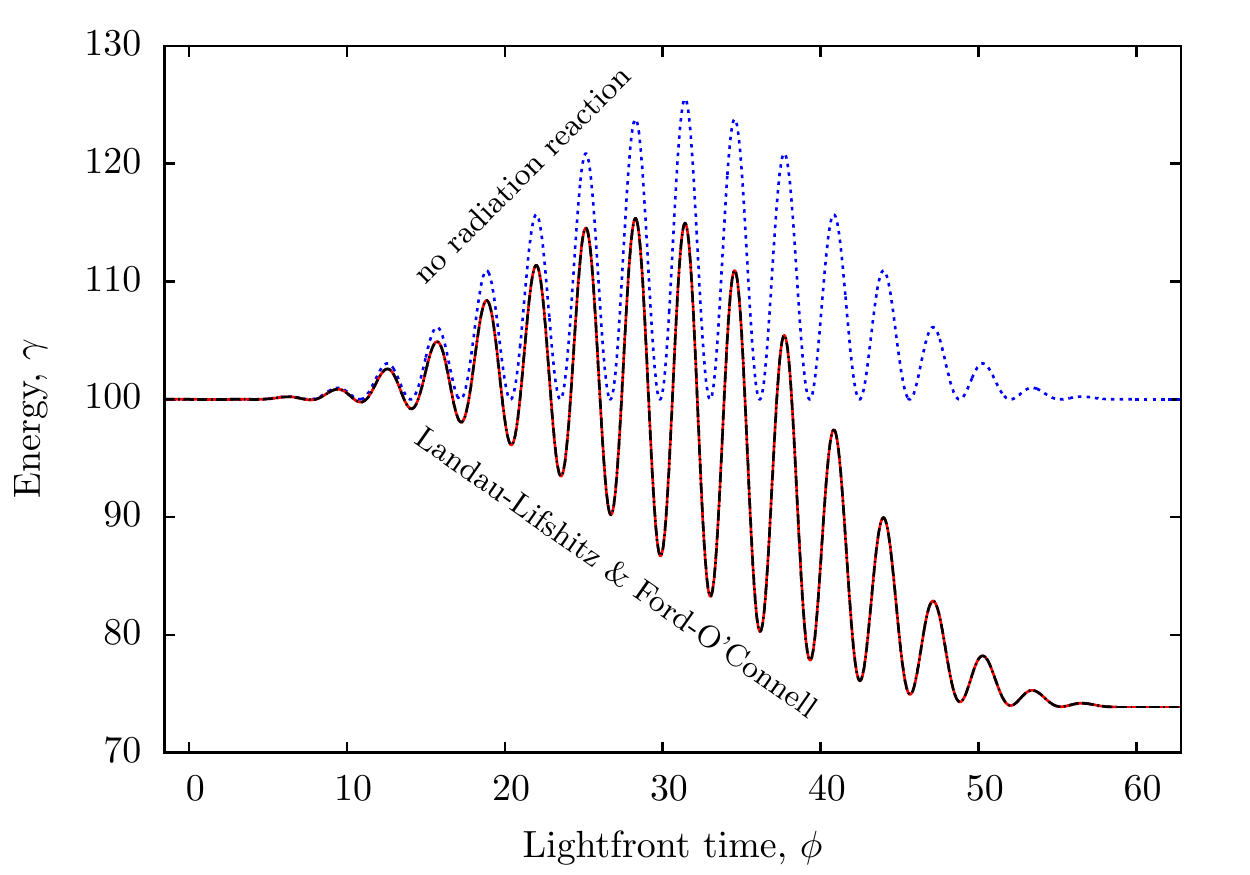}
\vspace{-0.5em}
\caption{\label{fig:3:gammaphilowfinite} 
Radiation reaction effects of a pulse with $a_0=100$ on an electron of initial energy $\gamma_\text{in}=100$: $\gamma$ as a function of $\phi$. Dotted blue curves without radiation reaction; solid red curves with Landau-Lifshitz radiation reaction; double-dotted black curves with Ford-O'Connell radiation reaction.}
\end{figure}

Again, considering the highest currently attainable laser intensities ($a_0=100$) and moderately high initial electron energies ($\gamma_\text{in}=100$) it is shown in Fig.~\ref{fig:3:phislowfinite}, ~\ref{fig:3:gammaphilowfinite} that radiation reaction has a significant effect, however we do not observe the breakdown of the Landau-Lifshitz approximation since the Landau-Lifshitz and Ford-O'Connell predictions are in excellent agreement, as expected.
%\vspace{-1em}

\begin{figure}
\centering
\includegraphics[width=0.85\textwidth]{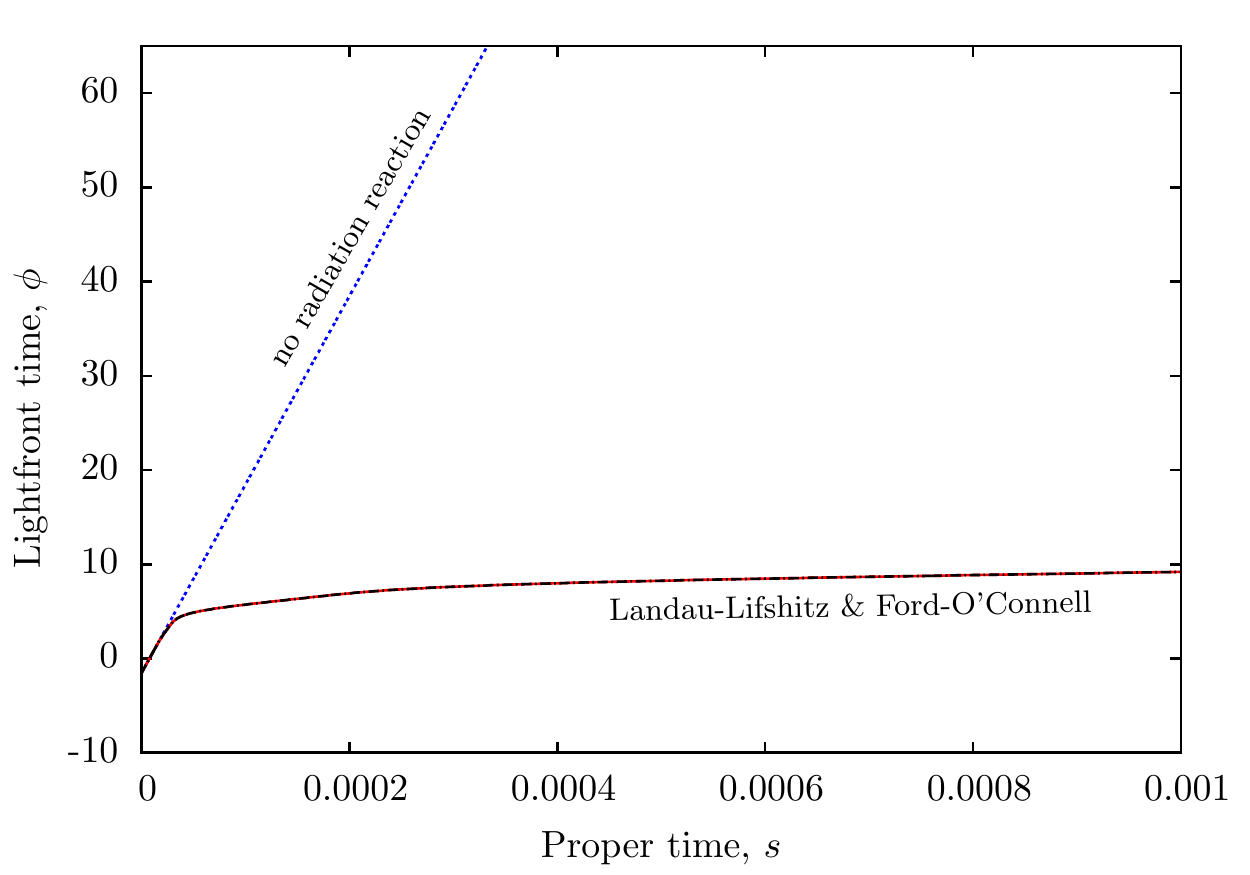}
\caption{\label{fig:3:phishighfinite}
Radiation reaction effects of a pulse with $a_0=1000$ on an electron of initial energy $\gamma_\text{in}=10^5$: $\phi$ as a function of $s$.} %Dotted blue curve without radiation reaction; solid red curves with LL radiation reaction; double-dotted black curves with FO radiation reaction.}
\end{figure}

\begin{figure}
\centering
\includegraphics[width=0.85\textwidth]{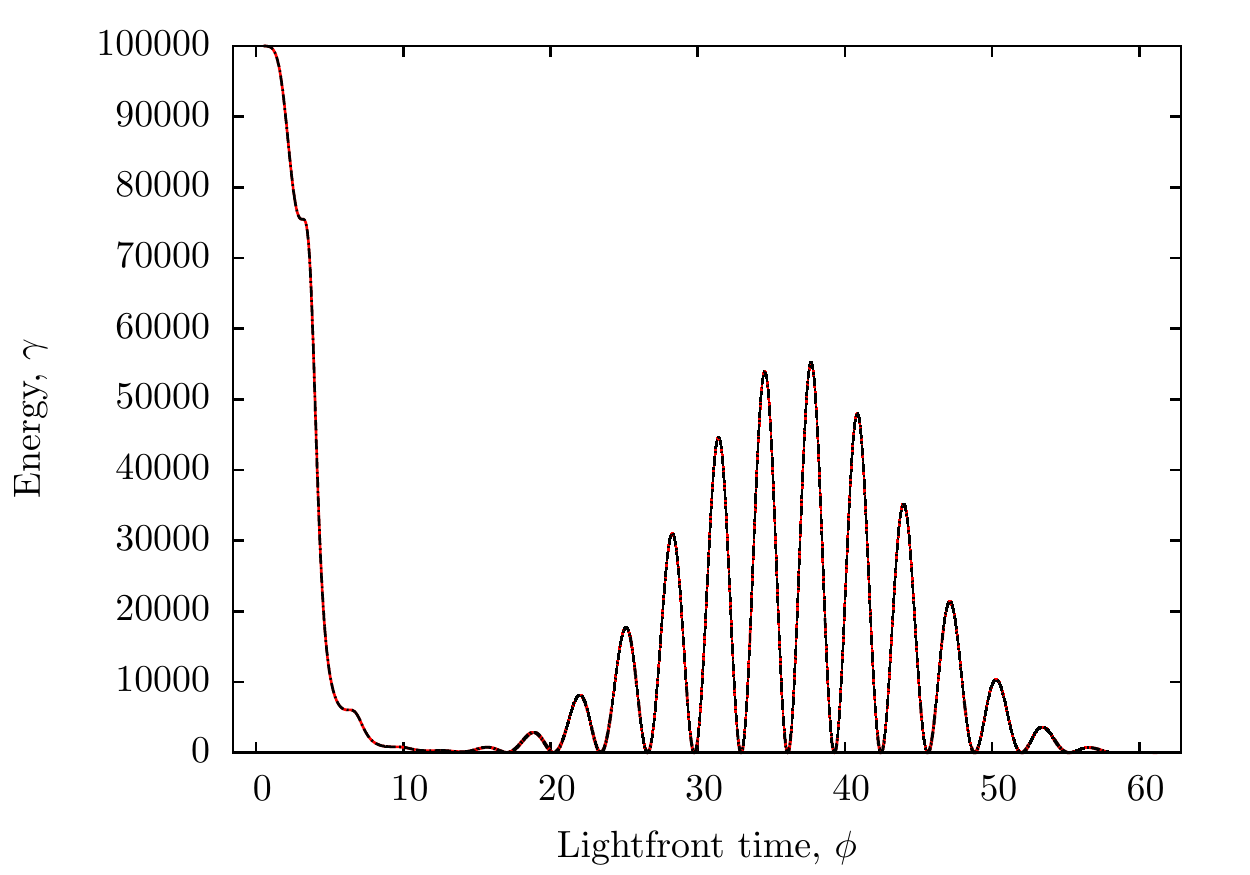}
\caption{\label{fig:3:gammaphihighfinite}
Radiation reaction effects of a pulse with $a_0=1000$ on an electron of initial energy $\gamma_\text{in}=10^5$: $\gamma$ as a function of $\phi$. Solid red curve with Landau-Lifshitz radiation reaction; double-dotted black curve with Ford-O'Connell radiation reaction.}
\end{figure}

Considering the most intense lasers under development ($a_0 = 1000$) and the highest energy electrons available ($\gamma_\text{in} = 10^5$), we appear to be in a regime where the condition (\ref{eq:3:Tcond}) is violated, and we would expect strong differences between Landau-Lifshitz and Ford-O'Connell corrections. However, as shown in Fig.~\ref{fig:3:phishighfinite}, ~\ref{fig:3:gammaphihighfinite}, although the dynamics is dominated by radiation reaction, agreement between the two theories remains strong despite the expected prediction, as for the infinite wave.


\begin{figure}
\centering
\includegraphics[width=0.85\textwidth]{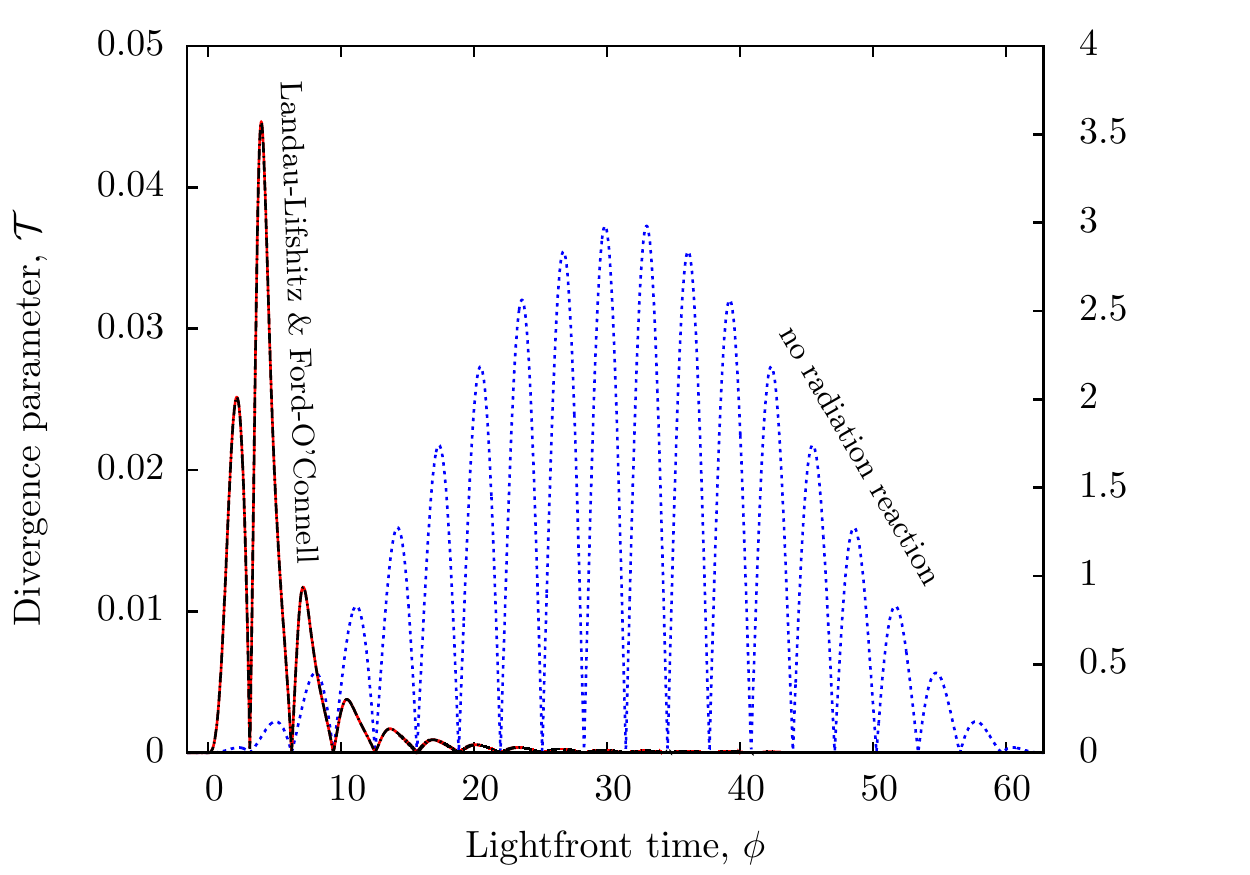}
\caption{\label{fig:det} 
$\T$ as a function  of $\phi$ with radiation reaction (left axis, solid red curve (Landau-Lifshitz), double-dotted black curve (Ford-O'Connell)), and without radiation reaction (right axis, dotted blue curve).}
\end{figure} 

Consider the $\mathcal{T}$ parameter, which is one possible quantitative measure for the divergence between the two approaches investigated. Because the particle begins in vacuum, initially $\T=0$, and from Fig.~\ref{fig:3:gammaphihighfinite}, it is clear that the electron loses almost all its energy to radiation in the first two cycles, while $\E_1\ll \omega a_0$. After this, the radiation becomes a small effect, and its evolution is well described by the Lorentz force alone, at a greatly reduced initial energy. As shown in Fig.~\ref{fig:det}, as the electron propagates through the laser pulse, its energy loss occurs at such a rate that $\T$ never approaches unity. Thus the Landau-Lifshitz equation remains a good description of radiation reaction phenomena for field strengths and electron energies far exceeding those currently proposed. Comparison with the values of $\T$ calculated for a (hypothetical) particle experiencing the Lorentz force alone demonstrates that the validity of the Landau-Lifshitz equation for such high energies is a direct consequence of radiation reaction itself (note the different scales in Fig.~\ref{fig:det}).

\section{Summary}

Radiation reaction can have a significant effect on the motion of a charged particle interacting with a laser pulse, even coming to dominate over the applied Lorentz force. Nonetheless, a high energy electron traversing an ultra-intense laser pulse loses most of its energy to radiation in the first few cycles. When it reaches the peak of the field, therefore, its energy is comparatively low. For field strengths and electron energies far exceeding those currently proposed, radiation reaction effects ensure that the instantaneous evolution of the particle's worldline can be accurately described by treating radiation reaction as a small correction, as in the prescription of Landau and Lifshitz.

\chapter{Interaction of a particle bunch with a laser pulse}\label{sec:benchmark}

\begin{figure}[H]
\vspace{-16em}
\centering
\includegraphics[width=0.9\textwidth]{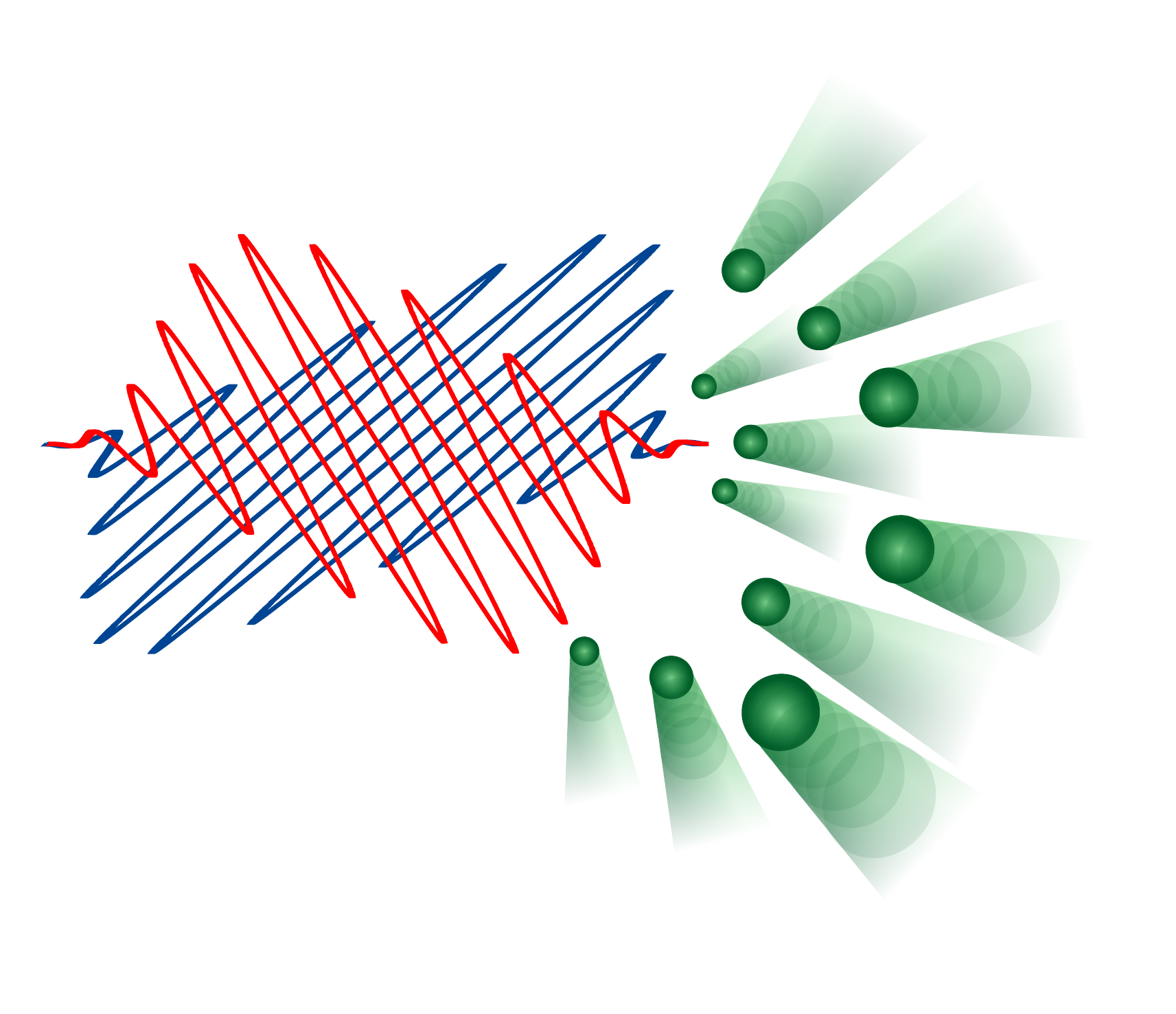}
\end{figure} 
\newpage 

In the previous Chapter we considered the interaction of a single particle with an intense laser pulse. However, realistic experiments involve a bunch of particles. For a bunch of particles we need to consider the evolution of bulk properties such as the average momentum and relative momentum spread. 

Many upcoming laser facilities, such as the Extreme Light Infrustructure (ELI) and Exawatt Center for Extreme Light Studies (XCELS) use all-optical setups, with electron bunches generated by \textit{laser wakefield acceleration} (LWFA). 

The concept of LWFA was proposed by Toshiki Tajima and John M. Dawson in 1979 \cite{Tajima}, where they showed that the ponderomotive force of an intense laser can cause charge separation, leaving a charged region in a previously neutral plasma. Particles injected into this region are accelerated, which provides a compact alternative to conventional accelerators. In typical LWFA experiments electron bunches are produced with charge $\sim 10 \enskip \text{pC}$ corresponding to $10^8$ particles, average energy $\sim$~ 1~ GeV~ \cite{Mangles, Geddes, Faure}, and relative energy spread $\sim$ 1$\%$ \cite{Wiggins}.

In this Chapter we extend the single particle model presented in Chapter 3 and study how radiation reaction influences the average momentum and momentum spread of a bunch of particles during their propagation through an intense laser pulse. We are particularly interested in how the evolution of the distribution depends on the length of a pulse and its total energy per unit area.

\section{Introduction of the particle distribution function}

After considering radiation reaction effects on the motion of a single particle colliding with an intense laser pulse we can extend our study further and look at the behaviour of a bunch of particles colliding with the laser pulse under the influence of radiation reaction. 

Because we consider a plane wave, the particles' spatial spread would only define the moment in time when the particular particle enters the pulse, so for simplicity we take all particles to originate from the same point. This is reasonable as we are primarily interested in the momentum distribution.

Considering the case with no spread in the transverse directions we are using the initial thermal Maxwellian particle distribution for longitudinal velocities $v$, which can be written as follows:
\begin{equation}
\label{eq:4:MJ_rest}
f\left(v, 0\right) = \frac{N_p}{\sqrt{2\pi\theta}} e^{-\frac{(v-\bar{v})^2}{2\theta}}\ ,
\end{equation}
where $\theta = \frac{\mathrm{k}T}{mc^2}$ is the thermal momentum spread, with $\mathrm{k}$ the Boltzman constant, and $N_p$ is the number of particles.

Although such a distribution is usually associated with a non-relativistic thermal momentum spread, whereas we investigate situations which are neither thermal nor non-relativistic it is a convenient distribution to illustrate evolution of the bulk properties of the particle bunch.

The average velocity $\bar{v}^a = \left(\sqrt{1+\bar{v}^2}, 0, 0, \bar{v}\right) = n^a/\sqrt{-n_b n^b}$, with 
\begin{equation}
\label{number_flux}
n^a = \int f\left(v\right)\frac{\dot{x}^a}{\sqrt{1 + v^2}}\mathrm{d}v\ ,
\end{equation}
where $n^a$ is the particle number current. 

We consider the \textit{number density}, as the $a = 0$ component of the \textit{number current} vector:  
\begin{equation}
\label{eq:4:number_flux}
n^0 = \int f\left(v\right)\frac{\dot{x}^0}{\sqrt{1 + v^2}}\mathrm{d}v = \int f\left(v\right)\mathrm{d}v\ ,
\end{equation}
To find the evolution of the distribution $f(v,\phi)$ we could solve the Vlasov equation, modified to include radiation reaction \cite{Hazeltine, Berezhiani, Hakim, kinetic, AdamSPIE, Tamburini}. However for computational efficiency we follow the evolution of a finite number of 500 particles, chosen to represent the distribution (\ref{eq:4:MJ_rest}). Because the bunch is moving relativistically we can neglect interparticle interactions over the timescale that the bunch experiences the pulse.

Typically one would sample the velocities of the particle distribution at random which would require a large number of particles to accurately represent the distribution. Instead, we determine the velocity spacing $\delta v$ between the particles from the initial distribution, by truncating the integral in (\ref{eq:4:number_flux}) so the particle number increases by 1:
\begin{equation}
\label{eq:4:number_density}
1 = \int\limits_{v-\frac{\delta v}{2}}^{v+\frac{\delta v}{2}} f\left(v\right)\mathrm{d}v \simeq f\left(v\right)\delta v\ ,
\end{equation}
leading to particles having initial velocities:
\begin{equation}
\dots\ , \enskip \bar{v} - \frac{1}{f\left(\bar{v}\right)} - \frac{1}{f\left(\bar{v} - \frac{1}{f\left(\bar{v}\right)}\right)}\ , \enskip \bar{v} - \frac{1}{f\left(\bar{v}\right)}\ , \enskip \bar{v}\ , \enskip \bar{v} + \frac{1}{f\left(\bar{v}\right)}\ , \enskip \bar{v} + \frac{1}{f\left(\bar{v}\right)} + \frac{1}{f\left(\bar{v} + \frac{1}{f\left(\bar{v}\right)}\right)}\ , \enskip \dots 
\end{equation}

Using these initial conditions we then integrate the Landau-Lifshitz equation and using the later spacing between particle velocities apply (\ref{eq:4:number_density}) in reverse to reconstruct the distribution, $f(v,\phi)$.

\section{Numerical simulations. Impact of the pulse length on the particle distribution}

As in the previous Chapter we consider an $N$-cycle pulse of central frequency $\omega$, modulated by a $\sin^2$-envelope. $a_0$ is the usual intensity parameter (so called ``normalised vector potential''). We use units such that $\omega=1$.
\begin{equation}
\label{eq:4:pulseshape}
{\cal E}= 
\begin{cases}
\omega a_0 \sin (\omega \phi) \sin^2(\omega \phi/ 2N) & \text{for } 0 < \phi < 2\pi N/\omega\ , \\
0 & \text{otherwise}\ .
\end{cases}
\end{equation}

Since we are interested in velocity rather than spatial distribution, all the particles originate at a single point in space in front of the laser pulse and are evaluated to the point of exit from the pulse, $\phi=2\pi N/\omega$.

The evolution is tracked using two different approaches. We consider the case with \textit{no radiation reaction} and that with the radiation reaction taken into account using the \textit{Landau-Lifshitz} correction. As shown in the previous Chapter for the case of interactions with a plane wave Ford-O'Connell and Landau-Lifshitz predictions agree, therefore there is no need to go beyond Landau-Lifshitz corrections in this Chapter.

Pulse parameters being varied between the simulations are: 
\vspace{-0.5em}
\begin{itemize}
\item number of oscillations $N$ of the pulse;
\vspace{-0.5em}
\item energy (per unit area) of the pulse $\sim N a^2_0$.
\end{itemize}
\vspace{-0.5em}
The same pulse parameters are considered for the initial average velocity $\bar{v}$ of $10^2$, $10^3$ and  $10^4$.

While evaluating cases with different pulse length, we keep the energy in the pulse per unit area constant. The energy in the pulse is given by $E = \int\limits_{\phi_{i}}^{\phi_{f}} \mathcal{E}^2\mathrm{d\phi}$, which for the pulse we are considering (\ref{eq:4:pulseshape}) is given by:
\begin{equation}
\label{eq:4:energyin}
E = \int\limits_{0}^{2\pi N} a^2_0 \sin^2 (\phi) \sin^4(\phi/ 2N) \mathrm{d\phi} = \frac{3\pi}{8} N a^2_0\ ,
\end{equation}
where $N$ is the number of oscillations in the pulse and $a_0$ is the peak intensity. Therefore, keeping $N a_0{}^2$ constant ensures the above requirement is met.

We are interested in establishing the impact of the pulse length on the width of the velocity distribution after the interaction. Therefore, we fix the initial distribution width to be $1\%$ of the initial average velocity of the distribution and compare with the spread after all the particles have passed through the pulse. The relative distribution width is calculated as:
\begin{equation}
\hat{\sigma} = \frac{\sqrt{\theta}}{\bar{v}}\ .
\end{equation}

Special attention in the following simulations is given to cases with initial average velocity $\bar{v}_i = 10^3$ as these correspond to average energy $\sim$ 1 GeV typically observed in LWFA experiments. The relative energy spread of the bunches used in these experiments can be $\sim$ 1$\%$, which justifies our choice of the initial distribution width.

\newpage

\subsection{Numerical results for a particle bunch with a central velocity of $\bar{v} = 10^2$}

All the particles start at a single point in space in front of the laser pulse and are evaluated to the point of exit from the pulse which has energy $E=\frac{3\pi}{8} \cdot 10^5$.

For this case we consider the following laser pulses of different length:
    \begin{enumerate}
    \vspace{-0.5em}
    \item Short laser pulse with peak $a_0 = 141.4$ and $N = 5$ oscillations
    \vspace{-0.5em}
    \item Laser pulse with peak $a_0 = 100$ and $N = 10$ oscillations
    \vspace{-0.5em}
    \item Long pulse with peak $a_0 = 44.7$ and $N = 50$ oscillations
    \end{enumerate}

\vspace{-0.5em}
The evolutions are tracked using two different approaches. We consider cases with \textit{no radiation reaction} and with the \textit{Landau-Lifshitz} radiation reaction force.

\begin{figure}[H]
\centering
\includegraphics[width=0.48\textwidth]{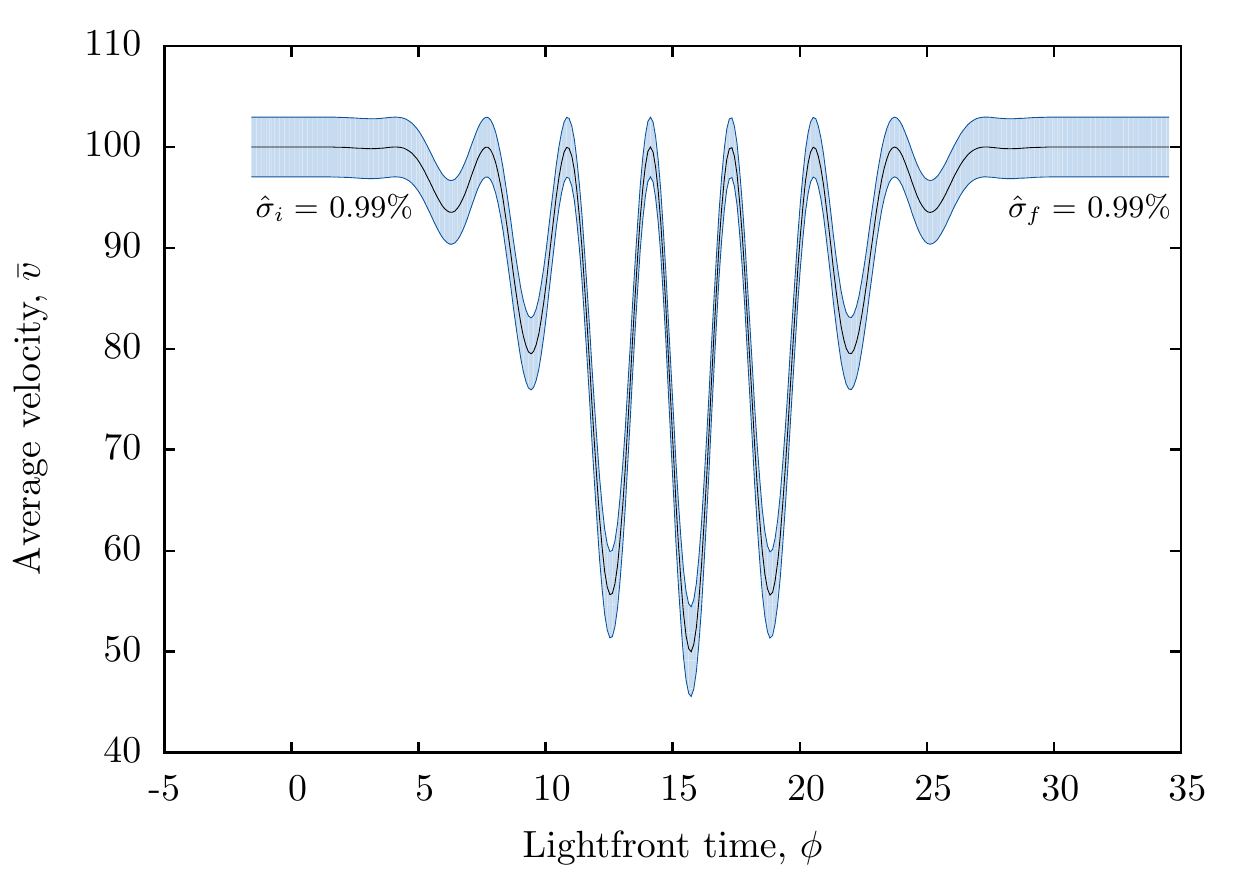}
\includegraphics[width=0.48\textwidth]{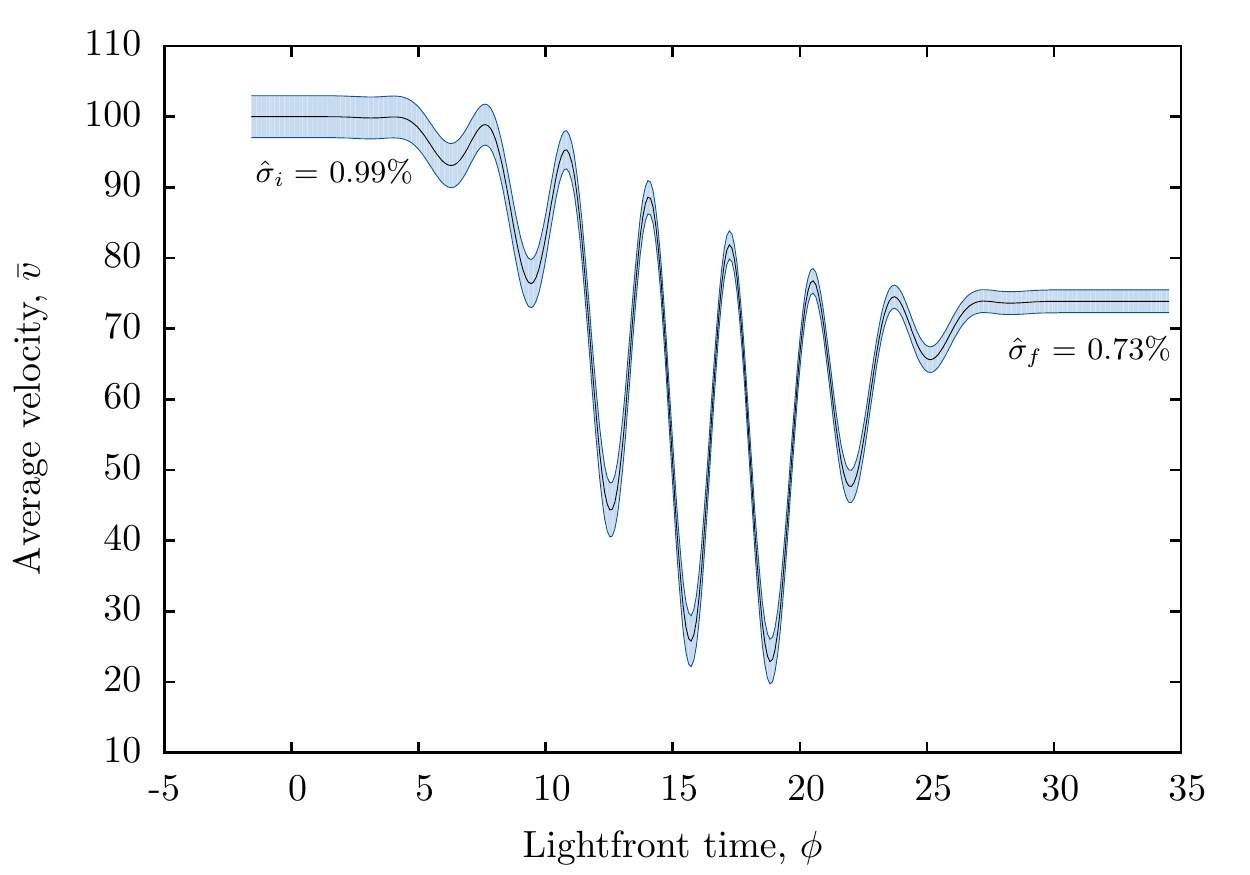}
\includegraphics[width=0.48\textwidth]{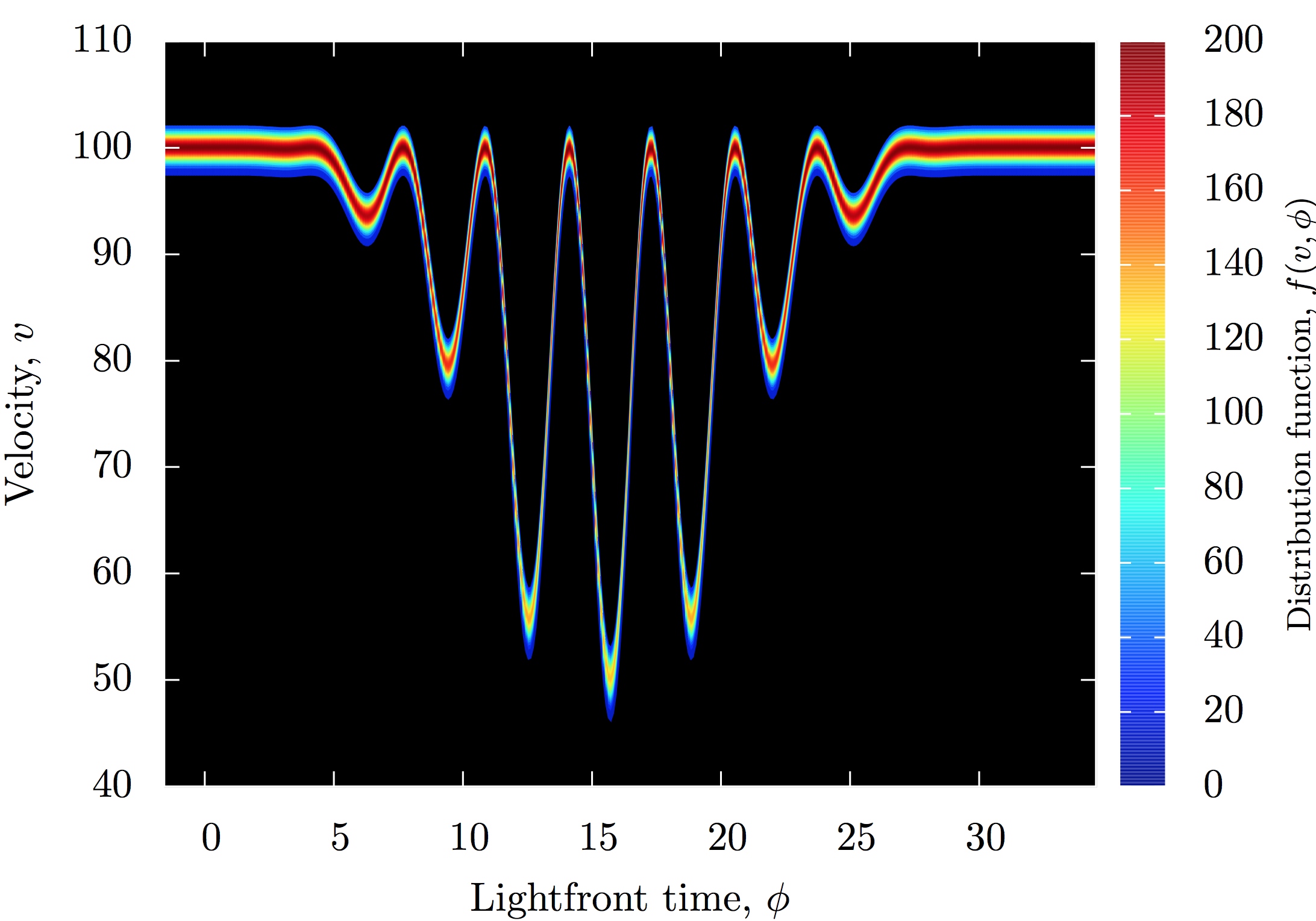}
\includegraphics[width=0.48\textwidth]{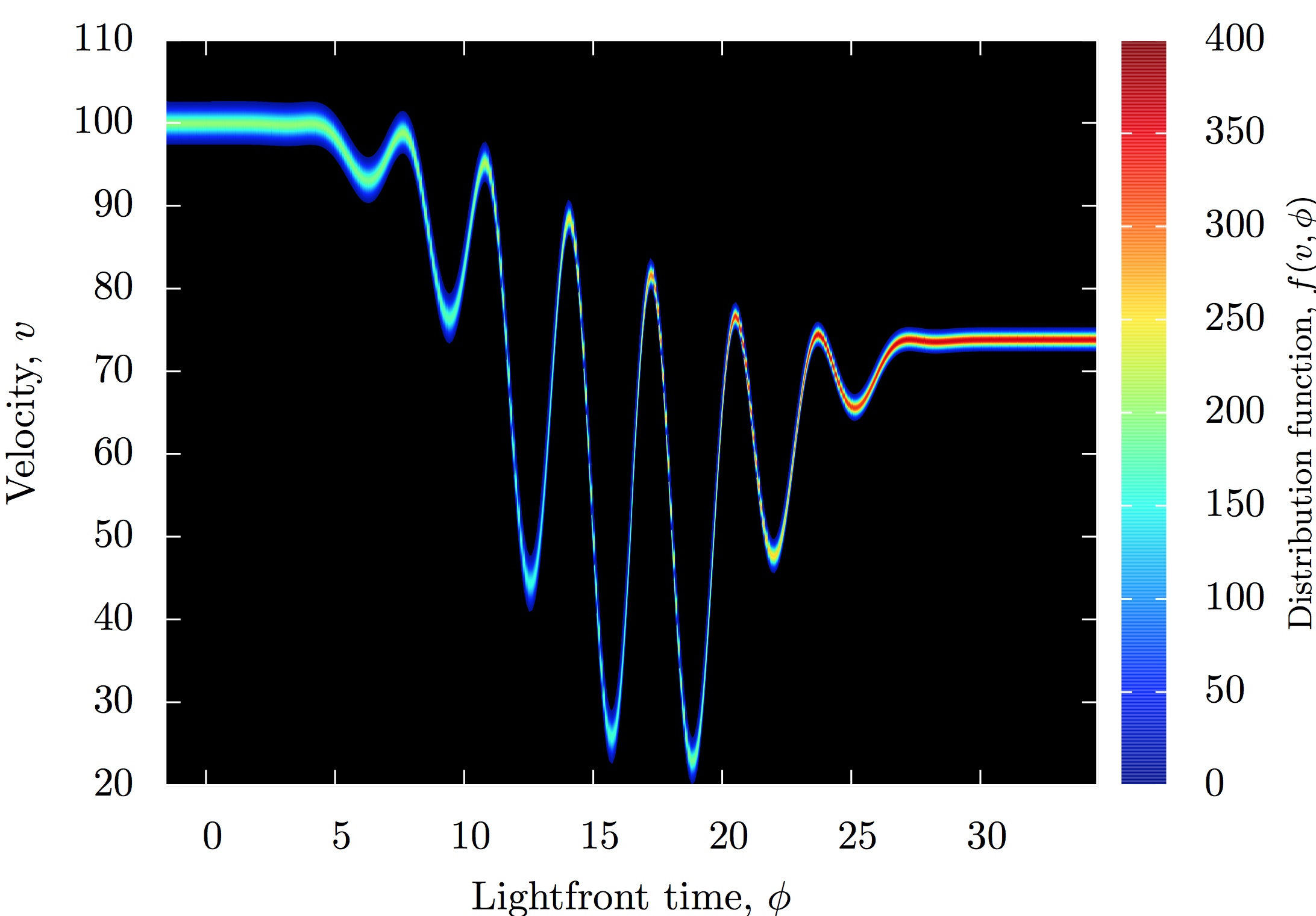}
\vspace{-1.5em}
\caption{\label{fig:dist_v100_N5} 
Distribution for $N = 5$ \textit{without} (left) and \textit{with} (right) radiation reaction.}
\end{figure} 
\vspace{-1.5em}
\begin{figure}[H]
\centering
\includegraphics[width=0.48\textwidth]{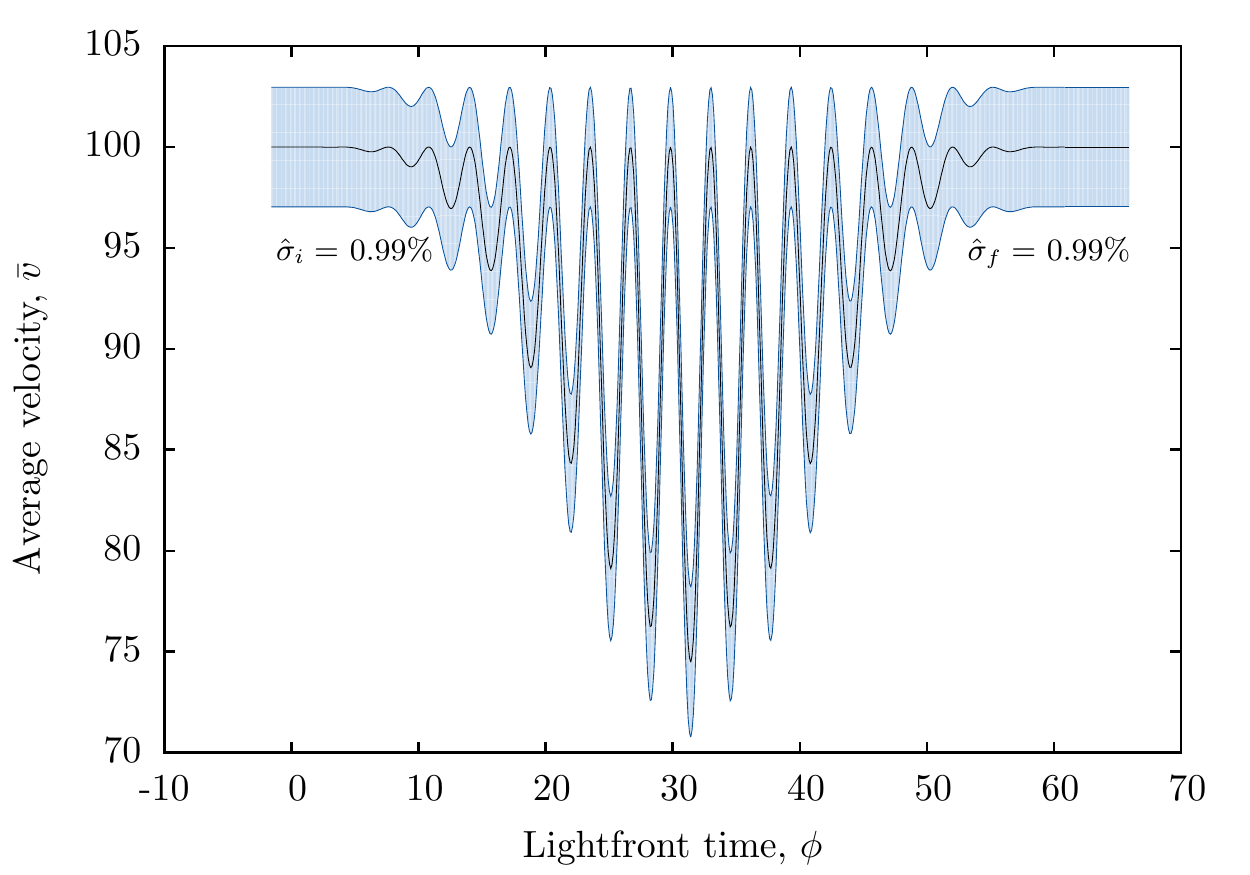}
\includegraphics[width=0.48\textwidth]{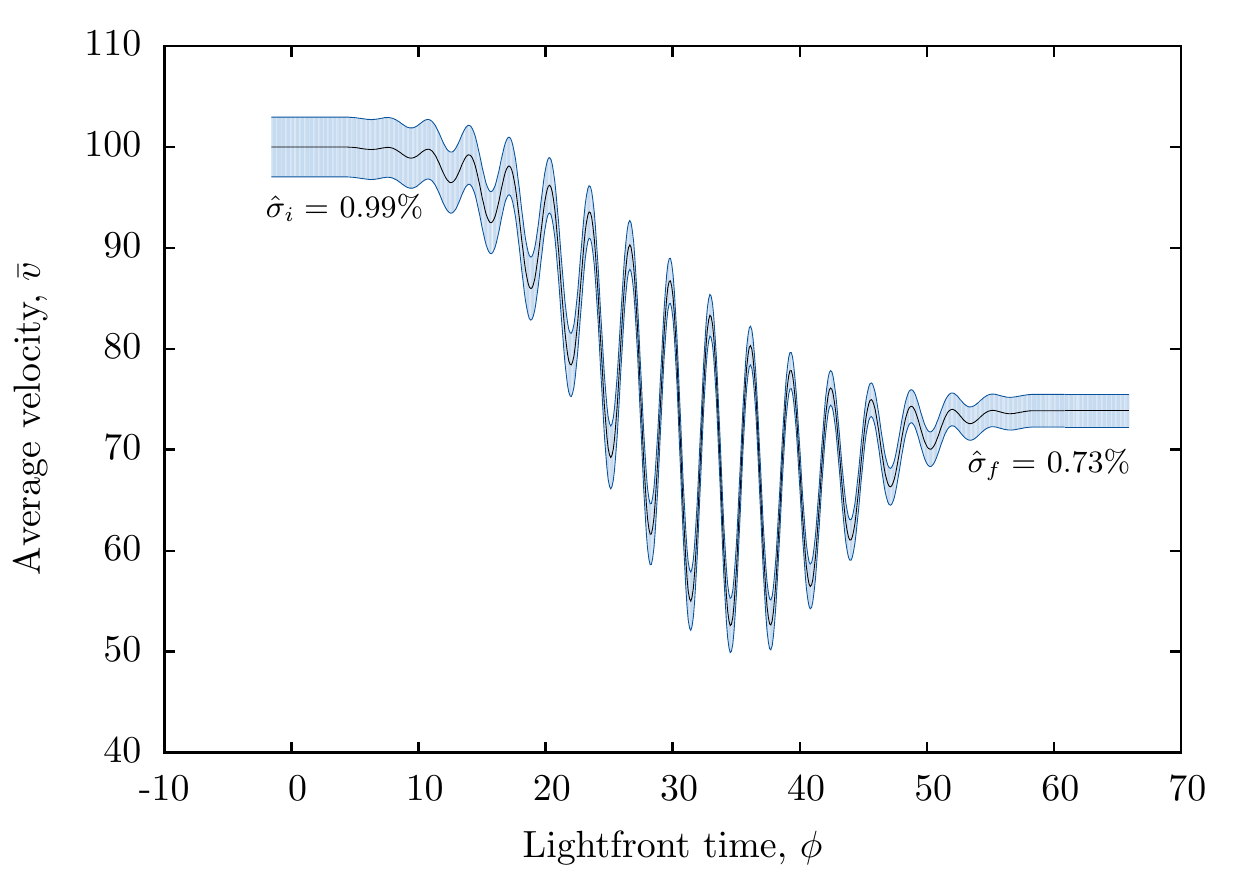}
\includegraphics[width=0.48\textwidth]{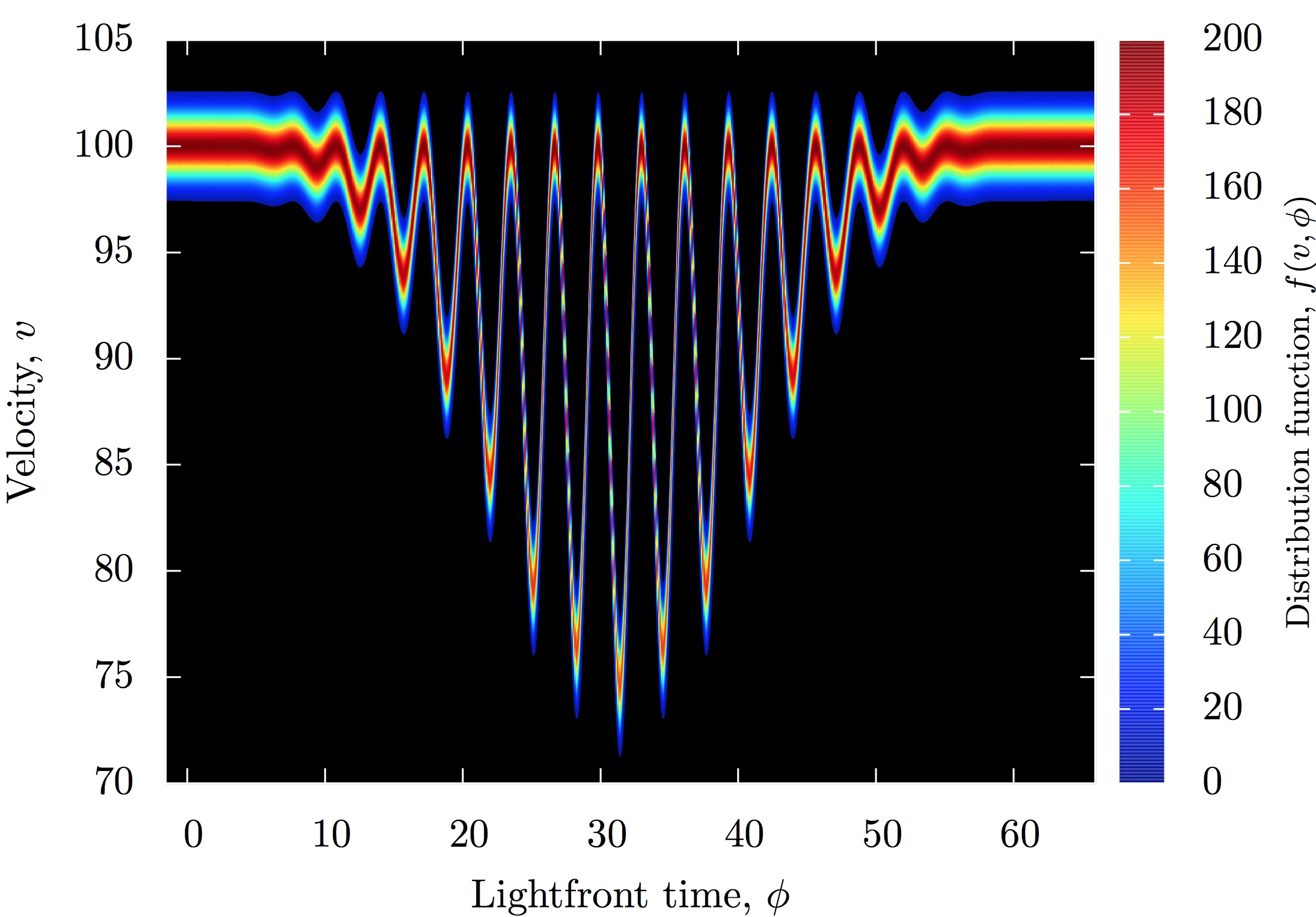}
\includegraphics[width=0.48\textwidth]{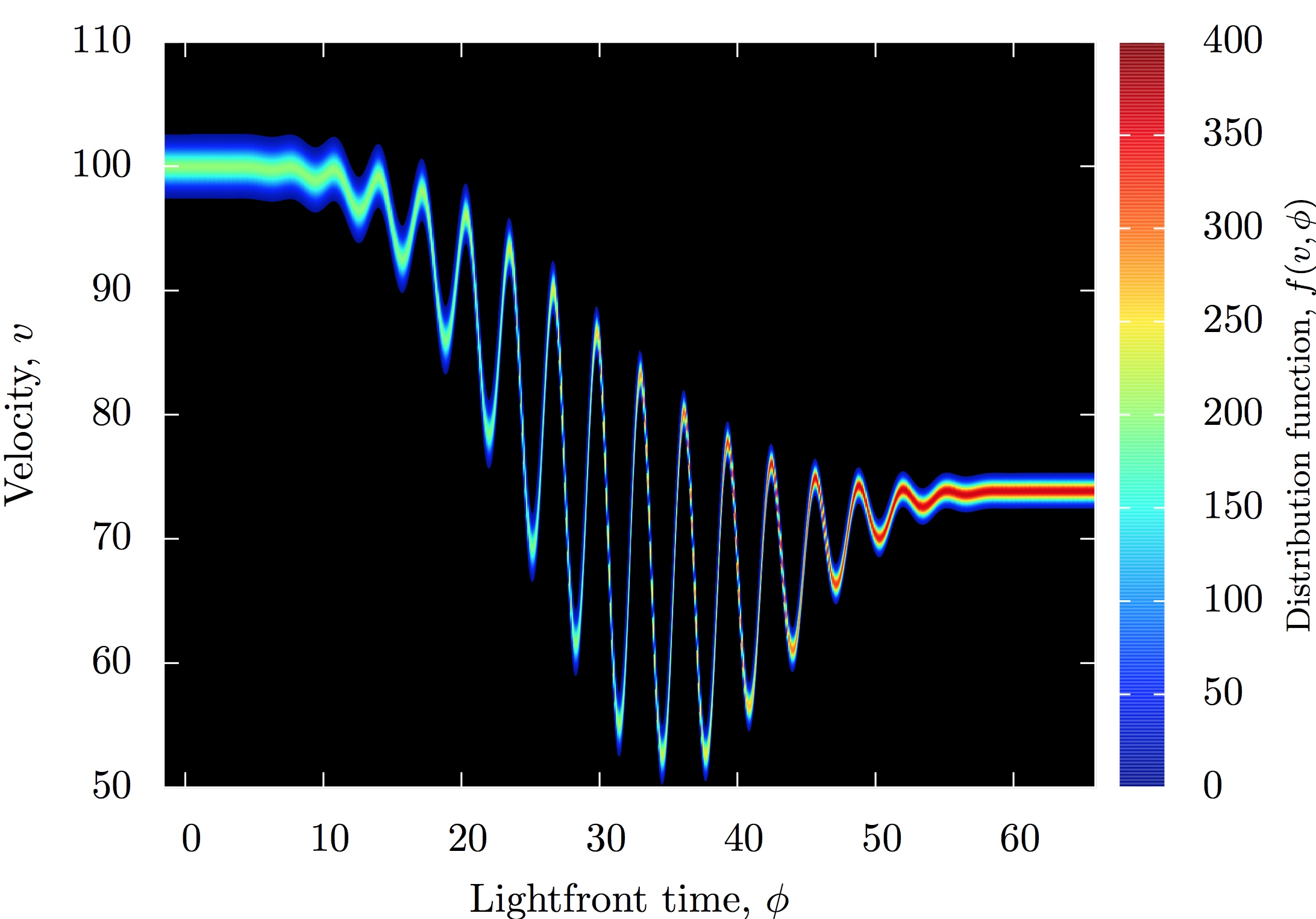}
\vspace{-1.5em}
\caption{\label{fig:dist_v100_N10} 
Distribution for $N = 10$ \textit{without} (left) and \textit{with} (right) radiation reaction.}
\end{figure} 

\vspace{-1.5em}
\begin{figure}[H]
\centering
\includegraphics[width=0.48\textwidth]{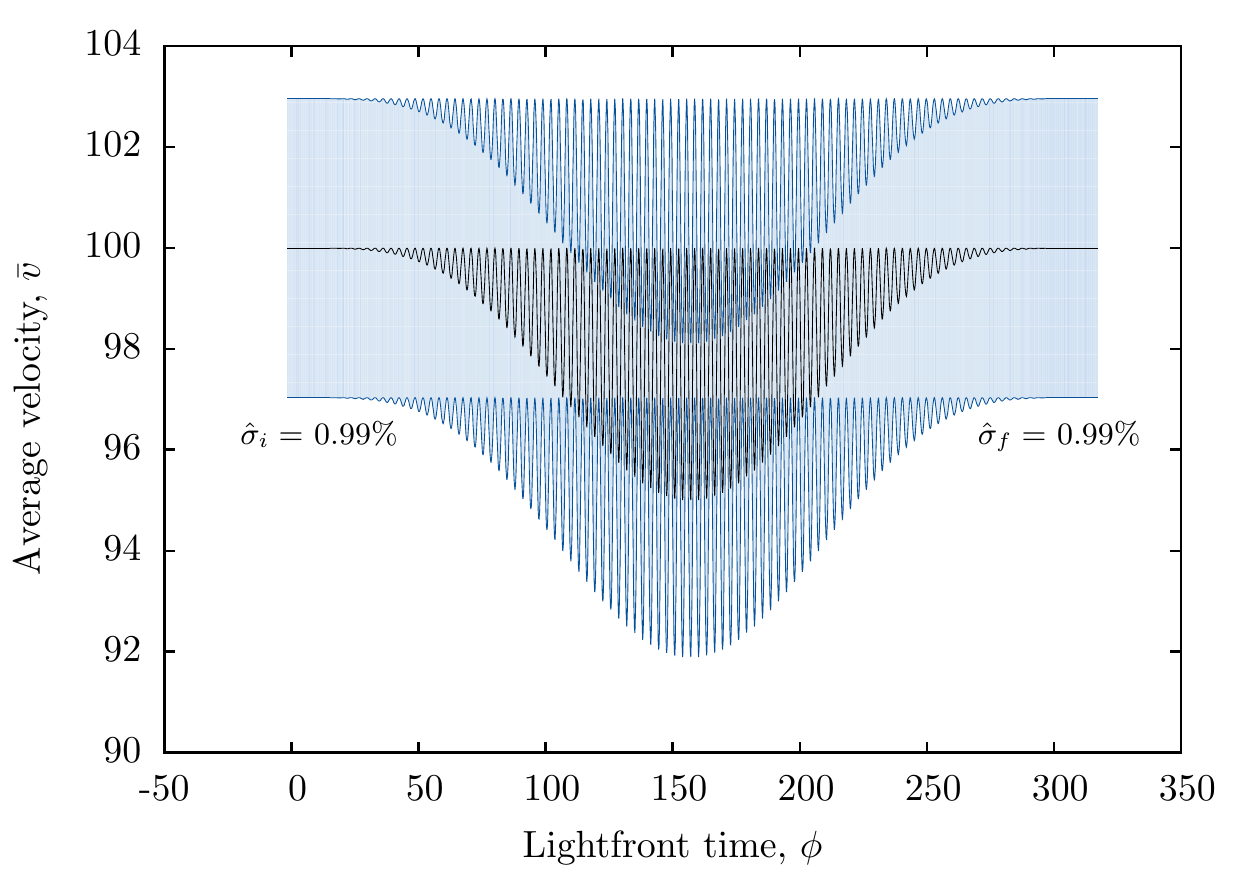}
\includegraphics[width=0.48\textwidth]{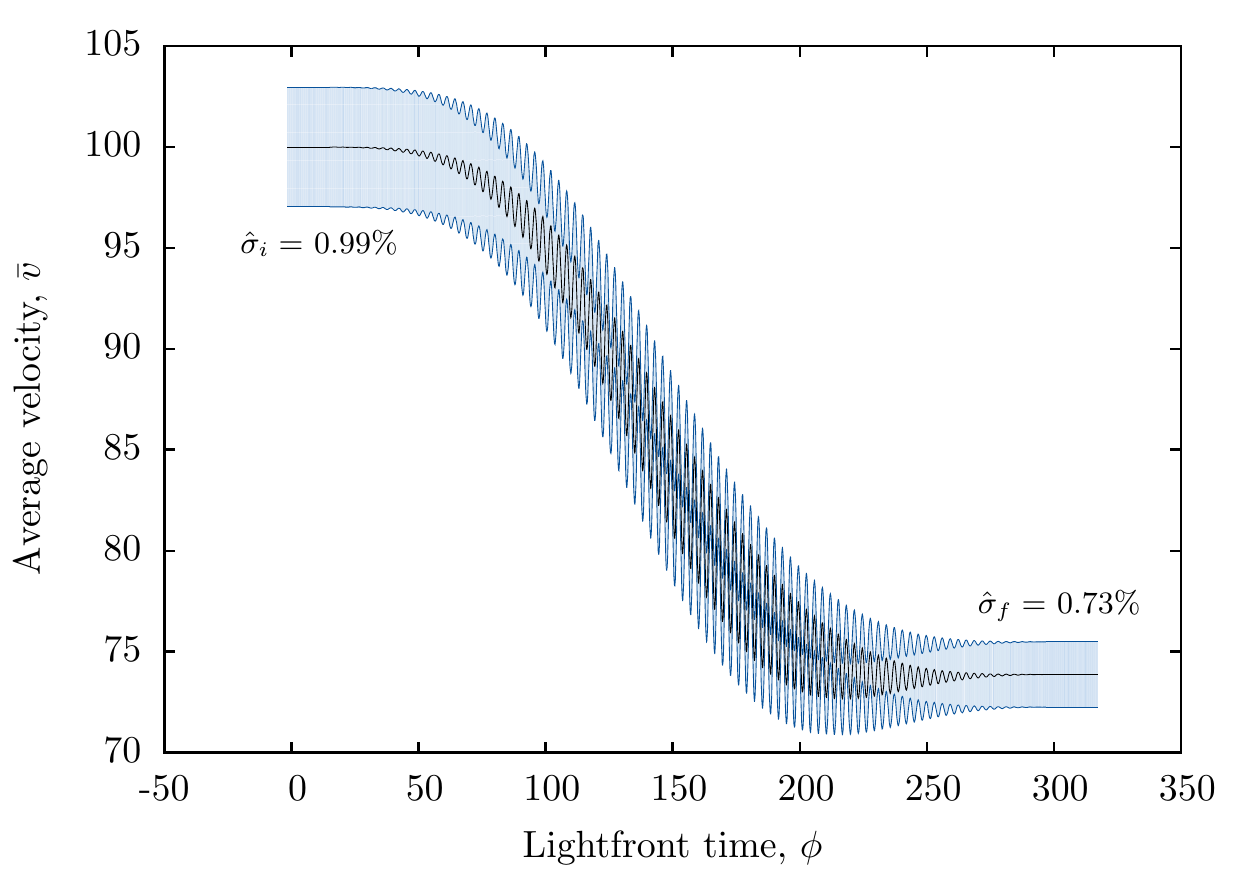}
\includegraphics[width=0.48\textwidth]{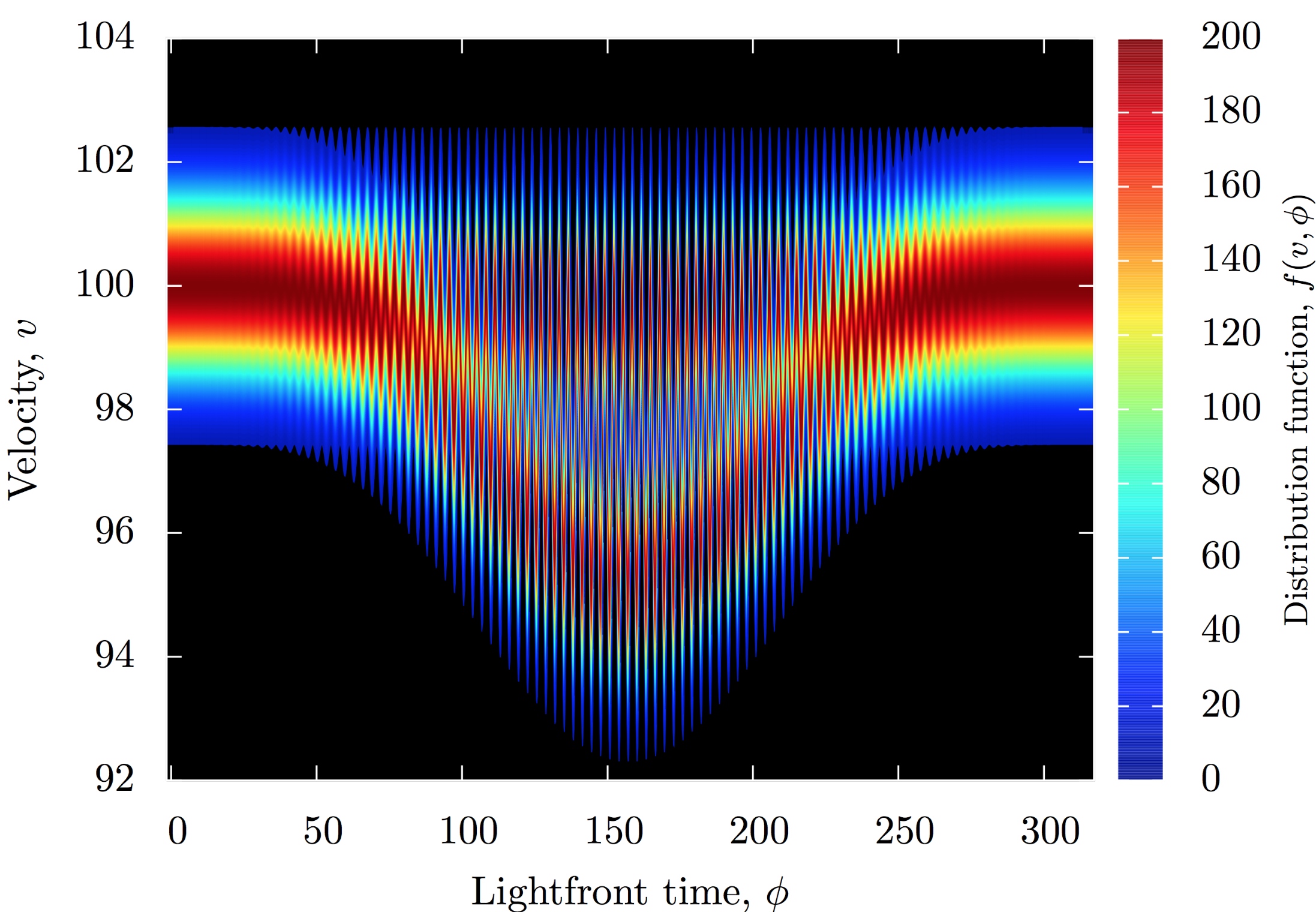}
\includegraphics[width=0.48\textwidth]{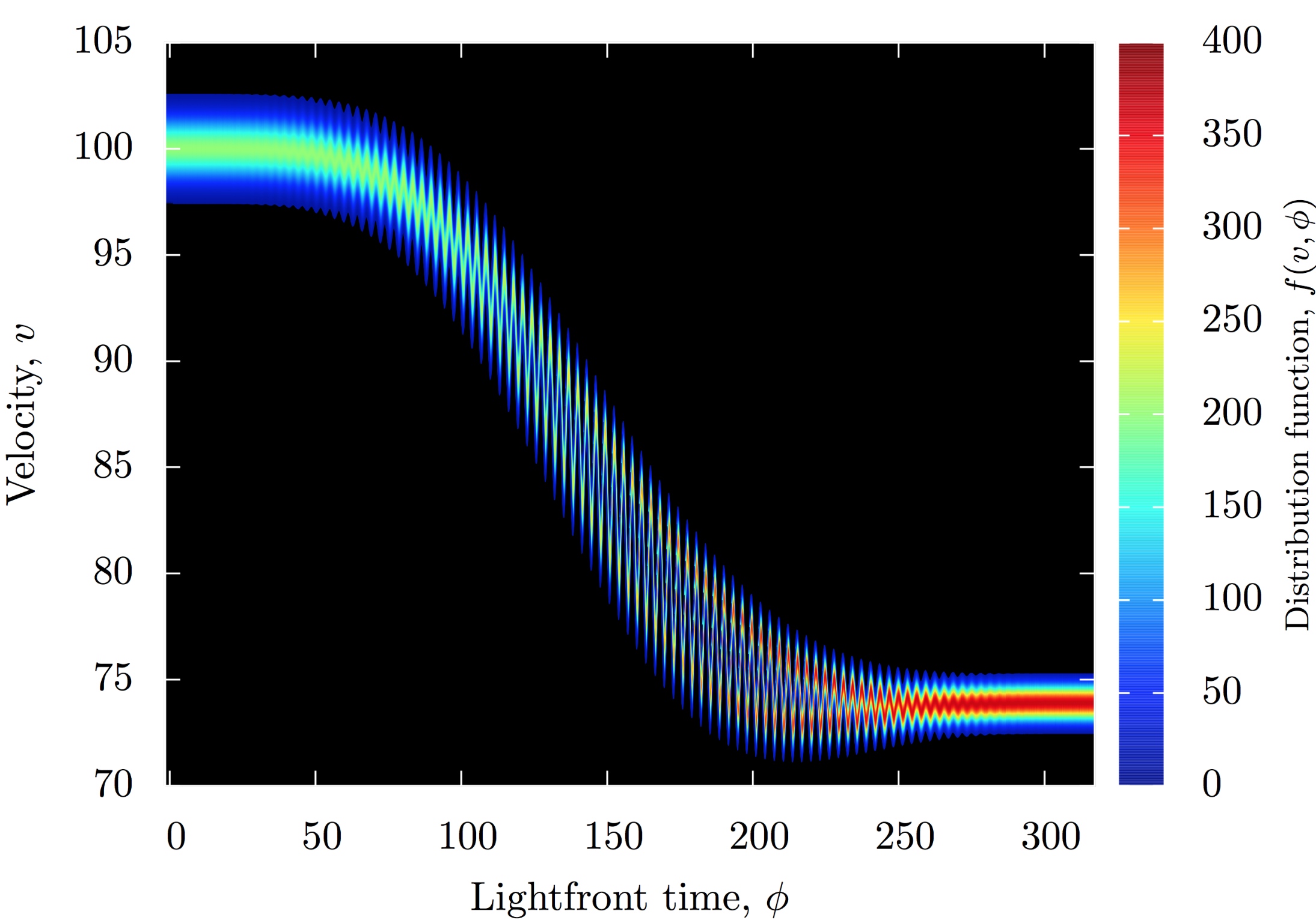}
\vspace{-1.5em}
\caption{\label{fig:dist_v100_N50} 
Distribution for $N = 50$ \textit{without} (left) and \textit{with} (right) radiation reaction.}
\end{figure} 

\subsection{Particle bunch with a central velocity of $\bar{v} = 10^3$}

All the particles start at a single point in space in front of the laser pulse and are evaluated to the point of exit from the pulse which has energy $E=\frac{3\pi}{8} \cdot 10^5$.

For this case we consider the following laser pulses of different length:
    \begin{enumerate}
    \vspace{-0.5em}
    \item Extremely short laser pulse with peak $a_0 = 223.6$ and $N = 2$ oscillations
    \vspace{-0.5em}
    \item Short laser pulse with peak $a_0 = 141.4$ and $N = 5$ oscillations
    \vspace{-0.5em}
    \item Laser pulse with peak $a_0 = 100$ and $N = 10$ oscillations
    \vspace{-0.5em}
    \item Laser pulse with peak $a_0 = 70.7$ and $N = 20$ oscillations
    \vspace{-0.5em}
    \item Long laser pulse with peak $a_0 = 44.7$ and $N = 50$ oscillations
    \vspace{-0.5em}
    \item Extremely long pulse with peak $a_0 = 31.6$ and $N = 100$ oscillations
    \end{enumerate}

\vspace{-0.5em}
The evolutions are tracked using two different approaches. We consider cases with \textit{no radiation reaction} and with the \textit{Landau-Lifshitz} radiation reaction force.

\begin{figure}[H]
\centering
\includegraphics[width=0.48\textwidth]{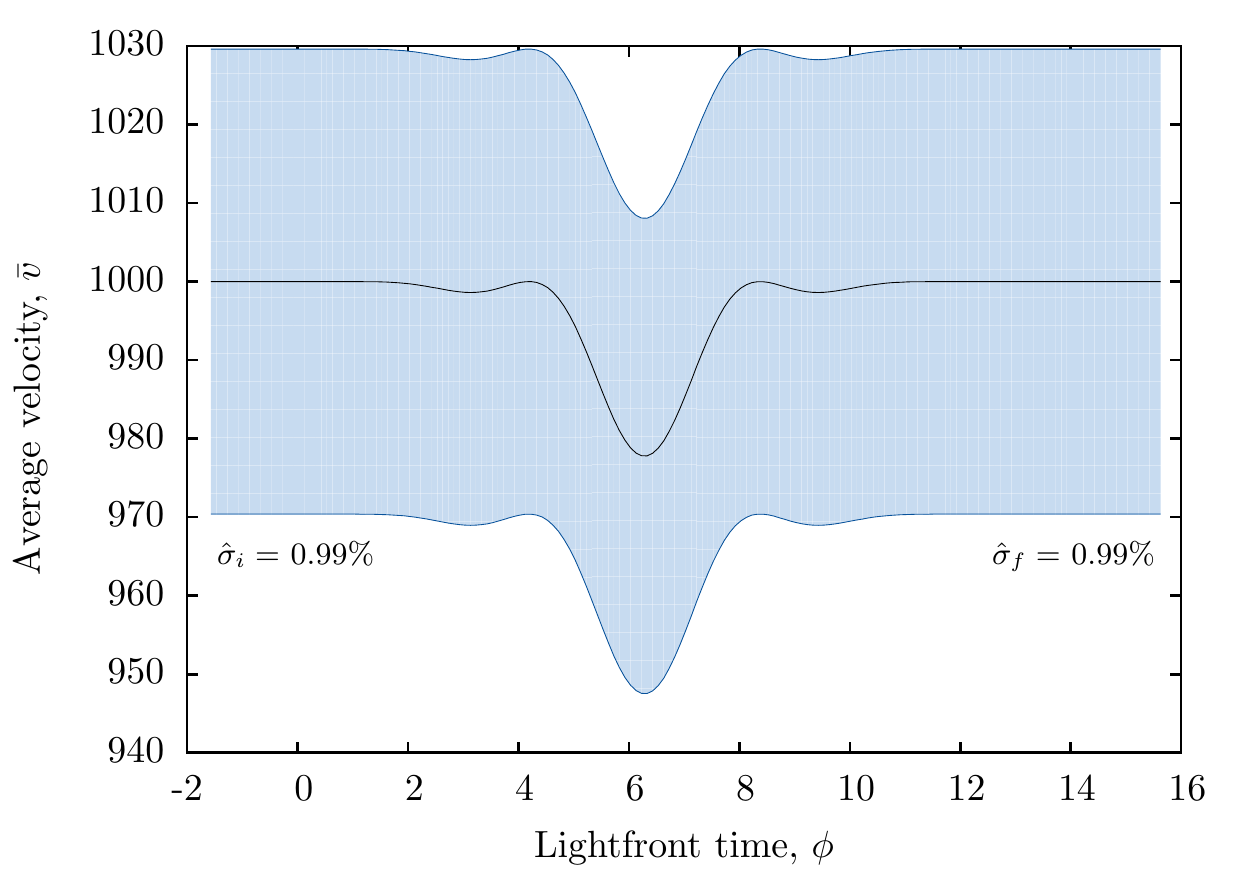}
\includegraphics[width=0.48\textwidth]{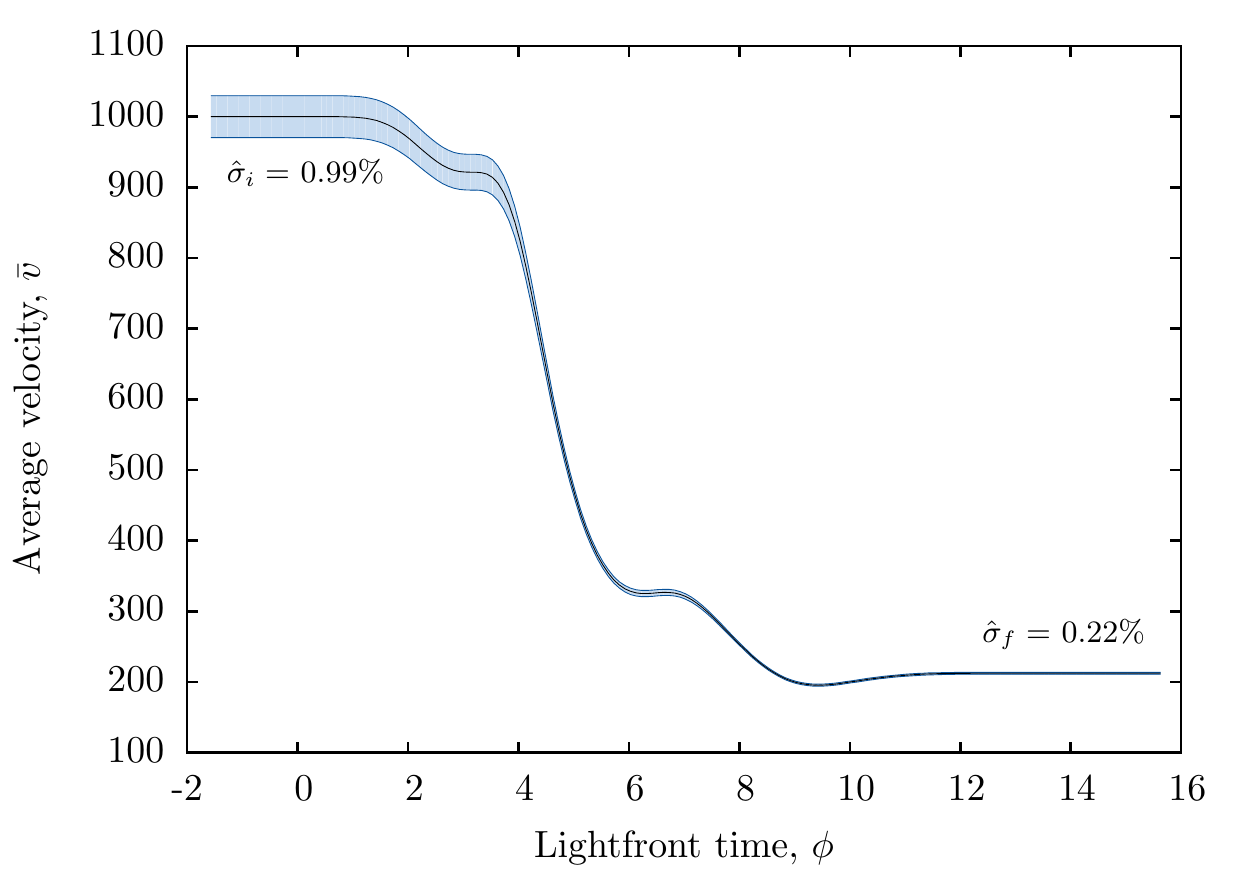}
\includegraphics[width=0.48\textwidth]{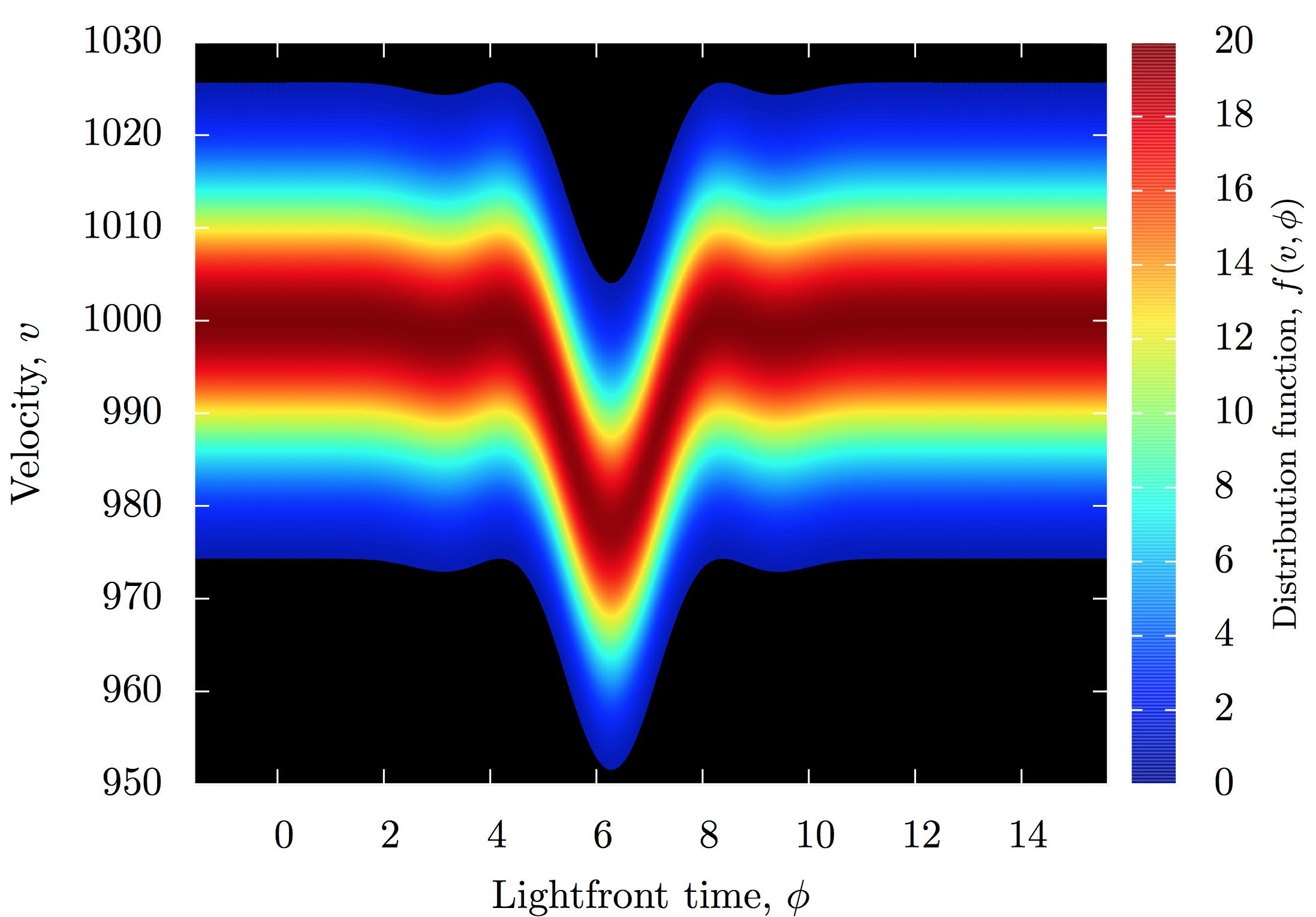}
\includegraphics[width=0.48\textwidth]{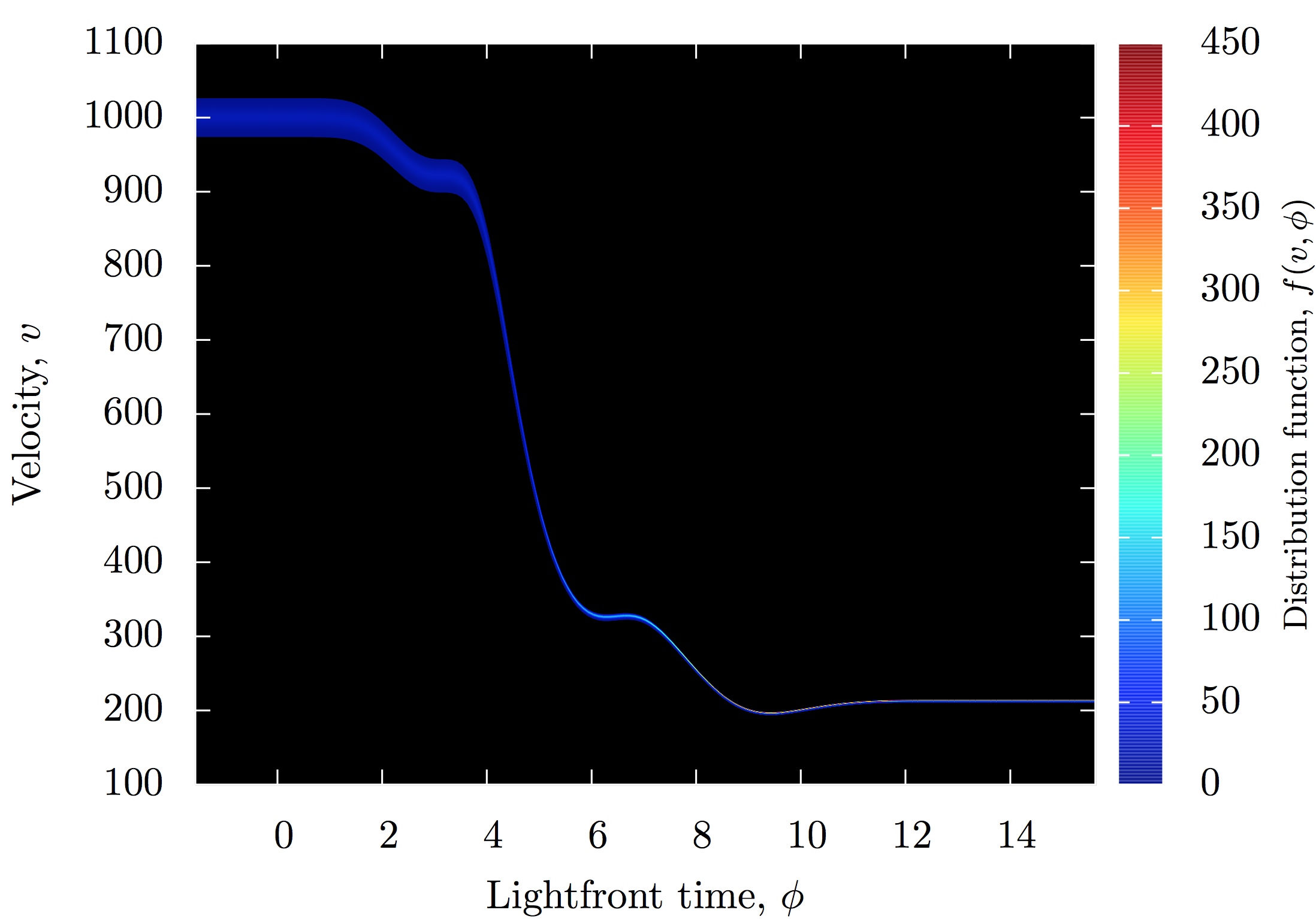}
\vspace{-1.5em}
\caption{\label{fig:dist_v1000_N2} 
Distribution for $N = 2$ \textit{without} (left) and \textit{with} (right) radiation reaction.}
\end{figure} 
\vspace{-1.5em}
\begin{figure}[H]
\centering
\includegraphics[width=0.48\textwidth]{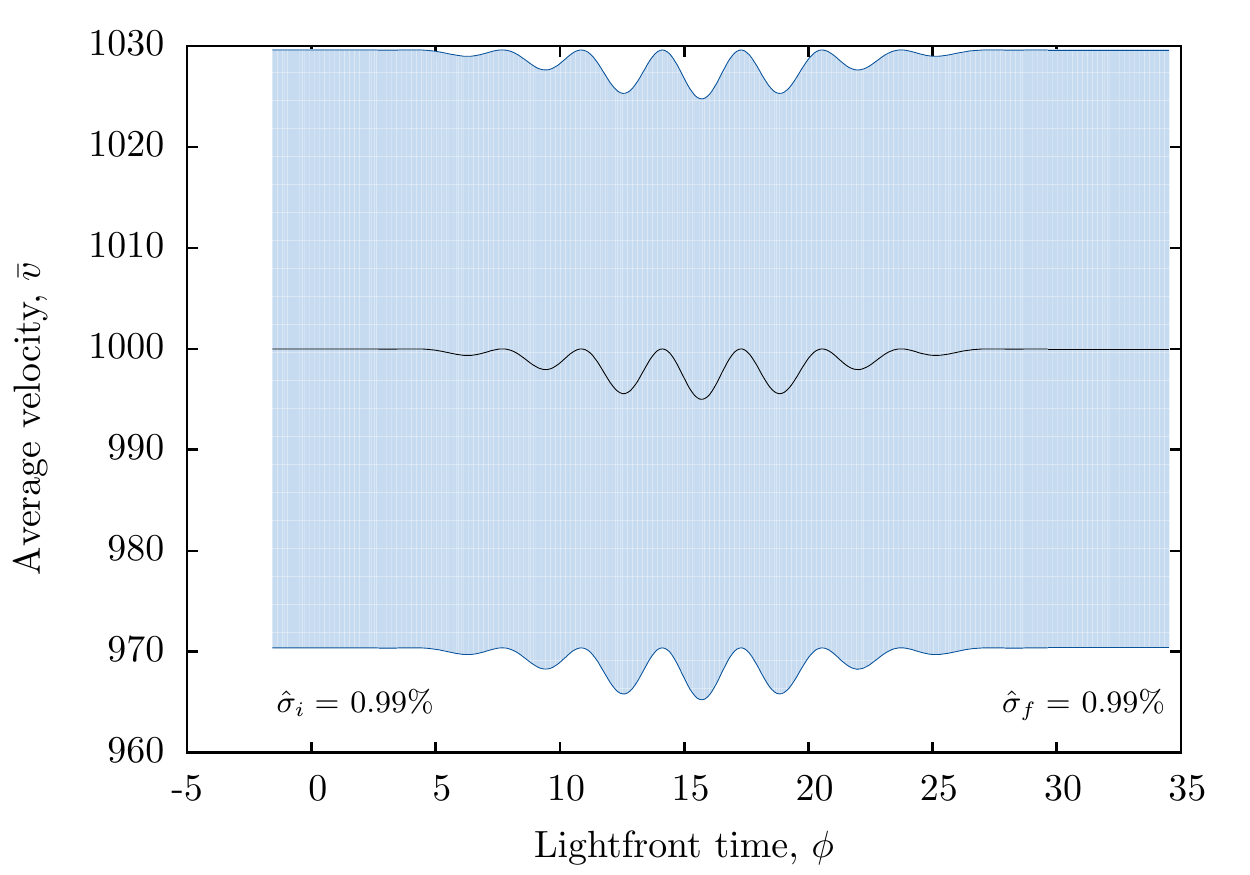}
\includegraphics[width=0.48\textwidth]{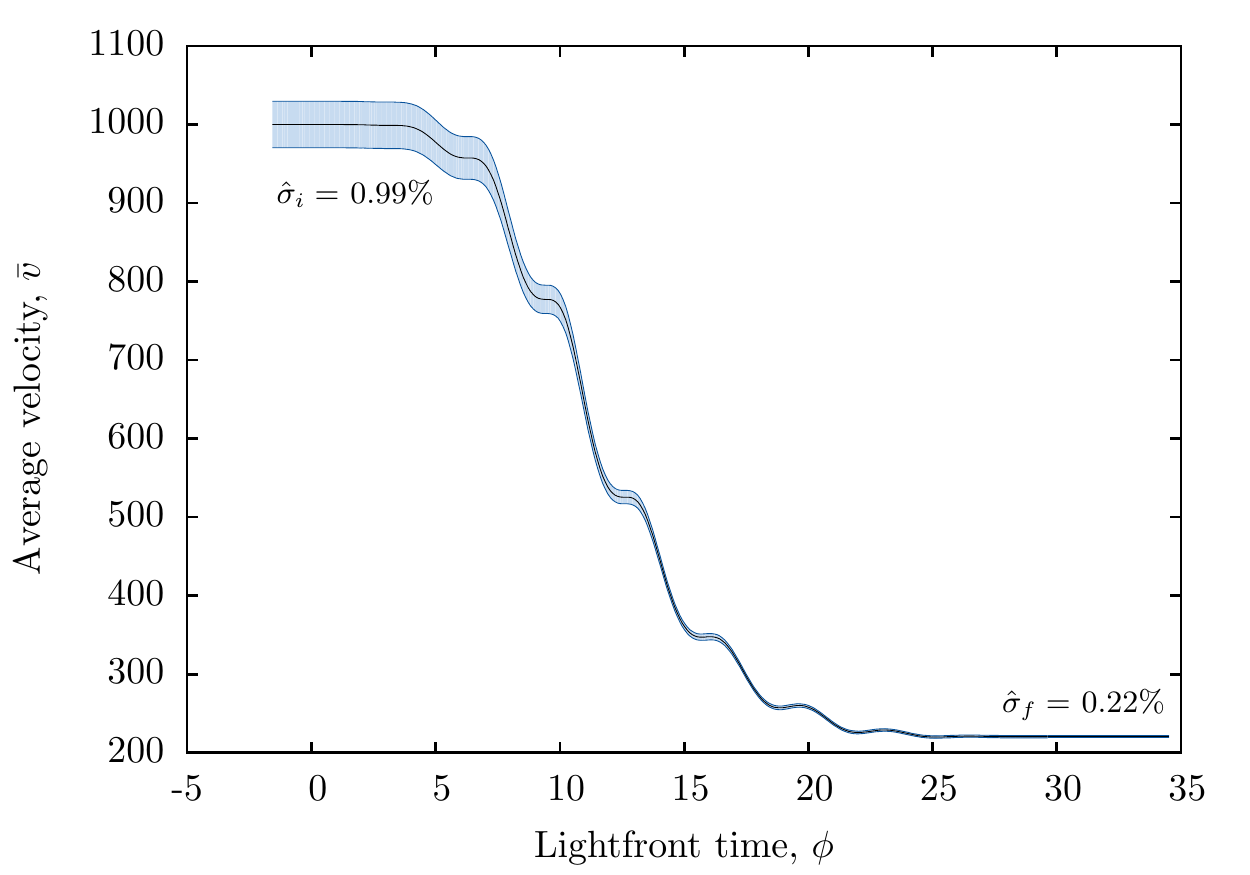}
\includegraphics[width=0.48\textwidth]{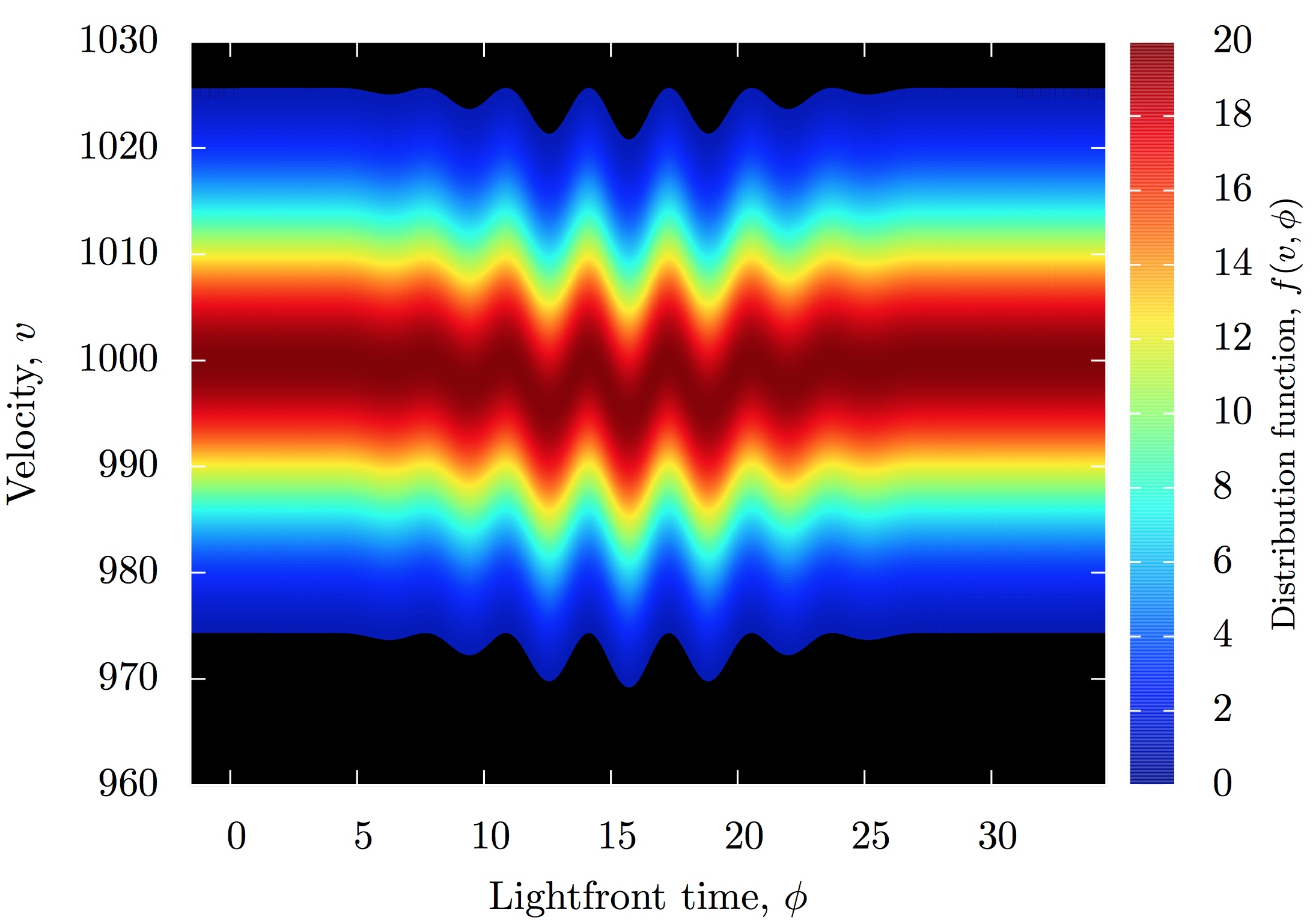}
\includegraphics[width=0.48\textwidth]{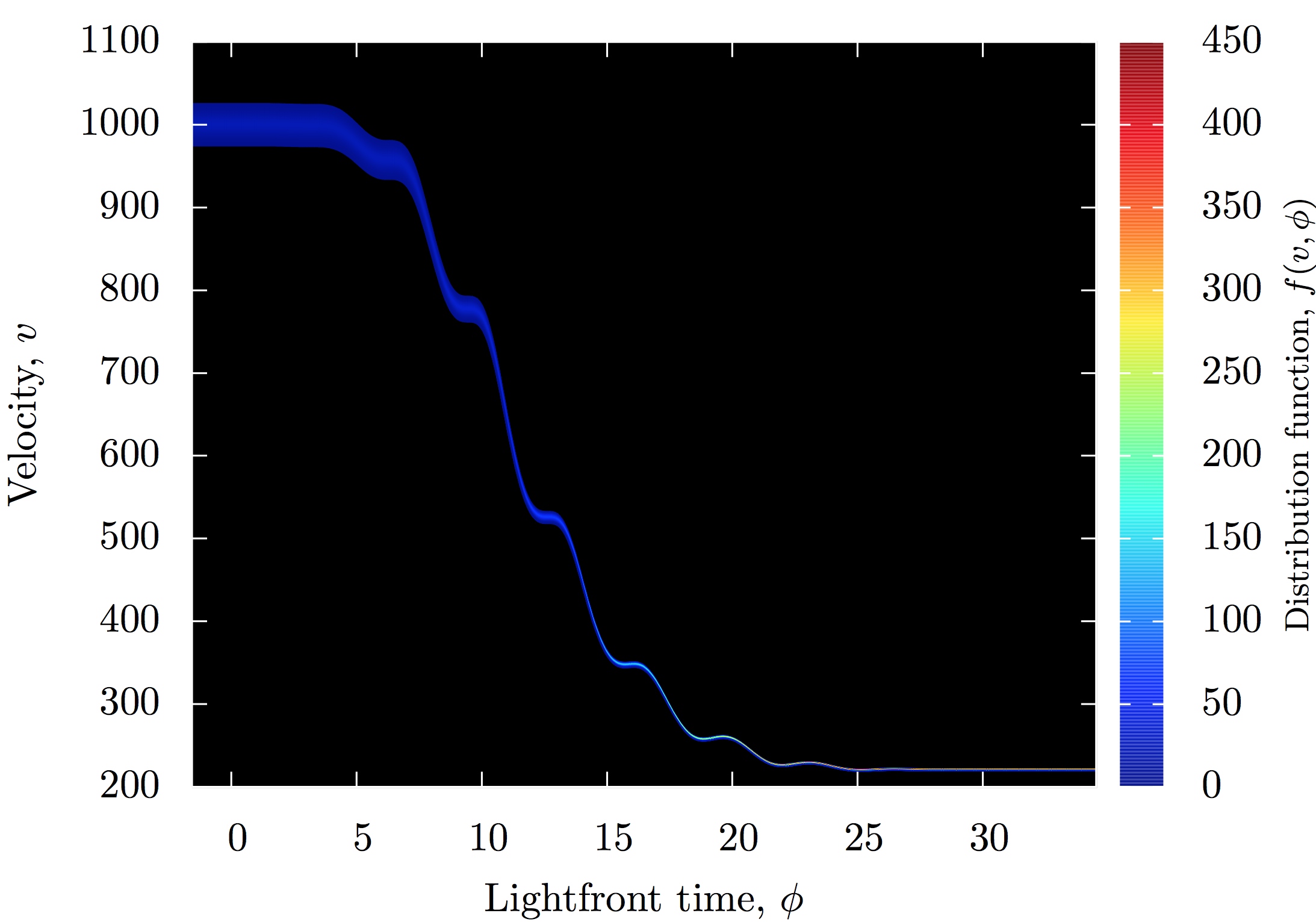}
\vspace{-1.5em}
\caption{\label{fig:dist_v1000_N5} 
Distribution for $N = 5$ \textit{without} (left) and \textit{with} (right) radiation reaction.}
\end{figure} 

\vspace{-1.5em}
\begin{figure}[H]
\centering
\includegraphics[width=0.48\textwidth]{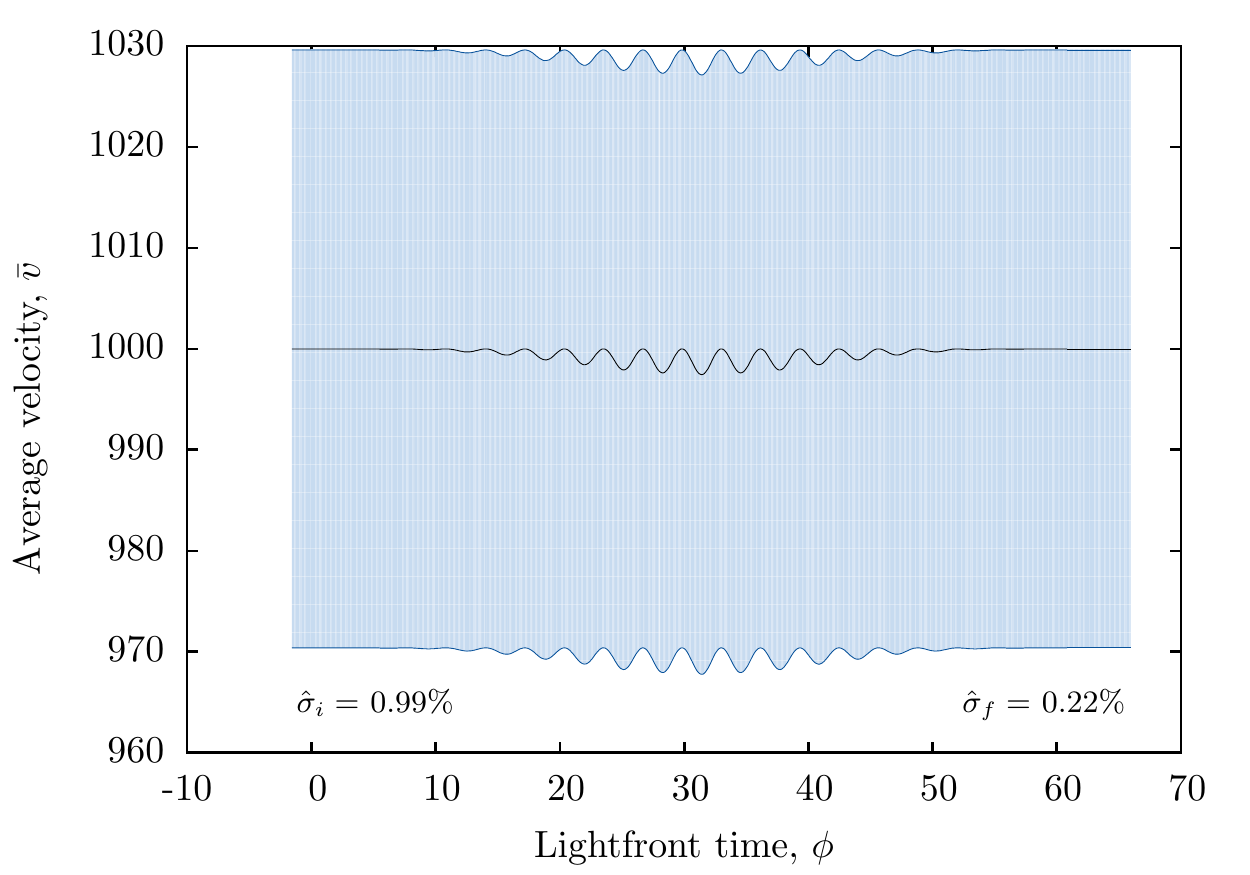}
\includegraphics[width=0.48\textwidth]{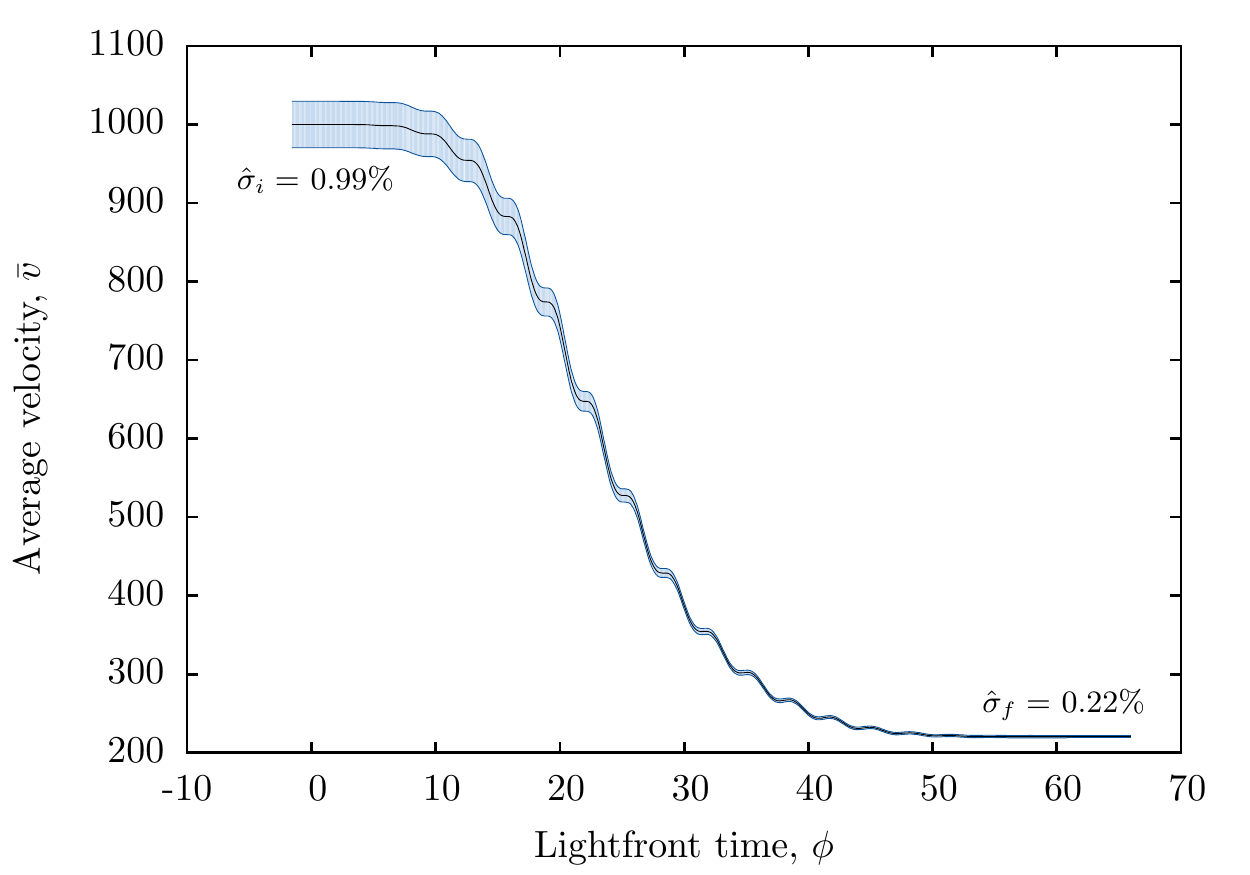}
\includegraphics[width=0.48\textwidth]{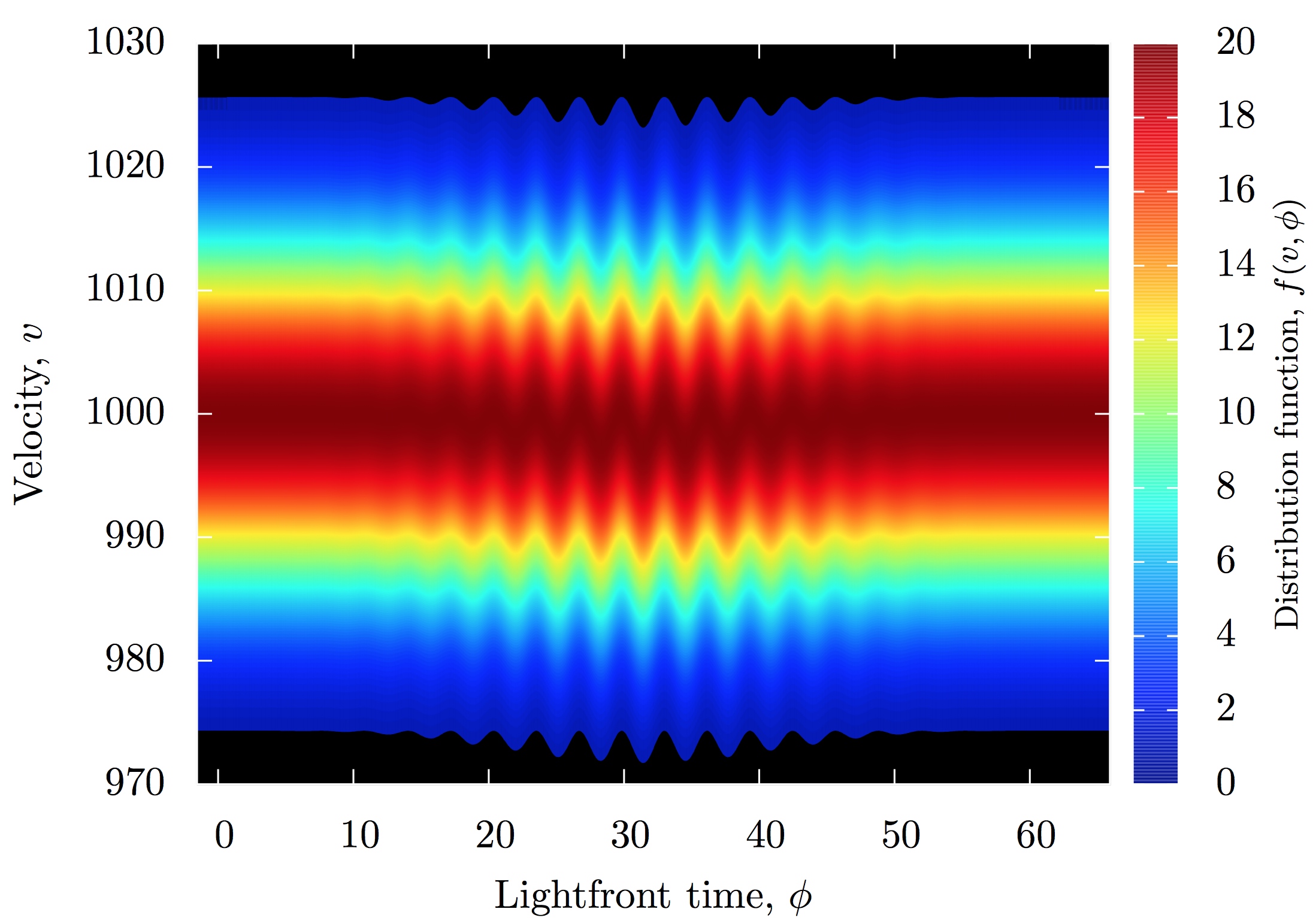}
\includegraphics[width=0.48\textwidth]{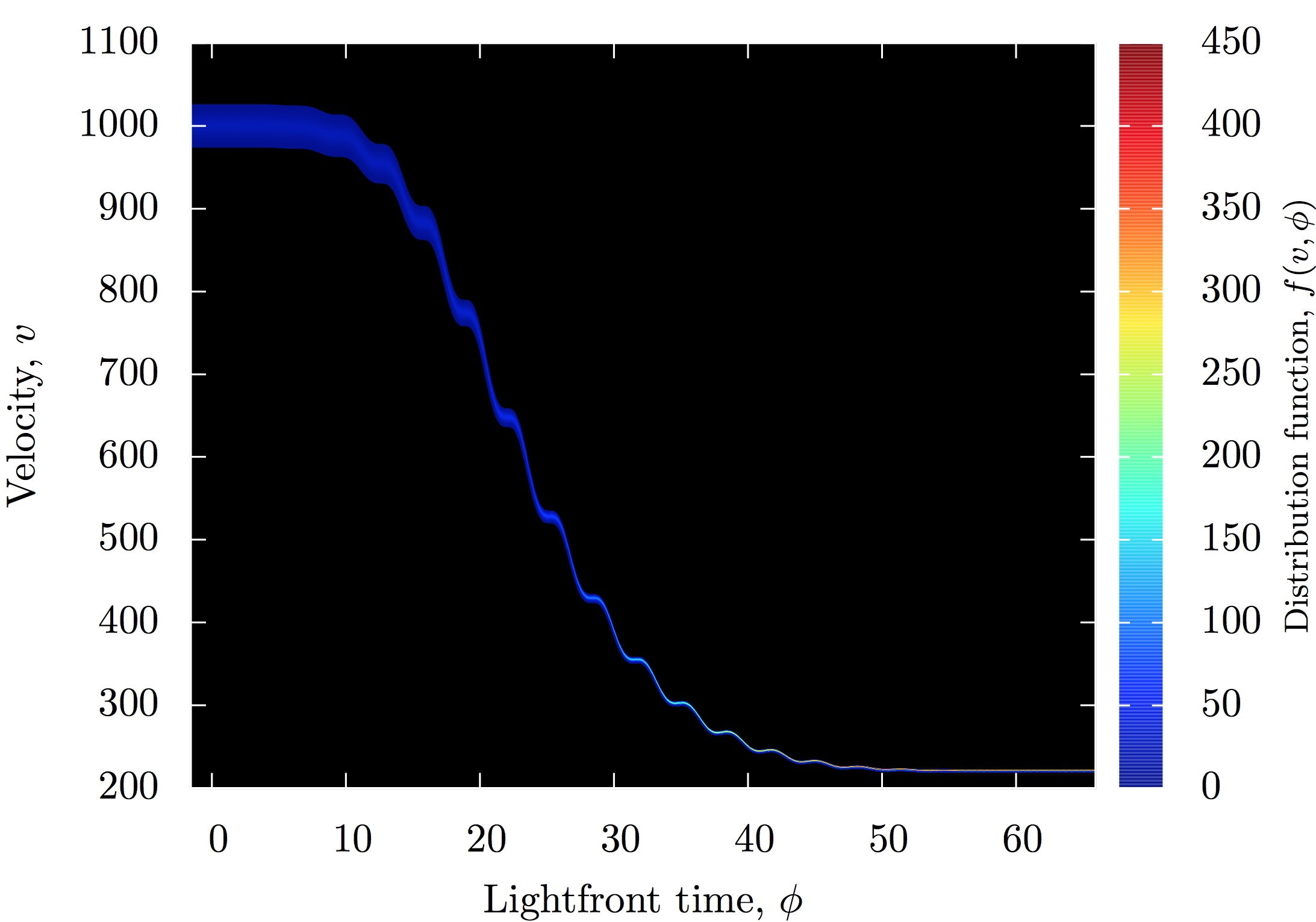}
\vspace{-1.5em}
\caption{\label{fig:dist_v1000_N10} 
Distribution for $N = 10$ \textit{without} (left) and \textit{with} (right) radiation reaction.}
\end{figure} 
\vspace{-1.5em}
\begin{figure}[H]
\centering
\includegraphics[width=0.48\textwidth]{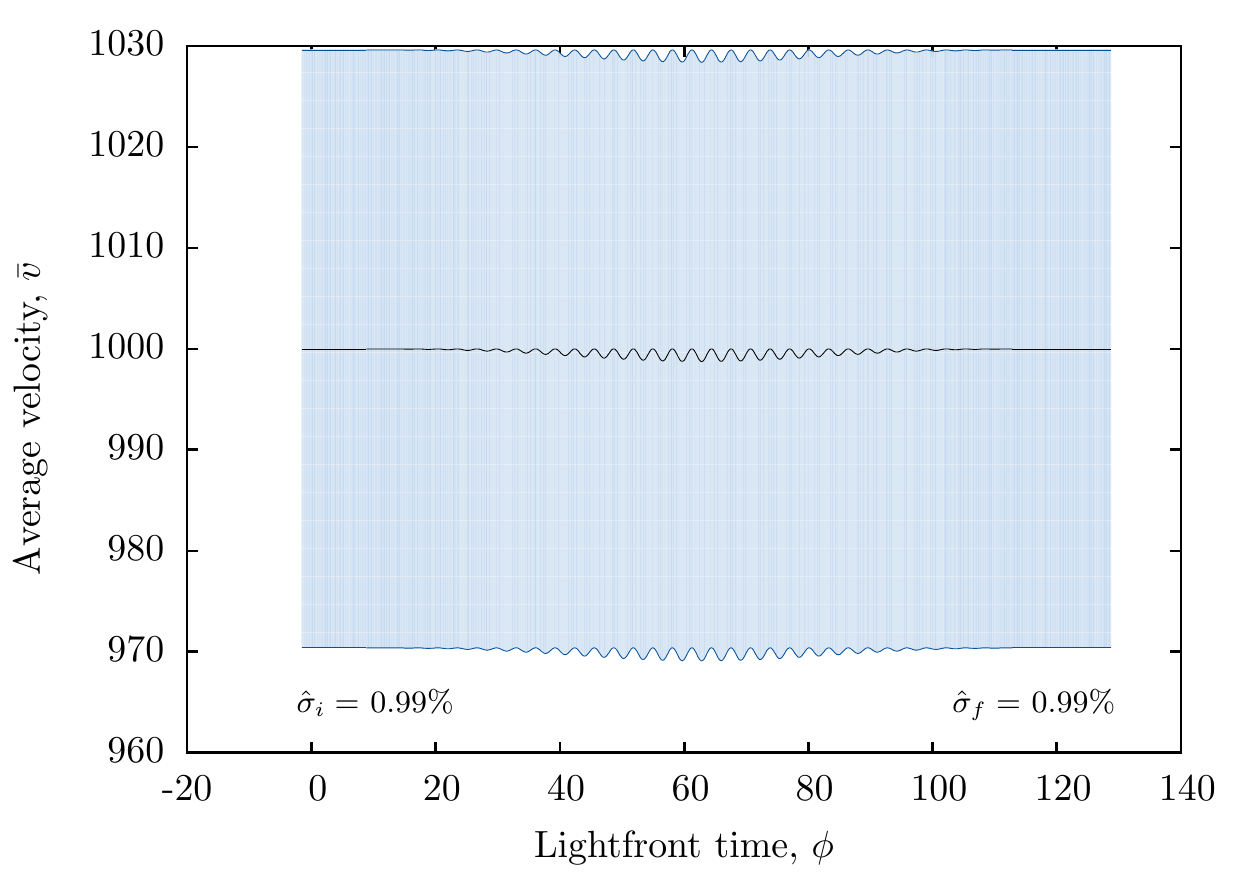}
\includegraphics[width=0.48\textwidth]{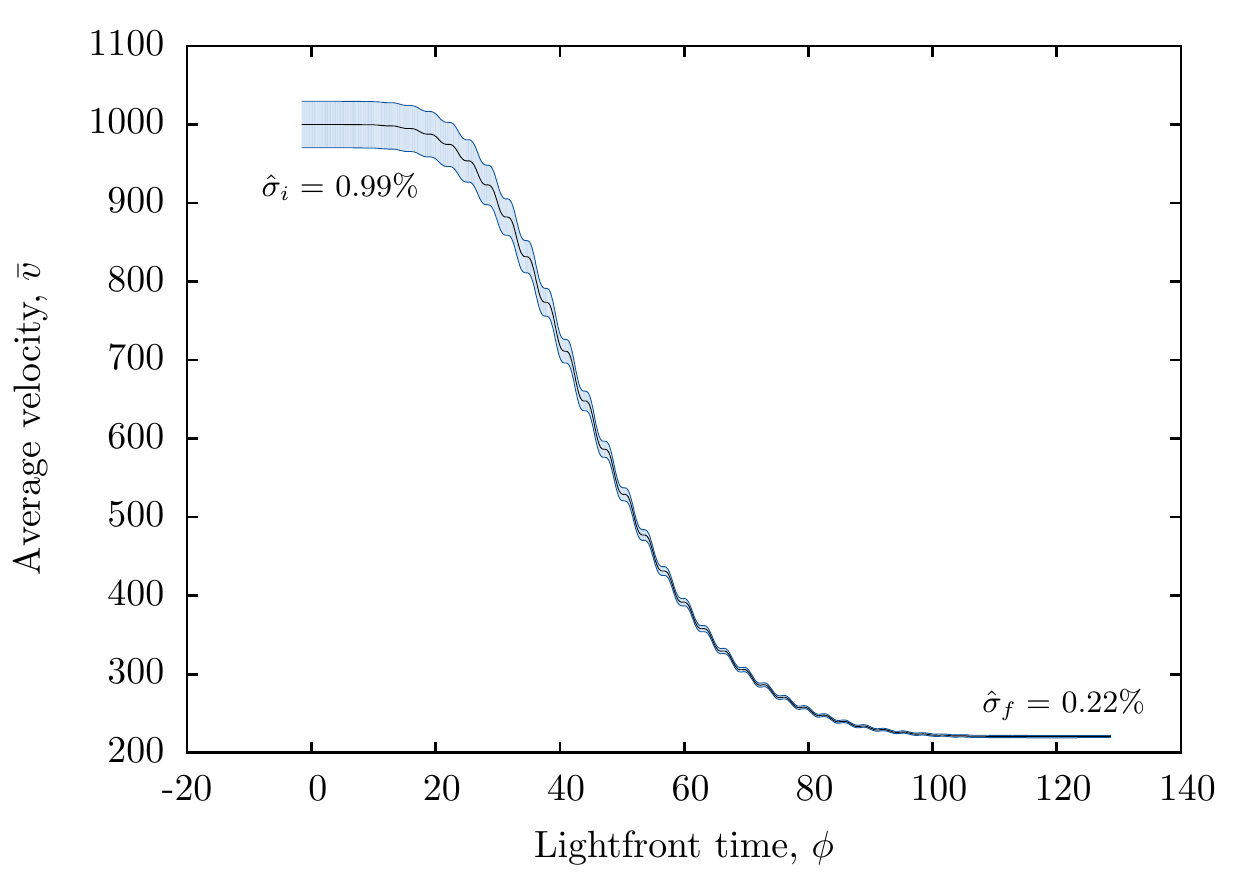}
\includegraphics[width=0.48\textwidth]{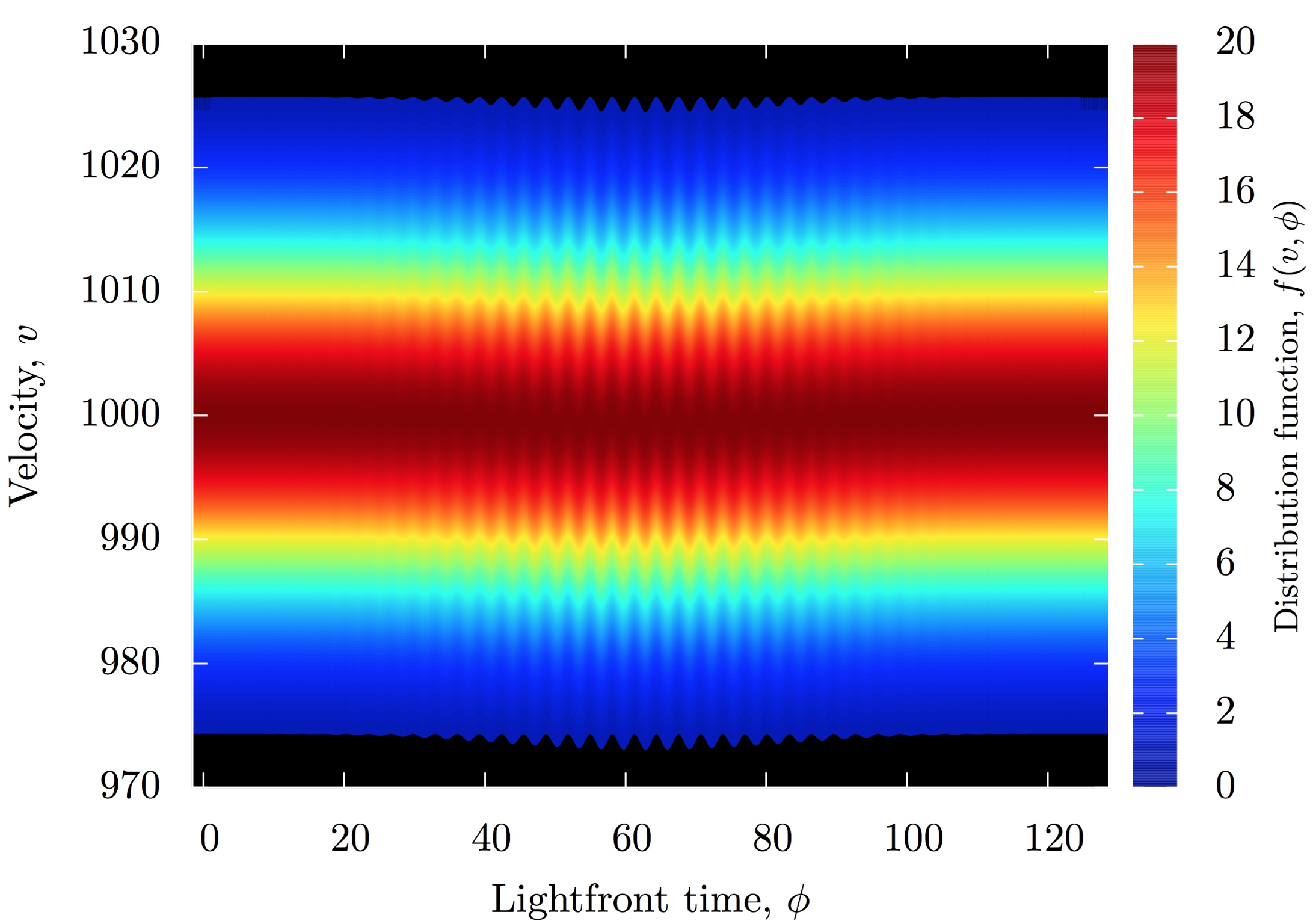}
\includegraphics[width=0.48\textwidth]{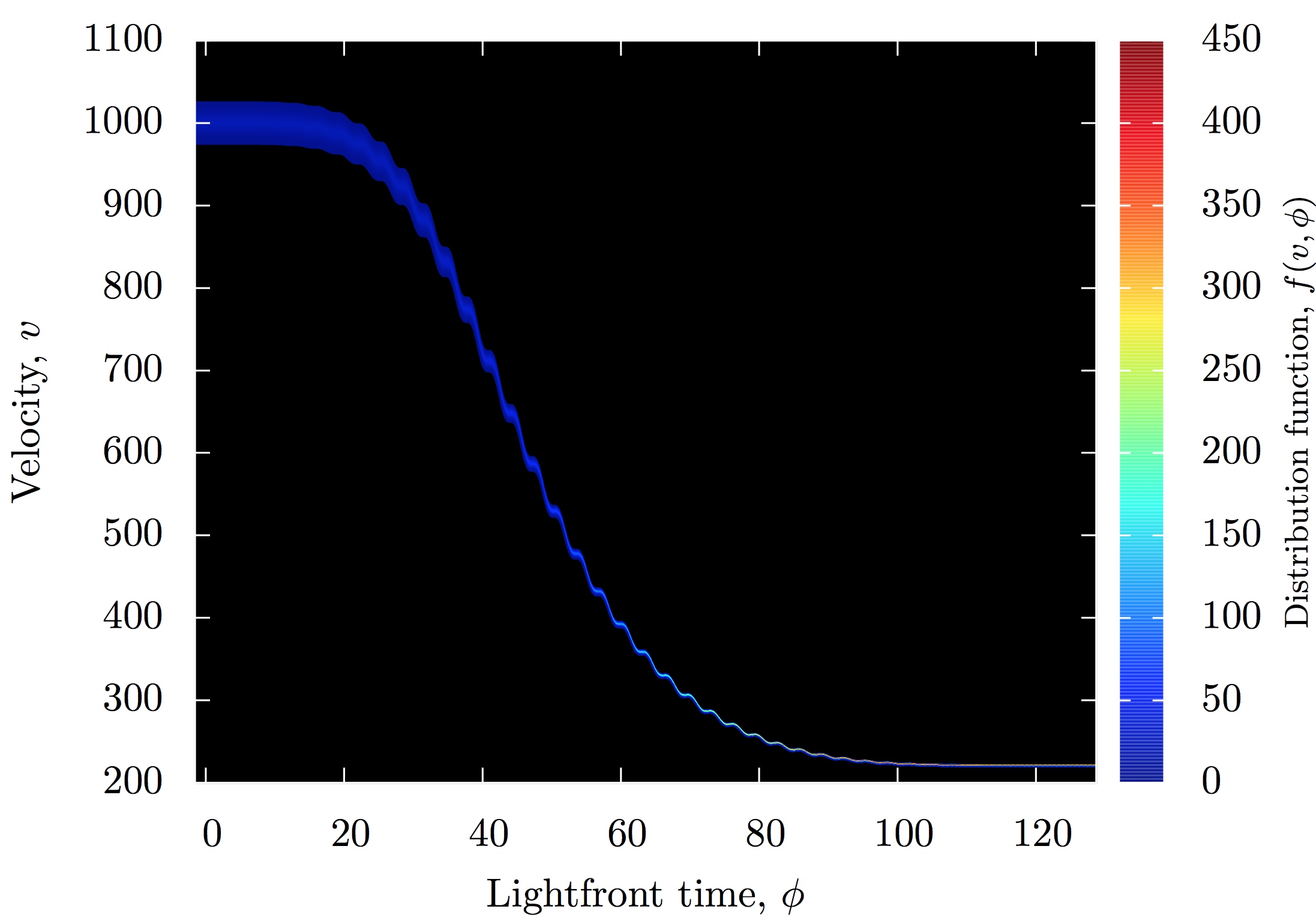}
\vspace{-1.5em}
\caption{\label{fig:dist_v1000_N20} 
Distribution for $N = 20$ \textit{without} (left) and \textit{with} (right) radiation reaction.}
\end{figure} 

\vspace{-1.5em}
\begin{figure}[H]
\centering
\includegraphics[width=0.48\textwidth]{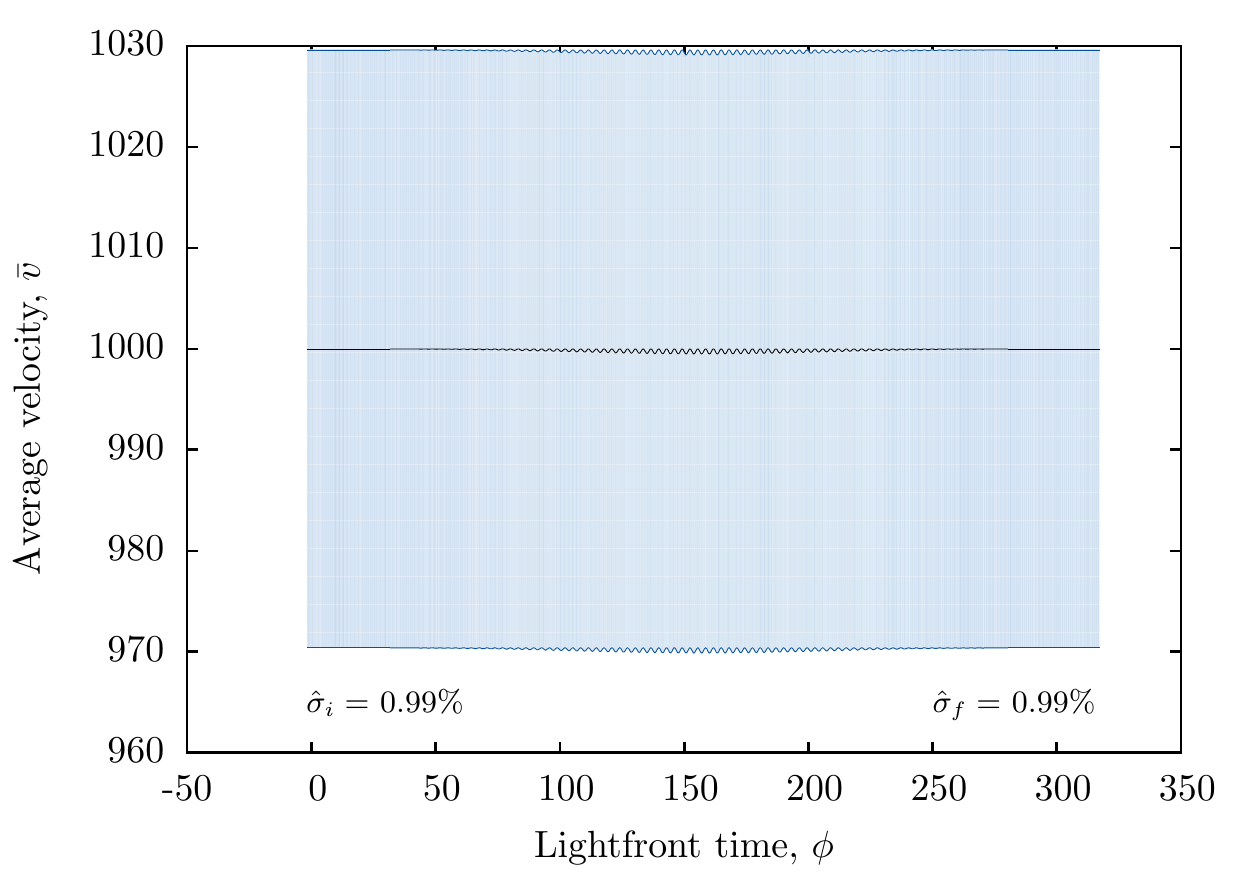}
\includegraphics[width=0.48\textwidth]{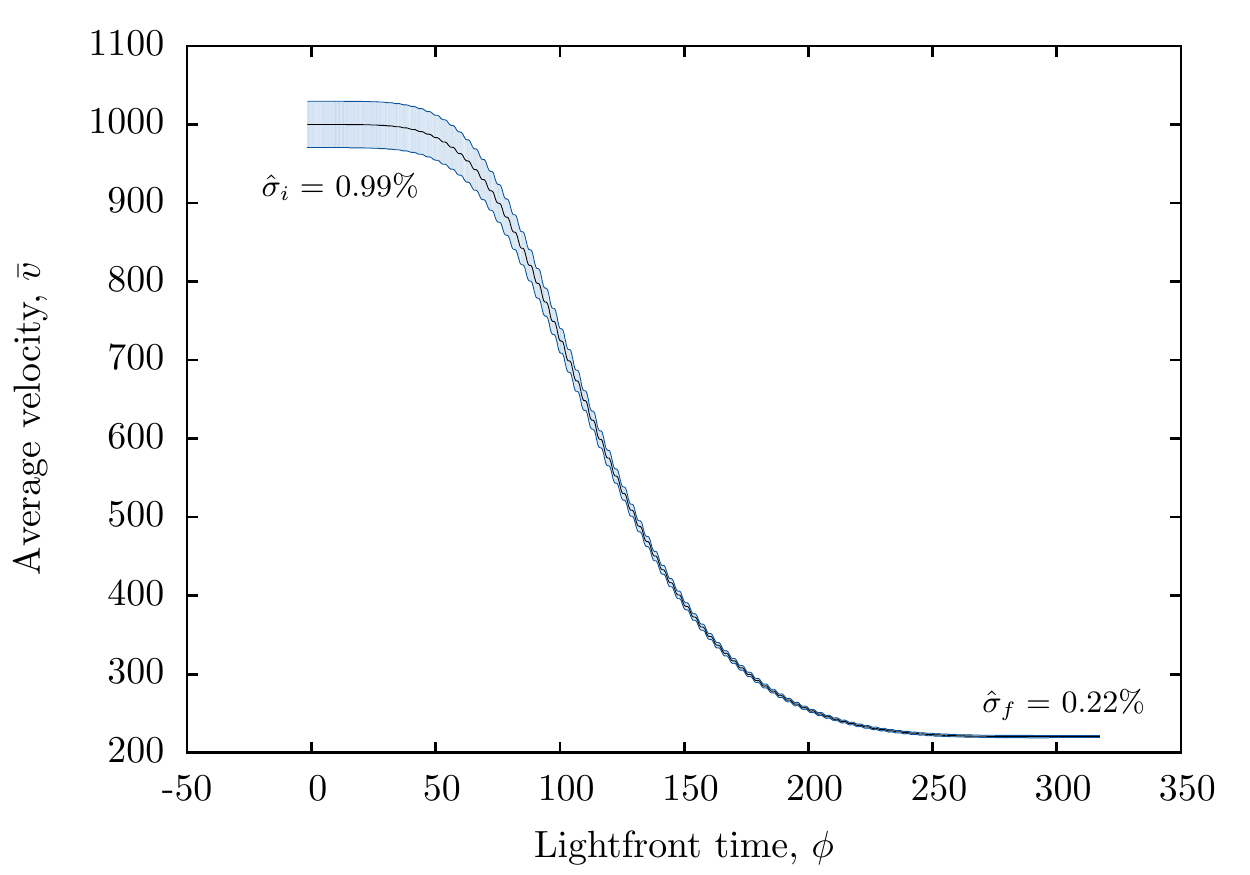}
\includegraphics[width=0.48\textwidth]{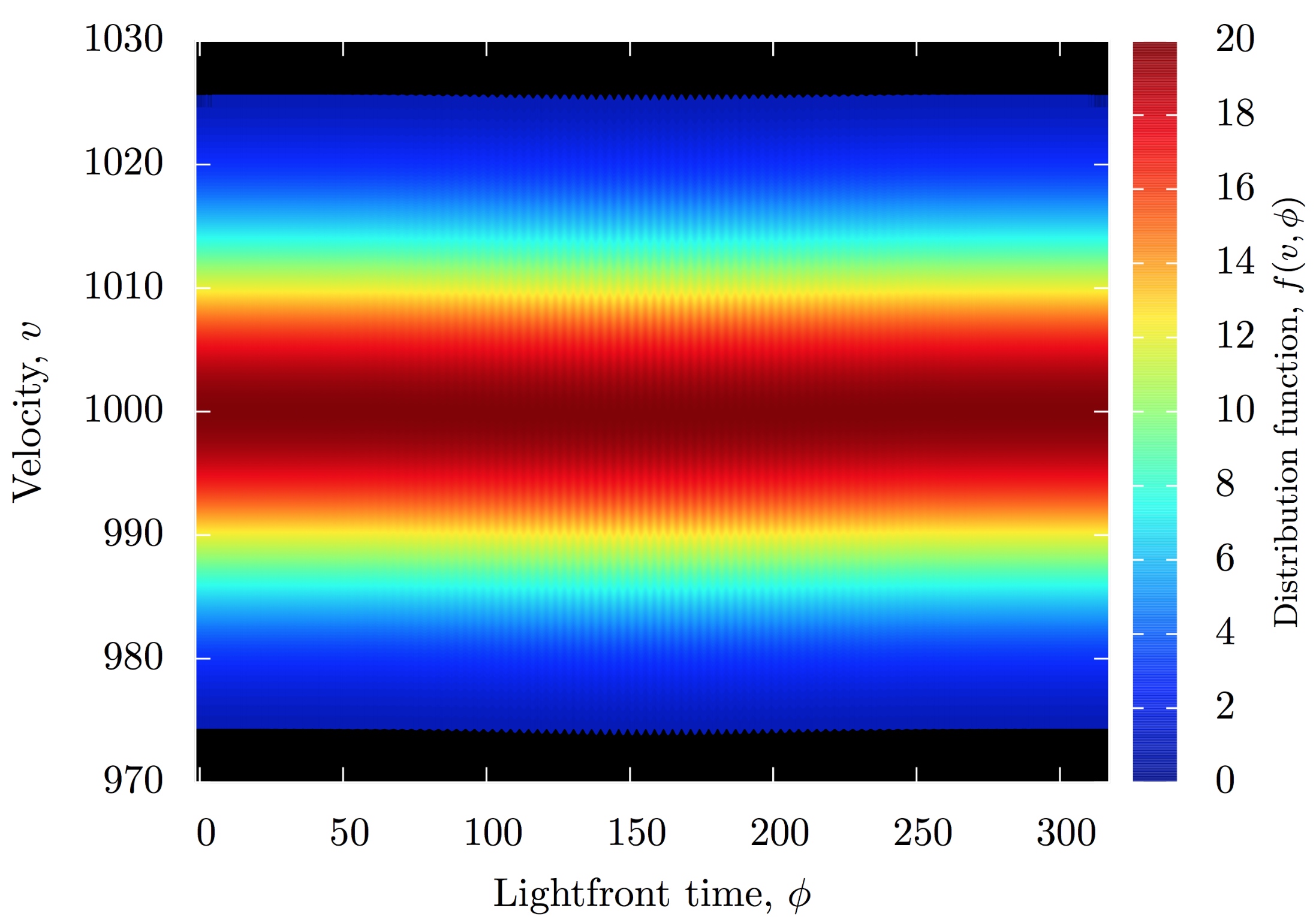}
\includegraphics[width=0.48\textwidth]{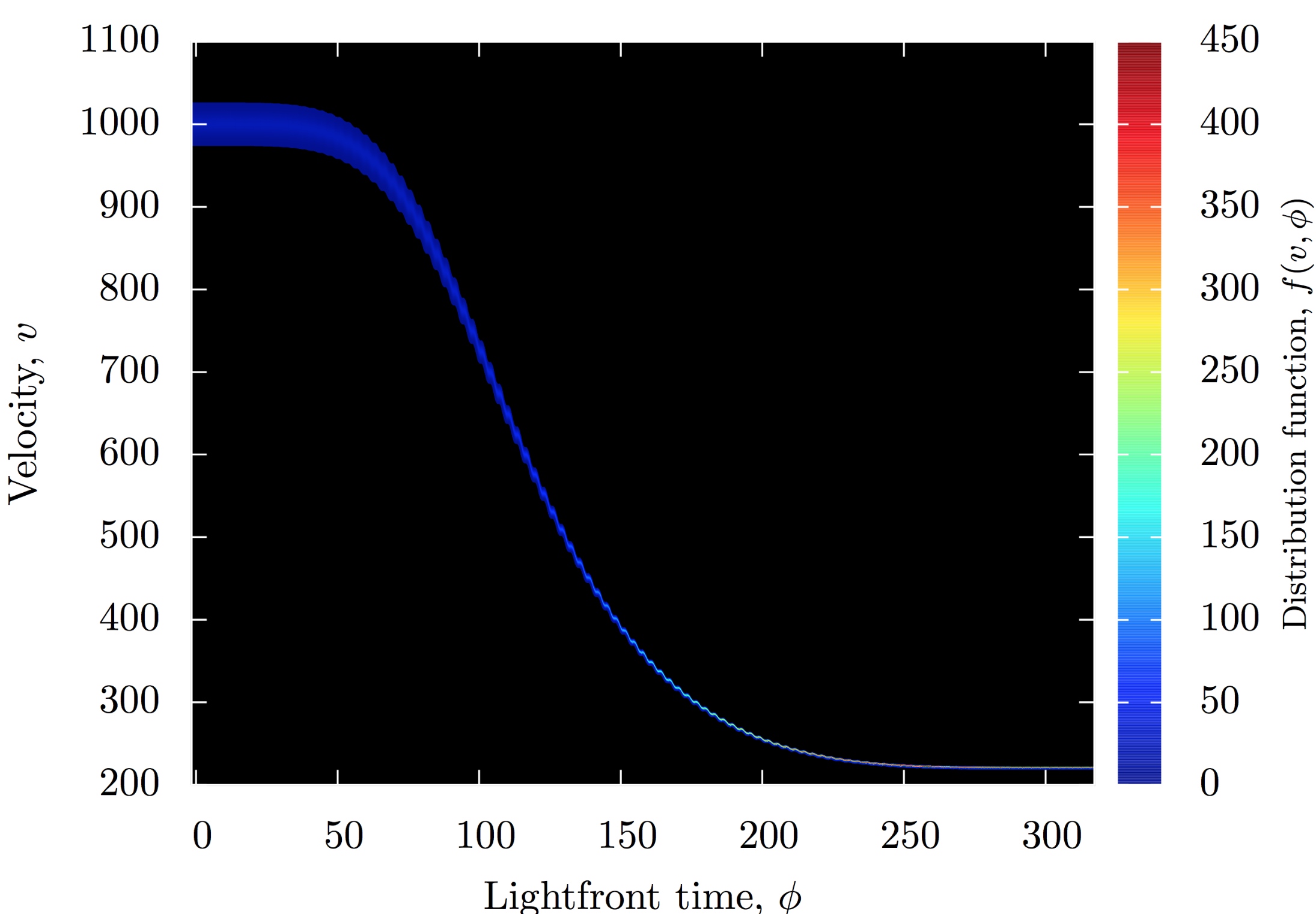}
\vspace{-1.5em}
\caption{\label{fig:dist_v1000_N50} 
Distribution for $N = 50$ \textit{without} (left) and \textit{with} (right) radiation reaction.}
\end{figure} 

\vspace{-1.5em}
\begin{figure}[H]
\centering
\includegraphics[width=0.48\textwidth]{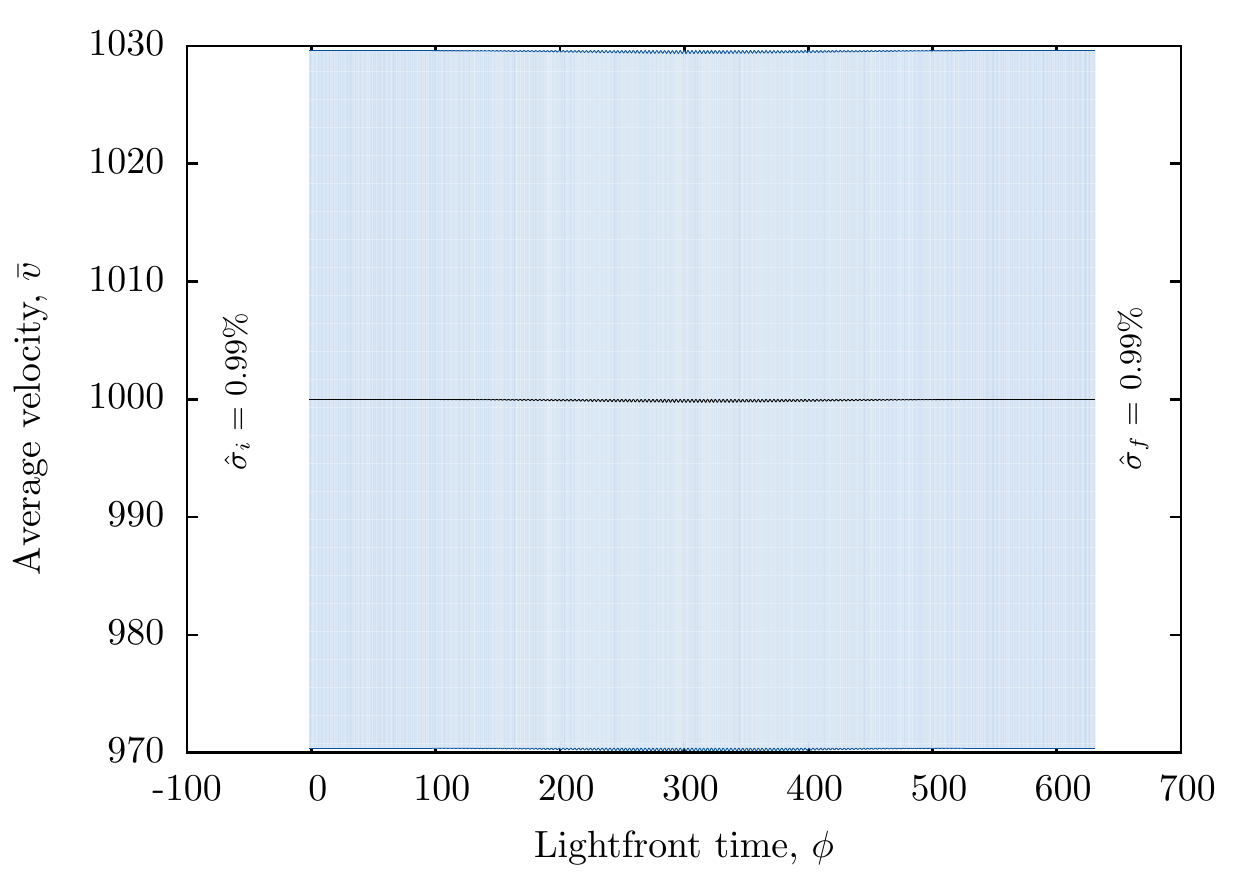}
\includegraphics[width=0.48\textwidth]{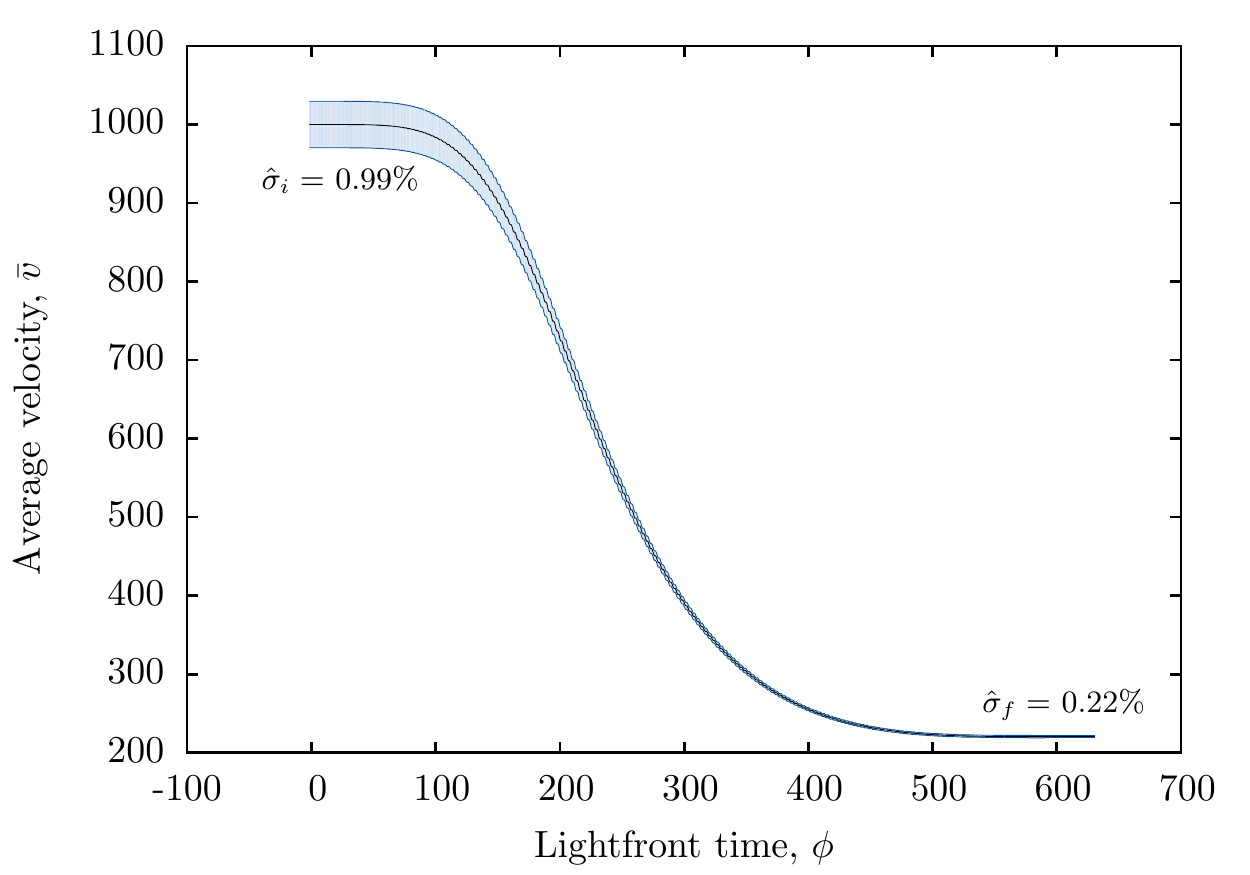}
\includegraphics[width=0.48\textwidth]{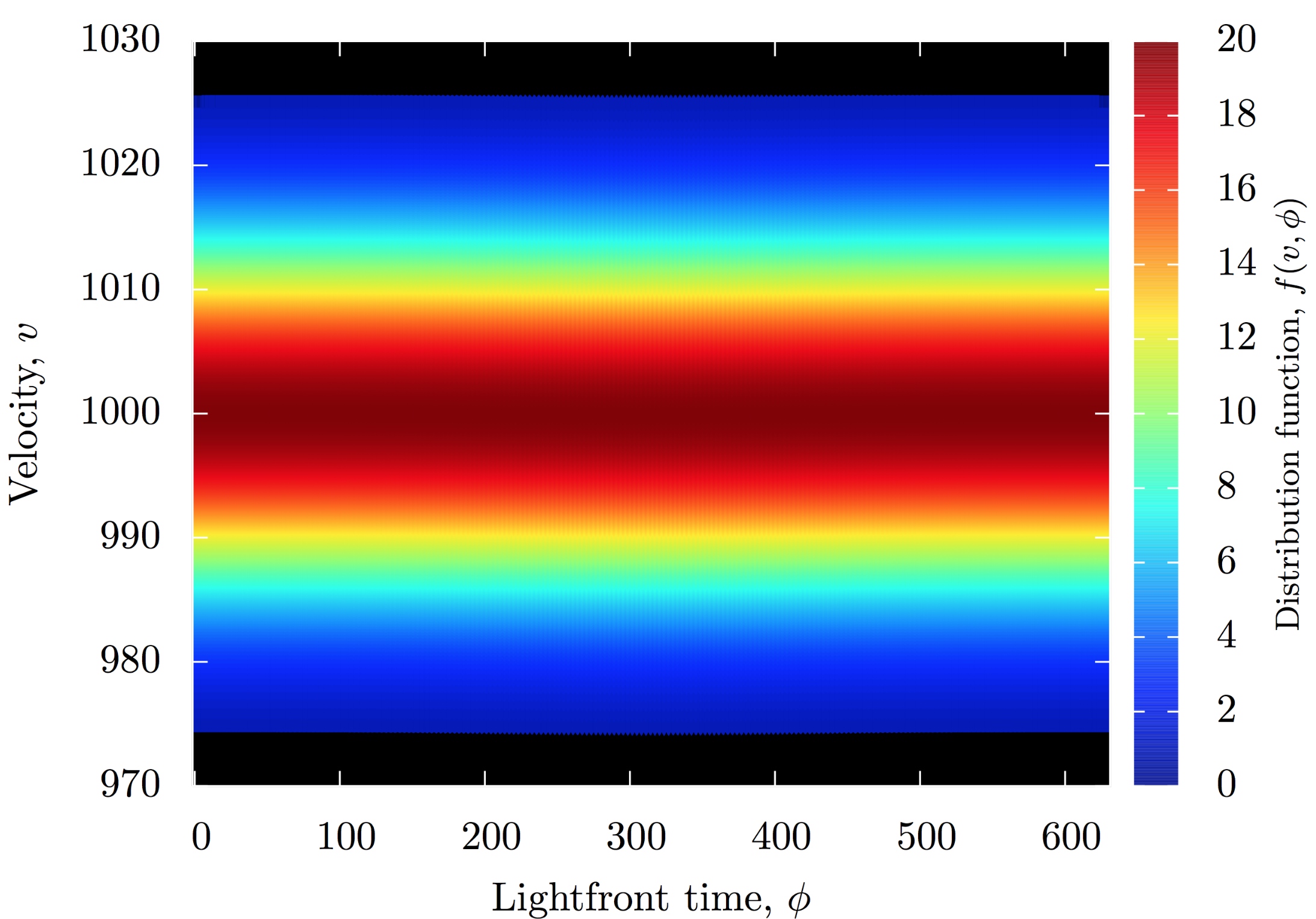}
\includegraphics[width=0.48\textwidth]{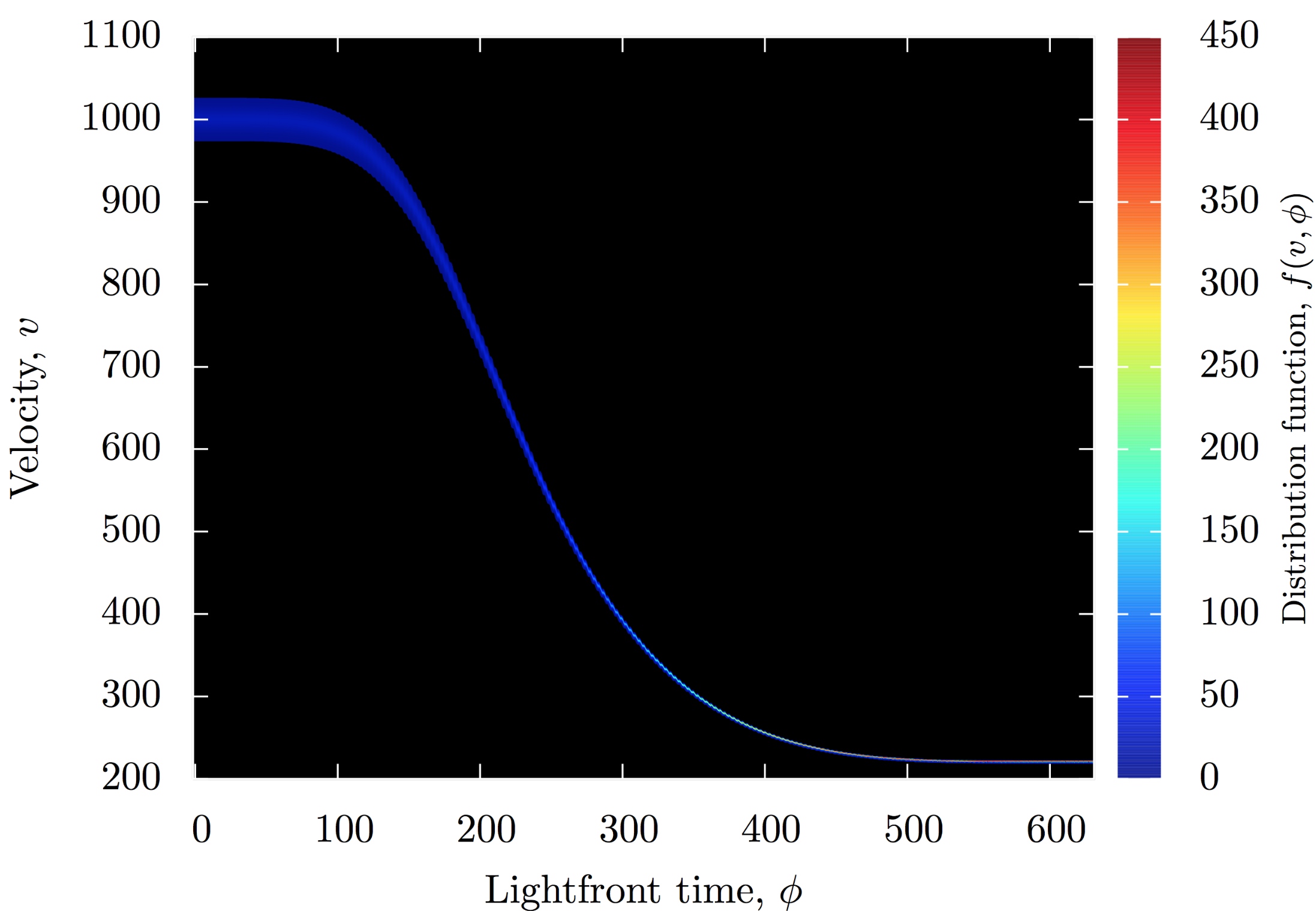}
\vspace{-1.5em}
\caption{\label{fig:dist_v1000_N100} 
Distribution for $N = 100$ \textit{without} (left) and \textit{with} (right) radiation reaction.}
\end{figure} 

\subsection{Particle bunch with a central velocity of $\bar{v} = 10^4$}

All the particles start at a single point in space in front of the laser pulse and are evaluated to the point of exit from the pulse which has energy $E=\frac{3\pi}{8} \cdot 10^5$.

For this case we consider the following laser pulses of different length:
    \begin{enumerate}
    \vspace{-0.5em}
    \item Short laser pulse with peak $a_0 = 141.4$ and $N = 5$ oscillations
    \vspace{-0.5em}
    \item Laser pulse with peak $a_0 = 100$ and $N = 10$ oscillations
    \vspace{-0.5em}
    \item Long pulse with peak $a_0 = 44.7$ and $N = 50$ oscillations
    \end{enumerate}

\vspace{-0.5em}
The evolutions are tracked using two different approaches. We consider cases with \textit{no radiation reaction} and with the \textit{Landau-Lifshitz} radiation reaction force.

\begin{figure}[H]
\centering
\includegraphics[width=0.48\textwidth]{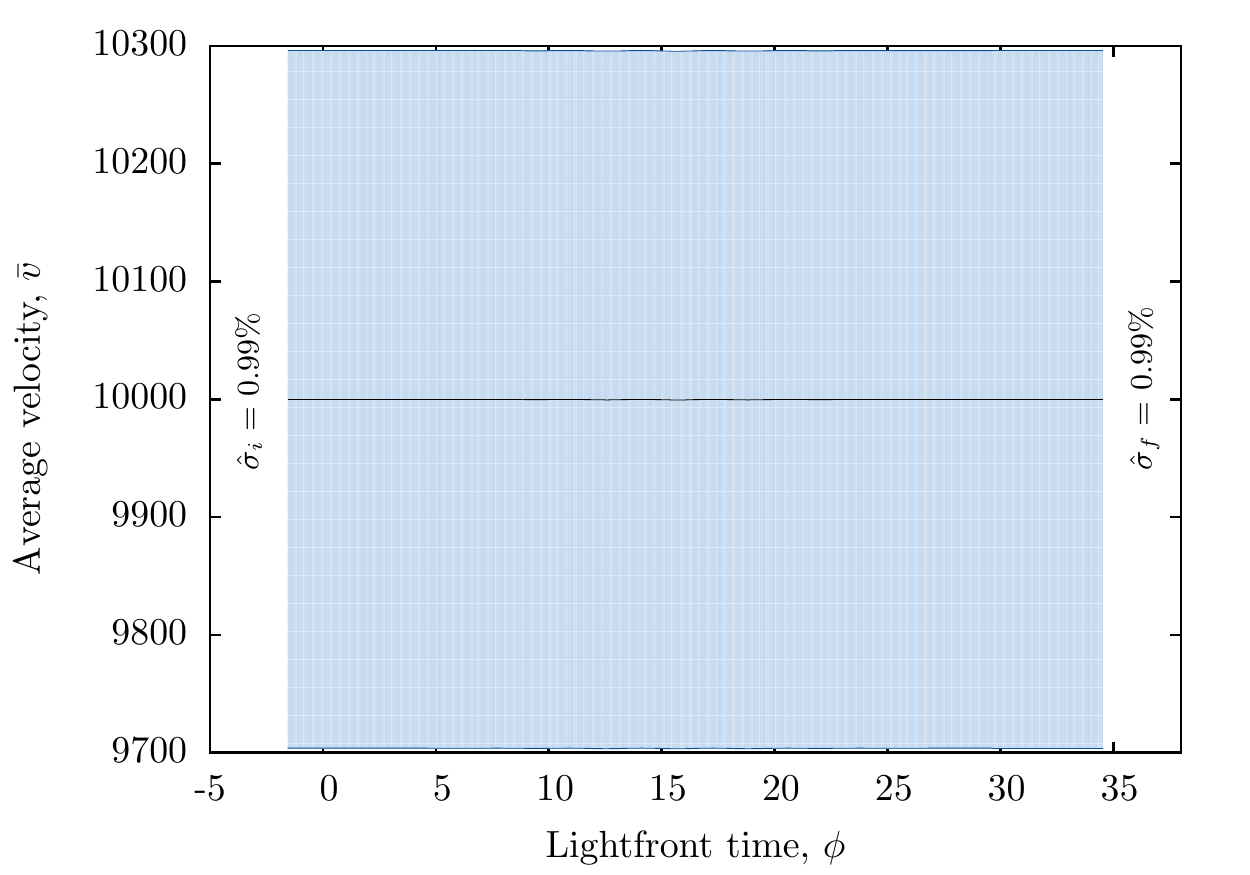}
\includegraphics[width=0.48\textwidth]{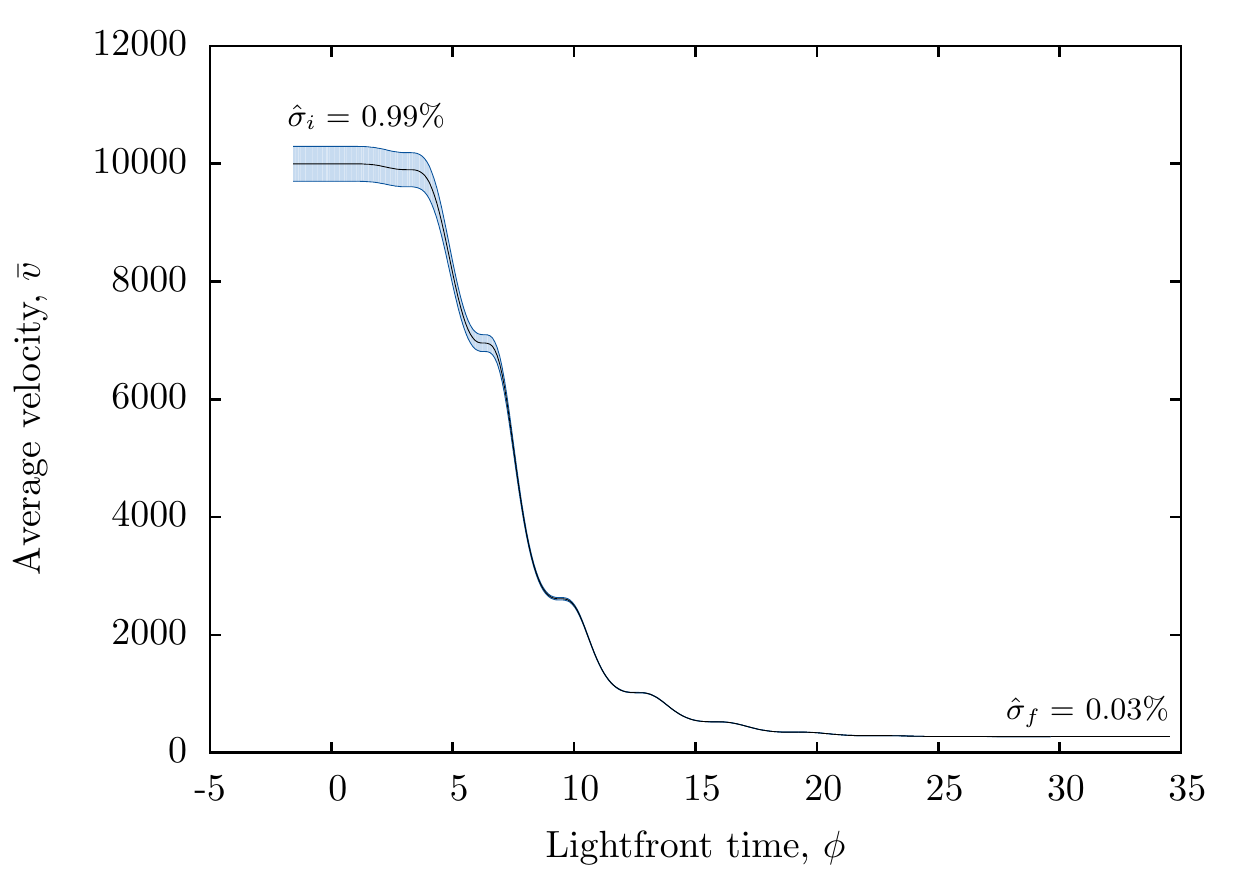}
\includegraphics[width=0.48\textwidth]{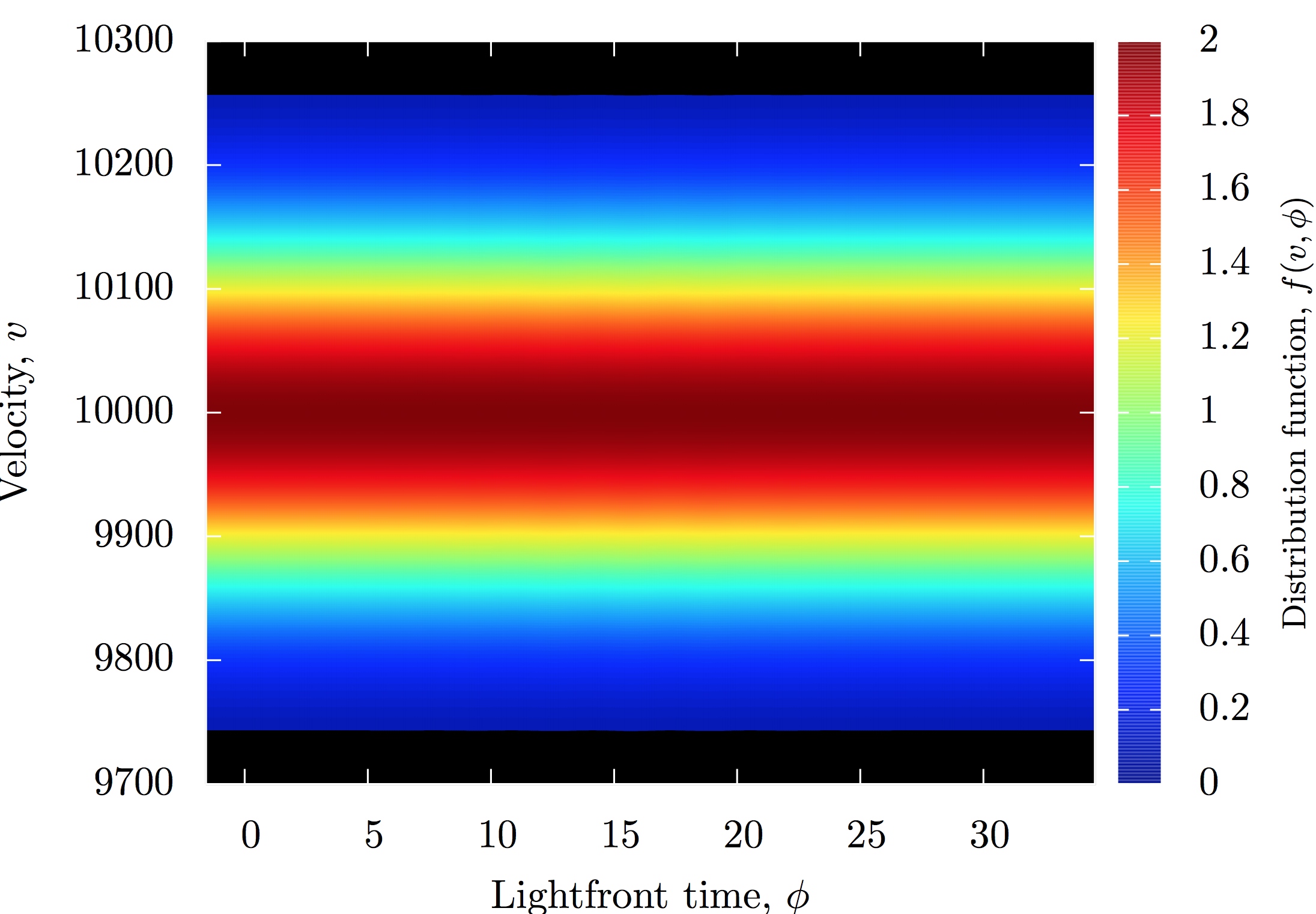}
\includegraphics[width=0.48\textwidth]{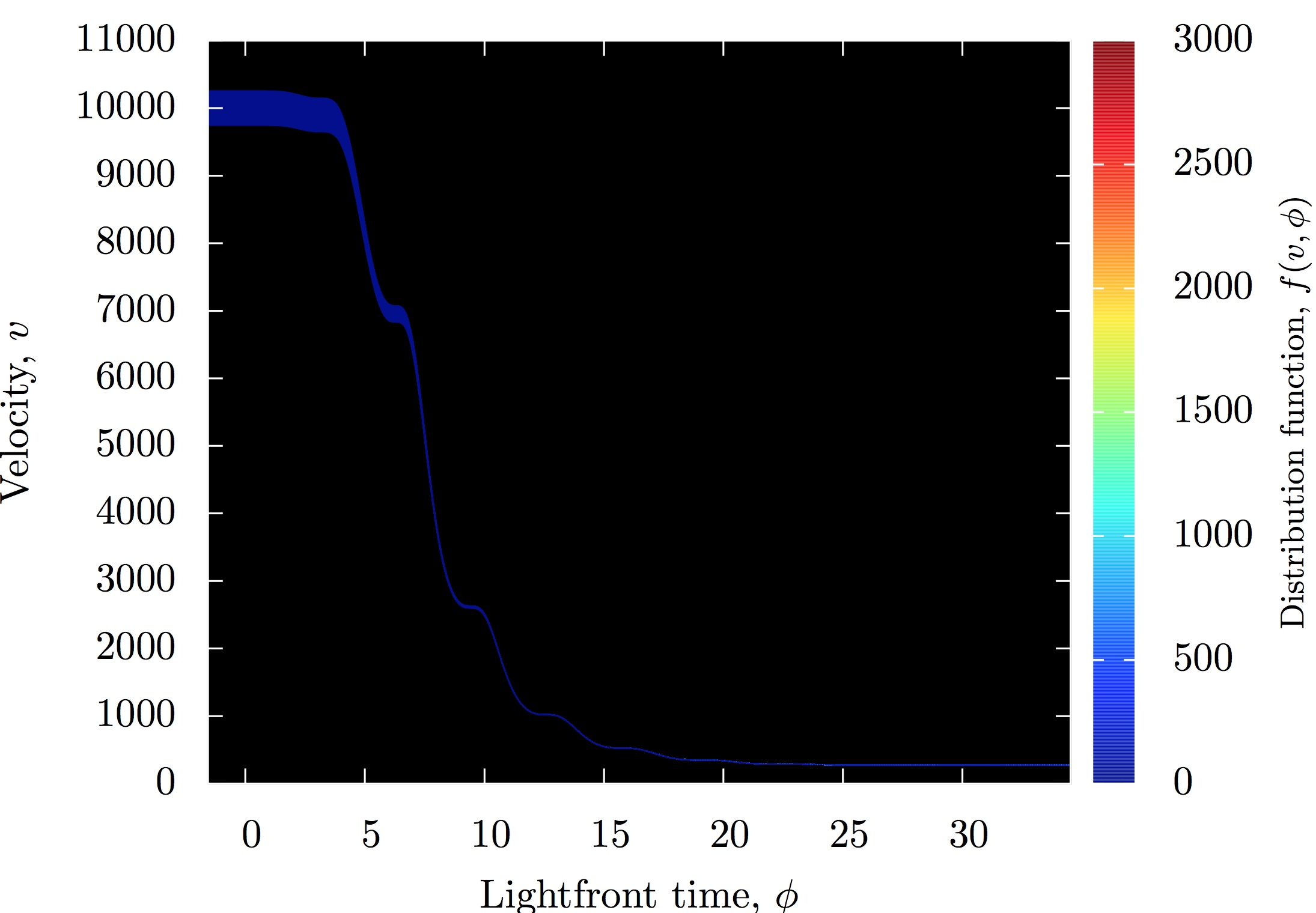}
\vspace{-1.5em}
\caption{\label{fig:dist_v10000_N5} 
Distribution for $N = 5$ \textit{without} (left) and \textit{with} (right) radiation reaction.}
\end{figure} 
\vspace{-1.5em}
\begin{figure}[H]
\centering
\includegraphics[width=0.48\textwidth]{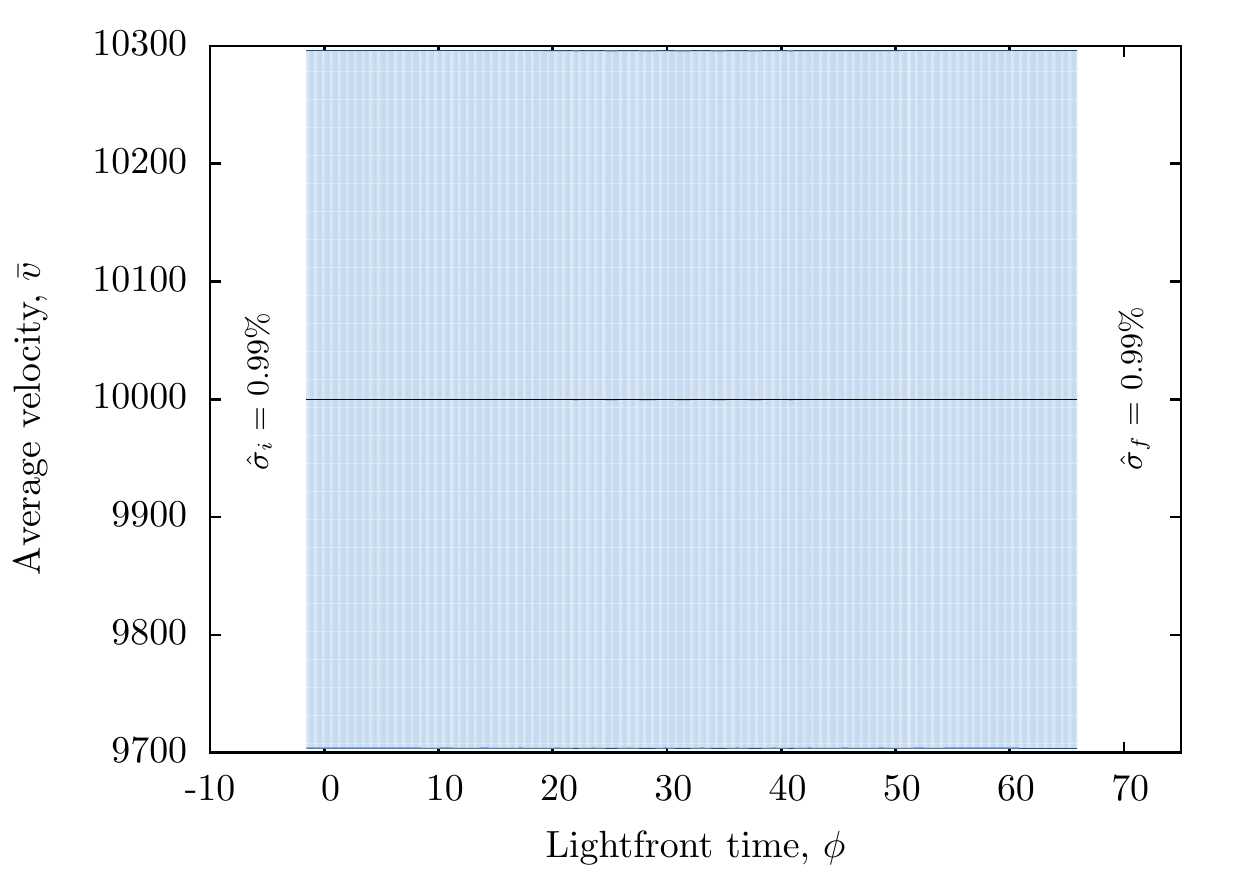}
\includegraphics[width=0.48\textwidth]{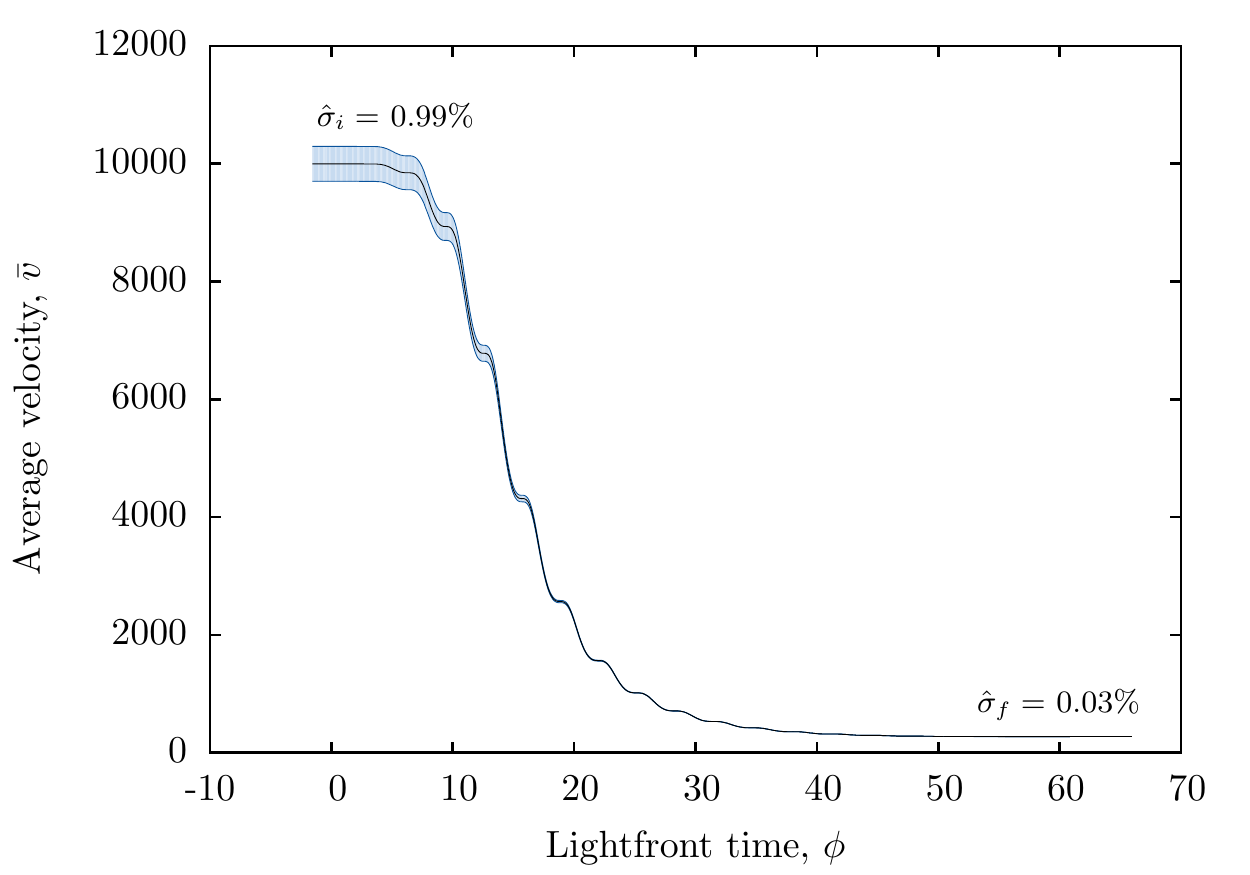}
\includegraphics[width=0.48\textwidth]{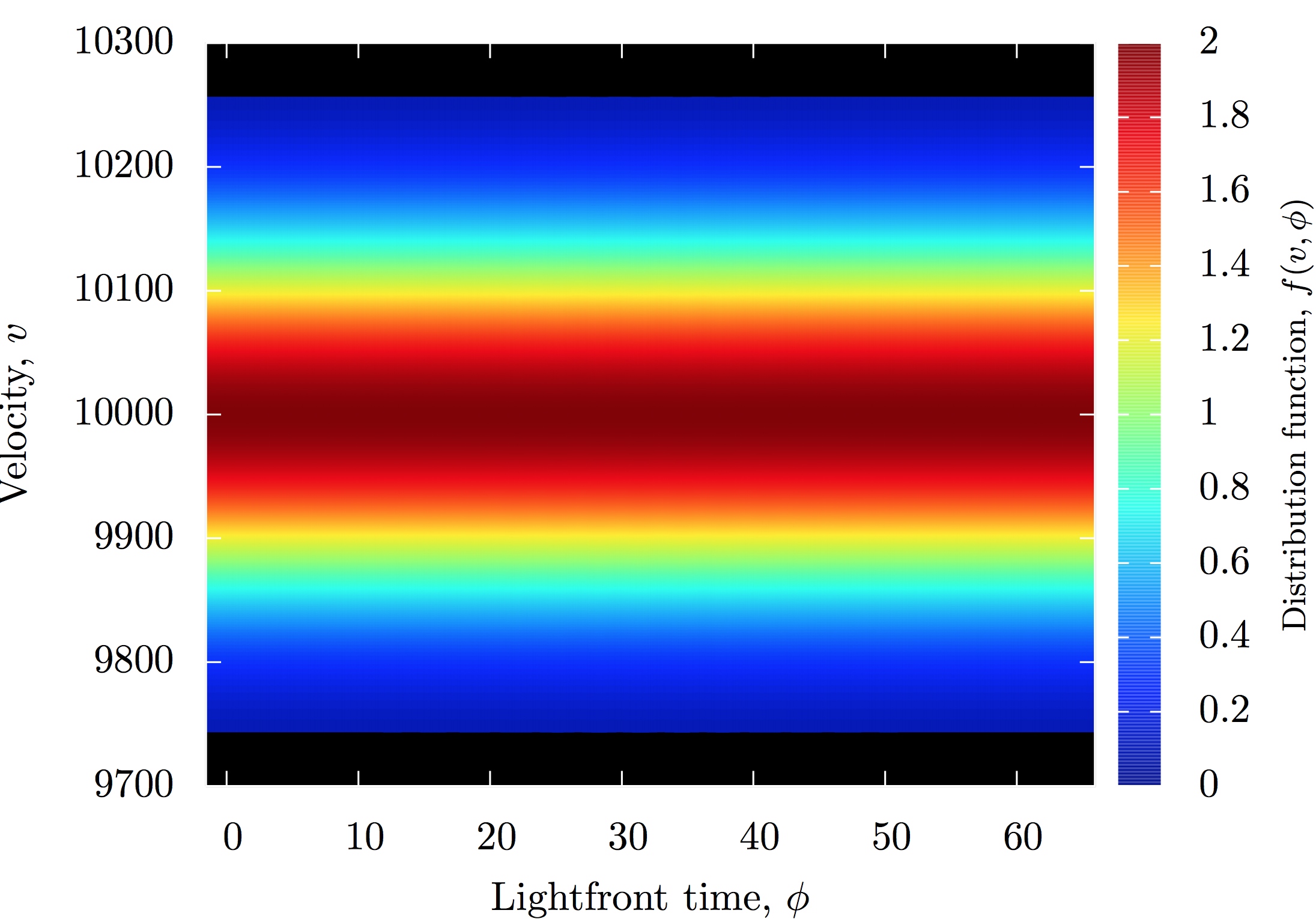}
\includegraphics[width=0.48\textwidth]{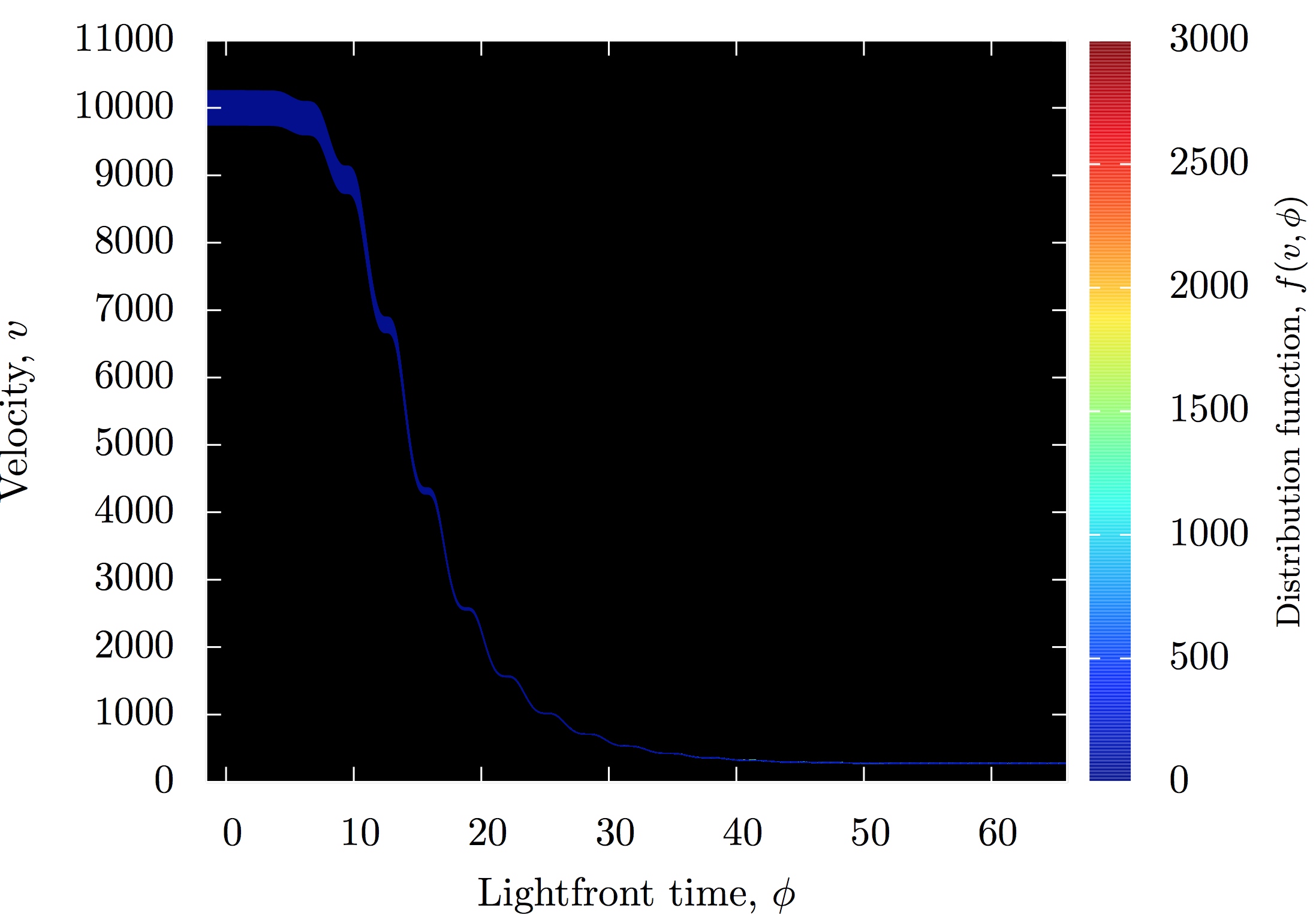}
\vspace{-1.5em}
\caption{\label{fig:dist_v10000_N10} 
Distribution for $N = 10$ \textit{without} (left) and \textit{with} (right) radiation reaction.}
\end{figure} 

\vspace{-1.5em}
\begin{figure}[H]
\centering
\includegraphics[width=0.48\textwidth]{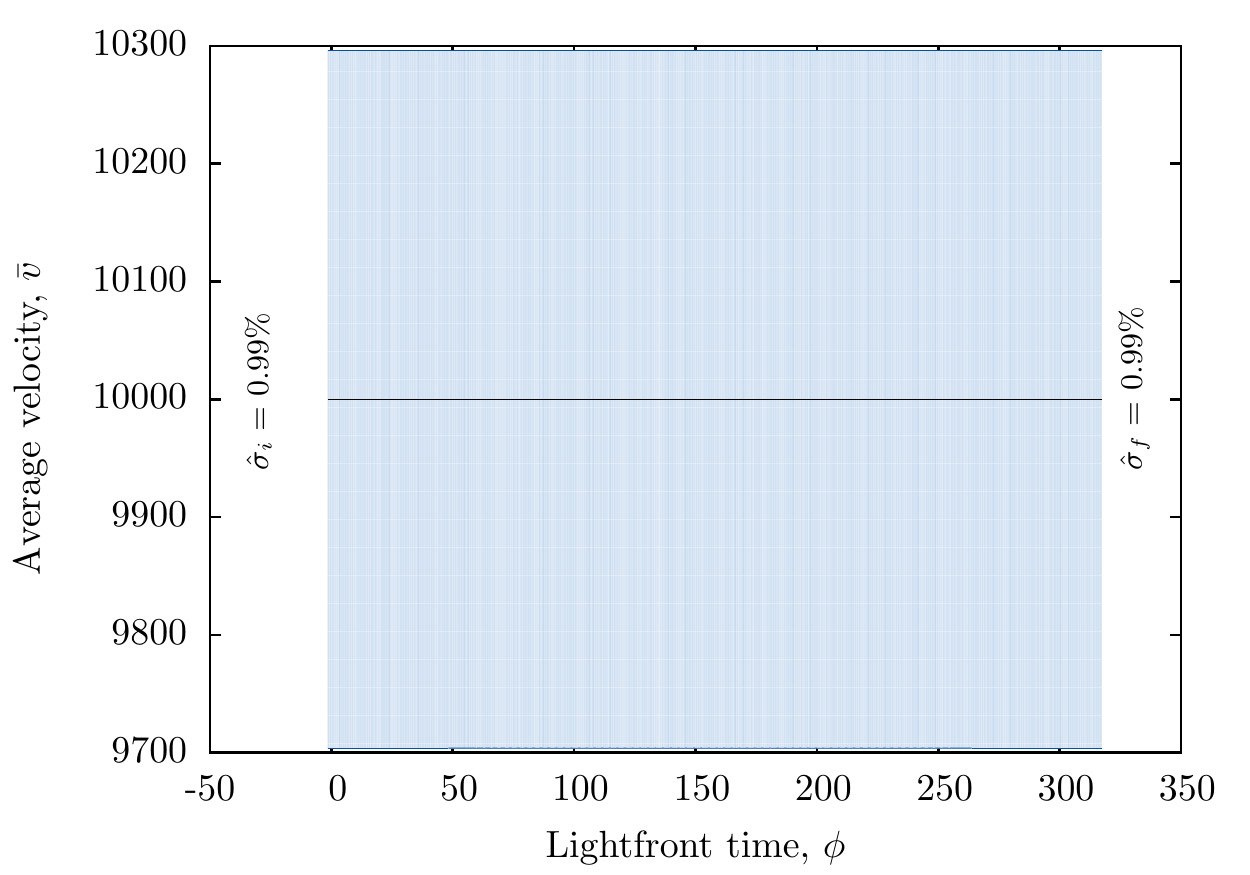}
\includegraphics[width=0.48\textwidth]{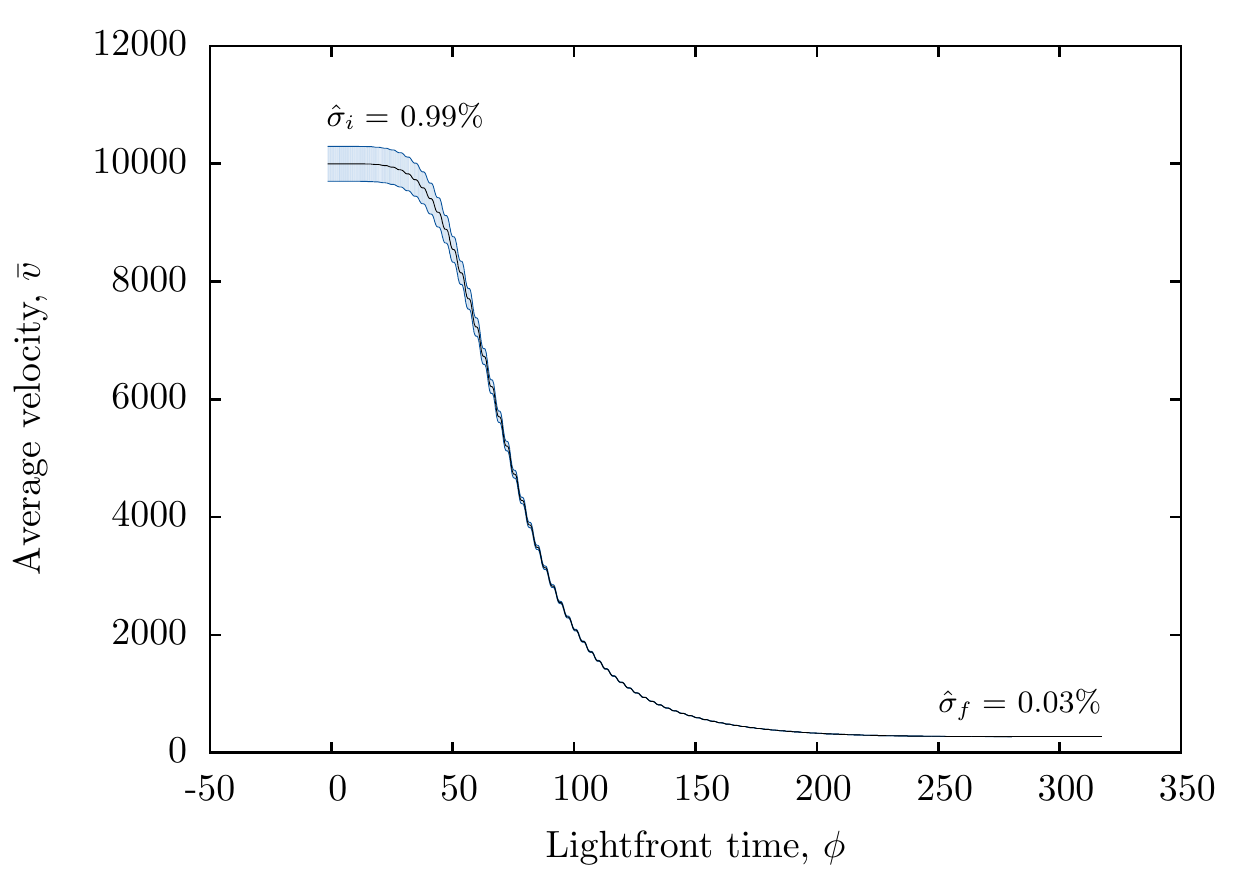}
\includegraphics[width=0.48\textwidth]{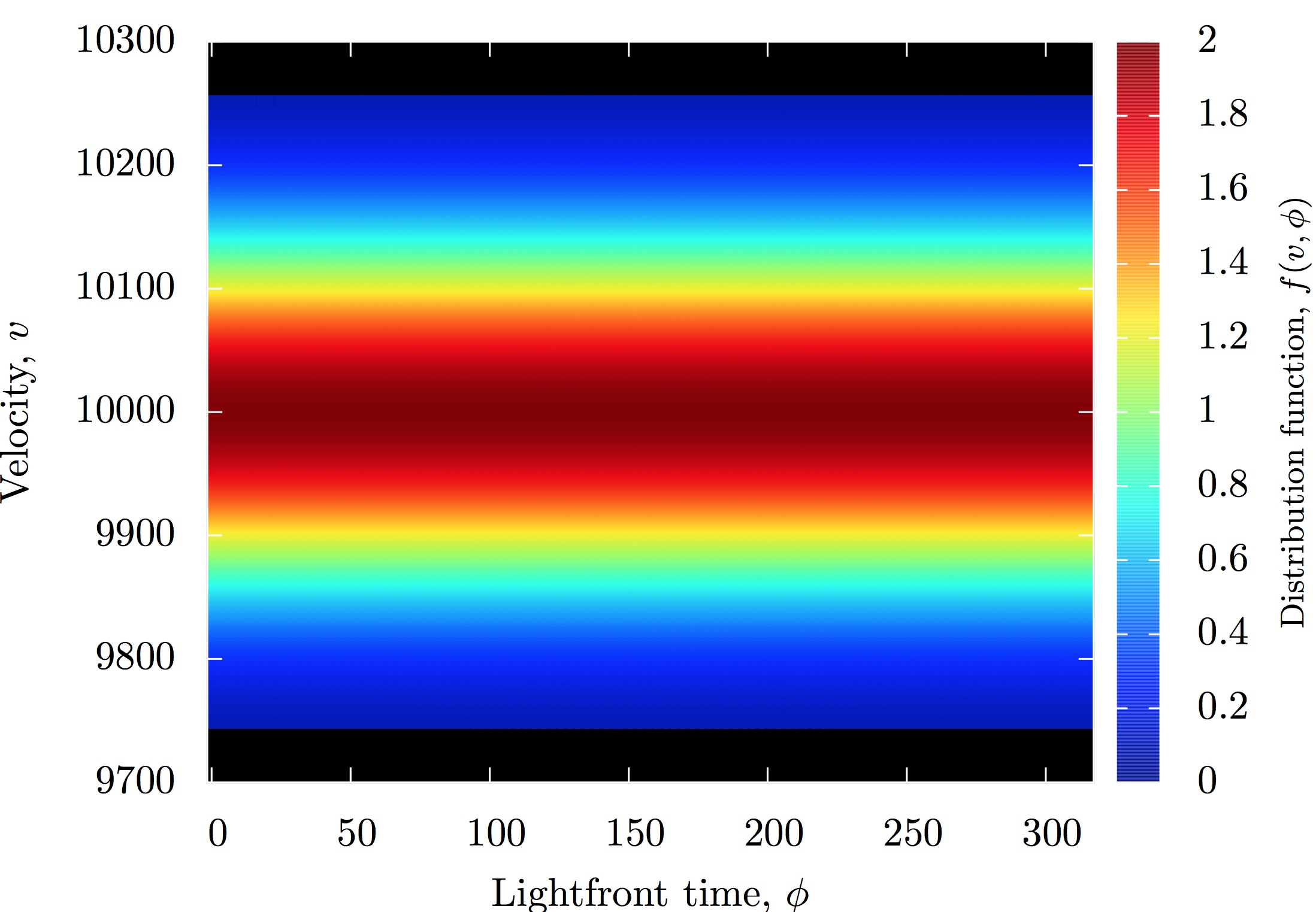}
\includegraphics[width=0.48\textwidth]{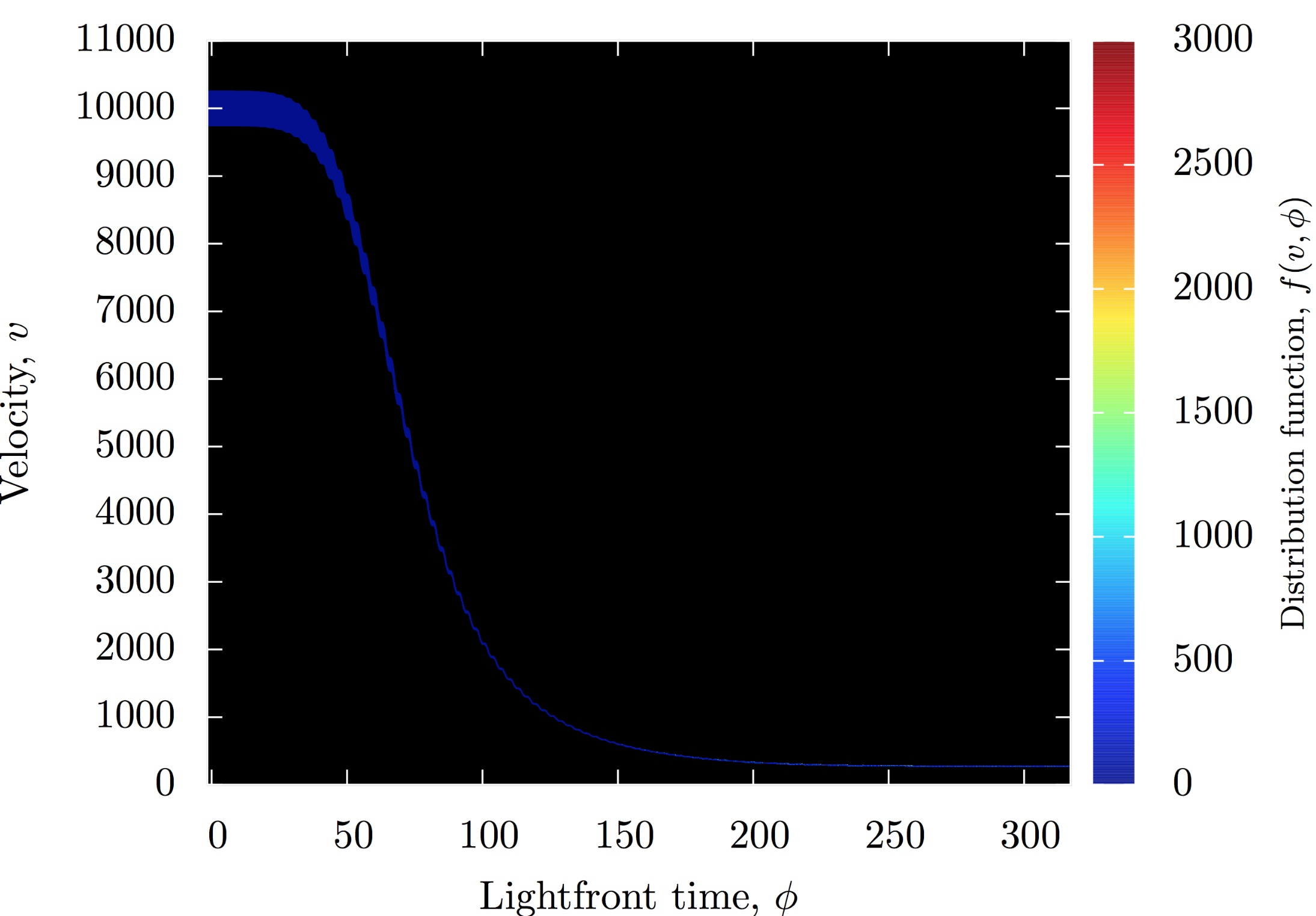}
\vspace{-1.5em}
\caption{\label{fig:dist_v10000_N50} 
Distribution for $N = 50$ \textit{without} (left) and \textit{with} (right) radiation reaction.}
\end{figure} 

Fig.~\ref{fig:dist_v100_N5}, ~\ref{fig:dist_v100_N10}, ~\ref{fig:dist_v100_N50} represent the evolution of the particle distribution with the initial average velocity $\bar{v} = 10^2$ passing through pulses of $N = 5, 10, 50$ oscillations respectively. They show the final distribution width $\hat{\sigma}_f$ for the case with \textit{no radiation reaction} remains the same as the initial distribution width $\hat{\sigma}_i$ as expected. Where radiation reaction is included, however, although the instantaneous distribution width depends on the pulse length, the initial width of $\hat{\sigma}_i = 0.99\%$ decreases to a final width of $\hat{\sigma}_f = 0.73\%$ irrespective of the pulse length. Similarly, the average velocity $\bar{v}$ decreases to 73.9.

If we now increase the central velocity of the particle bunch to $\bar{v} = 10^3$ we observe a respective decrease in the final distribution width to $\hat{\sigma}_f = 0.22\%$, as shown in Fig.~\ref{fig:dist_v1000_N2}, ~\ref{fig:dist_v1000_N5}, ~\ref{fig:dist_v1000_N10}, ~\ref{fig:dist_v1000_N20}, ~\ref{fig:dist_v1000_N50}, ~\ref{fig:dist_v1000_N100}, and these figures also demonstrate that the final distribution width still remains independent of the pulse length. This is accompanied by the decrease in average velocity to $\bar{v}_f = 220.5$. 

Moving to still higher average velocities, (the example of $\bar{v} = 10^4$ is presented in Fig.~\ref{fig:dist_v10000_N5}, ~\ref{fig:dist_v10000_N10}, ~\ref{fig:dist_v10000_N50}), the effect mentioned above still remains valid: we observe 30 times decrease in distribution width to $\hat{\sigma}_f=0.03\%$ which again depends only on the energy in the pulse, not on how it is distributed. 

\section{Numerical simulations. Impact of the pulse energy on the particle distribution}

In the previous section we have discussed the independence of the final distribution from the length of the pulse under the assumption of the fixed energy contained within these laser pulses. In this section we explore the impact of the laser pulse energy on the evolution of the distribution function. 

To achieve this goal we show scatter plots of the average final velocity $\bar{v}$ vs. energy contained in the pulse and the final distribution width $\hat{\sigma}_f$ vs. energy.

An analytical solution for the velocity evolution of the Landau-Lifshitz equation was presented in \cite{DiPiazza}. Although $\bar{v}$ is not a solution to Landau-Lifshitz, for a sufficiently narrow distribution it will approximate such a solution. According to \cite{DiPiazza} the velocity changes as: 
\begin{equation}
\label{eq:4:Antonino_Sol}
\bar{v} = \frac{\bar{v}_i}{h}-\frac{h^2-1}{2 h \dot{\phi}\left(0\right)} - \frac{I^2}{2 h \dot{\phi}\left(0\right)}\ ,
\end{equation}
where 
\begin{equation}
h = 1 + \tau \dot{\phi}\left(0\right) \int\limits_{\phi_{i}}^{\phi_{f}} \mathcal{E}^2\mathrm{d\phi} = 1 + \frac{3\pi}{8} N a^2_0 \tau \dot{\phi}\left(0\right) = 1 + \tau \dot{\phi}\left(0\right) E\ ,
\end{equation}
and
\begin{equation}
\label{eq:4:Idef}
I = - \int\limits_{\phi_{i}}^{\phi_{f}} h \mathcal{E} \mathrm{d\phi}\ . 
\end{equation}
This allows us to rewrite (\ref{eq:4:Antonino_Sol}) as:
\begin{equation}
\label{eq:4:Antonino_Sol_rewritten}
\bar{v} = \frac{\bar{v}_i}{1 + \tau \dot{\phi}\left(0\right) E}-\frac{\tau^2 \dot{\phi}\left(0\right)^2 E^2 + 2\tau \dot{\phi}\left(0\right) E}{2\dot{\phi}\left(0\right) \left[1 + \tau \dot{\phi}\left(0\right) E\right]} - \frac{I^2}{2\dot{\phi}\left(0\right) \left[1 + \tau \dot{\phi}\left(0\right) E\right]}\ .
\end{equation}
According to (\ref{eq:4:Idef}) $I^2$ is typically $\mathcal{O}\left(\tau^2\right)$. Considering that terms of $\mathcal{O}\left(\tau^2\right)$ in (\ref{eq:4:Antonino_Sol_rewritten}) are negligible due to $\tau$ being small and $\tau E \ll \bar{v}_i$ the analytical solution (\ref{eq:4:Antonino_Sol}) presented in \cite{DiPiazza} can be approximated as:
\begin{equation}
\label{eq:4:vf}
\bar{v}_f = \frac{\bar{v}_i}{1 + 2\tau \bar{v}_i E} = : g\left(\bar{v}_i, E\right)\ ,
\end{equation}
where $\bar{v}_i$ is the initial average velocity, $\bar{v}_f$ is the final average velocity and $E$ is the energy of the laser pulse.

Analogously, an approximate analytical solution for the final distribution width can be obtained. Consider (\ref{eq:4:vf}) with the replacement $\bar{v}_i \rightarrow \bar{v}_i\left(1 + \hat{\sigma}_i\right)$. This results in:
\begin{equation}
\bar{v}_f\left(1 + \hat{\sigma}_f\right) = g\left[\bar{v}_i\left(1 + \hat{\sigma}_i\right), E\right] \simeq g\left(\bar{v}_i, E\right) + \frac{\partial f}{\partial \bar{v}_i}\bar{v}_i\hat{\sigma}_i\ ,
\end{equation}
therefore 
\begin{equation}
\label{eq:4:hatsigmaf}
\hat{\sigma}_f = \frac{\partial f}{\partial \bar{v}_i}\frac{\bar{v}_i}{\bar{v}_f}\hat{\sigma}_i\ .
\end{equation}
Using (\ref{eq:4:vf}) the partial derivative and the $\bar{v}_i/\bar{v}_f$ ratio in (\ref{eq:4:hatsigmaf}) can be expressed as:
\begin{equation}
\label{eq:4:partial_der}
\frac{\partial f}{\partial \bar{v}_i} = \frac{1}{\left(1 + 2\tau \bar{v}_i E\right)^2}\ ,
\end{equation}
\begin{equation}
\label{eq:4:vel_ratio}
\frac{\bar{v}_i}{\bar{v}_f} = 1 + 2\tau \bar{v}_i E\ .
\end{equation}
Combining (\ref{eq:4:partial_der}) and (\ref{eq:4:vel_ratio}) into (\ref{eq:4:hatsigmaf}) the final distribution width change can be approximated as: 
\begin{equation}
\label{eq:4:sigmaf}
\hat{\sigma}_f = \frac{\hat{\sigma}_i}{1 + 2\tau\bar{v}_i E}\ ,
\end{equation}
where $\bar{v}_i$ is the initial average velocity, $\hat{\sigma}_i$ is the initial width of the distribution, $\hat{\sigma}_f$ is the final distribution width and $E$ is the energy contained in the laser pulse.

Plotting the simulation results for three different initial average velocities $\bar{v} = 10^2$, $10^3$ and $10^4$ it can be seen that these are in excellent agreement with the analytical approximations (\ref{eq:4:vf}, \ref{eq:4:sigmaf}).

\subsection{Particle bunch with central velocity of $\bar{v} = 10^2$}
\vspace{-1em}
\begin{figure}[H]
\centering
\includegraphics[width=0.85\textwidth]{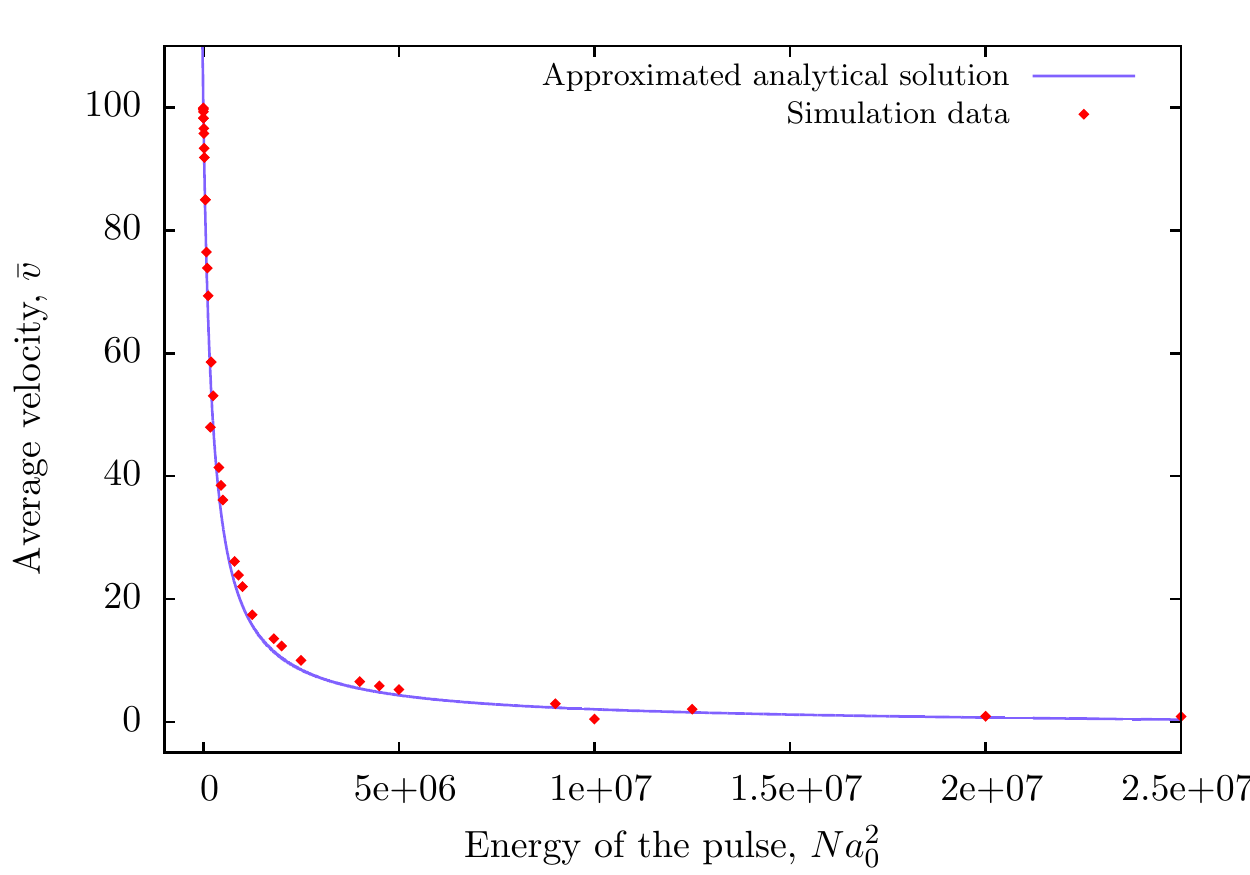}
\vspace{-1em}
\caption{\label{fig:dist_summary_g100_v} 
Final average velocity $\bar{v}_f$ as a function of energy in the pulse $E$: Approximate analytical solution and simulation data.}
\end{figure} 

\vspace{-1em}
\begin{figure}[H]
\centering
\includegraphics[width=0.85\textwidth]{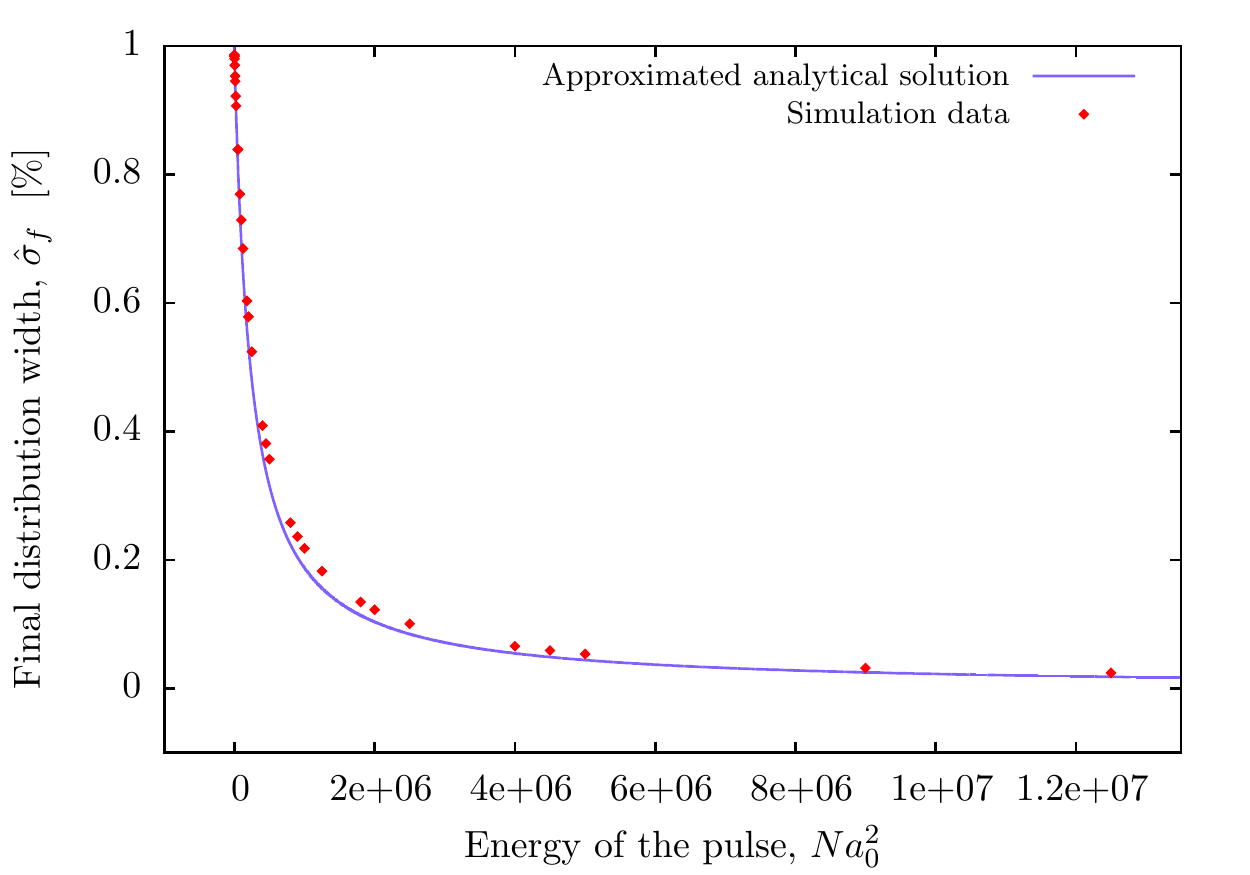}
\vspace{-1em}
\caption{\label{fig:dist_summary_g100_sigma} 
Final distribution width $\hat{\sigma}_f$ as a function of energy in the pulse $E$: Approximate analytical solution and simulation data.}
\end{figure} 

\subsection{Particle bunch with central velocity of $\bar{v} = 10^3$}
\vspace{-1em}
\begin{figure}[H]
\centering
\includegraphics[width=0.85\textwidth]{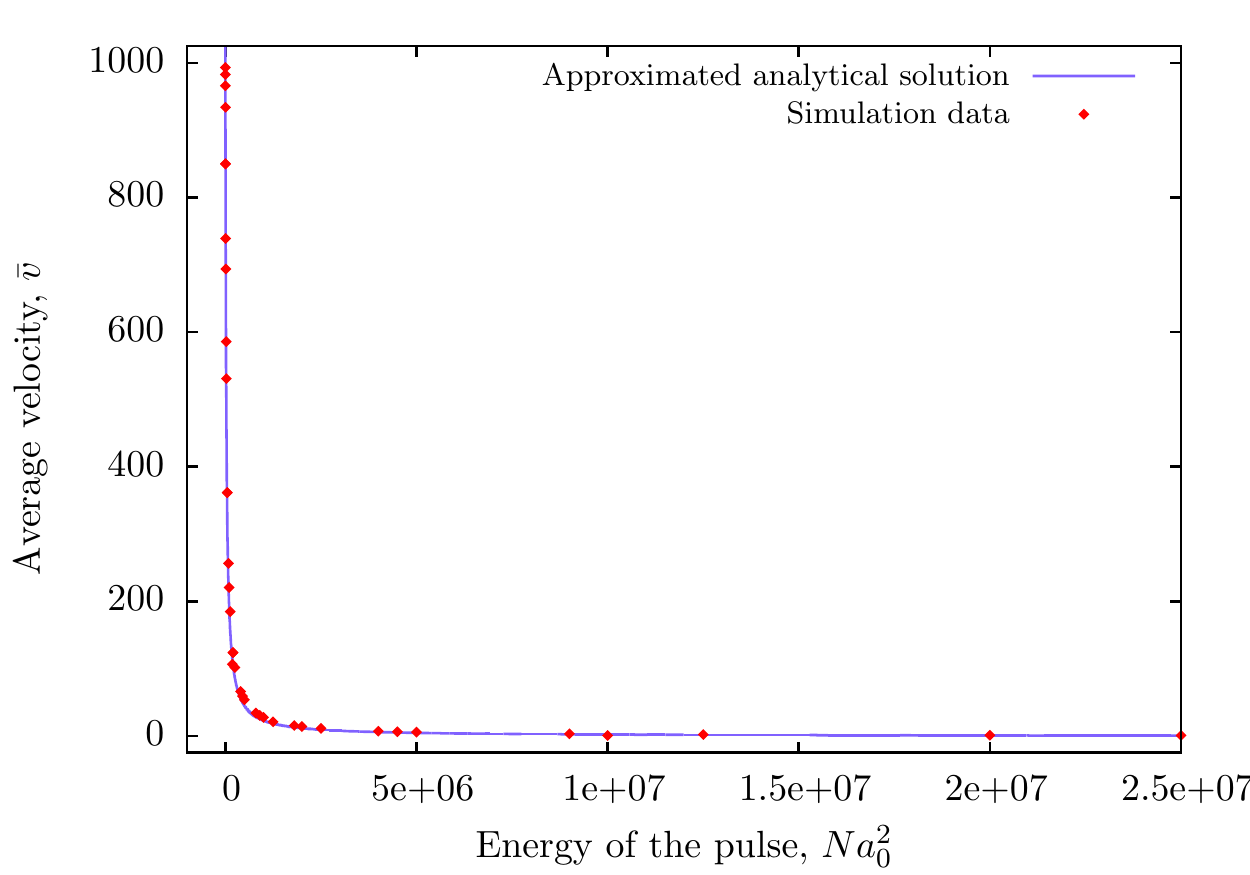}
\vspace{-1em}
\caption{\label{fig:dist_summary_g1000_v} 
Final average velocity $\bar{v}_f$ as a function of energy in the pulse $E$: Approximate analytical solution and simulation data.}
\end{figure} 

\vspace{-1em}
\begin{figure}[H]
\centering
\includegraphics[width=0.85\textwidth]{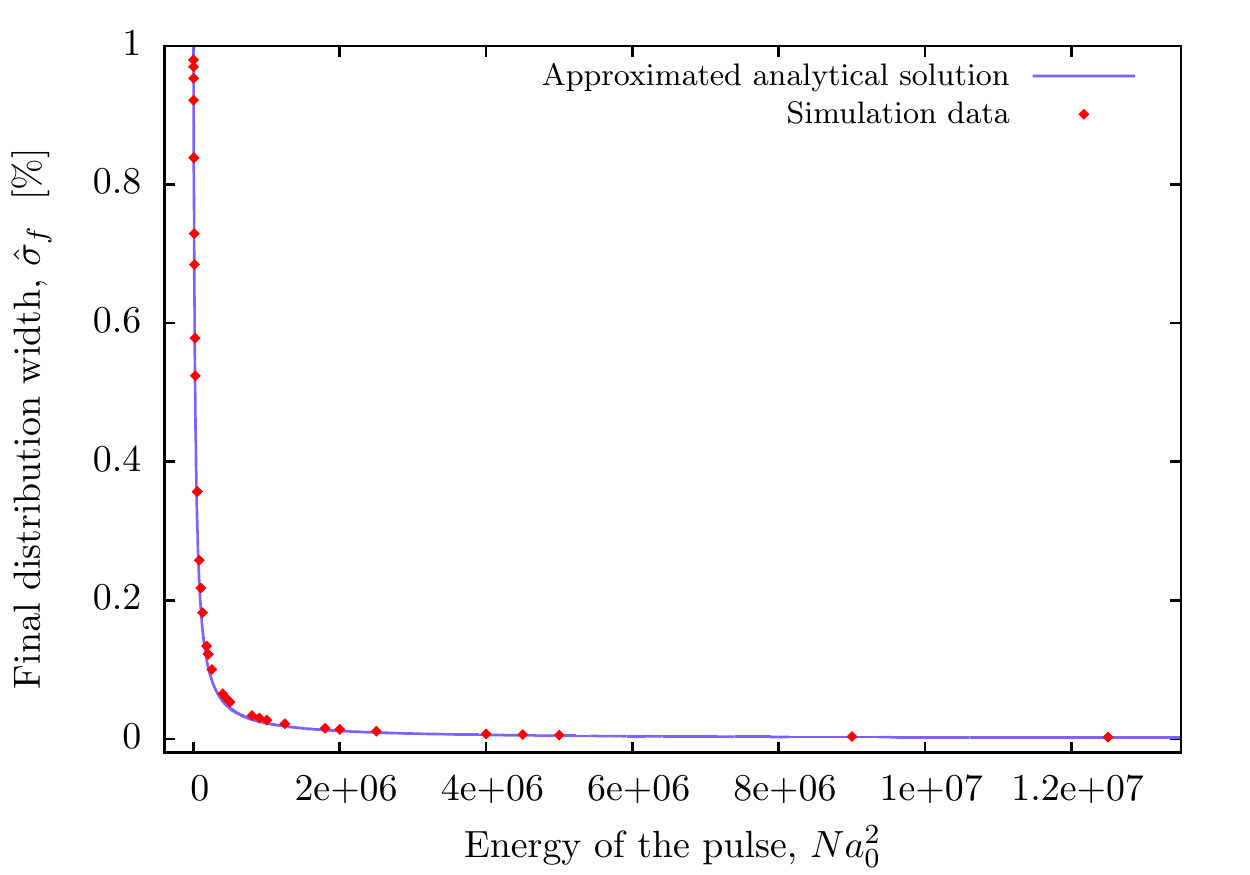}
\vspace{-1em}
\caption{\label{fig:dist_summary_g1000_sigma} 
Final distribution width $\hat{\sigma}_f$ as a function of energy in the pulse $E$: Approximate analytical solution and simulation data.}
\end{figure} 

\subsection{Particle bunch with central velocity of $\bar{v} = 10^4$}
\vspace{-1em}
\begin{figure}[H]
\centering
\includegraphics[width=0.85\textwidth]{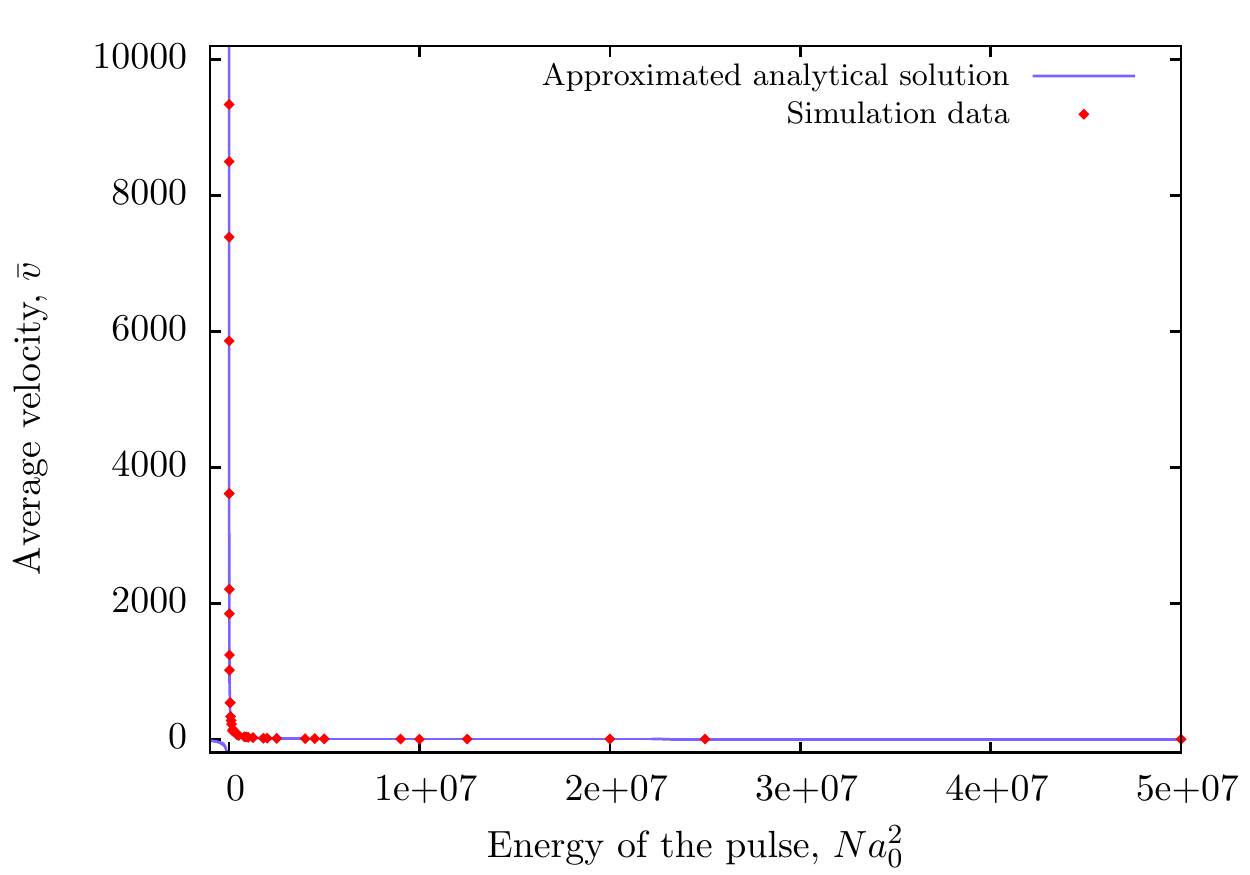}
\vspace{-1em}
\caption{\label{fig:dist_summary_g10000_v}
Final average velocity $\bar{v}_f$ as a function of energy in the pulse $E$: Approximate analytical solution and simulation data.}
\end{figure} 

\vspace{-1em}
\begin{figure}[H]
\centering
\includegraphics[width=0.85\textwidth]{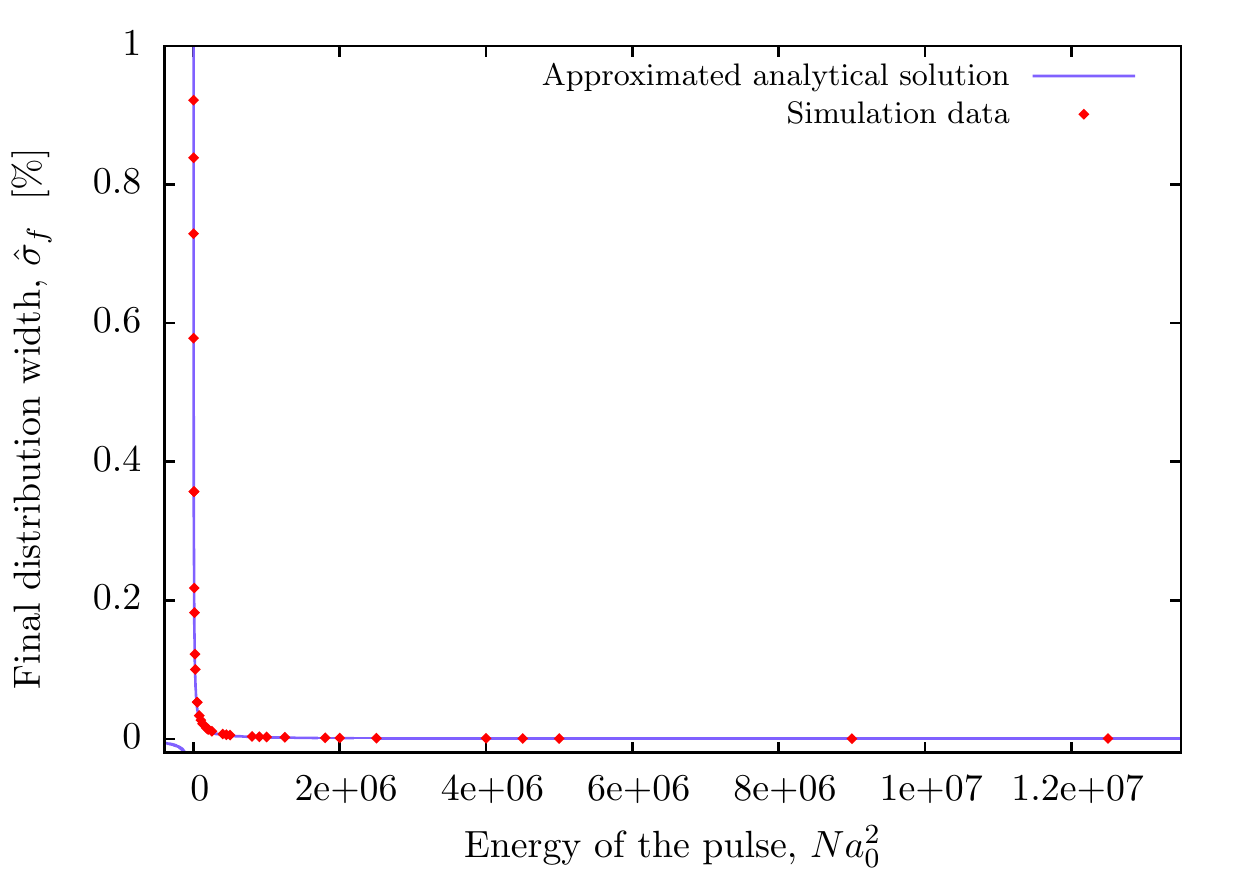}
\vspace{-1em}
\caption{\label{fig:dist_summary_g10000_sigma} 
Final distribution width $\hat{\sigma}_f$ as a function of energy in the pulse $E$: Approximate analytical solution and simulation data.}
\end{figure} 

In order to improve the plots' visibility some of the data points representing ``abnormal behaviour'' have not been plotted in Fig.~\ref{fig:dist_summary_g100_v}, ~\ref{fig:dist_summary_g100_sigma}, ~\ref{fig:dist_summary_g1000_v}, ~\ref{fig:dist_summary_g1000_sigma}, ~\ref{fig:dist_summary_g10000_v}, ~\ref{fig:dist_summary_g10000_sigma}. These points correspond to extreme values of $a_0 = 500, 1000$ where the final distribution width $\hat{\sigma}_f$ becomes large, indicating that although each individual particle obeys the Landau-Lifshitz equation, the average velocity $\bar{v}$ does not satisfy (\ref{eq:4:Antonino_Sol}).

\section{Summary}

Based on results presented in this Chapter we confirm that, while decreasing the energy of particles, radiation reaction also leads to a reduction in the momentum spread when a relativistic particle bunch passes through an intense laser pulse. Both analytical considerations and simulation results presented in this Chapter indicate that the change in average velocity and momentum spread of the particle distribution depends only on the total energy of the laser pulse, and remains completely independent of the way this energy is distributed. 

Both (\ref{eq:4:vf}) and (\ref{eq:4:sigmaf}) indicate that further increase of the initial average velocity $\bar{v}_i$ of the particle bunch leads to a unique final average velocity $\bar{v}_f = 1/2\tau E$ with zero velocity spread, $\hat{\sigma}_f = 0$. This can be interpreted as an effect of phase-space attractors of the Landau-Lifshitz equation \cite{Lehmann}.

These results remain valid in the classical theory, however it has recently been demostrated \cite{NeitzDiPiazza} that quantum radiation reaction may lead to a broadening of the distribution width.

\chapter{Scattering of an electron by a heavy nucleus}

\begin{figure}[H]
\vspace{-17em}
\centering
\includegraphics[width=\textwidth]{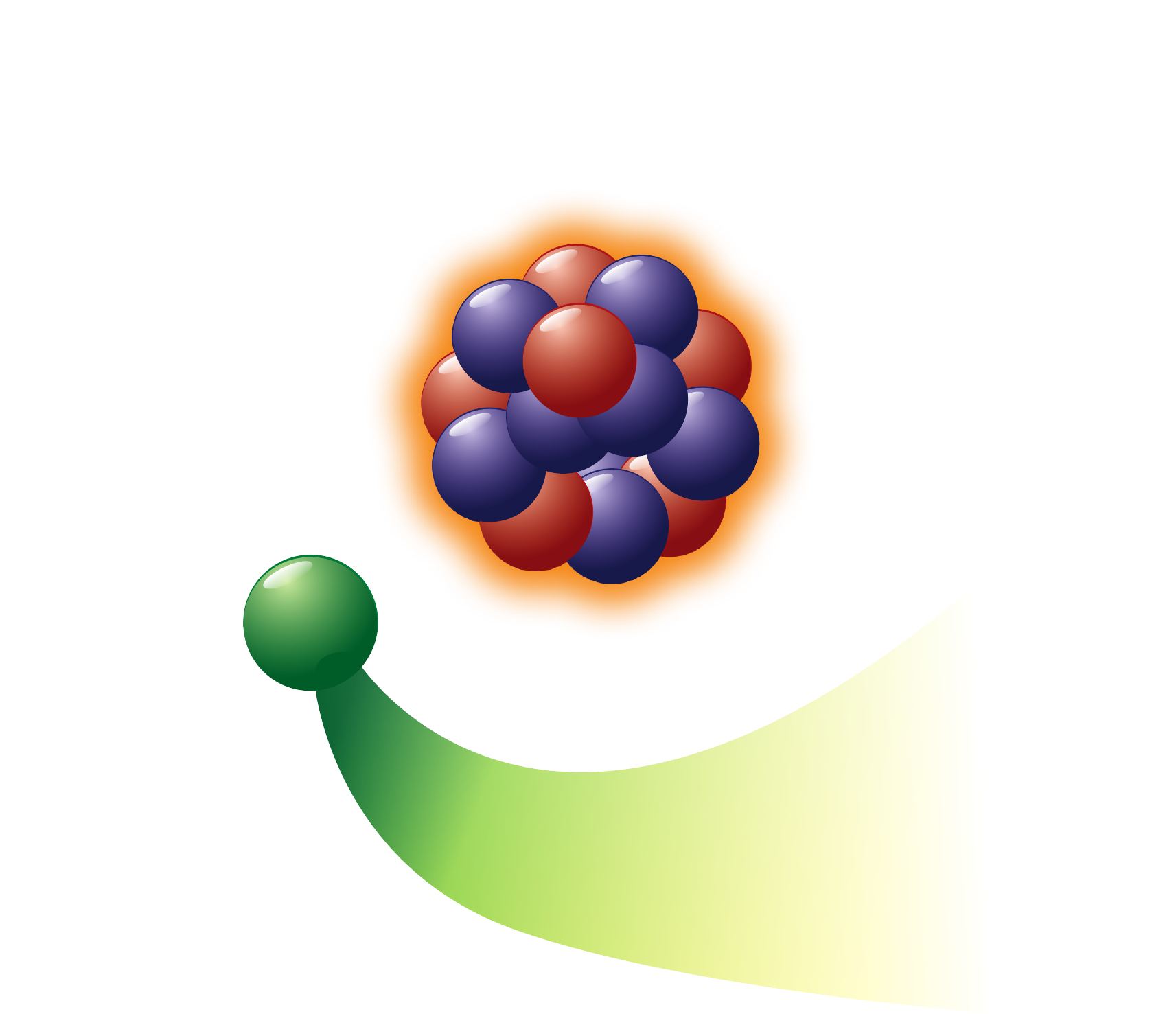}
\end{figure} 

\newpage 

In the previous Chapters we considered particles interacting with intense laser pulses. An alternative source of an extremely high electromagnetic field is the Coulomb field of an \textit{atomic nucleus} \cite{Rajeev}. The value of the electric field close to the surface of the nucleus can be as large as:
\begin{equation}
E = \frac{Ze}{4\pi\epsilon_{0}\boldsymbol{r}^2} \simeq 2.07 \cdot 10^{21}\  \text{V}\text{m}^{-1}\ ,
\end{equation}
for the $^{235}U$ nucleus, where $e = -q$ is the charge on the proton.

\begin{figure}[H]
\centering
\includegraphics[width=0.85\textwidth]{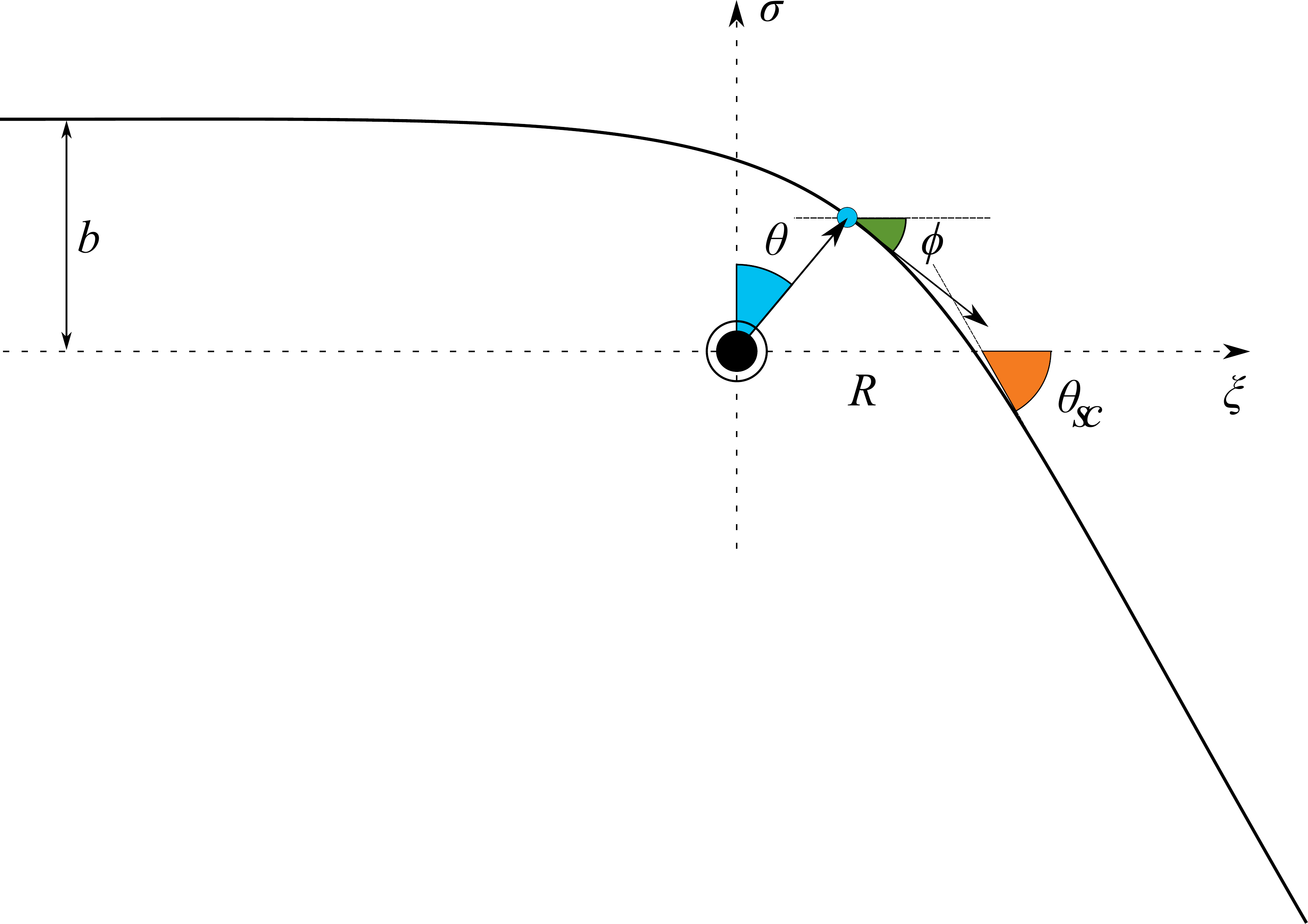}
\caption{\label{fig:5:setup} ``Experimental'' configuration}
\end{figure}

We consider a setup where the particle is fired at a stationary nucleus from a large (compared with nuclear scales) distance with impact parameter $b$ (see Fig.~\ref{fig:5:setup}).

A similar setup has previously played an important role in modern physics: in 1911, Ernest Rutherford performed an experiment in which he fired a beam of alpha particles at layers of gold leaf only a few atoms thick \cite{Rutherford}. He noted that while some of the particles passed through with little deflection a small fraction were deflected by very large angles. This result led Rutherford to postulate the existence of the atomic~nucleus.

This Chapter will be devoted to an investigation of effects of radiation reaction on the motion of a high energy particle scattered by a heavy nucleus. We are interested in how radiation reaction during the particle-nucleus interaction impacts its trajectory and energy evolution. Regions of deviation of Ford-O'Connell predictions from the Landau-Lifshitz ones are investigated and the importance of quantum effects during the interaction is discussed.

Previously, work has appeared in the literature on related problems. Eliezer \cite{Eliezer_nucleus} considered head-on collisions between an electron and a nucleus using the Lorentz-Abraham-Dirac equation, finding that this leads solely to runaway solutions. Huschilt and Baylis \cite{Baylis1, Huschilt} and Comay \cite{Comay} extended this result to show there is a minimum impact parameter below which there are no non-runaway solutions. This implies that the Lorentz-Abraham-Dirac equation cannot describe electron capture in the field of a nucleus. Rajeev \cite{Rajeev} solved the Landau-Lifshitz equation for a particle spiralling into the nucleus. These results are however limited to the non-relativistic case. 

\section{Theoretical model}

We investigate the effects of radiation reaction on the motion of a particle scattered by a heavy nucleus during a collision. To achieve this goal we consider the \textit{trajectory} and \textit{energy evolution} of a particle including radiation reaction described by Ford-O'Connell and compare the outcome with both Landau-Lifshitz predictions and the case with no radiation reaction.

Consider the orthonormal basis with 1 time-like vector, $\eta$, and 3 space-like vectors, $\epsilon$, $\lambda$, $\kappa$:
\begin{equation}
\label{cond1}
\eta^a\eta_a=-1\ , \quad \epsilon^a\epsilon_a=1\ , \quad \lambda^a\lambda_a=1\ , \quad \kappa^a\kappa_a=1\ , 
\end{equation}
and
\begin{equation}
\label{cond2}
\eta^a\epsilon_a=0\ , \quad \eta^a\lambda_a=0\ , \quad \epsilon^a\lambda_a=0\ , \quad \kappa^a\lambda_a=0\ , \quad \kappa^a\epsilon_a=0\ , \quad \kappa^a\eta_a=0\ .
\end{equation}
The relation between this basis and the one used in the previous Chapters is given by
\begin{equation} 
\eta = \frac{1}{2}\left(n + m\right)\ , \quad \kappa = \frac{1}{2}\left(n - m\right)\ .
\end{equation}
The metric tensor is then as follows:
\begin{equation}
g_{ab} = \epsilon_a \epsilon_b + \lambda_a \lambda_b + \kappa_a \kappa_b -\eta_a \eta_b\ .
\end{equation}

To study the problem of an electron scattering off a nucleus, we consider the physical setup shown in Fig.~\ref{fig:5:setup} which is described by the following parameters:
\vspace{-0.5em}
\begin{itemize}
\item $b$ -- impact parameter, 
\vspace{-0.5em}
\item $\theta_{sc}$ -- scattering angle, 
\vspace{-0.5em}
\item $\phi$ -- velocity angle (measured from $\xi$--axis), 
\vspace{-0.5em}
\item Longitudinal coordinate: $\xi = R\sin{\theta}$, 
\vspace{-0.5em}
\item Transverse coordinate: $\sigma = R\cos{\theta}$
\end{itemize}
\vspace{-0.5em}
and derive appropriate equations of motion. 

Introduce the coordinates such that, with $\alpha = - x^a \eta_a$, $\xi = x^a \epsilon_a$, $\sigma = x^a \lambda_a$ and $\zeta = x^a \kappa_a$, we have
 \begin{align}
 \label{eq:4:xadef}
  x^a = \alpha \eta^a + \xi\epsilon^a + \sigma\lambda^a + \zeta\kappa^a\ .
\end{align}
For a single particle the scattering occurs in a plane, so we take:
\begin{equation}
\zeta = x^a \kappa_a = 0\ ,
\end{equation}
without loss of generality. Note that this would not be true for the Lorentz-Abraham-Dirac equation, which requires three initial conditions, which can't be chosen to lie in a plane without loss of generality. 

Considering equation (\ref{eq:4:xadef}), the normalisation condition, $\dot{x}^a \dot{x}_a = -1$, requires:
\begin{equation}
\label{eq:4:normcond}
 \dot{\alpha}^2 = 1 + \dot{\xi}^2 + \dot{\sigma}^2\ .
\end{equation}
Since the time coordinate $\alpha$ always increases, its derivative satisfies the condition $\dot{\alpha}>0$, leaving us with:
\begin{equation}
 \dot{\alpha} = \sqrt{1 + \dot{\xi}^2 + \dot{\sigma}^2}\ .
\end{equation}

We consider the nucleus to be sufficiently massive that it can be treated as stationary. There are then two components of the Coulomb field of the nucleus acting on a particle in the $\zeta=0$ plane:
\begin{equation}
\label{eq:4:Fab}
- \frac{q}{m} F\indices{^a_b} = \mathcal{E}_1 \left(\eta^a \epsilon_b - \epsilon^a \eta_b \right) + \mathcal{E}_2 
 \left(\eta^a \lambda_b - \lambda^a \eta_b\right) \ ,
\end{equation}
where $(m/q)\mathcal{E}_i$ are the $\epsilon$ and $\lambda$ electric fields, with $i \in \{1, 2\}$, which can be further defined as:
\begin{equation}
\label{eq:5:E1E2cart}
{\cal E}_1 = \frac{\cal K}{\xi^2+\sigma^2}\frac{\xi}{\sqrt{\xi^2+\sigma^2}}\ , \qquad {\cal E}_2 = \frac{\cal K}{\xi^2+\sigma^2}\frac{\sigma}{\sqrt{\xi^2+\sigma^2}}\ ,
\end{equation}
where ${\cal K} = q q_n/4 \pi m = - 3Z\tau/2$, with $q_n = -Zq$ being the charge on the nucleus. Note that we use Heaviside-Lorentz units, where $\epsilon_0=1$.

Based on the definition of the sandwiched tensor, $G\indices{^a_b}$ then has the following form: 
\begin{align}
G^a{}_b=\nonumber &{\cal E}_1\epsilon^a\eta_b-{\cal E}_1\eta^a\epsilon_b+{\cal E}_2\lambda^a\eta_b-{\cal E}_2\eta^a\lambda_b-{\cal E}_1\dot{\alpha}\epsilon^a\xd_b-\left({\cal E}_1\dot{\xi}+{\cal E}_2\dot{\sigma}\right)\left[\eta^a\xd_b-\xd^a\eta_b\right]-\\
&{\cal E}_2\dot{\alpha}\lambda^a\xd_b+{\cal E}_1\dot{\alpha}\xd^a\epsilon_b+{\cal E}_2\dot{\alpha}\xd^a\lambda_b\ ,
\end{align}
which can be further expanded using the definition of $\xd^a$ from (\ref{eq:4:xadef}) as:
\begin{align}
G^a{}_b = \left({\cal E}_1 \dot{\sigma} - {\cal E}_2 \dot{\xi}\right)\left[\dot{\sigma}\left(\eta^a\epsilon_b - \epsilon^a\eta_b\right) - \dot{\xi}\left(\eta^a\lambda_b - \lambda^a\eta_b\right) - \dot{\alpha}\left(\epsilon^a\lambda_b - \lambda^a\epsilon_b\right) \right]\ .
\end{align}
The form of $G^a{}_b$ combined with the normalisation condition (\ref{eq:4:normcond}) allows us to rewrite the determinant as follows:
\begin{equation}
\label{eq:4:determinantnucleus}
\det M = 1 + \frac{\tau^2}{2} G^{ab} G_{ab} = 1+\tau^2\left({\cal E}_2\dot{\xi}-{\cal E}_1\dot{\sigma}\right)^2\ .
\end{equation}

If we now consider a head-on collision, the directions of the velocity and position vectors coincide, leading to the following relation:
\begin{equation}
\label{eq:5:1Drelation}
\frac{\sigma}{\xi} = \frac{\dot{\sigma}}{\dot{\xi}}\ .
\end{equation}
Substituting (\ref{eq:5:E1E2cart}) into (\ref{eq:4:determinantnucleus}) we obtain the expression for the determinant:
\begin{equation}
\label{eq:4:determinantnucleus_linear}
\det M = 1+\tau^2\left[\frac{\cal K}{\left(\xi^2+\sigma^2\right)^{3/2}}\left(\sigma\dot{\xi}-\xi\dot{\sigma}\right)\right]^2\ .
\end{equation}
Taking into account (\ref{eq:5:1Drelation}), the determinant (\ref{eq:4:determinantnucleus_linear}) reduces to 1, in keeping with our observation in Chapter 2 that Landau-Lifshitz and Ford-O'Connell equations for the case of linear motion are identical.

For more general collisions, the equation of motion that we wish to solve is:
\begin{equation}
\label{eq:4:generalmotion}
 \xdd = -\frac{q}{m} \left[ \Delta - \frac{\tau G - \tau^2 G^2}{\det{M}} \right] 
 \left[ F + \tau \dot{F} \right]\xd\ .
\end{equation}

Given the spherical symmetry of the field it is appropriate to use polar coordinates:
\begin{equation}
 \xi = R \sin{\theta} 
\end{equation}
\begin{equation}
\label{eq:4:zetadot}
 \dot{\xi} = \dot{R}\sin{\theta} + R \dot{\theta} \cos{\theta}  
\end{equation}
\begin{equation}
 \ddot{\xi} = \ddot{R} \sin{\theta} + 2 \dot{R} \dot{\theta} \cos{\theta} + R\ddot{\theta} \cos{\theta} - R\dot{\theta}^2 \sin{\theta} 
\end{equation}
and
\begin{equation}
 \sigma = R \cos{\theta}
\end{equation}
\begin{equation}
\label{eq:4:betadot}
 \dot{\sigma} = \dot{R}\cos{\theta} - R \dot{\theta} \sin{\theta}
\end{equation}
\begin{equation}
\ddot{\sigma} = \ddot{R} \cos{\theta} - 2 \dot{R} \dot{\theta} \sin{\theta} - R\ddot{\theta} \sin{\theta} - R\dot{\theta}^2 \cos{\theta}\end{equation}
 
The field is
\begin{equation}
\label{eq:4:E1E2}
 \mathcal{E}_1 = \frac{{\cal K}\sin{\theta}}{R^2}\ , \qquad \mathcal{E}_2 = 
 \frac{{\cal K}\cos{\theta}}{R^2}\ .
\end{equation}
The derivatives of the field are then:
\begin{equation}
\dot{\mathcal{E}}_1 = \dot{\theta}\mathcal{E}_2 - \frac{2 \dot{R}}{R} \mathcal{E}_1\ ,
\end{equation}
\begin{equation}
\dot{\mathcal{E}}_2 = -\dot{\theta}\mathcal{E}_1 - \frac{2 \dot{R}}{R} \mathcal{E}_2\ .
\end{equation}
In the polar coordinates defined above, the determinant (\ref{eq:4:determinantnucleus}) has the form
\begin{equation}
 \det{M} = 1 + \left( \frac{\tau {\cal K} \dot{\theta}}{R} \right)^2 \ .
\end{equation}

The left-hand side of the general equation of motion (\ref{eq:4:generalmotion}) can be expanded by taking into consideration the form of $\dot{x}^a$, which is as follows:
\begin{equation}
\xd^a=\dot{\alpha}\eta^a+\dot{\xi}\epsilon^a+\dot{\sigma}\lambda^a\ .
\end{equation}
Taking into account (\ref{eq:4:zetadot}, \ref{eq:4:betadot}) and the normalisation condition $-\dot{\alpha}^2+\dot{\xi}^2+\dot{\sigma}^2=-1$, we obtain $\dot{\alpha}$ in terms of polar coordinates:
\begin{equation}
\dot{\alpha}=\sqrt{1+\dot{R}^2+R^2\dot{\theta}^2}\ ,
\end{equation}
leading to an explicit form of $\xd^a$:
\begin{equation}
\label{xfirst}
\xd^a=\sqrt{1+\dot{R}^2+R^2\dot{\theta}^2}\eta^a+\dot{R}\sin\theta\epsilon^a+R\dot{\theta}\cos\theta\epsilon^a+\dot{R}\cos\theta\lambda^a-R\dot{\theta}\sin\theta\lambda^a\ .
\end{equation}
Differentiating (\ref{xfirst}) we obtain the equation for $\xdd^a$ and therefore the entire left hand side of the equation of motion:
\begin{align}
\label{eq:4:LHS}
\xdd^a=\nonumber &\left[\frac{R\dot{R}\dot{\theta}^2+\dot{R}\ddot{R}+R^2\dot{\theta}\ddot{\theta}}{\sqrt{1+\dot{R}^2+R^2\dot{\theta}^2}}\right]{\color{red}\eta^a}+\left[2\dot{R}\dot{\theta}\cos\theta-R\dot{\theta}^2\sin\theta+\ddot{R}\sin\theta+R\ddot{\theta}\cos\theta\right]{\color{red}\epsilon^a}+\\
&\left[-2\dot{R}\dot{\theta}\sin\theta-R\dot{\theta}^2\cos\theta+\ddot{R}\cos\theta-R\ddot{\theta}\sin{\theta}\right]{\color{red}\lambda^a}.
\end{align}

To complete the equation of motion we also need to convert the right-hand side of (\ref{eq:4:generalmotion}). Combining (\ref{eq:4:E1E2}) with (\ref{eq:4:Fab}) we obtain the form of the field in polar coordinates:
\begin{equation}
\frac{q}{m}F^a{}_b=\frac{{\cal K}\sin\theta}{R^2}\left(\epsilon^a\eta_b-\eta^a\epsilon_b\right)+\frac{{\cal K}\cos\theta}{R^2}\left(\lambda^a\eta_b-\eta^a\lambda_b\right)\ .
\end{equation}
Contracting the above equation with $\xd^b$ will give us the external force. If we also consider (\ref{cond1}, \ref{cond2}), the expression for the external force becomes:
\begin{equation}
\label{eq:4:extforce}
 - \frac{q}{m}F^a{}_b \dot{x}^b= \frac{{\cal K}\dot{R}}{R^2}\eta^a + \frac{{\cal K}\sin\theta}{R^2}\sqrt{1+\dot{R}^2+R^2\dot{\theta}^2}\epsilon^a + \frac{{\cal K}\cos\theta}{R^2}\sqrt{1+\dot{R}^2+R^2\dot{\theta}^2}\lambda^a.
\end{equation}

After some manipulation of the above equations we obtain three final sets of equations in polar coordinates describing particle motion in the field of the nucleus with no radiation reaction taken into account, with radiation reaction taken into account using the Landau-Lifshitz force and with the Ford-O'Connell force. These sets are listed below respectively. 

\noindent With no radiation reaction: 
\begin{align}
\ddot{R} = &\left[R\dot{\theta}^2 + \frac{{\cal K}\dot\alpha}{R^2} \right], \\
\ddot{\theta} = & -\frac{2 \dot{R} \dot{\theta}}{R}, \\
\ddot\alpha =& \frac{\mathcal{K} \dot{R}}{R^2}, 
 \end{align}
With Landau-Lifshitz radiation reaction correction: 
\begin{align}
\nonumber \ddot{R} = &\left[R\dot{\theta}^2 + \frac{{\cal K}\dot\alpha}{R^2} \right] 
   - \tau\frac{{\cal K} \dot{R}}{R^3} \left[ 2 \dot\alpha + 
     {\cal K}R\dot\theta^2 \right],\\
\nonumber \ddot{\theta} = & -\frac{2 \dot{R} \dot{\theta}}{R} 
   + \tau \frac{{\cal K} \dot{\theta}}{R^4} \left[R \dot\alpha - 
     {\cal K}\Big(1+ R^2 \dot{\theta}^2 \Big)\right],\\
\nonumber \ddot\alpha =& \frac{{\cal K} \dot{R}}{R^2} 
   + \tau \frac{{\cal K}}{R^3} \left[R^2 \dot\theta^2 - 2 \dot{R}^2 - {\cal K}R \dot{\theta}^2 
     \dot\alpha \right].
\end{align}
With Ford-O'Connell radiation reaction correction: 
\begin{align}
\nonumber \ddot{R} = &\left[R\dot{\theta}^2 + \frac{{\cal K}\dot\alpha}{R^2} \right] 
   - \tau\frac{{\cal K} \dot{R}}{D R^3} \left[ 2 D \dot\alpha + 
     {\cal K}R\dot\theta^2 \right]
   + \tau^2 \frac{{\cal K}^2 \dot{\theta}^2}{D R^4} \left[R + 3R \dot{R}^2 - 
     {\cal K}\dot\alpha \right]
   + \\
   &\tau^3 \frac{2{\cal K}^3 \dot{R} \dot{\theta}^2 
     \dot{\alpha}}{D R^5} \\
\nonumber \ddot{\theta} = & -\frac{2 \dot{R} \dot{\theta}}{R} 
   + \tau \frac{{\cal K} \dot{\theta}}{D R^4} \left[ D R \dot\alpha - 
     {\cal K}\Big(1+ R^2 \dot{\theta}^2 \Big)\right]
   + \tau^2 \frac{{\cal K}^2 \dot{R} \dot{\theta}}{D R^5} \left[2 + 3R^2 
     \dot\theta^2 \right] 
   - \\
   &\tau^3 \frac{{\cal K}^3 \dot\theta^3 \dot\alpha}{D R^5} \\
\nonumber \ddot\alpha =& \frac{{\cal K} \dot{R}}{R^2} 
   + \tau \frac{{\cal K}}{D R^3} \left[ 
     D \Big(R^2 \dot\theta^2 - 2 \dot{R}^2 \Big) - {\cal K}R \dot{\theta}^2 
     \dot\alpha \right]
   + \tau^2 \frac{{\cal K}^2 \dot{R} \dot{\theta}^2}{D R^4} 
     \Big[ 3R\dot\alpha - {\cal K} \Big] 
   + \\
   &\tau^3 \frac{{\cal K}^3 \dot{\theta}^2}{D R^5} \left[ 2\dot{R}^2 - R^2 
     \dot\theta^2 \right]
 \end{align}
where $D = \det{M} = 1 + \left(\frac{\tau{\mathcal K}\dot{\theta}}{R}\right)^2$.

\section{Significance of quantum effects}
When we approach ultrahigh fields quantum effects may become significant, therefore we need to estimate their importance. To do so we compare the fields the particle is interacting with, to the Schwinger limit \cite{Sauter, Schwinger}, at which electron-positron pair creation leads to nonlinearities in the electromagnetic field. We assume that quantum effects can be ignored provided that:
\begin{equation}
\label{eq:5:quantum}
\chi := \frac{\hat{E}}{E_S} \ll 1 \ ,
\end{equation}
where $\hat{E}^a = - F\indices{^a_b} \dot{x}^b$ is the electric field as seen by the particle and $E_S = m^2 / q\hbar \simeq 1.3 \cdot 10^{18}\  \text{V}\text{m}^{-1}$ is the Schwinger field. The $\chi$ parameter is a recognized measure of the significance of quantum effects \cite{BulanovSchwinger, HeinzlSchwinger, DiPiazza_review, Ritus}.

With the polar coordinates used, this parameter corresponds to:
\begin{equation}
\label{eq:5:quantum_parameter}
 \chi = \frac{\hbar {\cal K}}{m} \frac{\sqrt{1 + R^2 \dot\theta^2}}{R^2} \ .
\end{equation}

Throughout the simulations we trace the value $\chi$ to estimate the importance of quantum effects in the given regime. However, exploring the consequence of quantum effects is beyond the scope of this thesis. 

\section{Numerical simulations}

As we are interested in cases involving high fields we consider a particle with initial energy $\gamma_\text{in}$ shot at a highly charged Uranium nucleus with $Z=92$ with an impact parameter $b$. 

Throughout the simulations we study the evolution of the particle energy and particle trajectory. The evolution is tracked for the three different approaches: we consider the cases with \textit{no radiation reaction}, and with the radiation reaction taken into account considering both \textit{Landau-Lifshitz} and \textit{Ford-O'Connell} corrections. 

Additionally, for each set of parameters we focus on the evolution of the divergence parameter: 
\begin{equation}
\label{eq:5:tau_nucleus}
\mathcal{T} = \tau \mathcal{K}\frac{\dot{\theta}}{R}\ ,
\end{equation}
which is a quantitative measure of the difference between Ford-O'Connell and Landau-Lifshitz. The quantum parameter, $\chi$ (\ref{eq:5:quantum_parameter}) is also tracked so we have an understanding when classical predictions are still reliable. 

Parameters being varied between the simulations are: 
\vspace{-0.5em}
\begin{itemize}
\item initial energy $\gamma_\text{in}$ of the particle;
\vspace{-0.5em}
\item impact parameter $b$.
\end{itemize}

\vspace{-0.5em}
Impact parameters of $1$, $10^{-1}$, $10^{-2}$ and $10^{-3}$ are being considered, where distance is measured in \r{a}ngstr\"{o}ms, for the initial energies $\gamma_\text{in}$ of $10^2$, $10^3$ and $10^5$ respectively.

The maximum impact parameter $b = 1\mathring{\mathrm{A}}$ places us in the regime over which the atomic nucleus has a significant impact, while the minimum impact parameter $b = 10^{-3}\mathring{\mathrm{A}}$ corresponds approximately to the surface of the Uranium nucleus. 

\begin{figure}[H]
\vspace{+1em}
\centering
\includegraphics[width=0.95\textwidth]{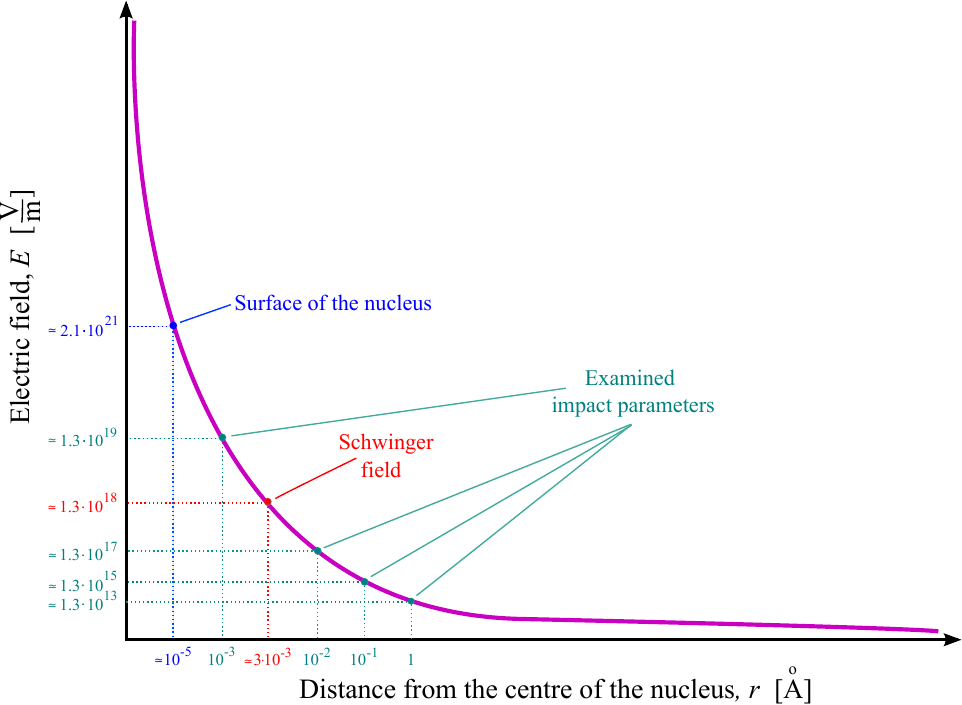}
\vspace{+1em}
\caption{\label{fig:5:ScatterDrawing} Relation of examined parameters and their respective electric fields to the Schwinger field and electric field at the surface of the nucleus.}
\end{figure}

The range of examined impact parameters with their corresponding electric fields along with the radial distance corresponding to the Schwinger field and that corresponding to the surface of the nucleus can be conveniently visualised in Fig.~\ref{fig:5:ScatterDrawing} (\textit{not} to scale).

\subsection{Particle with energy $\gamma_\text{in} = 10^5$ and impact parameter $b = 1\mathring{\mathrm{A}}$}
\begin{figure}[H]
\centering
\includegraphics[width=0.85\textwidth]{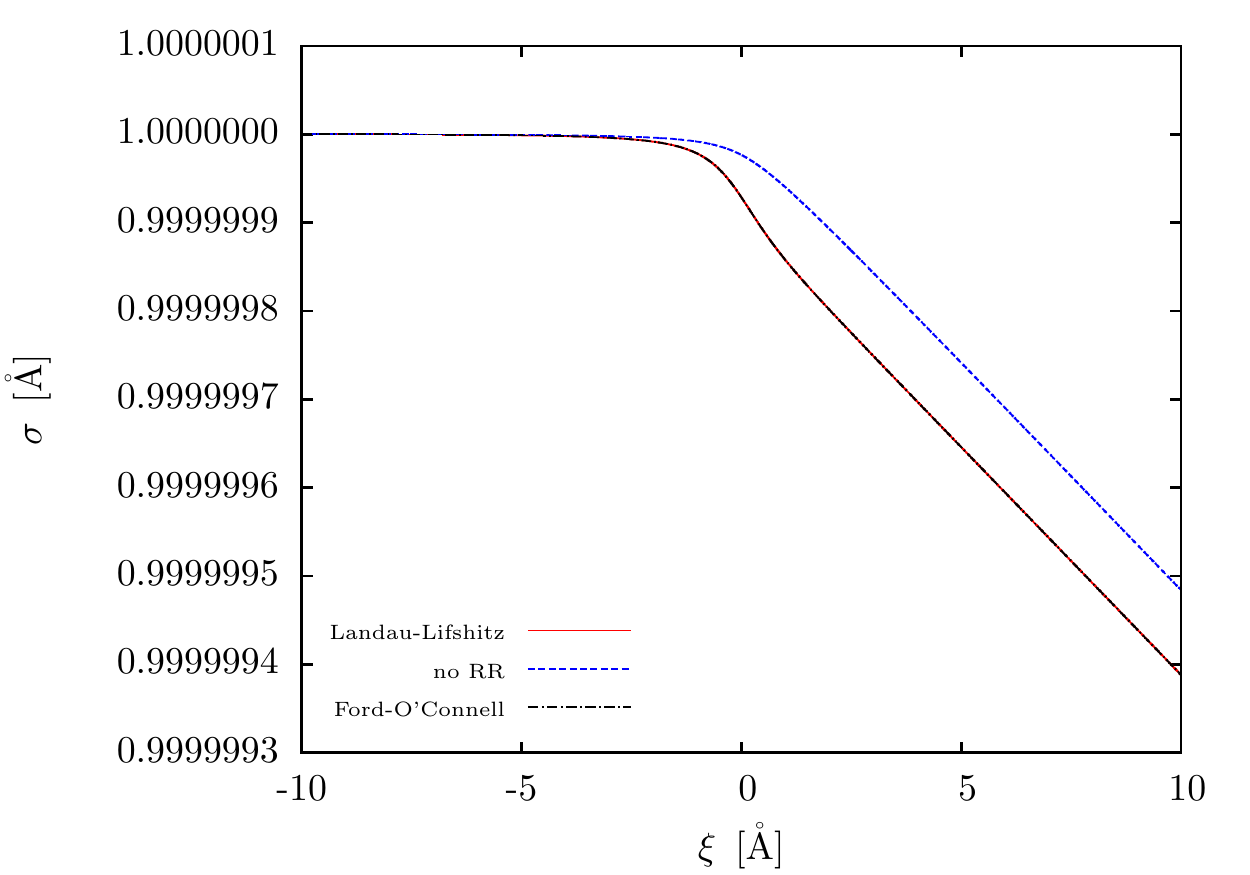}
\vspace{-1em}
\caption{\label{fig:5:trajectory_g5_i10} Trajectory of the particle with $\gamma_\text{in} = 10^5$ and $b = 1\mathring{\mathrm{A}}$. The \textit{Landau-Lifshitz} and \textit{Ford-O'Connell} predictions of the trajectory for this case coincide.}
\end{figure}

\vspace{-1em}
\begin{figure}[H]
\centering
\includegraphics[width=0.85\textwidth]{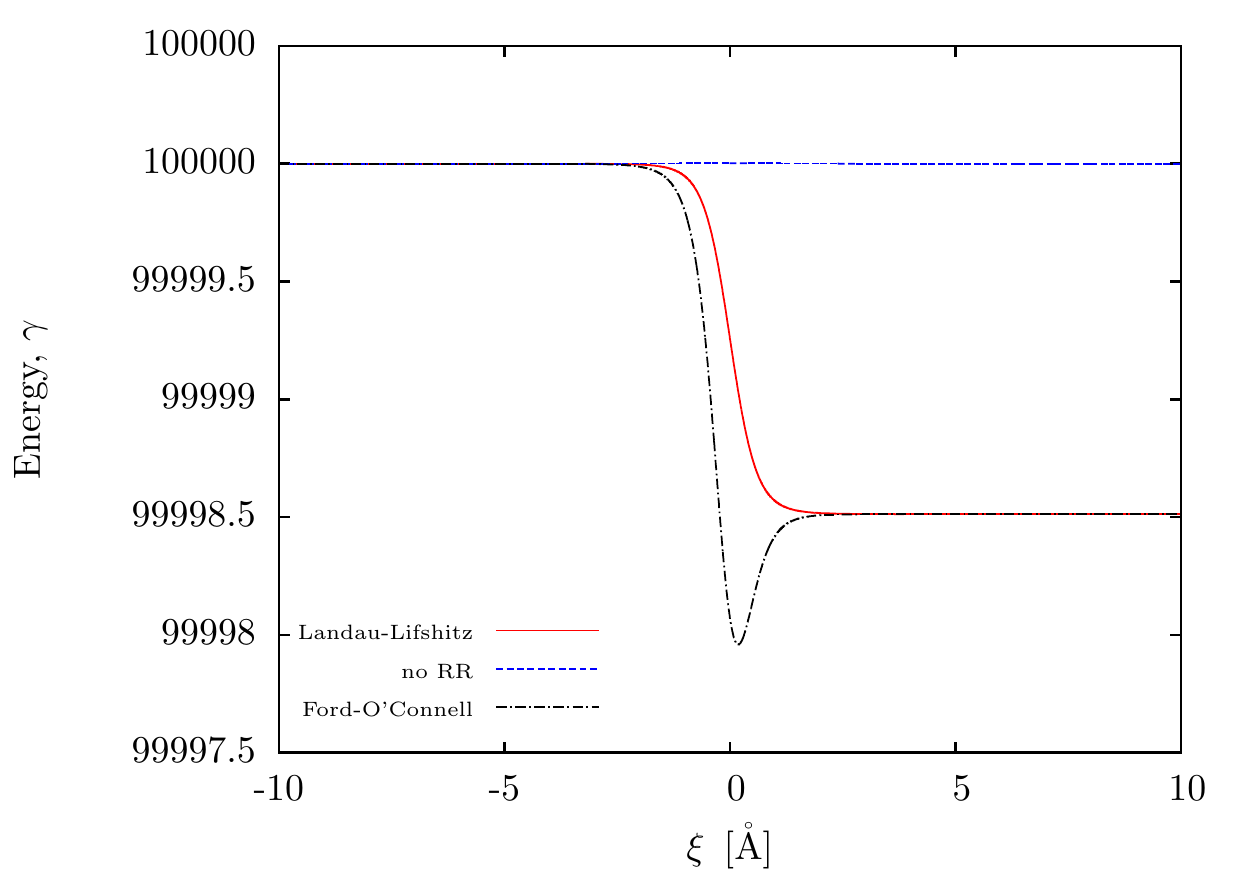}
\caption{\label{fig:5:energy_g5_i10} Energy evolution of the particle with $\gamma_\text{in} = 10^5$ and $b = 1\mathring{\mathrm{A}}$.}
\end{figure}

\vspace{-2em}
\begin{figure}[H]
\centering
\includegraphics[width=0.85\textwidth]{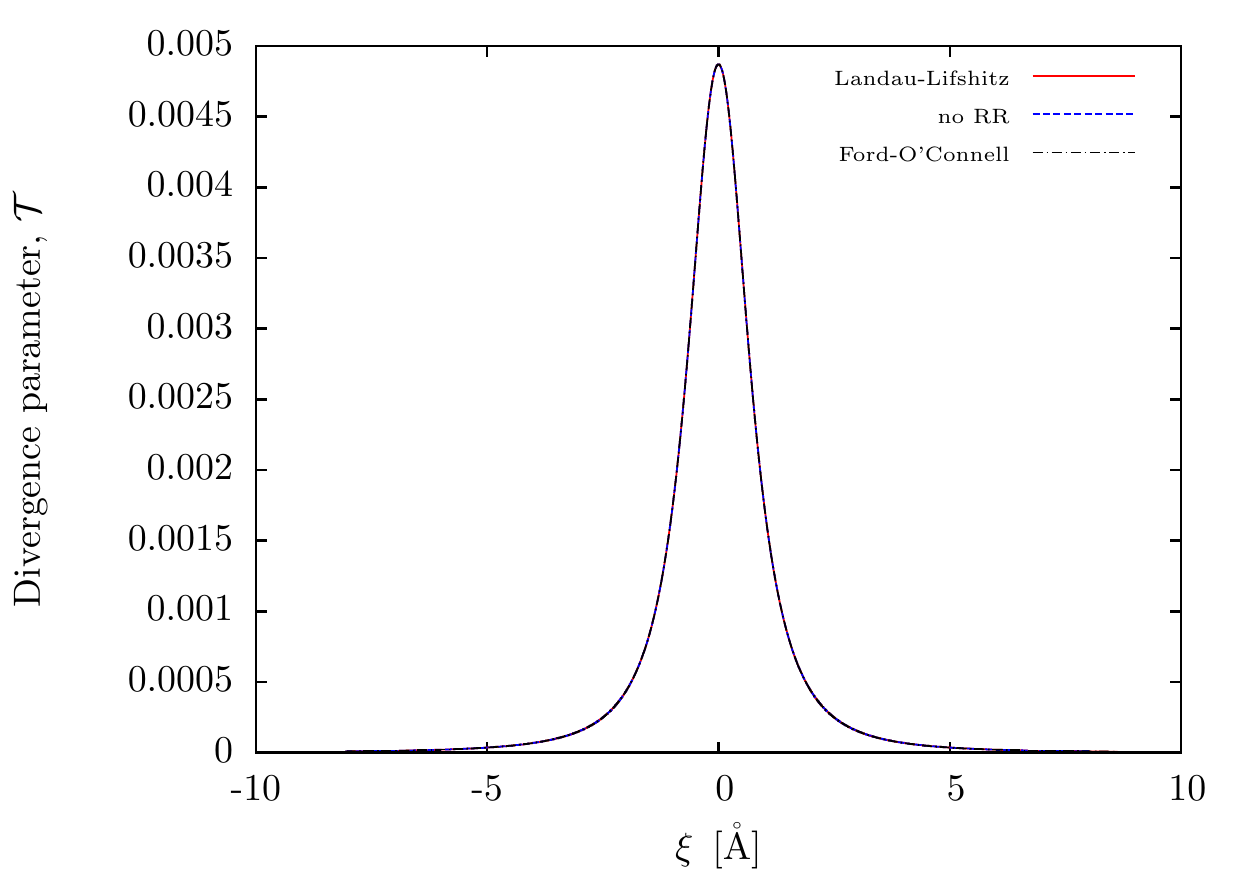}
\vspace{-1em}
\caption{\label{fig:5:divergence_g5_i10} Divergence parameter, $\mathcal{T}$ for the particle with $\gamma_\text{in} = 10^5$ and $b = 1\mathring{\mathrm{A}}$. The \textit{Landau-Lifshitz}, \textit{Ford-O'Connell} and \textit{no radiation reaction} predictions coincide.}
\end{figure}

\vspace{-2em}
\begin{figure}[H]
\centering
\includegraphics[width=0.85\textwidth]{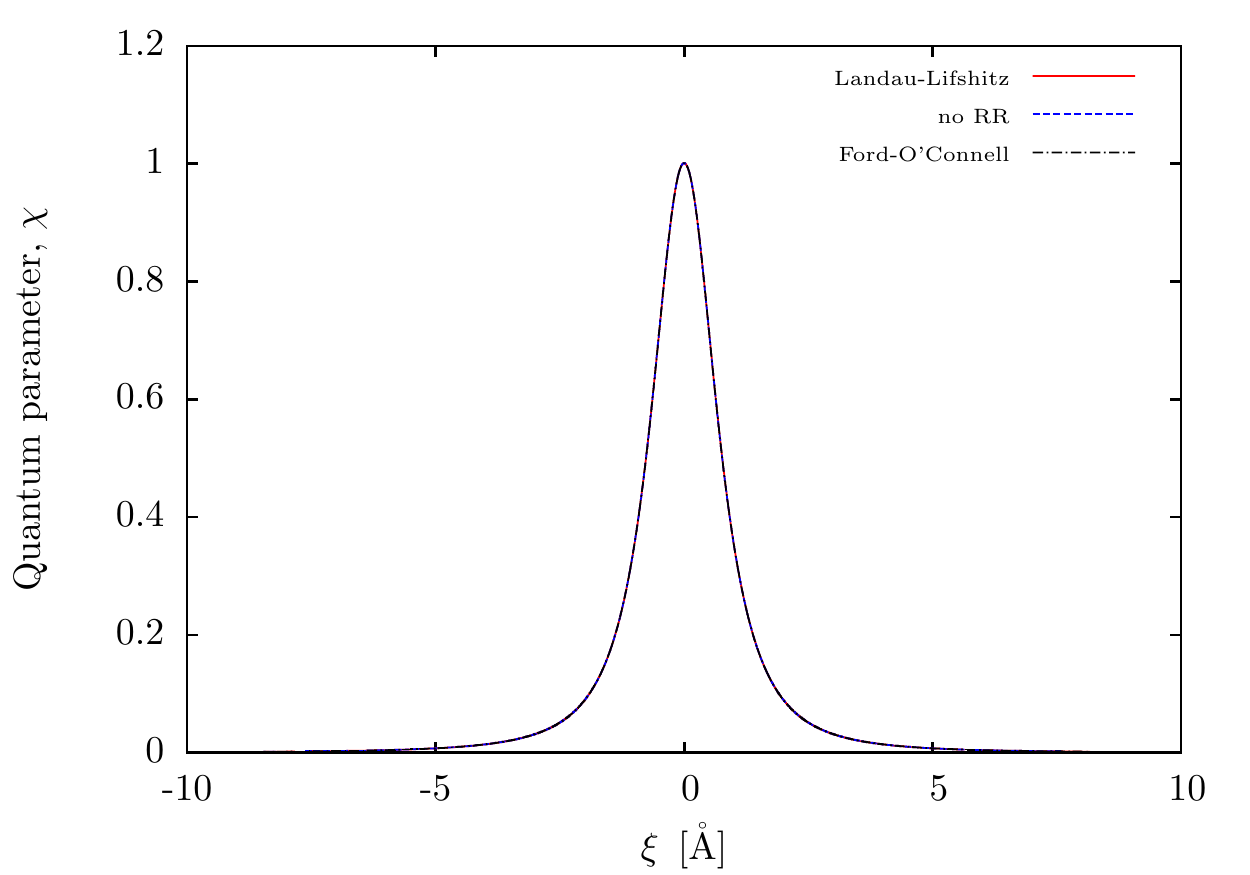}
\vspace{-1em}
\caption{\label{fig:5:quantum_g5_i10} Quantum parameter, $\chi$ for the particle with $\gamma_\text{in} = 10^5$ and $b = 1\mathring{\mathrm{A}}$. The \textit{Landau-Lifshitz}, \textit{Ford-O'Connell} and \textit{no radiation reaction} predictions of the trajectory for this case coincide.}
\end{figure}

\subsection{Particle with energy $\gamma_\text{in} = 10^5$ and impact parameter $b = 10^{-1}\mathring{\mathrm{A}}$}

\begin{figure}[H]
\centering
\includegraphics[width=0.85\textwidth]{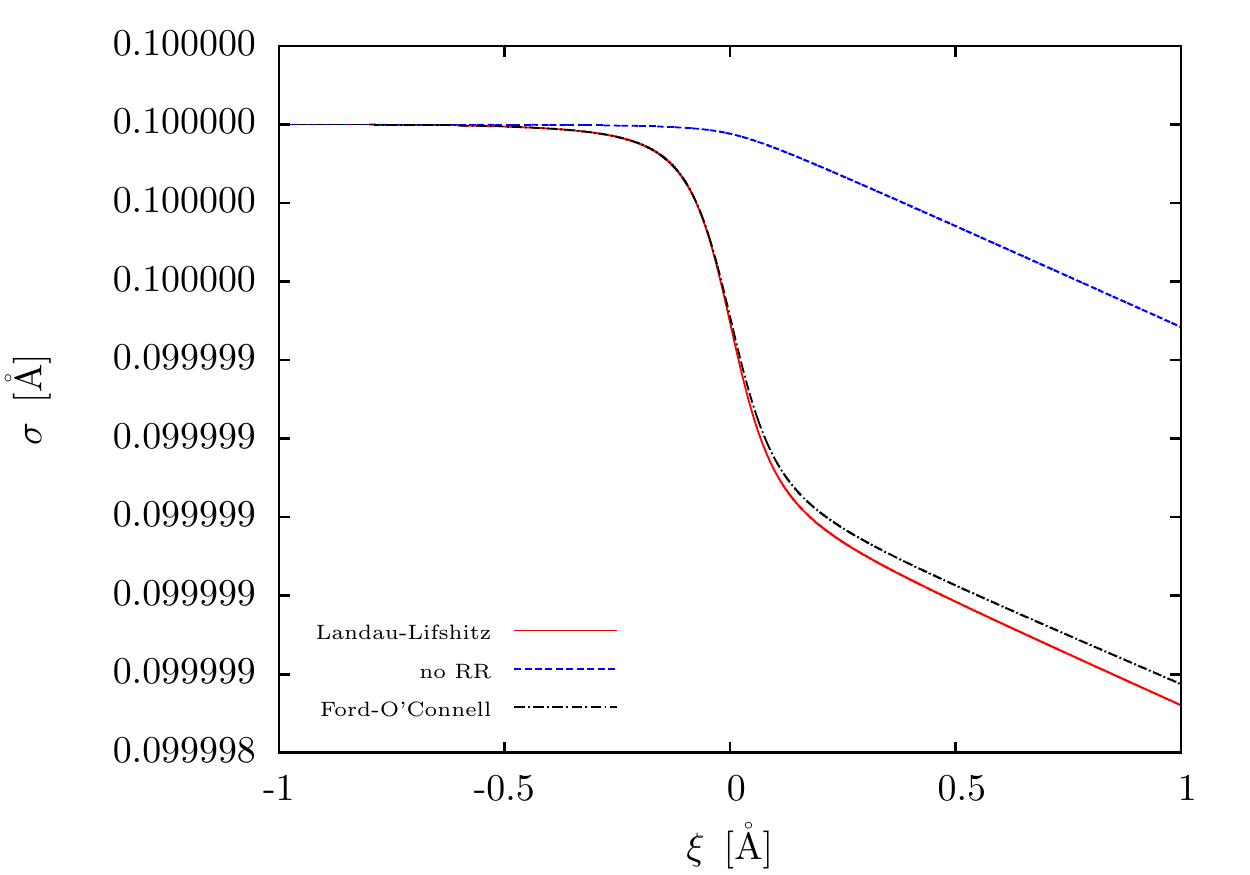}
\caption{\label{fig:5:trajectory_g5_i11} Trajectory of the particle with $\gamma_\text{in} = 10^5$ and $b = 10^{-1}\mathring{\mathrm{A}}$.}
\end{figure}

\begin{figure}[H]
\centering
\includegraphics[width=0.85\textwidth]{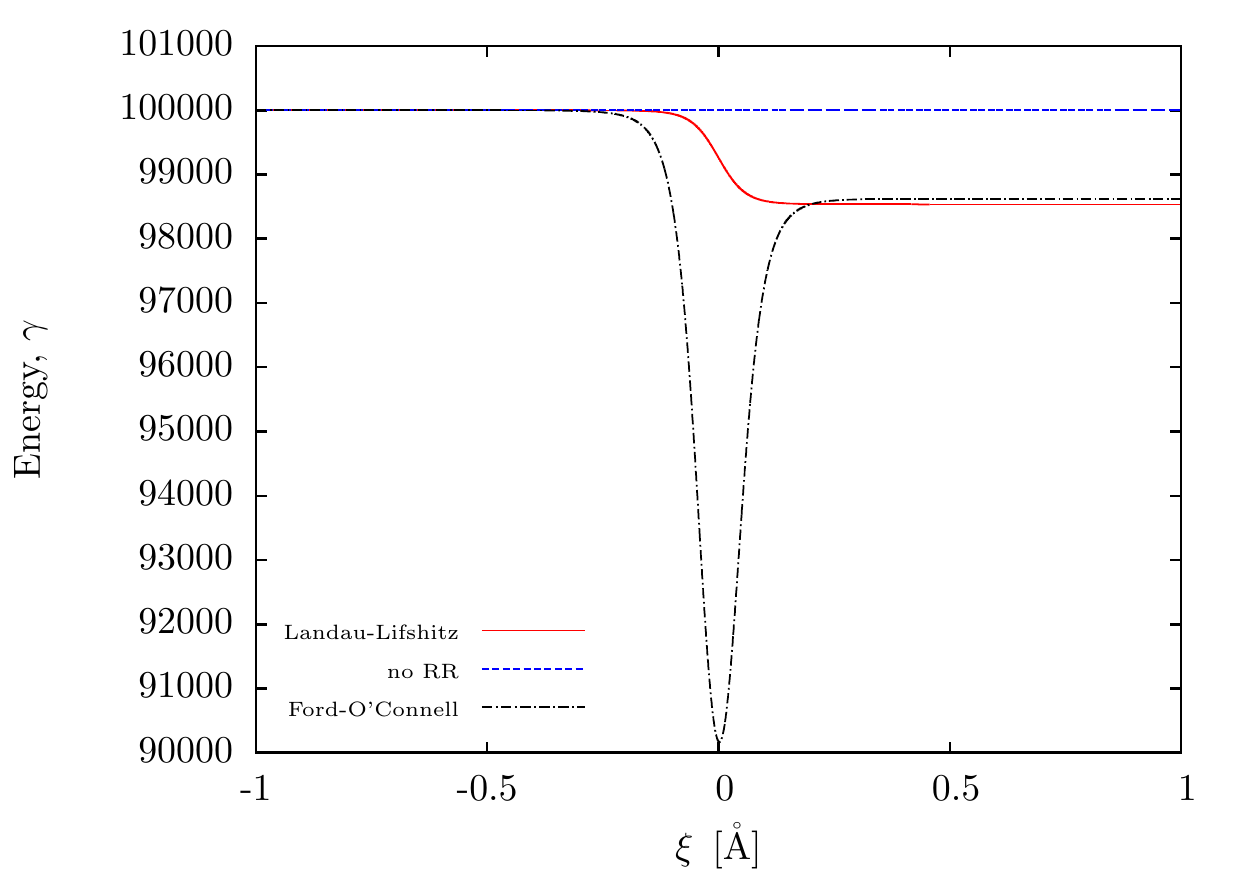}
\caption{\label{fig:5:energy_g5_i11} Energy evolution of the particle with $\gamma_\text{in} = 10^5$ and $b = 10^{-1}\mathring{\mathrm{A}}$.}
\end{figure}

\begin{figure}[H]
\centering
\includegraphics[width=0.85\textwidth]{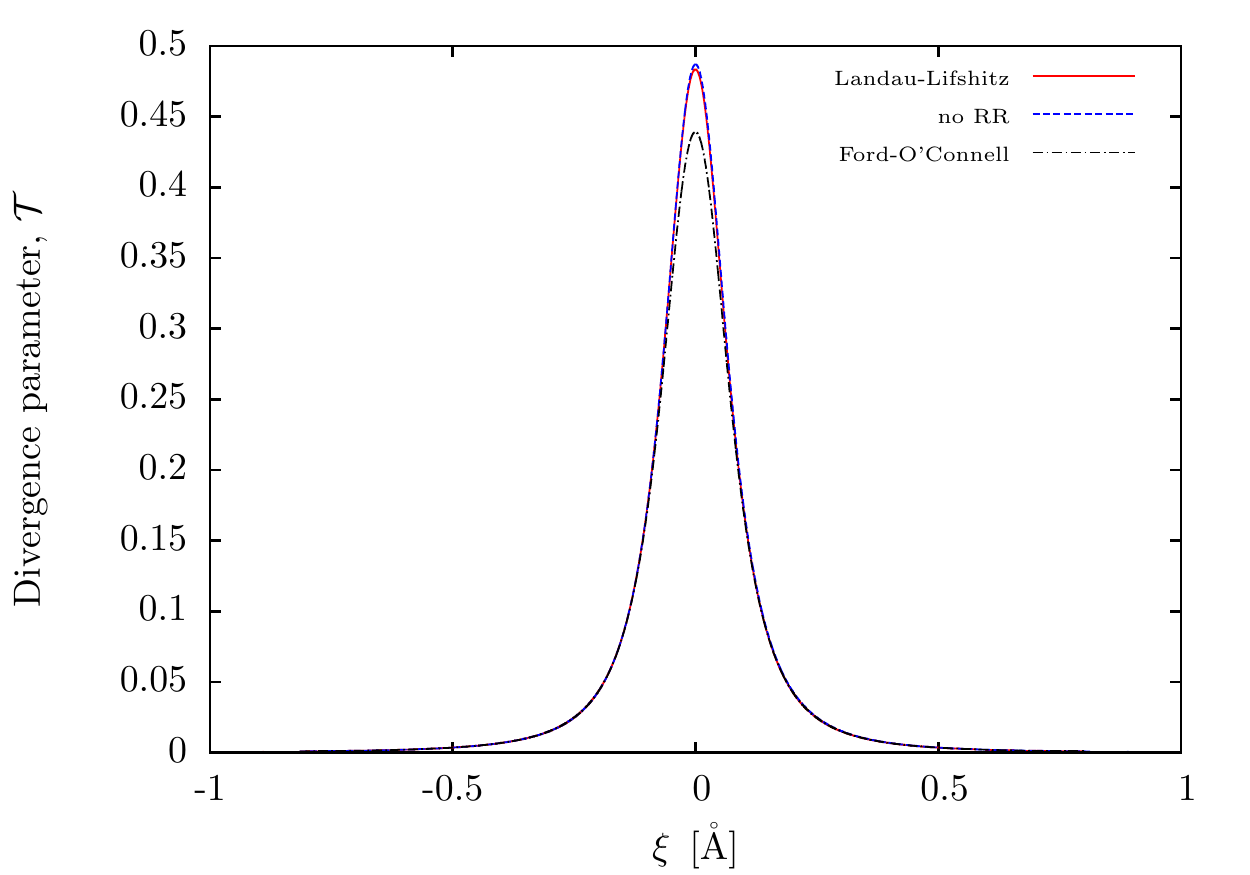}
\caption{\label{fig:5:divergence_g5_i11} Divergence parameter, $\mathcal{T}$ for the particle with $\gamma_\text{in} = 10^5$ and $b = 10^{-1}\mathring{\mathrm{A}}$.}
\end{figure}

\begin{figure}[H]
\centering
\includegraphics[width=0.85\textwidth]{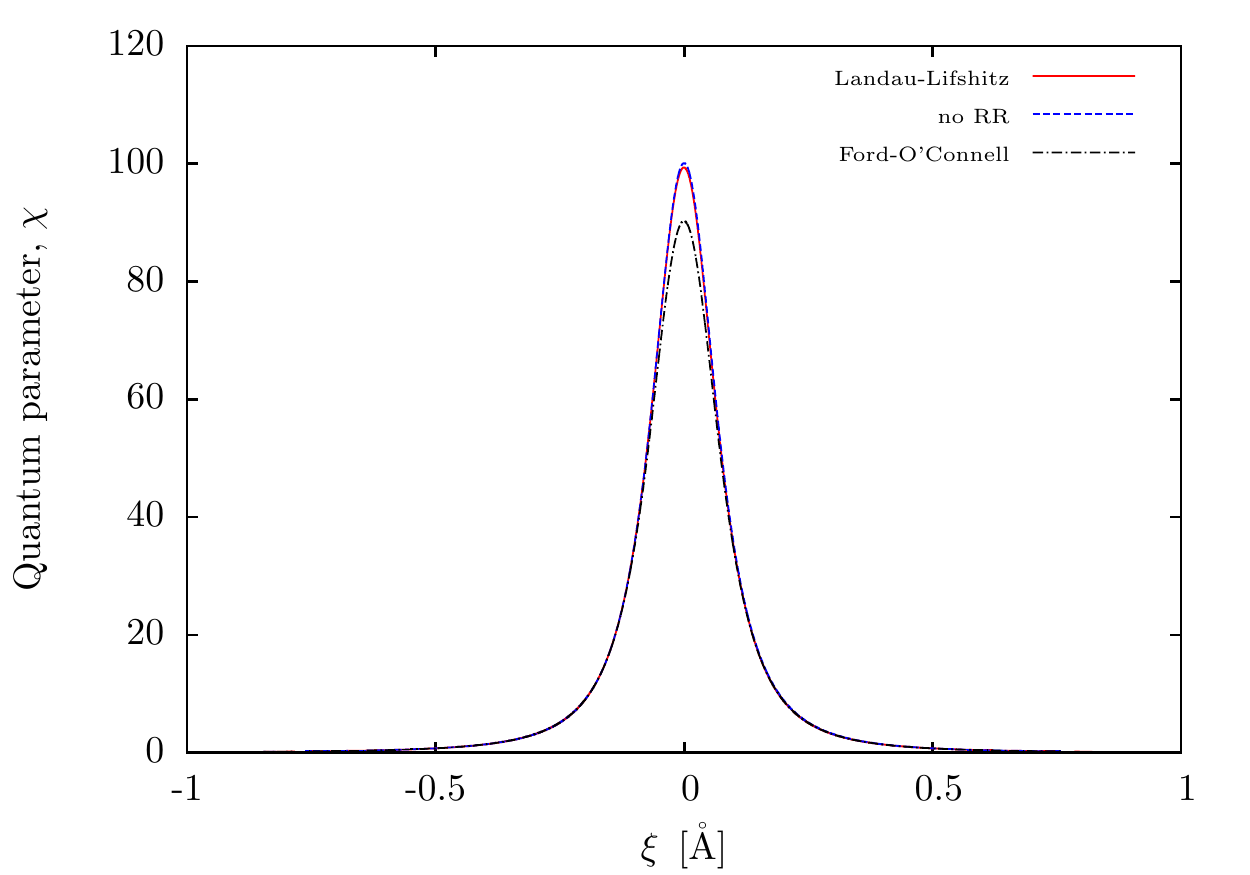}
\caption{\label{fig:5:quantum_g5_i11} Quantum parameter, $\chi$ for the particle with $\gamma_\text{in} = 10^5$ and $b = 10^{-1}\mathring{\mathrm{A}}$.}
\end{figure}

\subsection{Particle with energy $\gamma_\text{in} = 10^3$ and impact parameter $b = 10^{-2}\mathring{\mathrm{A}}$}

\begin{figure}[H]
\centering
\includegraphics[width=0.85\textwidth]{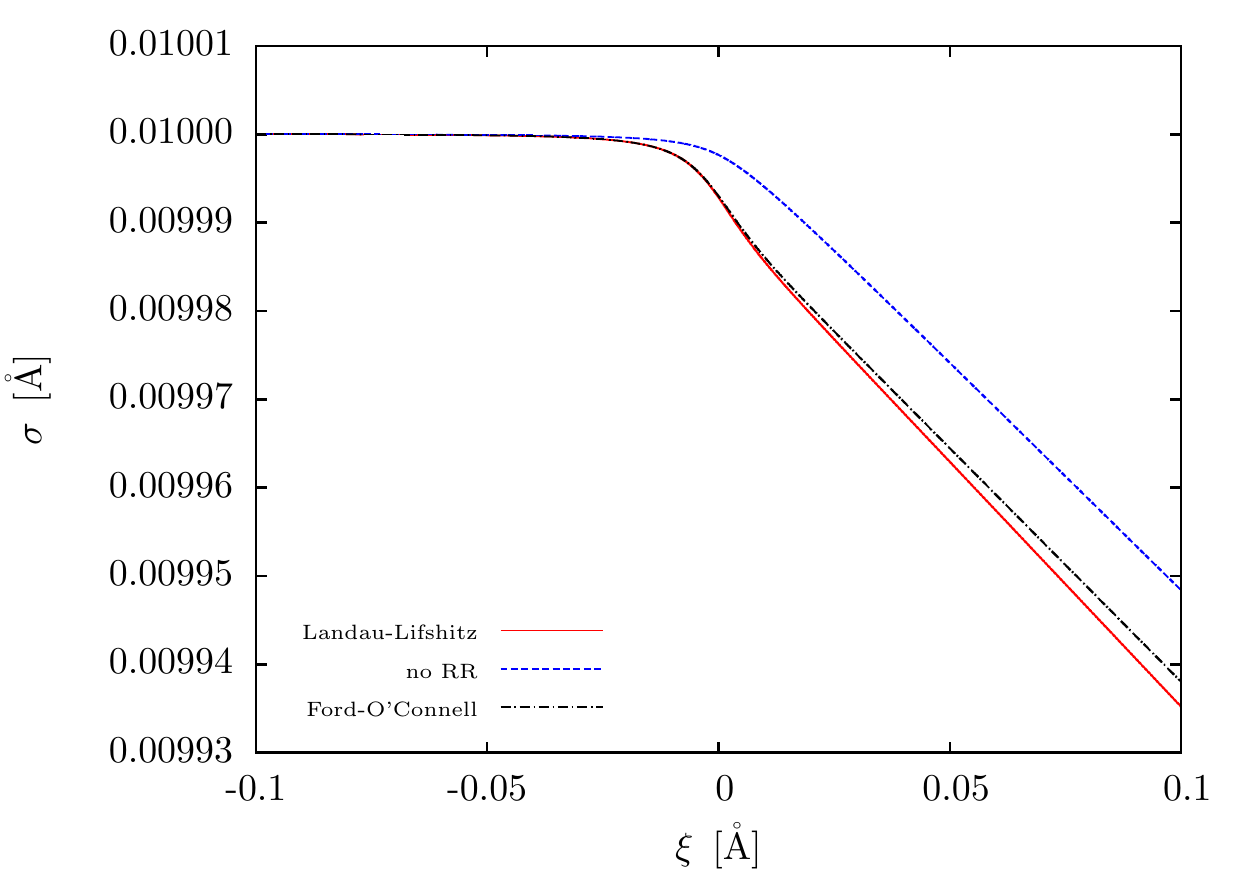}
\caption{\label{fig:5:trajectory_g3_i12} Trajectory of the particle with $\gamma_\text{in} = 10^3$ and $b = 10^{-2}\mathring{\mathrm{A}}$.}
\end{figure}

\begin{figure}[H]
\centering
\includegraphics[width=0.85\textwidth]{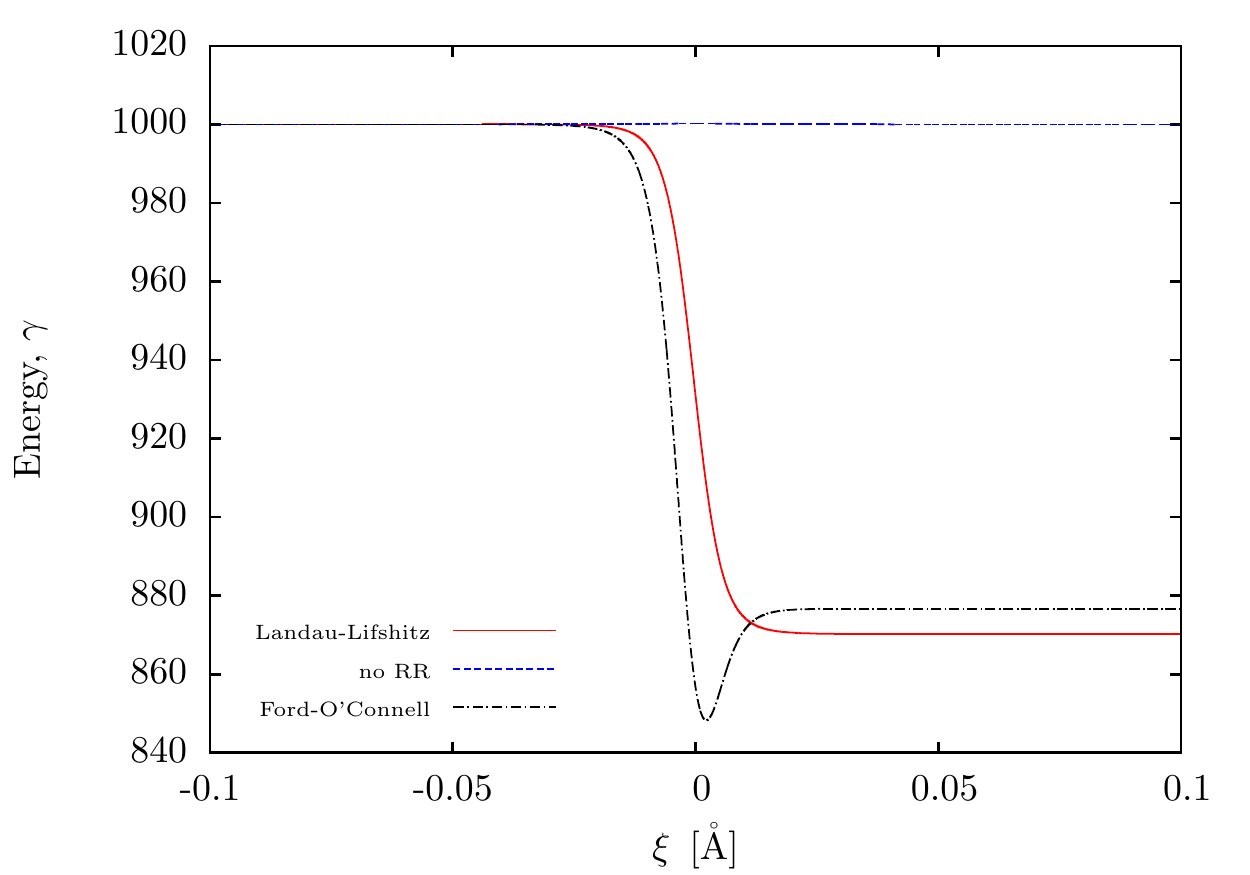}
\caption{\label{fig:5:energy_g3_i12} Energy evolution of the particle with $\gamma_\text{in} = 10^3$ and $b = 10^{-2}\mathring{\mathrm{A}}$.}
\end{figure}

\begin{figure}[H]
\centering
\includegraphics[width=0.85\textwidth]{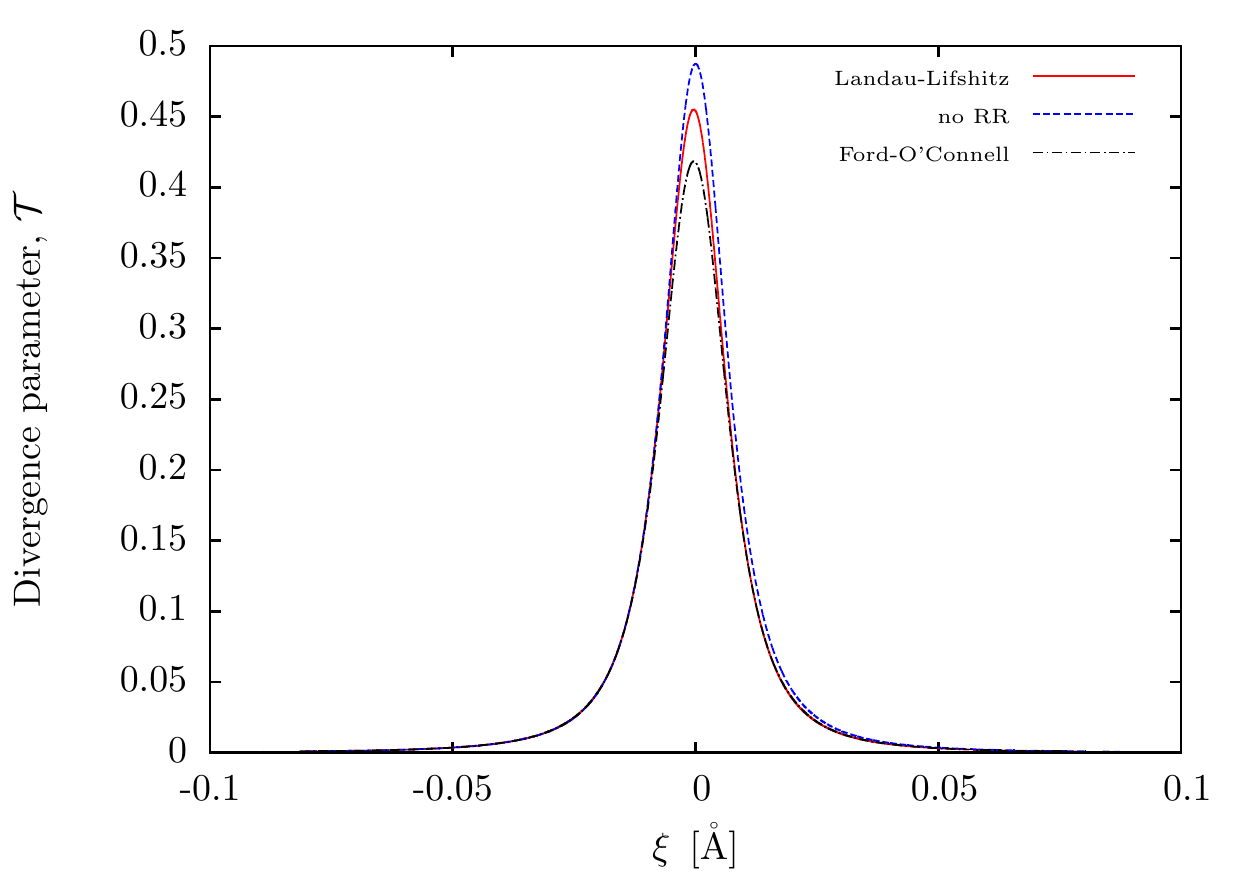}
\caption{\label{fig:5:divergence_g3_i12} Divergence parameter, $\mathcal{T}$ for the particle with $\gamma_\text{in} = 10^3$ and $b = 10^{-2}\mathring{\mathrm{A}}$.}
\end{figure}

\begin{figure}[H]
\centering
\includegraphics[width=0.85\textwidth]{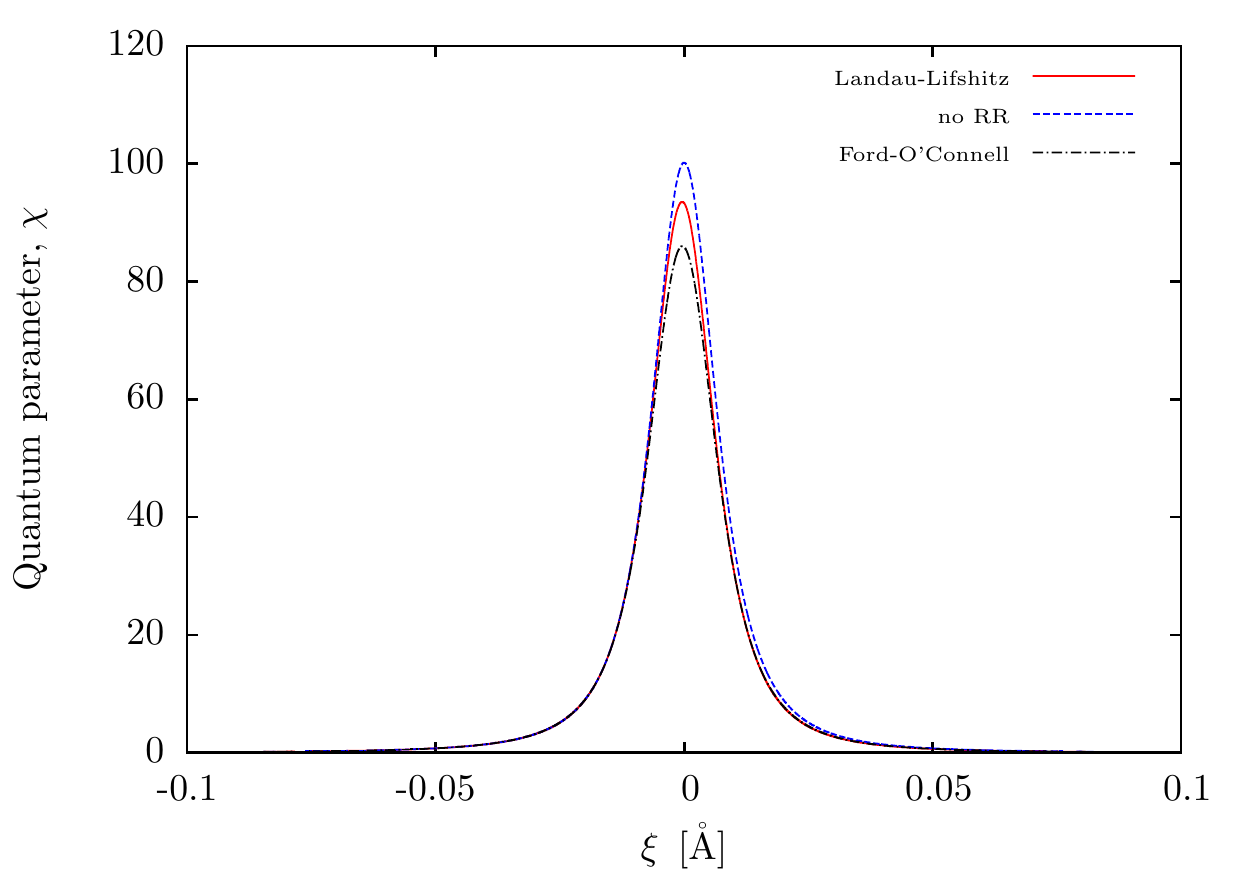}
\caption{\label{fig:5:quantum_g3_i12} Quantum parameter, $\chi$ for the particle with $\gamma_\text{in} = 10^3$ and $b = 10^{-2}\mathring{\mathrm{A}}$.}
\end{figure}

\subsection{Particle with energy $\gamma_\text{in} = 10^2$ and impact parameter $b = 10^{-2}\mathring{\mathrm{A}}$}

\begin{figure}[H]
\centering
\includegraphics[width=0.85\textwidth]{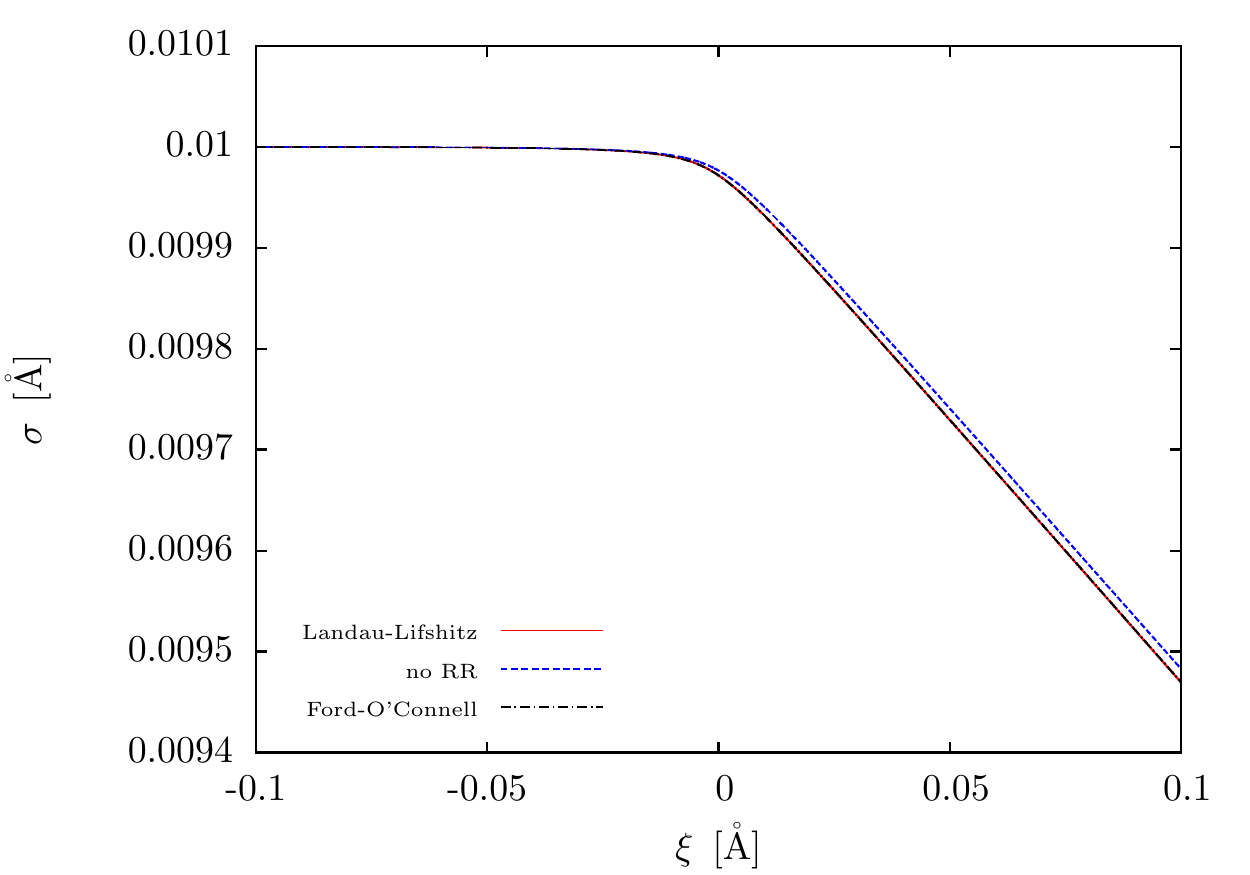}
\caption{\label{fig:5:trajectory_g2_i12} Trajectory of the particle with $\gamma_\text{in} = 10^2$ and $b = 10^{-2}\mathring{\mathrm{A}}$.}
\end{figure}

\begin{figure}[H]
\centering
\includegraphics[width=0.85\textwidth]{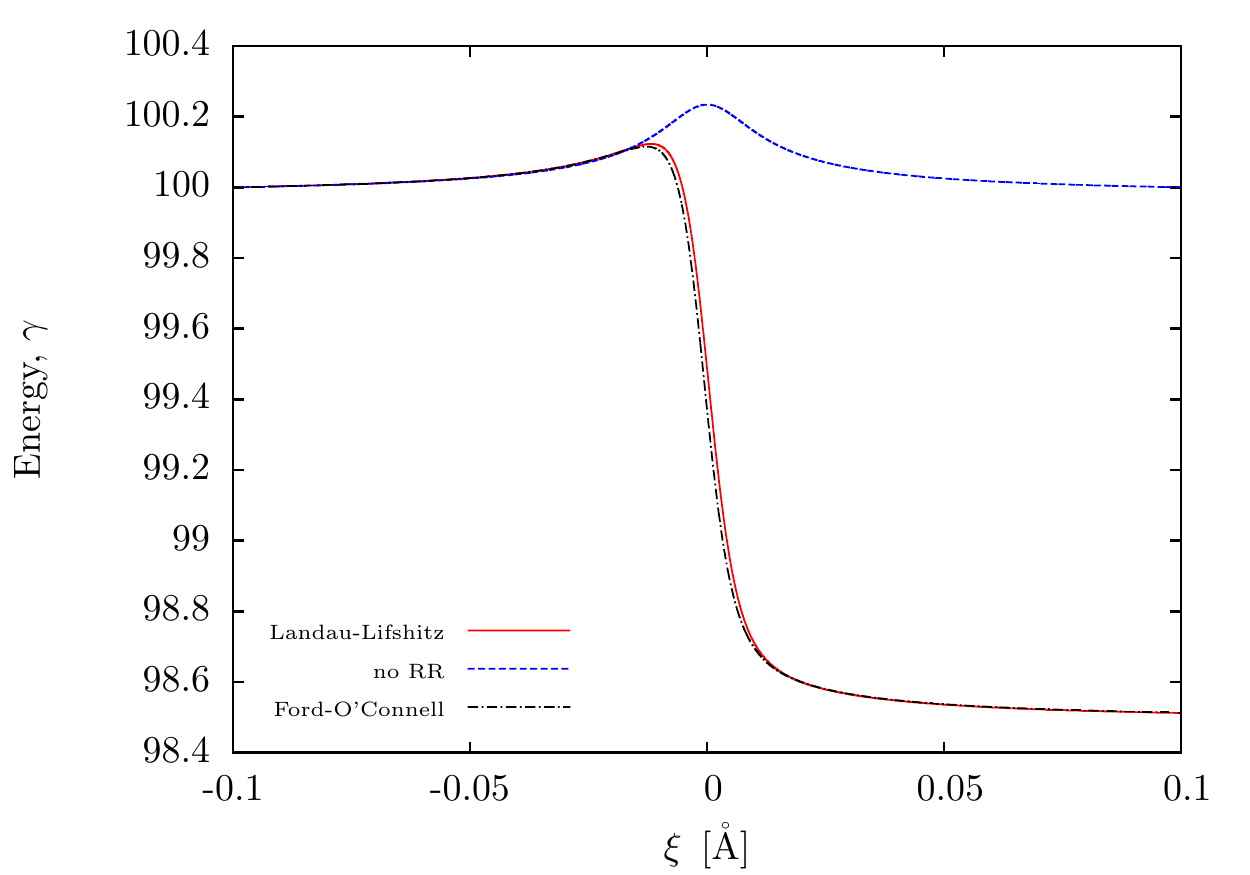}
\caption{\label{fig:5:energy_g2_i12} Energy evolution of the particle with $\gamma_\text{in} = 10^2$ and $b = 10^{-2}\mathring{\mathrm{A}}$.}
\end{figure}

\begin{figure}[H]
\centering
\includegraphics[width=0.85\textwidth]{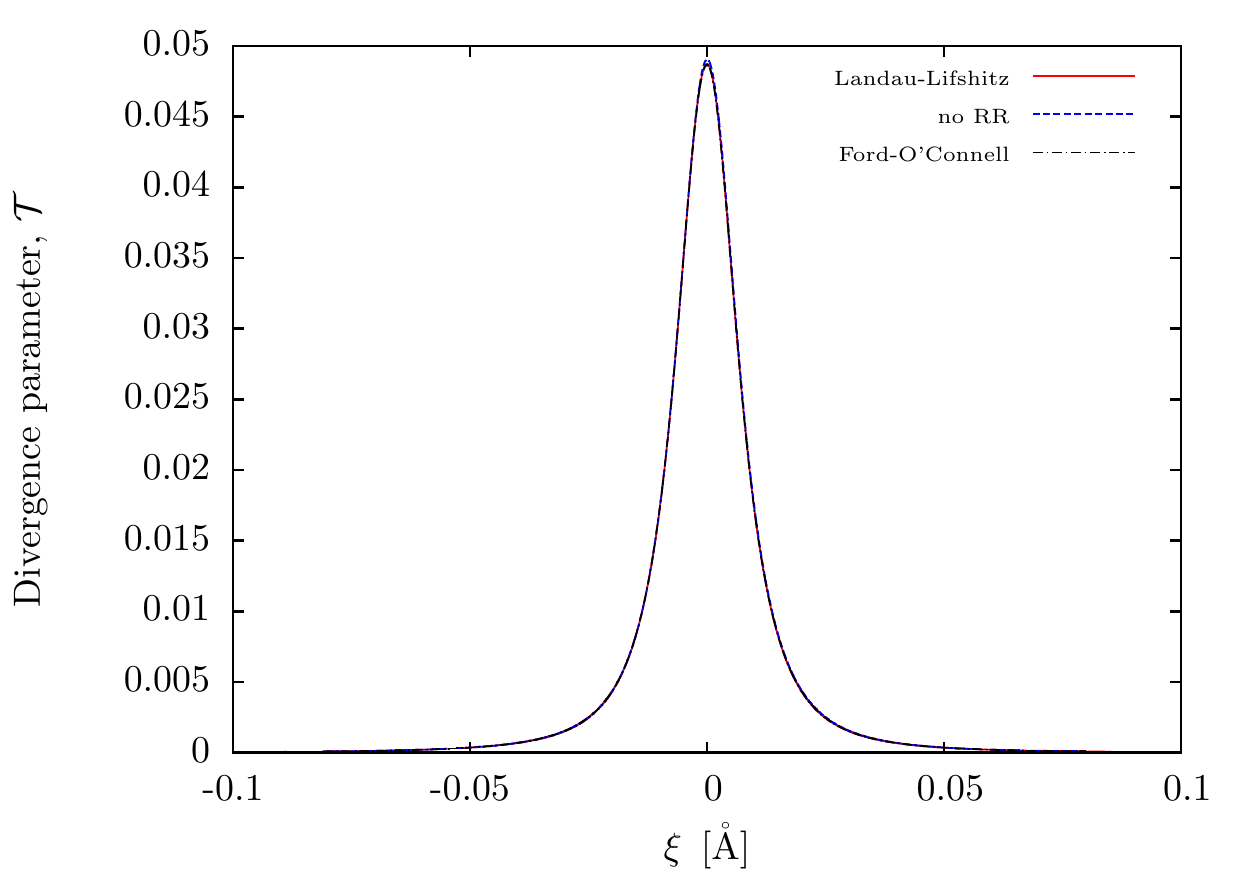}
\caption{\label{fig:5:divergence_g2_i12} Divergence parameter, $\mathcal{T}$ for the particle with $\gamma_\text{in} = 10^2$ and $b = 10^{-2}\mathring{\mathrm{A}}$.}
\end{figure}

\begin{figure}[H]
\centering
\includegraphics[width=0.85\textwidth]{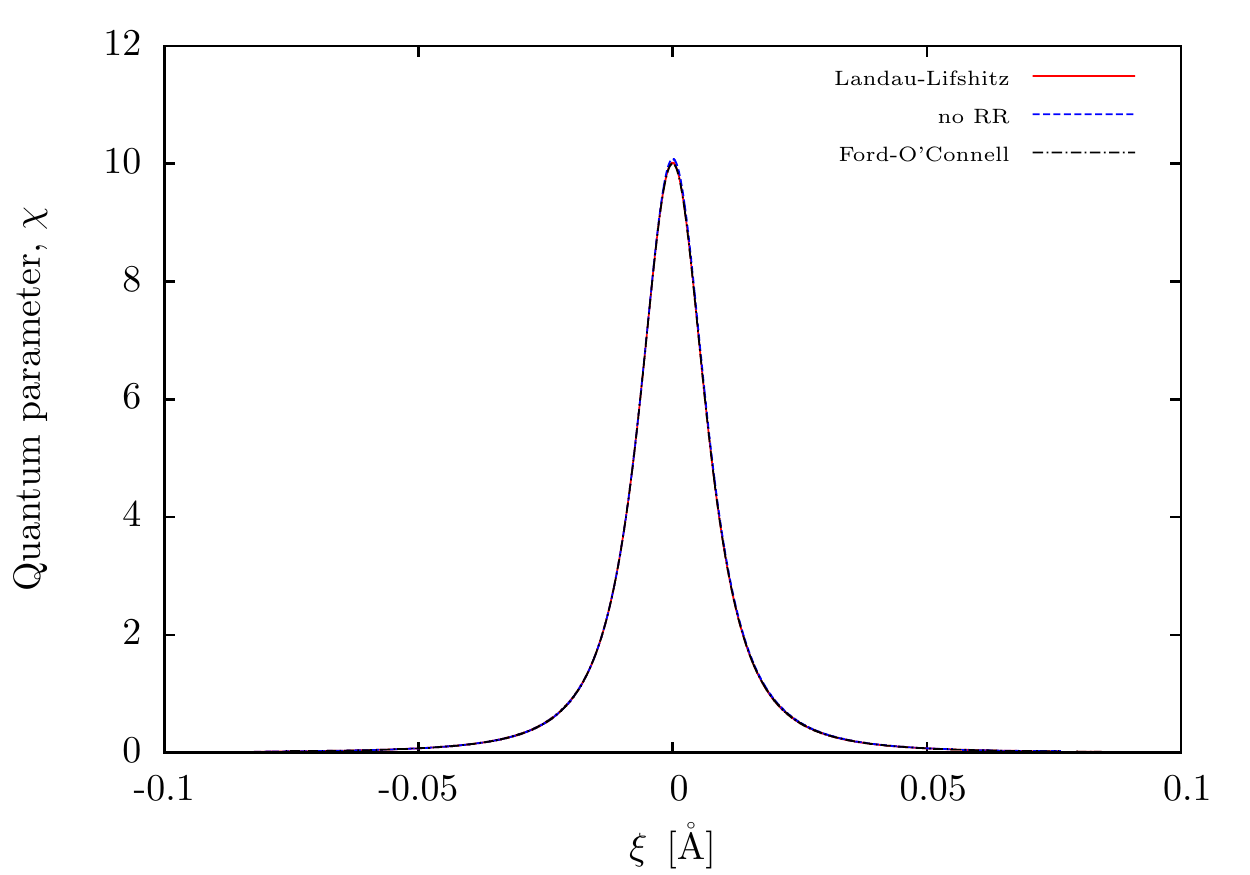}
\caption{\label{fig:5:quantum_g2_i12} Quantum parameter, $\chi$ for the particle with $\gamma_\text{in} = 10^2$ and $b = 10^{-2}\mathring{\mathrm{A}}$.}
\end{figure}

\subsection{Particle with energy $\gamma_\text{in} = 10^2$ and impact parameter $b = 10^{-3}\mathring{\mathrm{A}}$}

\begin{figure}[H]
\centering
\includegraphics[width=0.85\textwidth]{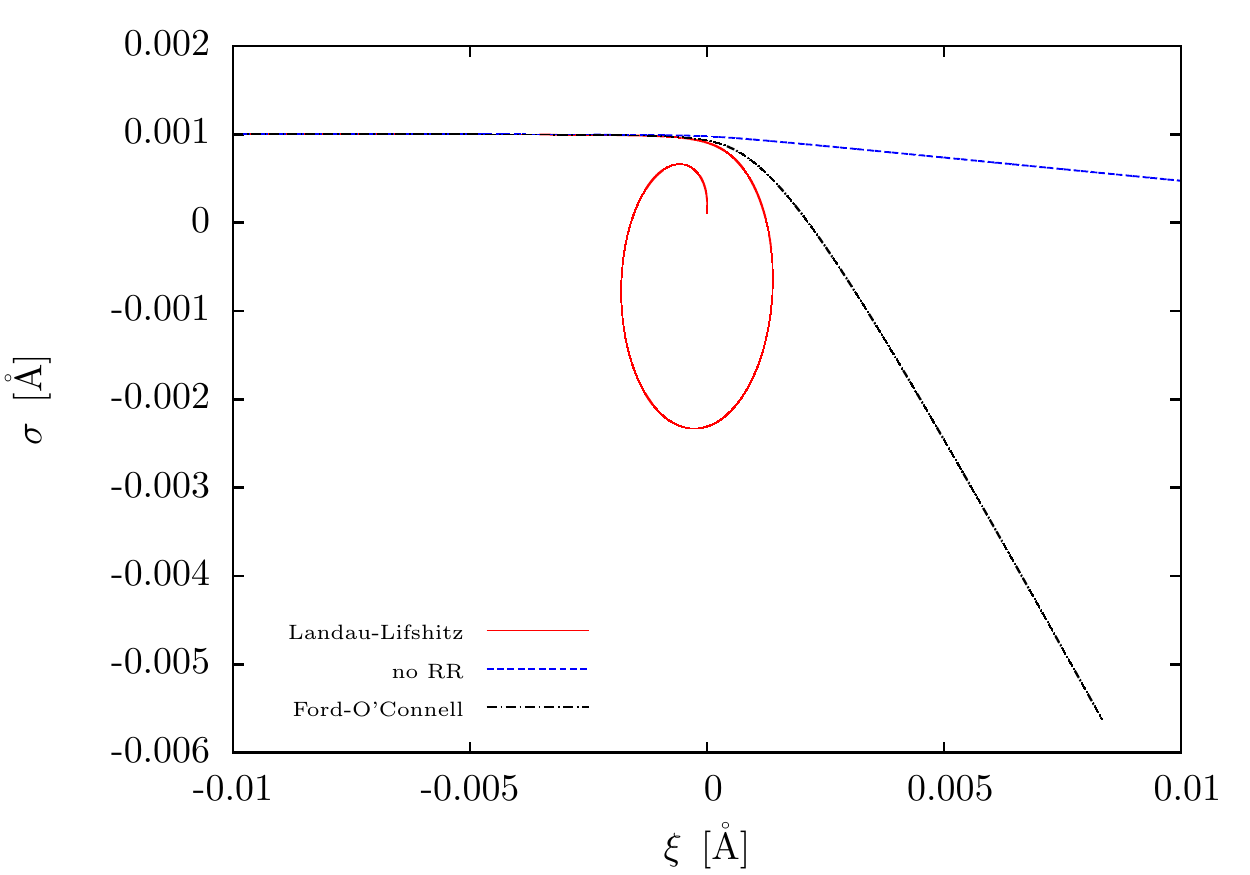}
\caption{\label{fig:5:trajectory_g2_i13} Trajectory of the particle with $\gamma_\text{in} = 10^2$ and $b = 10^{-3}\mathring{\mathrm{A}}$.}
\end{figure}

\begin{figure}[H]
\centering
\includegraphics[width=0.85\textwidth]{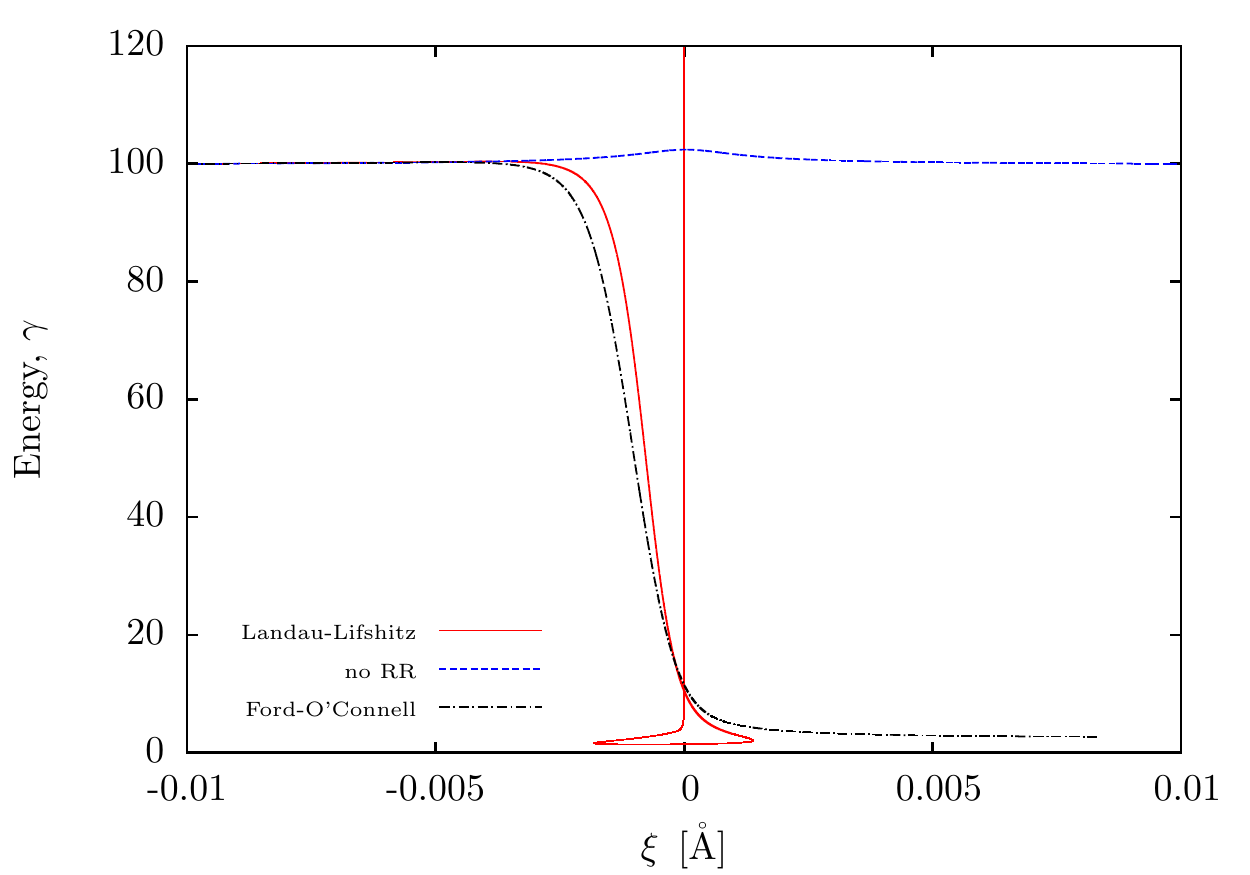}
\caption{\label{fig:5:energy_g2_i13} Energy evolution of the particle with $\gamma_\text{in} = 10^2$ and $b = 10^{-3}\mathring{\mathrm{A}}$.}
\end{figure}

\begin{figure}[H]
\centering
\includegraphics[width=0.85\textwidth]{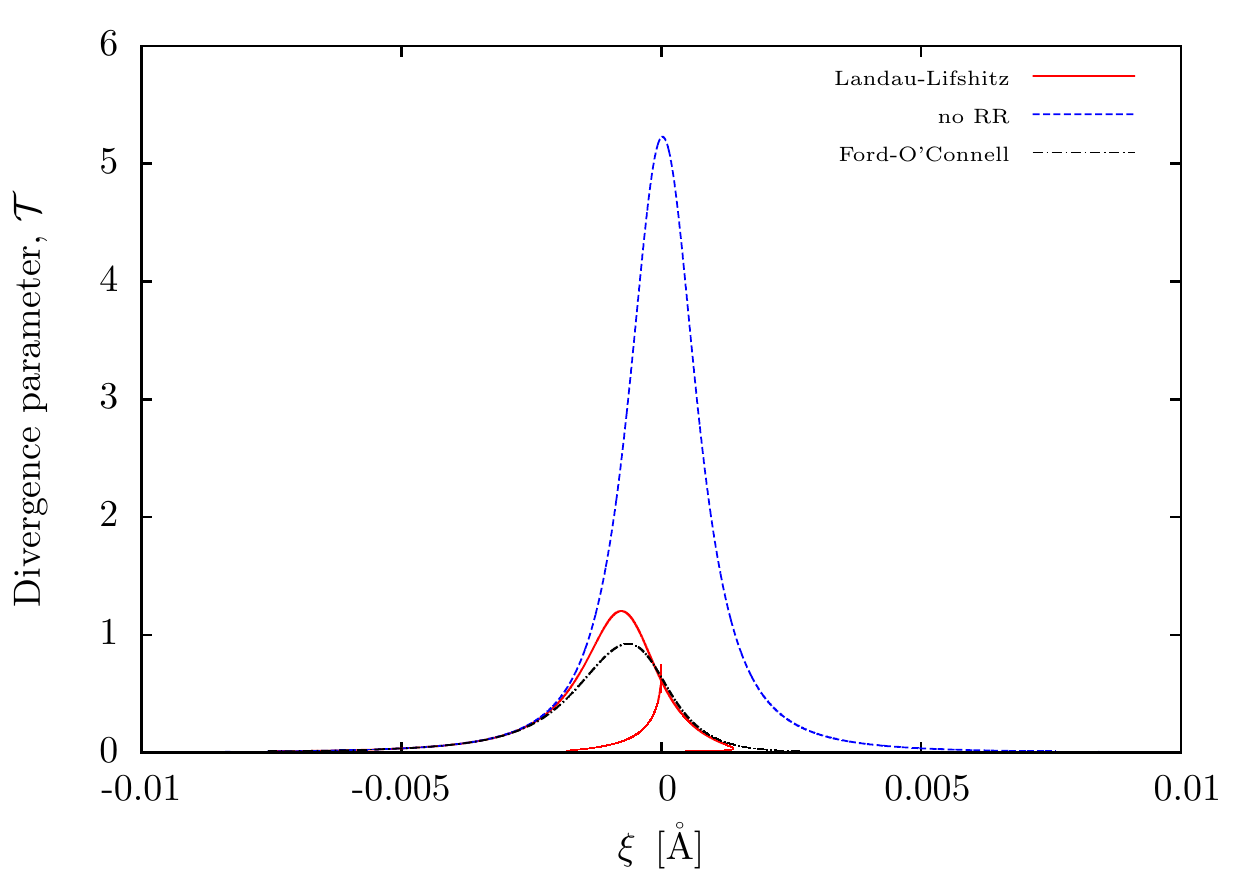}
\caption{\label{fig:5:divergence_g2_i13} Divergence parameter, $\mathcal{T}$ for the particle with $\gamma_\text{in} = 10^2$ and $b = 10^{-3}\mathring{\mathrm{A}}$.}
\end{figure}

\begin{figure}[H]
\centering
\includegraphics[width=0.85\textwidth]{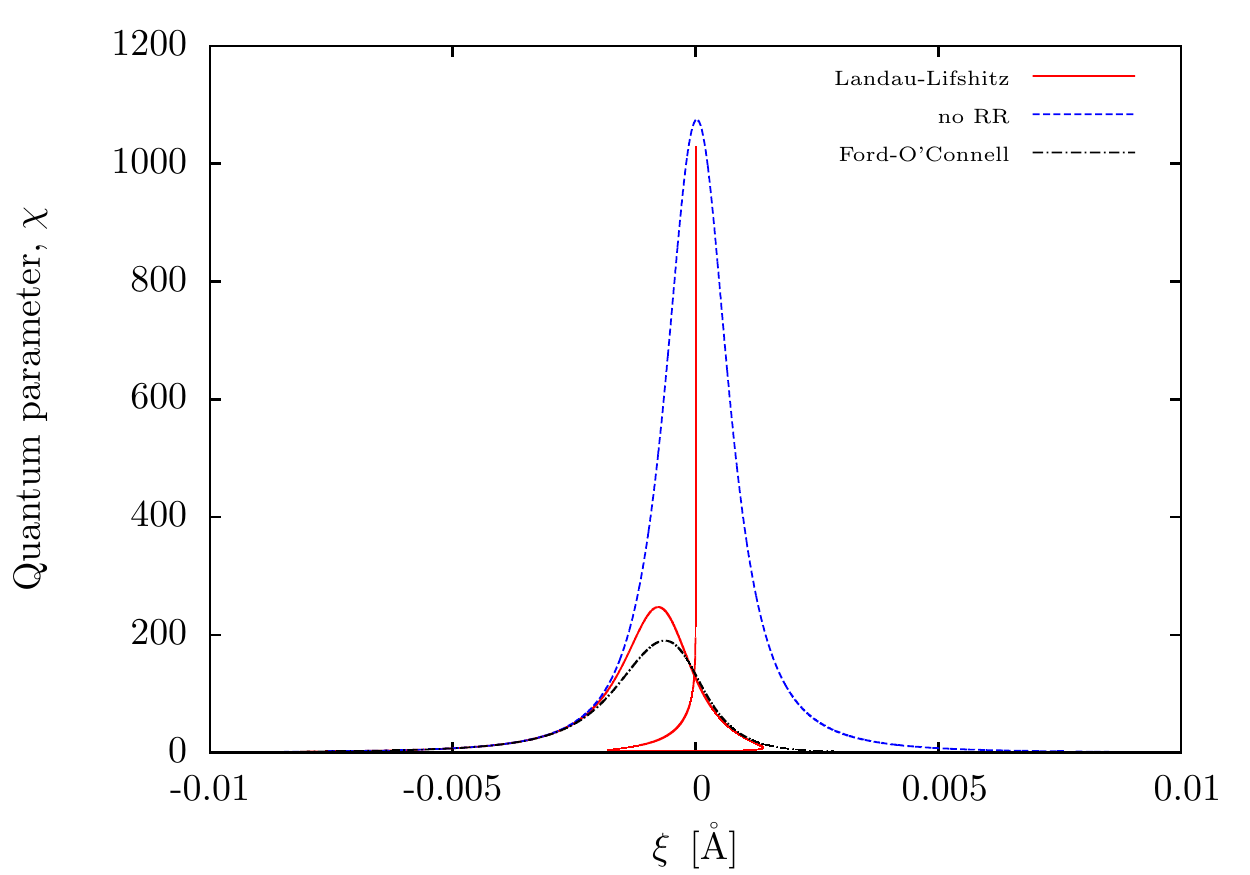}
\caption{\label{fig:5:quantum_g2_i13} Quantum parameter, $\chi$ for the particle with $\gamma_\text{in} = 10^2$ and $b = 10^{-3}\mathring{\mathrm{A}}$.}
\end{figure}

For the case of relatively high initial energy of the particle ($\gamma_\text{in} = 10^5$) incident on the nucleus with a relatively large impact parameter ($b = 1\mathring{\mathrm{A}}$) we observe negligible impact of radiation reaction on the particle trajectory compared to the no radiation reaction case, as shown in Fig.~\ref{fig:5:trajectory_g5_i10}. One can clearly see that the predictions of the Landau-Lifshitz and Ford-O'Connell approaches coincide. This also follows from the small instantaneous values of the divergence parameter $\mathcal{T}$, as shown in Fig.~\ref{fig:5:divergence_g5_i10}. If we now look at the energy evolution plot for this case, Fig.~\ref{fig:5:energy_g5_i10} it can be seen that Ford-O'Connell and Landau-Lifshitz instantaneous corrections differ, despite ultimately leading to the same final prediction. However, as can be seen in Fig.~\ref{fig:5:quantum_g5_i10} the quantum parameter $\chi$ has a value close to 1 in this case, indicating that quantum effects could play an important role in this setup.

Further decrease of the impact parameter $b$ leads to differences between Ford-O'Connell and Landau-Lifshitz predictions of both trajectory and energy evolution Fig.~\ref{fig:5:trajectory_g5_i11}, ~\ref{fig:5:energy_g5_i11}, which can be explained by the relatively high value of the divergence parameter $\mathcal{T}$ as shown on Fig.~\ref{fig:5:divergence_g5_i11}. However, evolution of the quantum parameter for this case Fig.~\ref{fig:5:quantum_g5_i11} indicates that we are in a strongly quantum regime.

Further decrease of both the initial energy of the particle $\gamma_\text{in}$ and impact parameter $b$ leads to more significant differences between Landau-Lifshitz and Ford-O'Connell, Fig.~\ref{fig:5:trajectory_g3_i12}--~\ref{fig:5:quantum_g2_i12}, however remaining in the quantum regime, see Fig.~\ref{fig:5:quantum_g3_i12}, ~\ref{fig:5:quantum_g2_i12}.

A shining example here is the case of initial energy $\gamma_\text{in} = 10^2$ and impact parameter $b = 10^{-3}\mathring{\mathrm{A}}$ where Ford-O'Connell predicts particle scattering, whereas Landau-Lifshitz predicts the electron will be captured by the nucleus, see Fig.~\ref{fig:5:trajectory_g2_i13}. The initially free particle retains this state if radiation reaction isn't taken into account, since its \textit{total energy} is conserved. However, radiation losses allow the particle's total energy to become negative, indicating that it becomes bound to the nucleus. Since in the Landau-Lifshitz case the particle spirals into the nucleus the $\xi$ coordinate is no longer single valued, as seen in Fig.~\ref{fig:5:trajectory_g2_i13}--\ref{fig:5:quantum_g2_i13}. Landau-Lifshitz predicts greater energy loss than Ford-O'Connell allowing the transition from the free state to the bound one, which is not the case with Ford-O'Connell. However, as can be seen in Fig.~\ref{fig:5:quantum_g2_i13}, in this case we remain in a strongly quantum regime.

If we now consider the ratio between the \textit{divergence parameter} (\ref{eq:5:tau_nucleus}) and the \textit{quantum parameter} (\ref{eq:5:quantum_parameter}) we get:
\begin{equation}
\label{eq:5:divquantratio}
\vspace{-0.5em}
\frac{\mathcal{T}}{\chi} = \frac{m\tau}{\hbar} \frac{R\dot{\theta}}{\sqrt{1 + R^2\dot{\theta}^2}}\ ,
\vspace{-0.5em}
\end{equation}
where $\frac{m\tau}{\hbar} = \frac{2}{3} \frac{q^2}{4\pi\hbar} = \frac{2}{3}\alpha \simeq 4.9\cdot 10^{-3}$, with $\alpha$ the fine structure constant.

On the other hand, the condition
\begin{equation}
\vspace{-0.5em}
\frac{R\dot{\theta}}{\sqrt{1 + R^2\dot{\theta}^2}} < 1\ ,
\end{equation}
is always fulfilled, therefore the ratio (\ref{eq:5:divquantratio}) is always small.

Since the divergence parameter $\mathcal{T}$ is responsible for differences between Ford-O'Connell and Landau-Lifshitz predictions, whereas the quantum parameter $\chi$ governs the significance of quantum effects, this suggests that we would not observe significant differences between Ford-O'Connell and Landau-Lifshitz predictions while quantum effects are still small enough not to be taken into account. It would be of interest to carry out a fully quantum mechanical treatment of this effect in future work.
\vspace{-0.5em}
\section{Summary}
\vspace{-0.5em}
It can be clearly seen that radiation reaction has an effect on particle motion in this physical scenario and one can see noticeable differences between the predictions of Ford-O'Connell and Landau-Lifshitz approaches, even when the divergence parameter $\mathcal{T}$ is small. This could lead to radiation reaction corrections to the Rutherford cross section (analogous to Dirac's radiation reaction correction to the Thomson cross section \cite{Dirac}).

For the extreme cases ($b = 10^{-3} \mathring{\mathrm{A}}$) significant differences in the particle's behaviour are observed (such as particle capture) in regions where $\mathcal{T}$ is no longer small.

However, we find that in these cases quantum effects appear to be very important and radiation reaction is not found to be significant while quantum effects remain small.

\chapter{Conclusions}

\begin{figure}[H]
\vspace{-15em}
\centering
\includegraphics[width=\textwidth]{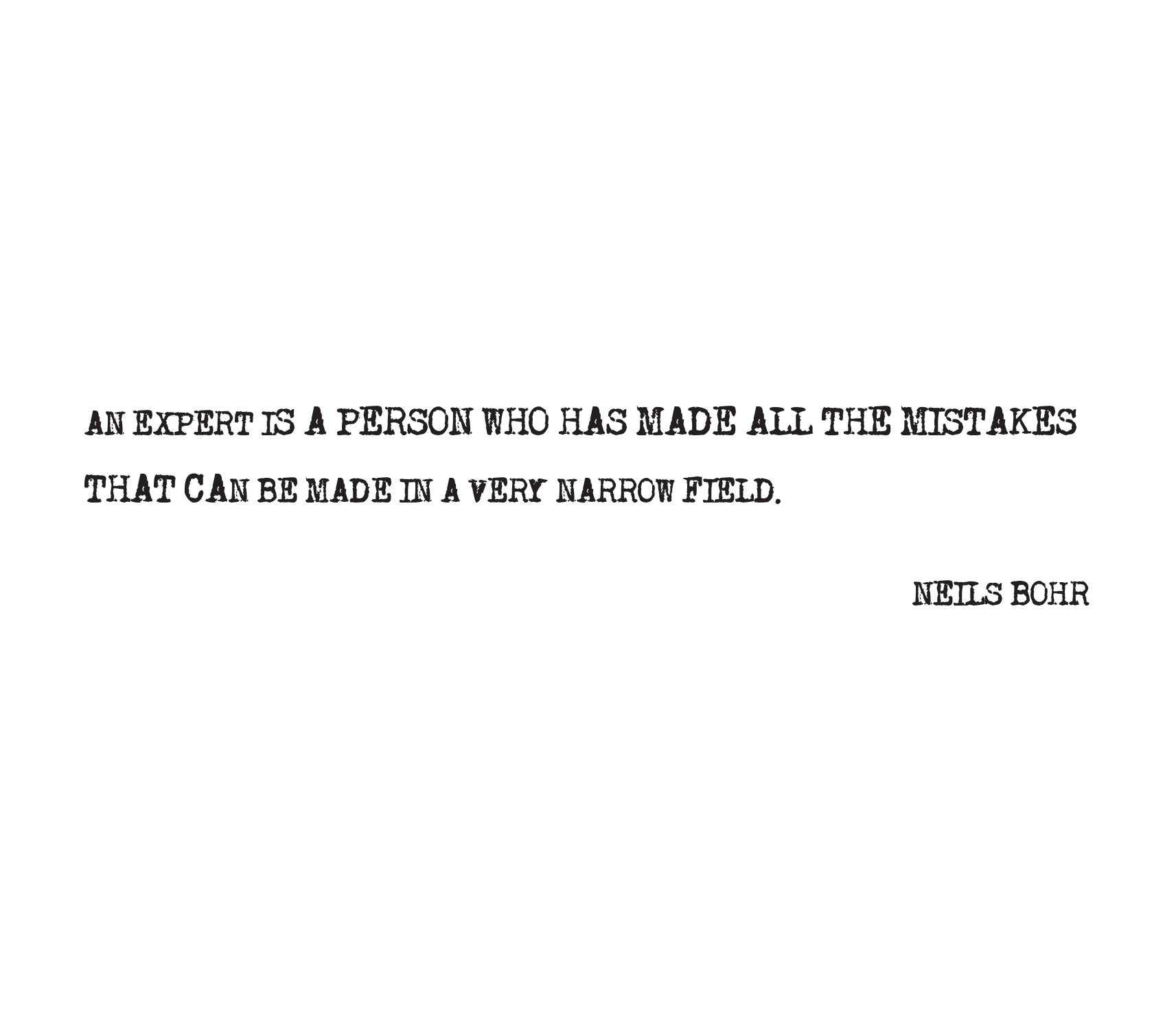}
\end{figure} 

\newpage 

\section{Summary}

The work presented in this thesis is focused on exploring the dynamics of charged particles in regimes where radiation reaction becomes an important effect. We have investigated a variety of physical setups, which have provided clarification of questions related to the significance of radiation reaction effects, and the validity of their theoretical descriptions. 

We started by discussing two common theoretical descriptions of radiation reaction, those of Lorentz, Abraham and Dirac, and of Landau and Lifshitz, respectively, emphasising the problems related to these models. For example, Lorentz-Abraham-Dirac suffers exponentially growing acceleration (“runaway solutions”) and violation of causality (pre–acceleration) while Landau-Lifshitz is pertubative, limiting its domain of validity. 

We then investigated an alternative model for classical radiation reaction based on the Ford-O'Connell equation. The main aim of this model is to address the issues mentioned above. We have derived a condition for the predictions of the Ford-O'Connell model to deviate from those of Landau-Lifshitz, which allows us to assess the validity of the latter. 
 
In subsequent Chapters we presented a study of the interaction of a high-energy particle with an intense laser pulse. By analysing this interaction we found that radiation reaction can have a significant effect on the particle motion and, in certain regimes, even dominates the applied Lorentz force. However, we have also found that the detailed interplay between the particle dynamics and the shape of the pulse ensures that, even with the most advanced technologies currently proposed, 
radiation reaction prevents the particle from accessing the regime where the Landau-Lifshitz approximation would break down. 

Extending this study we proceeded with an exploration of radiation reaction effects on the momentum spread of a bunch of particles during their propagation through an intense laser pulse. We confirm that, while decreasing the energy of particles, radiation reaction also leads to a reduction of the initial momentum spread when a relativistic particle bunch passes through an intense laser pulse. This leads to an improvement in the bunch quality at the expense of its energy. One of the main results of this study indicates that the change in momentum spread of the particle distribution depends only on the total energy of the laser pulse, and remains completely independent of the way this energy is distributed. This behviour is backed up by both analytical considerations and simulation results.

We have also shown that, with the increase of initial momentum, the particle bunch asymptotes to a constant final average momentum with zero momentum spread.

We finally considered an atomic nucleus as a source of high fields, which led us to explore a particle being scattered off a heavy nucleus. We presented an analytical model, which includes classical radiation reaction corrections and derived relevant equations of motion for this case and discussed the relevance of quantum effects during the interaction. 

We found that radiation reaction has an effect on the particle's motion and noticeable differences between the predictions of the Ford-O'Connell and Landau-Lifshitz approaches are evident, even when the divergence parameter $\mathcal{T}$ is small. An extreme case ($b = 10^{-3} \mathring{\mathrm{A}}$) leads to significantly different predictions in the particle's behaviour (such as particle capture), which is observed in regions where the divergence parameter $\mathcal{T}$ is no longer small.

However, one of the main conclusions of this work is that in these cases quantum effects appear to be very important and conversely radiation reaction is not found to be significant while quantum effects remain small. This implies the need to develop a quantum model in order to correctly interpret these cases.

\newpage 

\section{Outlook}

The work presented in this thesis has been restricted by certain limitations which could be addressed in future work. 

Chapter 3 was devoted to the exploration of the effects of radiation reaction for a single particle interacting with a plane wave. The results we obtained strongly indicated that while in the classical domain we do not anticipate strong differences between the Landau-Lifshitz and Ford-O'Connell approaches. However, realistic laser pulses have important transverse structure, and ponderomotive effects can eject the electron from the pulse. It would therefore be of interest to consider more realistic pulse shapes, which benefit from the transverse structure. This will require moving away from the plane wave approximation and would lead to a different condition for divergence between the Landau-Lifshitz and Ford-O'Connell approaches.

In Chapter 4 the evolution of the distribution function of a bunch of particles interacting with a plane wave was considered. One of the main limitations introduced there was the fact that all the particles of the distribution start at the same point in space, therefore having no spatial spread at all. Future extensions could include spatial as well as momentum spread and an investigation of the effects of radiation reaction also on the spatial spread. Consideration of an initial transverse spread, which is currently considered to be 0, and more realistic pulse structures are additional paths for future work.

As for the work in Chapter 5, we have clearly shown that in cases where Landau-Lifshitz and Ford-O'Connell approaches provide qualitatively different predictions, quantum effects appear to be very important and radiation reaction is not found to be significant while quantum effects remain small. This suggests the need to develop a quantum model in order to correctly analyse these cases.

\appendix
\chapter{Maxwell's equations for a plane wave}

\newpage

\noindent The aim of this Appendix is to demonstrate that, for the case of a plane wave, in order to satisfy Maxwell's equations the electric field depends on $x$ only through the coordinate $\phi$.

To achieve this goal consider Maxwell's equations written in terms of the electromagnetic field tensor:
\begin{equation}
\label{app:3:Max1}
\partial_a F^{ab} = 0\ ,
\end{equation}
and 
\begin{equation}
\label{app:3:Max2}
\partial_a F_{bc} + \partial_b F_{ca} + \partial_c F_{ab} = 0\ .
\end{equation}
Considering the form of the electromagnetic field tensor $F_{ab}$ for the plane wave case:
\begin{equation}
\label{app:3:planewave}
 \frac{q}{m} F\indices{_{ab}} = \mathcal{E}_1\left(x\right) \left( \epsilon_a n_b - 
 \epsilon_b n_a \right) + \mathcal{E}_2\left(x\right) \left( \lambda_a n_b - \lambda_b 
 n_a \right)\ ,
\end{equation}
where $\mathcal{E}_1$ and $\mathcal{E}_2$ correspond to electric fields of an arbitrary form in $\epsilon$ and $\lambda$ directions respectively, and
\begin{equation}
\label{app:3:assumption}
\mathcal{E}_1 = \mathcal{E}_1\left(\phi, \psi, \xi, \sigma\right)\ , \quad \mathcal{E}_2 = \mathcal{E}_2\left(\phi, \psi, \xi, \sigma\right)\ ,
\end{equation}
can depend on all 4 coordinates.

Based on (\ref{app:3:planewave}) the first Maxwell equation (\ref{app:3:Max1}) can be rewritten as:
\begin{equation}
\label{app:3:Max1_rewritten}
\partial_a F^{ab} = \partial_a \mathcal{E}_1 \epsilon^a n^b - \partial_a \mathcal{E}_1 \epsilon^b n^a + \partial_a \mathcal{E}_2 \lambda^a n^b - \partial_a \mathcal{E}_2 \lambda^b n^a\ .
\end{equation}
Using (\ref{app:3:assumption}) and $\phi = n_a x^a$, $\psi = m_a x^a$, $\xi = \epsilon_a x^a$, $\sigma = \lambda_a x^a$, the field derivatives in (\ref{app:3:Max1_rewritten}) can be expanded as:
\begin{align}
\label{app:3:der_rewritten}
\nonumber \partial_a \mathcal{E}_i & = \frac{\partial \mathcal{E}_i}{\partial \phi} \partial_a \phi + \frac{\partial \mathcal{E}_i}{\partial \psi} \partial_a \psi + \frac{\partial \mathcal{E}_i}{\partial \xi} \partial_a \xi + \frac{\partial \mathcal{E}_i}{\partial \sigma} \partial_a \sigma \\
& = \frac{\partial \mathcal{E}_i}{\partial \phi} n_a + \frac{\partial \mathcal{E}_i}{\partial \psi} m_a + \frac{\partial \mathcal{E}_i}{\partial \xi} \epsilon_a + \frac{\partial \mathcal{E}_i}{\partial \sigma} \lambda_a\ ,
\end{align}
where $i \in \{1,2\}$.

Substituting (\ref{app:3:der_rewritten}) into (\ref{app:3:Max1_rewritten}):
\begin{equation}
\left(\frac{\partial \mathcal{E}_1}{\partial \xi} + \frac{\partial \mathcal{E}_2}{\partial \sigma}\right) n^b + 2 \frac{\partial \mathcal{E}_1}{\partial \psi} \epsilon^b + 2 \frac{\partial \mathcal{E}_2}{\partial \psi} \lambda^b = 0
\end{equation}
leading to the following relations:
\begin{equation}
\label{app:3:Poisson}
\frac{\partial \mathcal{E}_1}{\partial \xi} = - \frac{\partial \mathcal{E}_2}{\partial \sigma}\ , 
\end{equation}
\begin{equation}
\label{app:3:cond12}
\frac{\partial \mathcal{E}_1}{\partial \psi} = 0\ ,
\end{equation}
\begin{equation}
\label{app:3:cond13}
\frac{\partial \mathcal{E}_2}{\partial \psi} = 0\ .
\end{equation}
Here equation (\ref{app:3:Poisson}) corresponds to Poisson's law, whereas equations (\ref{app:3:cond12}) and (\ref{app:3:cond13}) correspond to the Ampere-Maxwell law. Equations (\ref{app:3:cond12}) and (\ref{app:3:cond13}) indicate that the electric field components $\mathcal{E}_1$ and $\mathcal{E}_2$ for the case of the plane wave are independent of $\psi$. 

If we now analogously consider the second of Maxwell's equations (\ref{app:3:Max2}):
\begin{align}
\nonumber \partial_a F_{bc} + \partial_b F_{ca} + \partial_c F_{ab} = &\partial_a \mathcal{E}_1 \epsilon_b n_c - \partial_a \mathcal{E}_1 \epsilon_c n_b + \partial_a \mathcal{E}_2 \lambda_b n_c - \partial_a \mathcal{E}_2 \lambda_c n_b + \\
\nonumber &\partial_b \mathcal{E}_1 \epsilon_c n_a - \partial_b \mathcal{E}_1 \epsilon_a n_c + \partial_b \mathcal{E}_2 \lambda_c n_a - \partial_b \mathcal{E}_2 \lambda_a n_c + \\ &\partial_c \mathcal{E}_1 \epsilon_a n_b - \partial_c \mathcal{E}_1 \epsilon_b n_a + \partial_c \mathcal{E}_2 \lambda_a n_b - \partial_c \mathcal{E}_2 \lambda_b n_a\ .
\end{align}

Taking into account (\ref{app:3:cond12}) and (\ref{app:3:cond13}) and using the field derivative expansions (\ref{app:3:der_rewritten}) we obtain the second Maxwell's equation for the case of the plane wave:
\begin{align}
\label{app:3:Maxwell2_rewritten}
\nonumber \partial_a F_{bc} + \partial_b F_{ca} + \partial_c F_{ab} = &\left(\frac{\partial \mathcal{E}_1}{\partial \sigma} - \frac{\partial \mathcal{E}_2}{\partial \xi}\right)\Big[ \lambda_a \epsilon_b n_c - \lambda_a n_b \epsilon_c - \epsilon_a \lambda_b n_c + \\
& \qquad \qquad \qquad \quad \enskip \epsilon_a n_b \lambda_c + n_a \lambda_b \epsilon_c - n_a \epsilon_b \lambda_c\Big] = 0\ .
\end{align}

Equation (\ref{app:3:Maxwell2_rewritten}) implies
\begin{equation}
\label{app:3:cond_new}
\frac{\partial \mathcal{E}_1}{\partial \sigma} = \frac{\partial \mathcal{E}_2}{\partial \xi}\ .
\end{equation}
For the case of linearly polarised plane wave the second component of electric field $\mathcal{E}_2=0$. Taking this into account equations (\ref{app:3:Poisson}) and (\ref{app:3:cond_new}) indicate that $\mathcal{E}_1$ must be independant of $\xi$ and $\sigma$, therefore implying that $\mathcal{E}_1$ is a function of $\phi$ only.
 
For the general case of the plane wave we further differentiate (\ref{app:3:cond_new}) and (\ref{app:3:Poisson}):
\begin{equation}
\label{app:3:cond_new_diffed}
\frac{\partial^2 \mathcal{E}_1}{\partial \sigma^2} = \frac{\partial^2 \mathcal{E}_2}{\partial \sigma \partial \xi}\ ,
\end{equation}
\begin{equation}
\label{app:3:Poisson_diffed}
\frac{\partial^2 \mathcal{E}_1}{\partial \xi^2} = - \frac{\partial^2 \mathcal{E}_2}{\partial \sigma\partial \xi}\ . 
\end{equation}
If we now add (\ref{app:3:cond_new_diffed}) and (\ref{app:3:Poisson_diffed}) together:
\begin{equation}
\label{app:3:transverse_lagr}
\frac{\partial^2 \mathcal{E}_1}{\partial \sigma^2} + \frac{\partial^2 \mathcal{E}_1}{\partial \xi^2} = \nabla^2_{\perp} \mathcal{E}_1 = 0\ .
\end{equation}

Equation (\ref{app:3:transverse_lagr}) has two possible solutions: one implying constant $\mathcal{E}_1$ in the transverse directions $\xi$ and $\sigma$, and the other one leading to infinite values of $\mathcal{E}_1$ at infinite distances. 
Since the second one is unphysical we are only interested in the first solution, leading to the conclusion that for a general plane wave the $\mathcal{E}_1$ component of the electric field $\mathcal{E}$ depends solely on $\phi$. Analogously the same can be proven for the $\mathcal{E}_2$ component.

\chapter{Parameters describing the laser pulse}

\newpage
\noindent A laser pulse is characterised by a number of properties such as {\it frequency}, {\it pulse duration}, {\it energy} and {\it intesity}. For convenience, in this thesis we have been using the ``intensity parameter'' $a_0 = q\mathcal{E}/m\omega c$ and the number of cycles $N$. In this Appendix we show how these relate to physically measurable parameters. 

\begin{itemize}
\item Frequency
\end{itemize}

Since we are working in units such that $\omega = 1$, the relation between $\tau$ used throughout the thesis and the SI value $\tau_{si}$ gives us the frequency $\omega$ in SI units: 
\begin{equation}
\omega\tau_{si} = \tau\ ,
\end{equation}
where $\tau = 1.5 \cdot 10^{-8}$.

Therefore 
\begin{equation}
\omega = \frac{1.5 \cdot 10^{-8}}{6.3 \cdot 10^{-24}} = 2.4 \cdot 10^{15} \enskip \text{rad s$^{-1}$}\ .
\end{equation}

\begin{itemize}
\item Pulse duration
\end{itemize}

Given the above value of the frequency, the pulse duration can then be expressed as:
\begin{equation}
T = \frac{2\pi N}{\omega} = 2.6 \cdot N \enskip \text{fs}\ ,
\end{equation}
where $N$ is the number of oscillations in the pulse. 

\begin{itemize}
\item Energy per unit transverse area
\end{itemize}

The energy per unit transverse area (fluence) in the pulse can be expressed as:
\begin{equation}
E = \epsilon_0 c \int\limits^{2\pi N/\omega}_{0} \mathcal{E}^2 d\phi = \frac{3\pi}{8} \frac{m^2 c^3 \epsilon_0 \omega}{q^2} N a_0^2 = 2.2 \cdot 10^{7} \cdot N a_0^2 \enskip \text{Jm$^{-2}$}
\end{equation}

\newpage
\begin{itemize}
\item Intensity
\end{itemize}

The intensity is described by:
\begin{align}
\label{app:in}
\nonumber I & = \epsilon_0 c <\mathcal{E}^2> = \epsilon_0 c \frac{\omega}{2\pi N} \int\limits^{2\pi N/\omega}_{0}\mathcal{E}^2 \mathrm{d}\phi = \frac{3}{16} \frac{m^2 c^3 \omega^2 \epsilon_0}{q^2}  a_0^2 \\
& = 8.2 \cdot 10^{21} \cdot a_0^2 \enskip \text{Wm$^{-2}$}.
\end{align}
Intensities are typically measured in the hybrid units Wcm$^{-2}$, therefore (\ref{app:in}) corresponds to \mbox{$I = 8.2 \cdot 10^{17} \cdot a_0^2 \enskip \text{Wcm$^{-2}$}$}.

%\bibliographystyle{unsrt}
%\bibliography{bibtexfile}

\begin{thebibliography}{99}

\bibitem{Adam9} J. D. Jackson, ``\textit{Classical Electrodynamics}'', 3rd ed., Wiley \& Sons, Chichester, (1999).

\bibitem{Burton2} D. A. Burton and A. Noble, ``\textit{Aspects of electromagnetic radiation reaction in strong fields}'', Contemporary Physics {\bf 55}, 110 (2014).

\bibitem{Erber} T. Erber, ``\textit{The Classical Theories of Radiation Reaction}'', Fortschritte der Physik {\bf 9}, 343 (1961).

\bibitem{Rohrlich2} F. Rohrlich, ``\textit{Classical charged particles}'', World Scientific, Singapore (2007).

\bibitem{ELI} http://www.extreme-light-infrastructure.eu/

\bibitem{Lorentz} H. A. Lorentz, ``{\it The theory of electrons and its applications to the phenomena of light and radiant heat}'', Stechert, New York, (1916).

\bibitem{Abraham} M. Abraham, ``{\it The classical theory of electricity and magnetism}'', Blackie, London, (1932).

\bibitem{Dirac} P. A. M. Dirac, ``{\it Classical Theory of Radiating Electrons}'',  Proc. R. Soc. A {\bf 167}, 148 (1938).

\bibitem{Landau} L. D. Landau and E. M. Lifshitz, ``{\it The Classical Theory of Fields}'', Pergamon, London, (1962).

\bibitem{OConnell} R. F. O'Connell, ``{\it The equation of motion of an electron}'', Phys. Lett. A {\bf 313}, 491 (2003).

\bibitem{FO1} G. W. Ford and R. F. O'Connell, ``{\it Radiation reaction in electrodynamics and the elimination of runaway solutions}'', Phys. Lett. A {\bf 157}, 217 (1991).

\bibitem{FO2} G. W. Ford and R. F. O'Connell, ``{\it Relativistic form of radiation reaction}'', Phys. Lett. A {\bf 174}, 182 (1993).

\bibitem{ourPRE} Y. Kravets, A. Noble, D. Jaroszynski, ``{\it Radiation reaction effects on the interaction of an
electron with an intense laser pulse}'', Phys. Rev. E. {\bf 88}, 011201(R) (2013).

\bibitem{ourEPS} Y. Kravets, A. Noble and D. Jaroszynski, ``{\it Validity of the Landau-Lifshitz approximation in an ultra-high intensity laser pulse}'', Proc. EPS {\bf P4.214} (2013).

\bibitem{ourSPIE} Y. Kravets, A. Noble and D. Jaroszynski, ``{\it Energy losses due to radiation reaction in an intense laser pulse}'', Proc. SPIE {\bf 8779}, 87791X (2013).

\bibitem{ourCLF} A. Noble, Y. Kravets, S. Yoffe and D. Jaroszynski, ``{\it Radiation damping of an electron in an intense laser pulse}'', Central Laser Facility Annual Report (2012--2013).

\bibitem{Schott} G. A. Schott, ``{\it Electromagnetic radiation and the mechanical reactions arising from it}'', University Press, Cambridge, (1912).

\bibitem{Ferris} M. R. Ferris and J. Gratus, ``{\it The origin of the Schott term in the electromagnetic self force of a classical point charge}'', J. Math. Phys. {\bf 52}, 092902 (2011).

\bibitem{Bhabha} H. J. Bhabha, ``{\it Classical Theory of Mesons}'', Proc. R. Soc. A {\bf 172}, 384 (1939).

\bibitem{Wheeler} J. A. Wheeler and R. P. Feynman, ``{\it Interaction with the Absorber as the Mechanism of Radiation}'', Rev. Mod. Phys. {\bf 17}, 157 (1945).

\bibitem{Rohrlich1} F. Rohrlich, ``{\it Solution of the Classical Electromagnetic Self-Energy Problem}'', Phys. Rev. Lett. {\bf 12}, 375 (1964).

\bibitem{Teitelboim} C. Teitelboim, ``{\it Splitting of the Maxwell Tensor: Radiation Reaction without Advanced Fields}'', Phys. Rev. D {\bf 1}, 1572 (1970).

\bibitem{Barut} A. O. Barut, ``{\it Electrodynamics in terms of retarded fields}'', Phys. Rev. D {\bf 10}, 3335 (1974).

\bibitem{Gratus} D. A. Burton, J. Gratus and R. W. Tucker, ``{\it Asymptotic analysis for ultra-relativistic charge}'',  Ann. Phys. {\bf 322}, 599 (2006).

\bibitem{Baylis2} W. E. Baylis and J. Huschilt, ``\textit{Nonuniqueness of physical solutions to the Lorentz-Dirac equation}'', Phys. Rev. D. {\bf 13}, 3237 (1976).

\bibitem{Hammond1} R. T. Hammond, ``{\it Radiation reaction at ultrahigh intensities}'', Phys. Rev. A {\bf 81}, 062104 (2010).

\bibitem{Hammond2} R. T. Hammond, ``\textit{Relativistic particle motion and radiation reaction in electrodynamics}'', EJTP {\bf 7}, 221 (2010).

\bibitem{Spohn} H. Spohn, ``\textit{The critical manifold of the Lorentz-Dirac equation}'', Europhys. Lett. {\bf 50}, 287 (2000).

\bibitem{Griffiths} D. J. Griffiths, T. C. Proctor and D. F. Schroeter, ``{\it Abraham-Lorentz versus Landau-Lifshitz}'', Am. J. Phys. {\bf 78}, 391 (2010).

\bibitem{Bulanov} S. V. Bulanov {\it et al.}, ``{\it Lorentz-Abraham-Dirac versus Landau-Lifshitz radiation friction force in the ultrarelativistic electron interaction with electromagnetic wave (exact solutions)}'', Phys. Rev. E {\bf 84}, 056605 (2011).

\bibitem{Pandit} R. R. Pandit and Y. Sentoku, ``{\it Higher order terms of radiative damping in extreme intense laser-matter interaction}'', Phys. Plasmas {\bf 19}, 073304 (2012).

\bibitem{Galley} C. R. Galley, A. K. Leibovich and I. Z. Rothstein, ``\textit{Finite size corrections to the radiation reaction force in classical electrodynamics}'', Phys. Rev. Lett. {\bf 105}, 094802 (2010).

\bibitem{Gron} \O. Gr\o n, ``\textit{The significance of the Schott energy for energy-momentum conservation of a radiating charge obeying the Lorentz-Abraham-Dirac equation}'', Am. J. Phys., {\bf 79}, 115 (2011).

\bibitem{MoPapas} T. C. Mo and C. H. Papas, ``{\it New Equation of Motion for Classical Charged Particles}'', Phys. Rev. D {\bf 4}, 3566 (1971).

\bibitem{Baylis3} J. Huschilt and W. E. Baylis, ``\textit{Solutions to the `new' equation of motion for classical charged particles}'', Phys. Rev. D. {\bf 9}, 2479 (1976).

\bibitem{Sokolov} I. V. Sokolov {\it et al.}, ``{\it Dynamics of emitting electrons in strong laser fields}'', Phys. Plasmas {\bf 16}, 093115 (2009).

\bibitem{Eliezer} C. J. Eliezer, ``\textit{On the classical theory of particles}'', Proc. R. Soc. Lond. A {\bf 194}, 543 (1948).

\bibitem{XCELS} http://www.xcels.iapras.ru

\bibitem{Meyer} J. W. Meyer, ``{\it Covariant Classical Motion of Electron in a Laser Beam}'', Phys. Rev. D {\bf 3}, 621 (1971).

\bibitem{Heinzl} T. Heinzl and A. Ilderton, ``{\it A Lorentz and gauge invariant measure of laser intensity}'', Opt. Commun. {\bf 282}, 1879 (2009).

\bibitem{Hadad} Y. Hadad, L. Labun, J. Rafelski, N. Elkina, C. Klier, and H. Ruhl, ``{\it Effects of radiation reaction in relativistic laser acceleration}'', Phys. Rev. D {\bf 82}, 096012 (2010).

\bibitem{Harvey} C. Harvey, T. Heinzl and M. Marklund, ``{\it Symmetry breaking from radiation reaction in ultra-intense laser fields}'', Phys. Rev. D {\bf 84}, 116005 (2011).

\bibitem{DiPiazza} A. Di Piazza, ``{\it Exact Solution of the Landau-Lifshitz Equation in a Plane Wave}'', Lett. Math. Phys. {\bf 83}, 305 (2008).

\bibitem{Heinzl2} T. Heinzl, A. Ilderton and M. Marklund, ``{\it Finite size effects in stimulated laser pair production}'', Phys. Lett. B {\bf 692}, 250 (2010).

\bibitem{Mackenroth} F. Mackenroth A. Di Piazza and C. H. Keitel, ``{\it Determining the Carrier-Envelope Phase of Intense Few-Cycle Laser Pulses}'', Phys. Rev. Lett. {\bf 105}, 063903 (2010).

\bibitem{Tajima} T. Tajima and J.M. Dawson, ``\textit{Laser Electron Accelerator}'', Phys. Rev. Lett. {\bf 43}, 267 (1979).

\bibitem{Mangles} S. P. D. Mangles et al., ``\textit{Monoenergetic beams of relativistic electrons from intense laser-plasma interactions}'', Nature {\bf 431}, 535 (2004).

\bibitem{Geddes} C. G. R. Geddes et al., ``\textit{High-quality electron beams from a laser wakeﬁeld accelerator using plasma-channel guiding}'', Nature {\bf 431}, 538 (2004).

\bibitem{Faure} J. Faure et al., ``\textit{A laser-plasma accelerator producing monoenergetic electron beams}'', Nature  {\bf 431}, 541 (2004).

\bibitem{Wiggins} S. M. Wiggins et al., ``\textit{High quality electron beams from a laser wakefield accelerator}'', Plasma Phys. Control. Fusion {\bf 52}, 124032 (2010).

\bibitem{Hazeltine} R. D. Hazeltine and S. M. Mahajan, ``\textit{Radiation reaction in fusion plasmas}'', Phys. Rev. E {\bf 70}, 046407 (2004).

\bibitem{Berezhiani} V. I. Berezhiani, R. D. Hazeltine, and S. M. Mahajan, ``\textit{Radiation reaction and relativistic hydrodynamics}'', Phys. Rev. E {\bf 69}, 056406 (2004).

\bibitem{Hakim} R. Hakim and A. Mangeney, ``\textit{Relativistic Kinetic Equations Including Radiation Effects. I. Vlasov Approximation}'', J. Math. Phys. {\bf 9}, 116 (1968).

\bibitem{kinetic} A. Noble, D. A. Burton, J. Gratus, and D. A. Jaroszynski, ``{\it A kinetic model of radiating electrons}'', J. Math. Phys. {\bf 54}, 043101 (2013).

\bibitem{AdamSPIE} A. Noble et al., ``{\it Kinetic treatment of radiation reaction effects}'', Proc. SPIE {\bf 8079}, 80790L (2011).

\bibitem{Tamburini} M. Tamburini et al., ``\textit{Radiation reaction effects on electron nonlinear dynamics and ion acceleration in laser-solid interaction}'', Nucl. Instrum. Methods Phys. Res. A {\bf 653}, 181 (2011).

\bibitem{Lehmann} G. Lehmann and K. H. Spatschek, ``\textit{Phase-space contraction and attractors for ultrarelativistic electrons}'', Phys. Rev. E {\bf 85}, 056412 (2012).

\bibitem{NeitzDiPiazza} N. Neitz and A. Di Piazza, ``\textit{Stochasticity effects in quantum radiation reaction}'', Phys. Rev. Lett. {\bf 111}, 054802 (2013).

\bibitem{Rajeev} S. G. Rajeev, ``\textit{Exact solution of the Landau-Lifshitz equations for a radiating charged particle in the Coulomb potential}'', Ann. Phys. {\bf 323}, 2654 (2008).

\bibitem{Rutherford} E. Rutherford, ``\textit{The Scattering of $\alpha$ and $\beta$ rays by Matter and the Structure of the Atom}'', Philos. Mag. {\bf 6}, 21 (1911).

\bibitem{Eliezer_nucleus} C. J. Eliezer, ``\textit{The hydrogen atom and the classical theory of radiation}'', Proc. Cam. Phil. Soc. {\bf 39}, 173 (1943).

\bibitem{Baylis1} W. E. Baylis and J. Huschilt, ``\textit{Numerical solutions to two-body problems in classical electrodynamics: Head-on collisions with retarded fields and radiation reaction. II. Attractive case}'', Phys. Rev. D {\bf 13}, 3262 (1976).

\bibitem{Huschilt} J. Huschilt and W. E. Baylis, ``\textit{Rutherford scattering with radiation reaction}'', Phys. Rev. D {\bf 17}, 985 (1978).

\bibitem{Comay} E Comay, ``\textit{Solutions of the Lorentz-Dirac equation in the ultrarelativistic domain}'', J. Phys. A: Math. Gen. {\bf 29}, 2111 (1996).

\bibitem{Sauter} F. Sauter, ``\textit{\"{U}ber das Verhalten eines Elektrons im homogenen elektrischen Feld nach der relativistischen Theorie Diracs}'', Zeitschrift für Physik {\bf 82}, 742 (1931).

\bibitem{Schwinger} J. Schwinger, ``\textit{On Gauge Invariance and Vacuum Polarization}'', Phys. Rev. {\bf 82}, 664 (1951). 

\bibitem{BulanovSchwinger} S. S. Bulanov et al., ``\textit{On the Schwinger limit attainability with extreme power lasers}'', Phys. Rev. Lett. {\bf 105}, 220407 (2010).

\bibitem{HeinzlSchwinger} T. Heinzl, ``\textit{Strong-Field QED and High Power Lasers}'', Plenary talk QFEXT11 Benasque Conference.

\bibitem{DiPiazza_review} A. Di Piazza, C. M\"{u}ller, K. Z. Hatsagortsyan and C. H. Keitel, ``\textit{Extremely high-intensity laser interactions with fundamental quantum systems}'', Rev. Mod. Phys. {\bf 84}, 1177 (2012).

\bibitem{Ritus} V. I. Ritus, ``\textit{Quantum effects of the interaction of elementary particles with an intense electromagnetic field}'', (1979).

\end{thebibliography}

\end{document}